**Е. И. Жмуриков, И.А. Бубненков, В.В. Дрёмов, С.И. Самарин, А. С. Покровский, Д. В. Харьков**

# ГРАФИТ В НАУКЕ И ЯДЕРНОЙ ТЕХНИКЕ

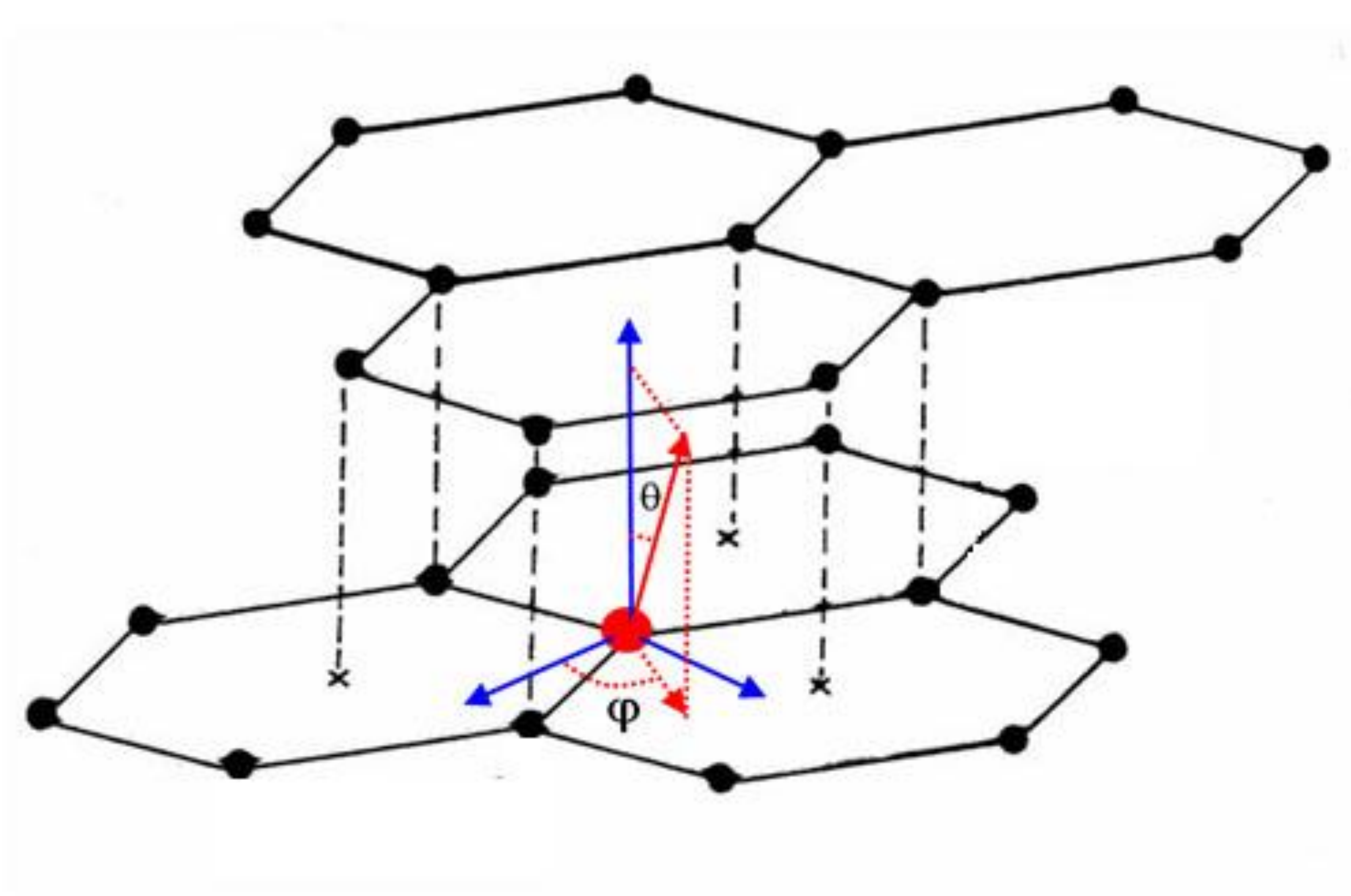

Монография посвящена вопросам применения графита и графитовых композитов в науке и технике. Рассмотрены структура и электрофизические свойства, технологические аспекты получения высокопрочных искусственных графитов, динамика разрушения графитов, традиционно используемых в атомной промышленности. Особое внимание уделено особенностям графитации и свойствам графитовых композитов на основе изотопа углерода $^{13}C$. Книга основана, как правило, на оригинальных результатах, и сосредоточена на актуальных проблемах применения и испытания графитовых материалов в современной ядерной физике, в научных и технических приложениях.

Для научных сотрудников и инженеров, специализирующихся в физике конденсированного состояния, для бакалавров, магистрантов и аспирантов физических специальностей университетов.

***

**GRAPHITE IN SCIENCE AND NUCLEAR TECHNOLOGY**

**E.I.Zhmurikov, I.A. Bubnenkov, V.V. Dremov, S.I. Samarin,**

**A.S. Pokrovsky, D. V.Harkov,**

The monograph is devoted to the application of graphite and graphite composites in the science and technology. The structure and electrical properties, technological aspects of the high-strength artificial graphite producing and dynamics of its destruction are considered. This type of graphite are traditionally used in the nuclear industry, so authors concentrate on the actual problems of the application and testing of graphite materials in modern science and technology. Chapter 4 and 5 of the monograph is focused on the characteristics of graphitization and properties of graphite composites based on the carbon isotope $^{13}C$. The book relay on the original results and concentrates on actual problems of application and testing of graphite.

For scientists and engineers specializing in condensed matter; for undergraduate, graduate and post-graduate students of physics specialties of universities.

# ВВЕДЕНИЕ

Графит обладает тем уникальным набором качеств, которые делают его незаменимым для задач ядерной физики и энергетики. Природный углерод — это смесь двух стабильных изотопов: $^{12}C$ (98,892%) и $^{13}C$ (1,108 %). Из четырех радиоактивных изотопов ($^{10}C$, $^{11}C$, $^{14}C$ и $^{15}C$) долгоживущим является только изотоп $^{14}C$ с периодом полураспада 5730 лет [1]. Это чистый низкоэнергетический ß-излучатель с максимальной энергией частиц 156 кэВ относится к числу глобальных радионуклидов, однако радиационный порог образования этого радиоизотопа достаточно высокий, поэтому образуется он главным образом при ядерных взрывах либо при взаимодействии вторичных нейтронов космического излучения с ядрами азота по реакции $^{14}N$ (n, p) =>$^{14}C$. Роль других реакций в образовании изотопа $^{14}C$ крайне незначительна.

К другим особенностям графита как материала для ядерной физики, прежде всего, относится малое эффективное сечение σ фотоядерных реакций для углерода в области гигантского резонанса, связанного с возбуждением γ-квантами собственных колебаний протонов относительно нейтронов (дипольные колебания). Нуклоны могут покидать ядро не только в процессе дипольных колебаний, но и после их затухания.

Таким образом, для чистого графита при облучении даже достаточно высокоэнергетичным (до 50МэВ) протонным пучком вторичная радиация сравнительно невелика вследствие малого сечения поглощения образующихся вторичных нейтронов в реакции с ядрами углерода - менее 4,5 микробарн для графита высокой чистоты [2]. При этом большая часть столкновений нейтронов с ядрами углерода происходит по механизму упругого рассеяния, последнее обстоятельство обусловило эффективное использование графита в качестве замедлителя или поглотителя нейтронов. В частности, для атомного реактора, работающего на обогащённом уране, графит как замедлитель по эффективности идёт вслед за бериллием и тяжёлой водой. В этом случае используется графит повышенной чистоты, где общее содержание примесей не превышает $1\times10^{-3}$ %. Для использования в полупроводниковой технике созданы графиты ещё более чистые, с содержанием примесей не выше $1\times10^{-6}$ %.

Графит является хорошим конструкционным материалом, его применение во многом основано на том, что благодаря очень высокой температуре сублимации графит остаётся твёрдым вплоть до температур порядка $4000^0$С. В то же самое время графит при невысокой плотности является материалом не только достаточно прочным, но и пластичным, легко обрабатывается механически, имеет низкое давление насыщенных паров в вакууме даже при повышенной температуре. Кроме того, графит обладает высокой теплопроводностью и теплоёмкостью, не обязательно обладая при этом высокой электропроводностью. Прочность и пластичность графита заметно возрастает с температурой, вплоть до ~ $2500^o$С [2, стр.215]. Графит, кроме того, за счёт высокой пористости устойчив как к тепловому шоку, так и к высокому градиенту температур, способен отдавать избыточное тепло переизлучением в ИК и оптическом диапазоне. А коррозионная и химическая стойкость в сочетании с антифрикционными свойствами делают его незаменимым в целом ряде научных и практических применений.

На воздухе графит не окисляется до температуры $400^{\circ}$C, в двуокиси углерода до $500^{\circ}$C. При более высоких температурах изделия из графита необходимо использовать в защитной среде либо в вакууме.

В то же самое время, графит как конструкционный материал, изучен совершенно недостаточно. В частности, не ясны причины сильных разбросов физико-механических и теплофизических свойств графита для различных марок графита и даже в пределах одной промышленной марки. Не вполне понятна причина сильной анизотропии для хорошо графитированных материалов, недостаточно изучены свойства этого материала в сложных условиях эксплуатации, в частности, в условиях повышенной радиации и/или высокой температуры.

Прочность графита значительно изменяется в зависимости от метода его изготовления, поэтому графиты с одинаковой плотностью, но разных марок, отличаясь структурой, могут иметь различную прочность. Общим правилом является то, что более тонко структурированный графитовый композит обладает, как правило, большей прочностью и большим временем жизни.

Следует оговориться, что по мере возможности авторы старались придерживаться терминологии, используемой, например, в обзоре Виргильева Ю.С. «Графиты для реакторостроения» [3] или в «Толковом терминологическом словаре углеродных материалов» [4]. Тем не менее, нужно понимать, что по ряду причин не всегда удаётся придерживаться строгих терминов, и, в данной книге, как и в другой научной литературе, слово «графит» часто употребляется расширительно, для краткости изложения, взамен строгого определения «искусственный графит, полученный по электродной технологии с использованием в качестве наполнителя углеродных порошков».

Кроме того, словарь терминов слово «композит» определяет как армированный материал или материал с волокнистым наполнителем, однако определение для композита [4, стр.226] не противоречит, как нам кажется, его употреблению также и для углеродных материалов с дисперсным углеродным наполнителем. Поэтому в главе 5 употребляется слово «композит» для материала, созданного на основе мелкодисперсного изотопного порошка $^{13}$C, хотя этот термин в данном случае и не является строгим. Для большей ясности в конце книги приведён список основных используемых сокращений и терминов.

Авторы выражают искреннюю признательность: сотрудникам ФГУП НИИграфит, д.т.н. Виргильеву Ю.С.; к.т.н. Калягиной И.П. за проведение рентгеноструктурного анализа изотопа $^{13}$C при исследовании его графитации; д.ф.м.н. Котосонову А.С. за исследование диамагнитной восприимчивости изотопа графита $^{13}$C и проведение электронной микроскопии.

Группе профессора Романенко А.И. (ИНХ СО РАН) за комплекс электрофизических измерений; группе профессора Цыбули С.В. (ИК СО РАН) за рентгенографические и электронно-микроскопические измерения; д.ф.-м.н. Окотруб А.В. и д.ф.-м.н. Булушевой Л.Г. (ИНХ СО РАН) за рентгенофлюоресцентные измерения и квантовохимическое моделирование изотопа графита $^{13}C$; д.ф.м.н. Станкус С.В. и его сотрудникам (ИТ СО РАН) за комплекс теплофизических измерений; к.ф.-м.н. Е.Б. Бургиной за ИК-измерения и измерения КРС; к.т.н. Титову А.Т. (ИГ СО РАН) за электронно-растровые измерения и элементный анализ графитовых образцов; М.Г. Голковскому за комплекс работ по облучению графитовых образцов электронным пучком и неизменное внимание к работе.

А также лаборантам, техникам и инженерам ИЯФ СО РАН, в частности, Жуль И.Е., Кот Н.Х., Болховитянову Д.Ю., всем другим, кто помогал в создании макетов для измерений, в проведении измерений и испытаний графитовых образцов; Ю.П Шаркееву (ИФП и М СО РАН) за неизменное внимание и интерес к работе.

ГЛАВА 1

# ОСНОВНЫЕ ПРЕДСТАВЛЕНИЯ О СТРУКТУРЕ И СВОЙСТВАХ ГРАФИТА. ДИНАМИКА РАЗРУШЕНИЯ, ПРОЧНОСТЬ И ДОЛГОВЕЧНОСТЬ ГРАФИТОВЫХ КОМПОЗИТОВ

Графит является термодинамически стабильной аллотропной модификацией углерода – элемента 4-ой группы главной подгруппы 2-го периода периодической системы Менделеева, порядковый номер 6, атомная масса природной смеси изотопов 12,0107 г/моль [5].

Следует сказать, что углерод существует в нескольких аллотропных модификациях и в различных изоморфных состояниях с очень разнообразными физическими свойствами [6-9]. Разнообразие аллотропных модификаций обусловлено способностью углерода к образованию химических связей различного типа. Электронные орбитали атома углерода могут иметь различную геометрию, в зависимости от степени их гибридизации. Существует три основных геометрии атома углерода: тетраэдрическая, тригональная и дигональная. Тетраэдрической геометрии атома углерода соответствуют аллотропные модификации углерода: алмаз и лонсдейлит. Такой же гибридизацией обладает углерод, например, в метане и других углеводородах.

Тригональная модификация образуется при смешении одной *s*- и двух *p*-электронных орбиталей ($sp^2$-гибридизация). Атом углерода имеет три равноценные $\boldsymbol{\sigma}_{x,y}$-связи, расположенные в одной плоскости под углом 120° друг к другу. Не участвующая в гибридизации $p_z$-орбиталь, расположенная перпендикулярно плоскости σ-связей, используется для образования **π**-связи с другими атомами. Именно такая геометрия углерода характерна для графита.

Углерод с дигональной геометрией атома образует особую аллотропную модификацию – карбин, однако в синтезированных к настоящему времени материалах длина прямых карбиновых без изгибов и межцепочечных связок составляет менее сотни атомов [10]. При обычных температурах углерод химически инертен, при достаточно высоких соединяется со многими элементами, проявляя сильные восстановительные свойства. Фазовая диаграмма углерода Банди представлена на рис. 1.1, различные аллотропные модификации и изоморфизмы углерода показаны на рис.1.2.

Следует сказать, что данная фазовая диаграмма не всегда представляется бесспорной, в частности, до сих пор поднимается вопрос о температуре плавления графита [11, 12], поскольку, по мнению авторов [12], имеется «…значительное расхождение данных, относящихся к температуре плавления графита». Тем не менее, большинство авторов придерживается фазовой диаграммы Банди 1996г., которая на данный момент не вызывает особых сомнений, и речь идёт, как правило, об уточнениях диаграммы в той или иной области давлений и температур [13, 14].

## 1.1. Структурные и электрофизические свойства графитов

### 1.1.1. Кристаллическая решётка углерода

В графите каждый атом углерода ковалентно связан с тремя другими окружающими его атомами углерода. Различают гексагональную и ромбоэдрическую модификации графита [5], которые различаются упаковкой слоёв (рис.1.3). У гексагонального графита половина атомов каждого слоя располагается над и под центрами шестиугольника (мотив ABABABA…), в то время как у ромбоэдрического каждый четвёртый слой повторяет первый. Ромбоэдрический графит удобно представлять в гексагональных осях, чтобы показать его слоистую структуру. В чистом виде ромбоэдрический графит не наблюдается, так как является метастабильной фазой. Однако в природных графитах содержание ромбоэдрической фазы может достигать 30 %.

Гексагональная решётка графита относится к пространственной группе C6/mmc- $D^4_{6h}$ с четырьмя атомами, приходящимися на элементарную ячейку. Параметр *a* гексагональной ячейки составляет 2,46Å, параметр ***c***=6,71Å, теоретическая плотность подобного кристалла равна 2,267 г/см$^3$. В каждой плоскости углеродные атомы образуют сетку правильных шестигранников с расстоянием между атомами 1,42Å. Связи внутри слоёв, имеющие ковалентный характер, представляют тригональные гибриды ($2s$, $2p_x$, $2p_y$) [5].

Если связь в слое осуществляется за счёт ковалентных σ-связей, то связь между слоями обеспечивают относительно слабые **π** связи (иногда не совсем точно пишут, что связь между слоями осуществляется силами Ван-дер-Ваальса, при некотором вкладе поляризационных сил).

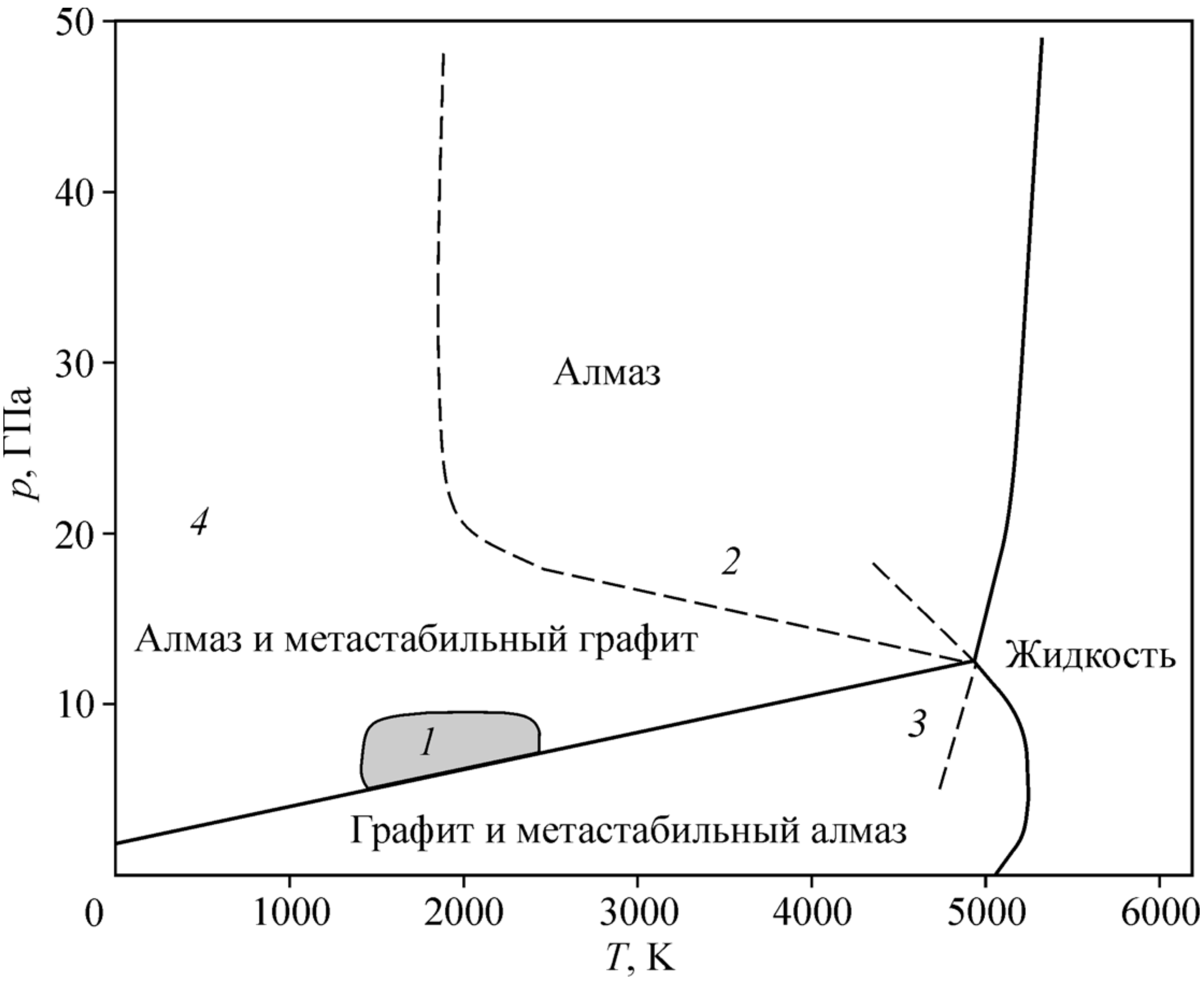

**Рис. 1.1.** Фазовая диаграмма углерода *Bundy et al.*, 1996г. [15, 16].

1.1.2. <u>Рентгенодифракционный метод исследования поликристаллических графитов</u> основан на том, что согласно условию Вульфа - Брэгга максимумы дифракционной картины возникают при отражении рентгеновских лучей от системы параллельных кристаллографических плоскостей, когда лучи, отражённые разными плоскостями этой системы, имеют разность хода, равную целому числу длин волн:

$$2d\ \sin\theta = m\,\lambda, \qquad (1.1)$$

где $d$ — межплоскостное расстояние, $\theta$ — угол скольжения, т. е. угол между отражающей плоскостью и падающим лучом, $\lambda$ - длина волны рентгеновского излучения и $m$ - так называемый, порядок отражения, т. е. положительное целое число.

Исследование структуры поликристаллических материалов с помощью дифракции рентгеновских лучей основано на методе Дэбая-Шерера [17, стр.43]. Металлы, сплавы, кристаллические порошки состоят из множества мелких монокристаллов данного вещества, для их исследования используют монохроматическое излучение.

В данном методе узкий параллельный пучок монохроматических рентгеновских лучей, падая на поликристаллический образец и отражаясь от кристалликов, из которых он состоит, даёт ряд коаксиальных, т. е. имеющих одну общую ось, дифракционных конусов. Осью этих конусов служит направление первичного пучка рентгеновских лучей, вершины их лежат внутри исследуемого объекта, а углы раствора определяются согласно условию Вульфа-Брэгга.

Угол раствора конуса равен учетверённому углу отражения $\theta$, и измерение углов раствора дифракционных конусов позволяет определить межплоскостные расстояния $d$. В некоторых случаях этих данных, в совокупности с измерением интенсивности лучей в каждом дифракционном конусе, достаточно для полного определения структуры кристаллической решётки.

Рентгенограмма (дебаеграмма) поликристаллов представляет собой несколько концентрических колец, в каждое из которых сливаются отражения от определённой системы плоскостей различно ориентированных монокристаллов. Дебаеграммы различных веществ имеют индивидуальный характер и широко используются для идентификации соединений (в том числе и в смесях).

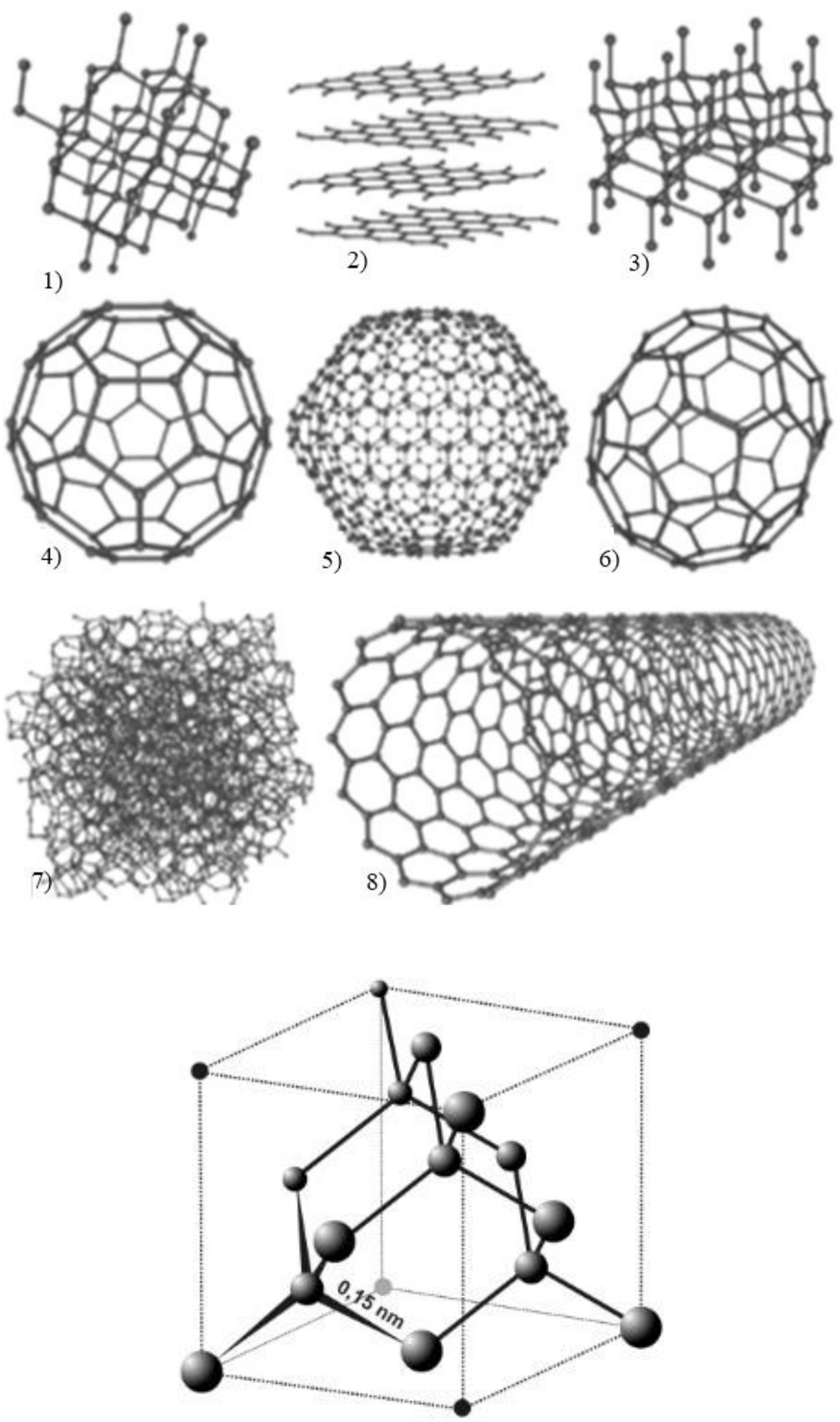

**Рис. 1.2,** а) **-** Схемы строения различных модификаций углерода:1) алмаз, 2) графит, 3) лонсдейлит 4) фуллерен - $C_{60}$, 5) фуллерен − $C_{540}$, 6) фуллерен − $C_{70}$ 7) аморфный углерод, 8) углеродная нанотрубка.

б) **-** Элементарная ячейка алмаза.

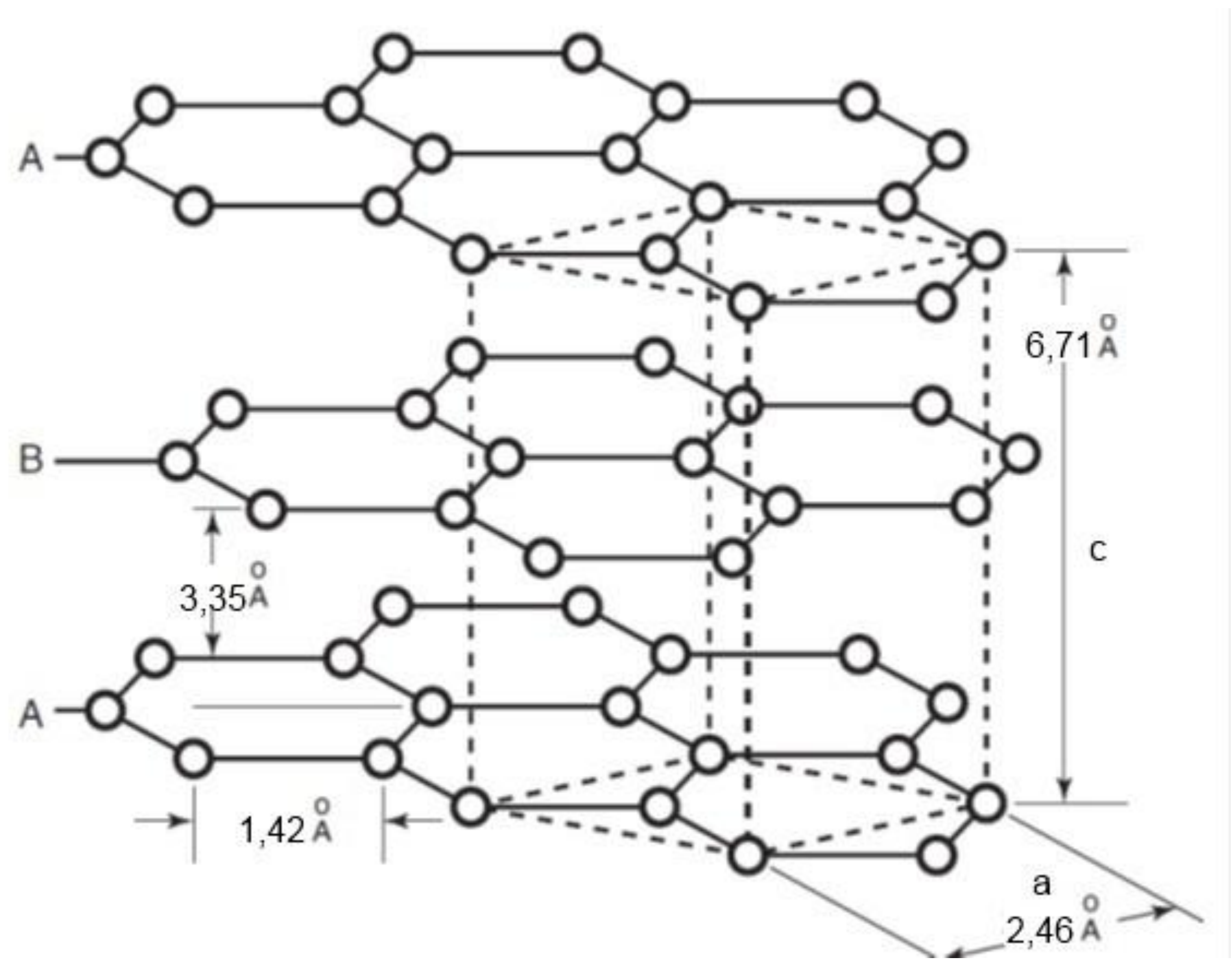

**Рис. 1.3.** Кристаллографическая решётка гексагонального графита [5].

Кристаллическая решётка двуслойная с чередованием слоёв в направлении оси С: …АВА-ВАВ…; Z=4; координационное число - 3; параметры элементарной ячейки: $a$ = 2,46 Å, $c$ =6,71 Å; длина σ – связи – 1,42 Å, π связи – 3,35 Å; энергия: σ – связи 418 – 461 кДж/моль; π связи – 16,8 – 41,9 кДж/моль; расчётная плотность – 2,267 г/см$^3$.

Рентгеноструктурный анализ поликристаллов позволяет определять фазовый состав образцов, устанавливать размеры и преимущественную ориентацию (текстурирование) зёрен в веществе, осуществлять контроль за напряжениями в образце и решать другие задачи. Чем ниже степень упорядоченности атомного строения материала, тем более размытый, диффузный характер будет иметь рассеянное им рентгеновское излучение.

Число $N$ достаточно крупных кристаллитов, участвующих в отражении рентгеновских лучей, определяется числом $\boldsymbol{n}$ точечных рефлексов, составляющих дебаевское кольцо рентгенограммы:

$$N = 2n/\alpha \cos\vartheta, \qquad (1.2)$$

где $\alpha$ — постоянная величина (параметр аппаратуры), $\vartheta$ — брэгговский угол.

Размеры мелких кристаллитов $L$ можно оценить по ширине дифракционных максимумов, используя формулу Селякова-Шеррера [18, 19]:

$$L = L_0 \cdot \lambda / (\beta \cdot \cos\vartheta), \qquad (1.3)$$

где $L_0$·- постоянная, зависящая от формы частиц;

ß·- ширина линий на половине высоты максимума;

$\lambda$ – длина волны рентгеновского отражения;

$\vartheta$ - брэгговский угол дифракции.

Рассматривая форму кристаллитов углеродных материалов в качестве дисковых образований, можно по ширине линии hkо определить диаметр кристаллитов $L_a$, высоту кристаллитов $L_c$ определяют по ширине линии 00l. Величину микроискажений кристаллической решётки можно оценить, используя соотношение дифракционных максимумов с индексами Миллера 002 и 004, однако как размеры кристаллитов, так и величина микроискажений могут быть получены с определённой достоверностью лишь при изучении дифракционных линий методами Фурье-анализа [20].

1.1.3. Дефекты структуры в графите. Типы дефектов

Дефекты в графите согласно [21, 22] можно разделить на два типа: дефекты, относящиеся к нарушениям между слоями, и дефекты связи в сетках. К первым относятся дефекты упаковки слоев, характеризующиеся нарушением порядка упаковки параллельных слоев гексагональных сеток. Так, углерод, состоящий из достаточно совершенных гексагональных сеток, но с нарушенным порядком в последовательности упаковки слоёв, называют обычно турбостратным. В такой структуре графитоподобные сетки смещены друг относительно друга случайным образом (со случайным вектором смещения одного слоя относительно другого).

Второй вид нарушений структуры в графите - дефекты в связях углеродной решетки. К ним относятся вакансии и их группы, атомы примесей, внедренные в гексагональный слой, дефекты изомерных связей, когда часть атомов имеет гибридизацию $sp^3$, краевые дефекты, и т.д.

*Краевые дефекты* возникают, когда связь С-С не может образоваться, например, если одна макромолекула не находится в плоскости своих ближайших соседей. С краевыми дефектами часто вступают во взаимодействие инородные атомы, например Н или группы типа – ОН, =О, -О- [22]. Кроме того, атомы углерода на краевых дефектах могут образовывать слабые связи с атомами соседних макромолекул, которые обычно распадаются при нагревании, обусловливая электронные свойства.

*"Клещевидные" дефекты*, или дефекты «расщепления», когда при разрушении связей образуются пустоты или разрывы в гексагональной сетке углеродных атомов. У клещевидных дефектов могут возникать винтовые дислокации или другие искривления гексагональной сетки.

*Дефекты изомерных связей*: для сеточных конфигураций предполагалось, что атомы углерода сохраняют $sp^2$ – гибридизацию, однако это не так, поскольку часть атомов могут быть присоединены к атомам углерода с гибридизацией $sp^3$. В первую очередь это относится к очень маленьким кристаллитам, у которых велика доля краевых дефектов.

*Дефекты двойникования*, когда на линии двойникования возникают чередующиеся кольца, состоящие из четырех и восьми атомов. Следует иметь в виду, что имеются два типа двойникования: в первом случае это происходит в виде срастания, называемого базисным, ось которого параллельна оси «*с*» графитовой решетки. При этом наблюдаются кристаллические образования из двух или нескольких одинаковых по составу и строению, но не одинаковых по форме и величине частей, закономерно расположенных относительно друг друга. Закономерность состоит в том, что решетка одной части совмещается с другой поворотом в двойниковой оси.

Второй вид двойникование – внебазисный, обусловлен отражением в двойниковой плоскости либо совмещением поворота и отражения.

Внебазисное двойникование предполагает наличие плоскости симметрии (плоскости зеркального отражения, называемой плоскостью двойникования) и проявляется как изгиб системы графитовых плоскостей на определенный угол.

Для гексагональной структуры графита с последовательностью упаковки слоев ABAB существуют только два угла (48°18' и 35°12') истинного двойникования. В остальных случаях законы симметрии нарушаются, т.е. последовательность слоев слева и справа от границы раздела не сохраняется даже при симметричной границе наклона. Отсюда, наклонные границы кристаллов, симметричные и особенно несимметричные должны сопровождаться многочисленными разрывами связей и достаточно развитой системой краевых дислокаций.

*Химические дефекты* предствляют из себя включения инородных атомов в углеродную сетку, и прямо связаны со смещением атомов из своих нормальных положений в решетке. Так, в решётку графита могут встраиваться атомы O, S, Se, As, N, P и другие. При этом можно ожидать, что атомы, находящиеся в периодической системе элементов левее углерода будут служить электронными акцепторами, а лежащие правее атомы окажутся донорами. В частности, атомы бора являются акцепторами, приводя к положительному температурному коэффициенту проводимости.

*К слоевым дефектам* относятся вакансии и их группы, это могут быть дефекты по Френкелю и по Шоттки, а также краевые и винтовые дислокации**.** Эти дефекты могут возникать, например, под действием нейтронного облучения. Под действием радиации атомы углерода смещаются из своих нормальных положений, образуя замыкающие цепи между углеродными макромолекулами. В сильно облучённых кристаллах графита может быть увеличено межплоскостное расстояние, это нарушение снимается отжигом при ~ $1500^0$C.

1.1.4. Электронная структура, электрические и тепловые свойства поликристаллического графита

Каждый атом углерода в графите имеет четыре валентных электрона, три из которых образуют прочные ковалентные σ-связи в гексагональном графеновом слое. Оставшийся электрон остаётся в π-состоянии, обеспечивая слабую ковалентную связь между соседними графеновыми слоями. Согласно многочисленным расчётам и ренгеноспектральным исследованиям [18] три связывающие и три разрыхляющие молекулярные орбитали (МО) [23, 24] создают заполненные и пустые σ-зоны, которые разделены промежутком в ~5эВ и этот промежуток включает уровень Ферми и π-зоны. На основе правил построения обратной решётки [25] и пренебрегая в первом приближении взаимодействием между графеновыми слоями можно получить первую зону Бриллюэна графита в виде прямой призмы с гексагональным основанием [26].

На рис.1.4 показана приведённая зона Бриллюэна для графита. Здесь $k_x$, $k_y$, $k_z$ есть проекции волновых векторов, на рис.1.4 *б* приведена двумерная зона Бриллюэна. Эта зона построена для поверхности в обратном пространстве (или в пространстве квазиимпульсов), при условии $k_z = 0$. Используемые в расчётах модельные волновые функции, не имеющие составляющей в направлении *c* (или в направлении *z* в обратном пространстве) условно обозначаются как σ, а волновая функция, содержащая лишь *z*-компоненту обозначается соответственно как π. Соответствующие зоны носят название σ - и π – зон.

Строго говоря, разделение на σ- и π – зоны справедливо только для высокосимметричных точек и направлений подзоны ΓALHKM, составляющей всего лишь 1/24 части всей зоны Бриллюэна графита. Для двумерных расчетов энергетической дисперсии в графите (когда в первом приближении можно пренебречь взаимодействием между слоями) обычно используется одноэлектронное и адиабатическое приближение. Пример такого расчёта зависимости дисперсии энергии от волнового вектора представлен на (рис.1.5).

Зонная структура графита впервые была посчитана в работе [27] в приближении сильно связанных электронов. Если пренебречь взаимодействием между графеновыми слоями, то, как уже было сказано, зона Бриллюэна сводится к гексагону (рис.1.6), а элементарная ячейка содержит два атома. Зависимость плотности электронных состояний от энергии согласно [28] представлена на рис.1.7.

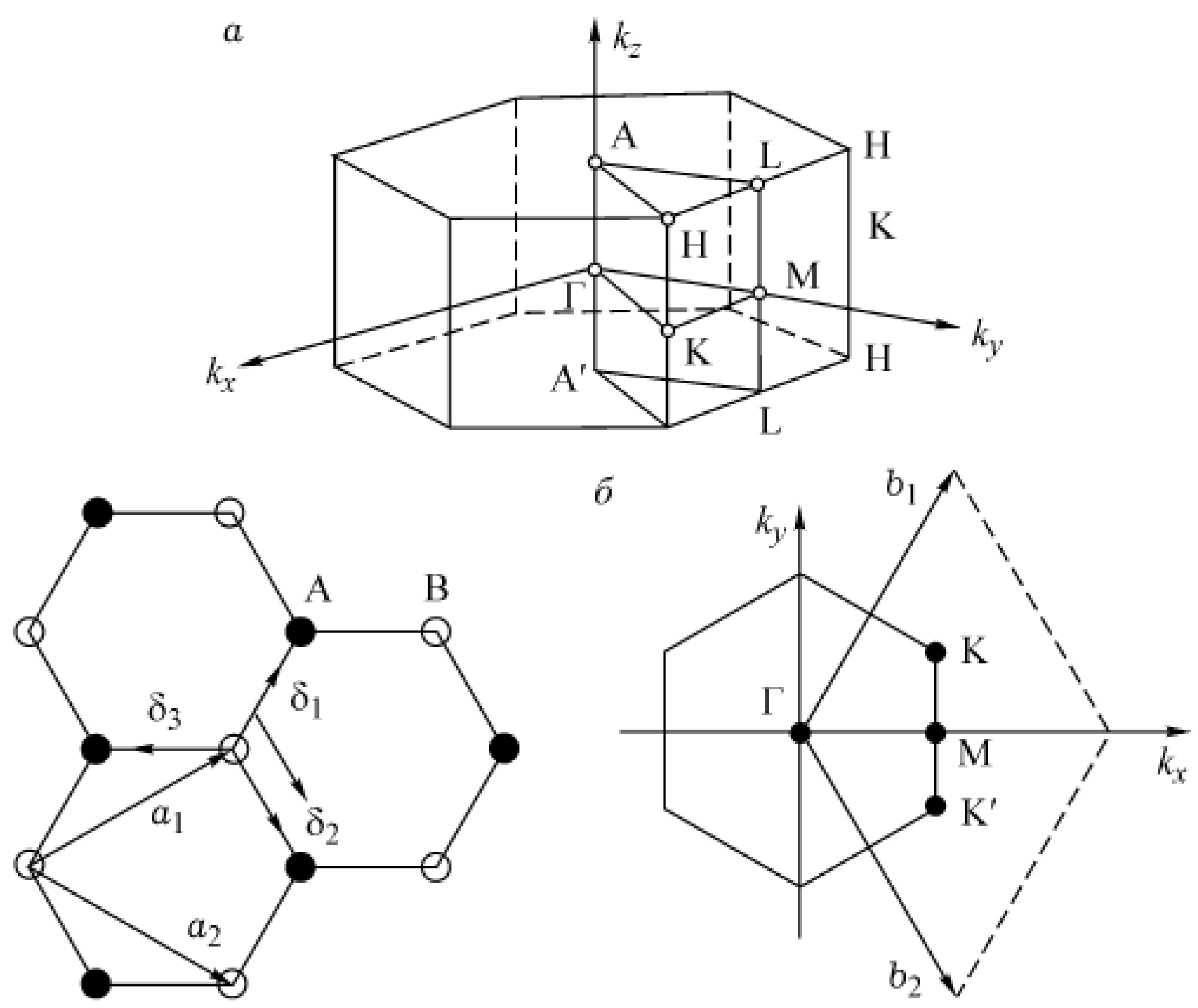

**Рис. 1.4** а) Элементарная ячейка и первая зона Бриллюэна гексагонального графита. б) Гексагональная решетка графена с векторами трансляции $a_1$, $a_2$ и первая зона Бриллюэна с векторами обратной решетки b1, b2 [29].

В приближении сильно связанных электронов интеграл перекрытия $\gamma_0$, то есть сила взаимодействия между слоями быстро спадает на межатомных расстояниях. Другими словами - основной вклад в формирование зонной структуры вносят ковалентные σ-связи центрального атома и трёх ближайших соседей.

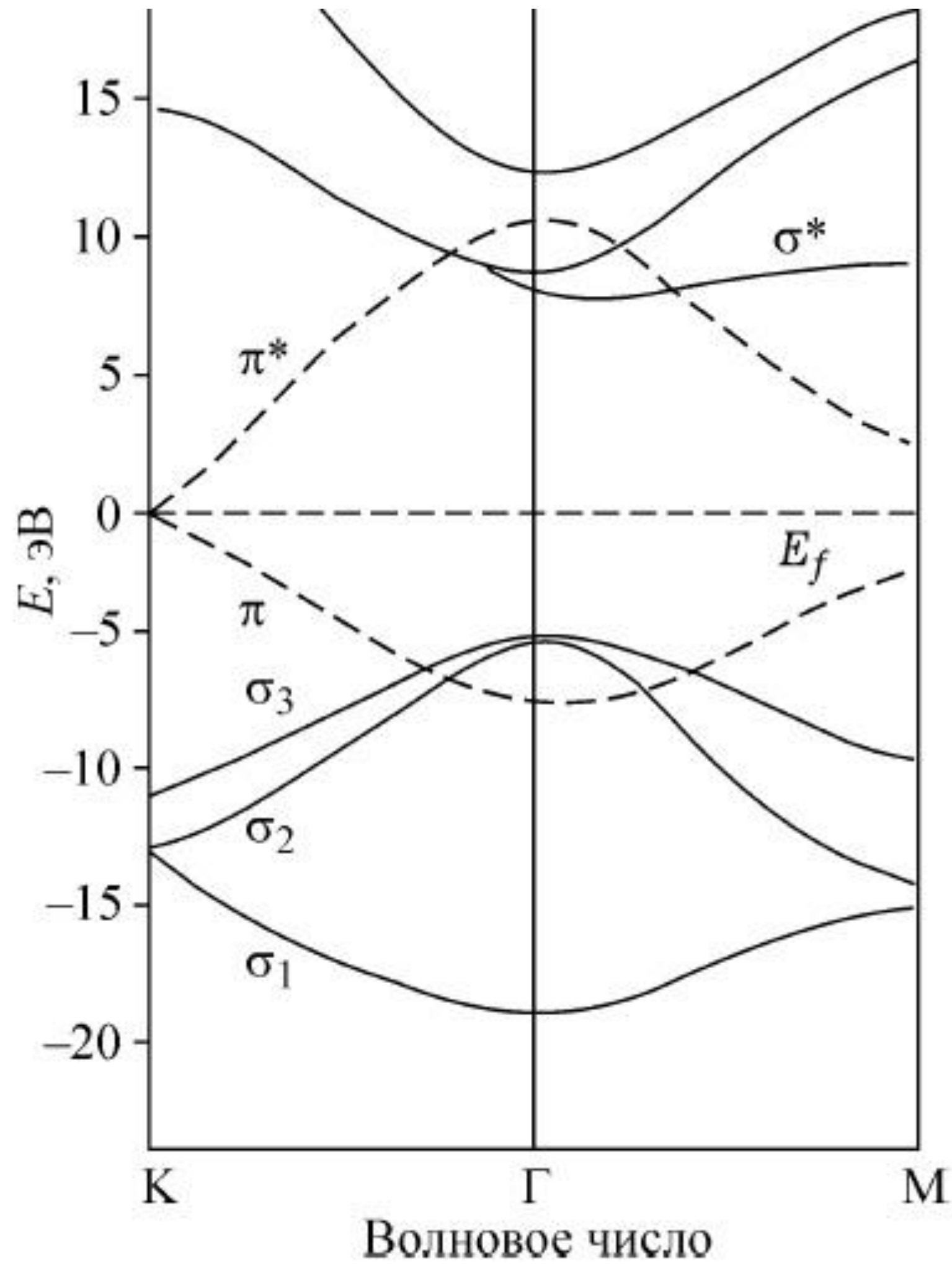

**Рис. 1.5.** Результаты двумерного расчёта дисперсии энергии вдоль двух характерных направлений зоны Бриллюэна [26].

Вблизи точек соприкосновения валентной зоны и зоны проводимости (K и K') закон дисперсии для носителей заряда (электронов) в графене имеет вид как показано на рис. 1.8. При этом согласно [29]:

$$\boldsymbol{E} = \hbar \upsilon_F k, \qquad (1.4),$$

где: $v_F$ - скорость Ферми (экспериментальное значение ($\upsilon_F = 10^6$ м/с);

$k$ - модуль волнового вектора в двумерном пространстве с компонентами $k_x$ и $k_y$ отсчитанного от K или K' точек Дирака;

$\hbar$ - постоянная Планка.

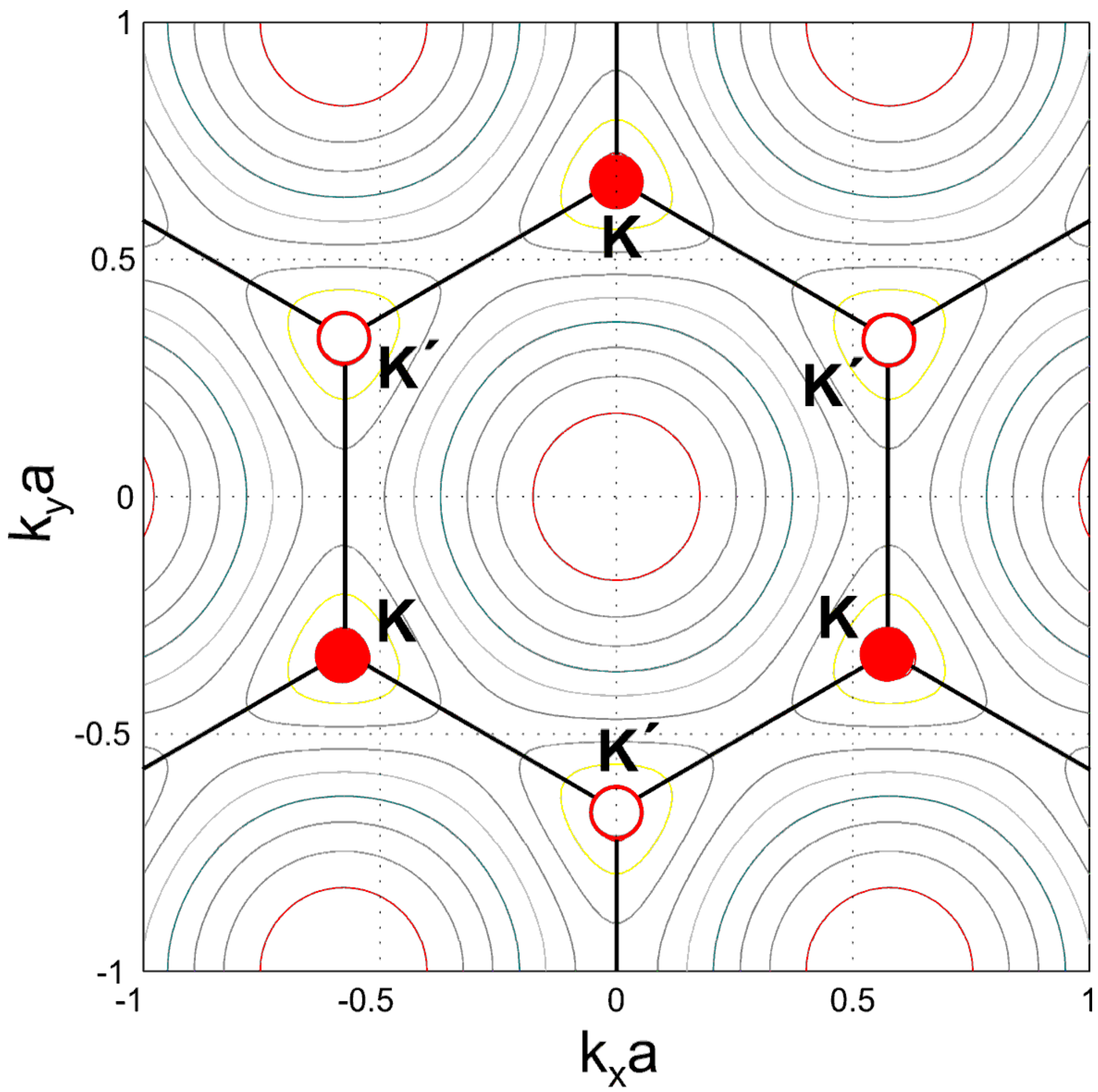

**Рис. 1.6.** Изолинии постоянной энергии для графенового слоя. Жирный чёрный шестиугольник — первая зона Бриллюэна. Внутренние окружности соответствуют краям первой зоны Бриллюэна, где закон дисперсии носителей линеен. **K** и **K'** обозначают две долины в **k**-пространстве с неэквивалентными волновыми векторами.

Линейный закон дисперсии приводит к линейной зависимости плотности состояний от энергии, в отличие от обычных двумерных систем с параболическим законом дисперсии, где плотность состояний не зависит от энергии. Плотность состояний на единицу площади равна [29]:

$$\nu(E) = g_s g_v |E|/2\pi\hbar v_F^2 \qquad (1.5),$$

где $g_s$ и $g_v$— двухкратное спиновое и четырёхкратное долинное вырождение соответственно, а модуль энергии появляется, чтобы описать электроны и дырки одной формулой. Отсюда видно, что при нулевой энергии плотность состояний равна нулю, то есть отсутствуют носители (при нулевой температуре).

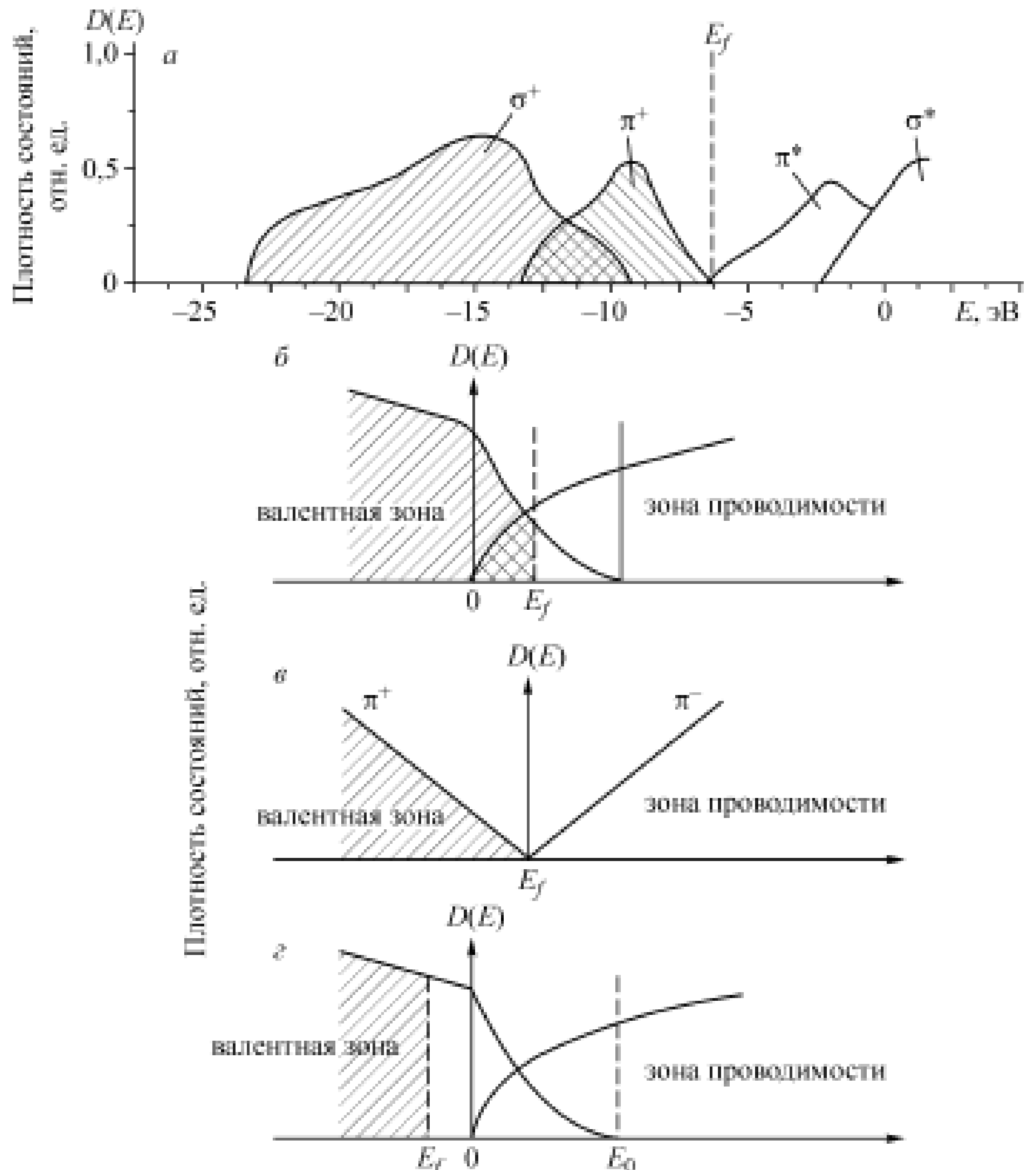

**Рис. 1.7.** а) Зависимость плотности электронных состояний в кристаллическом графите от энергии. Нулевое значение энергии отсчитано: *а* – от уровня вакуума, *б* – от дна зоны антисвязывающей π – зоны. Плотность состояний: *в* - для графеновой плоскости *г* - для кристаллического графита, легированного бором [28].

Вследствие малого взаимодействия между слоями зонная структура трёхмерной модели отличается от двухмерной только в окрестности соприкосновения валентной зоны и зоны проводимости. Однако это небольшое взаимодействие качественно меняет поверхность Ферми. Большинство электронных свойств, и, в частности, электропроводность, могут быть поняты только по трёхмерной модели.

Учёт межслоевых взаимодействий приводит к появлению зависимости энергии π-электронов от волнового вектора $k_z$ вдоль ребра НКН (рис. 1.4). Наиболее важным следствием межслоевых взаимодействий является перекрытие валентной зоны и зоны проводимости на величину~ 0,03÷0,04 эВ и искажение линейного закона дисперсии в интервале энергий от 0 ~ 0,5 эВ вблизи ребра НКН зоны (рис.1.4). Перекрытие зон приводит к равной концентрации электронов и дырок ~ $3 \cdot 10^{18}$ см$^{-3}$ при температуре 0 K, т.е. монокристалл графита является типичным полуметаллом [18].

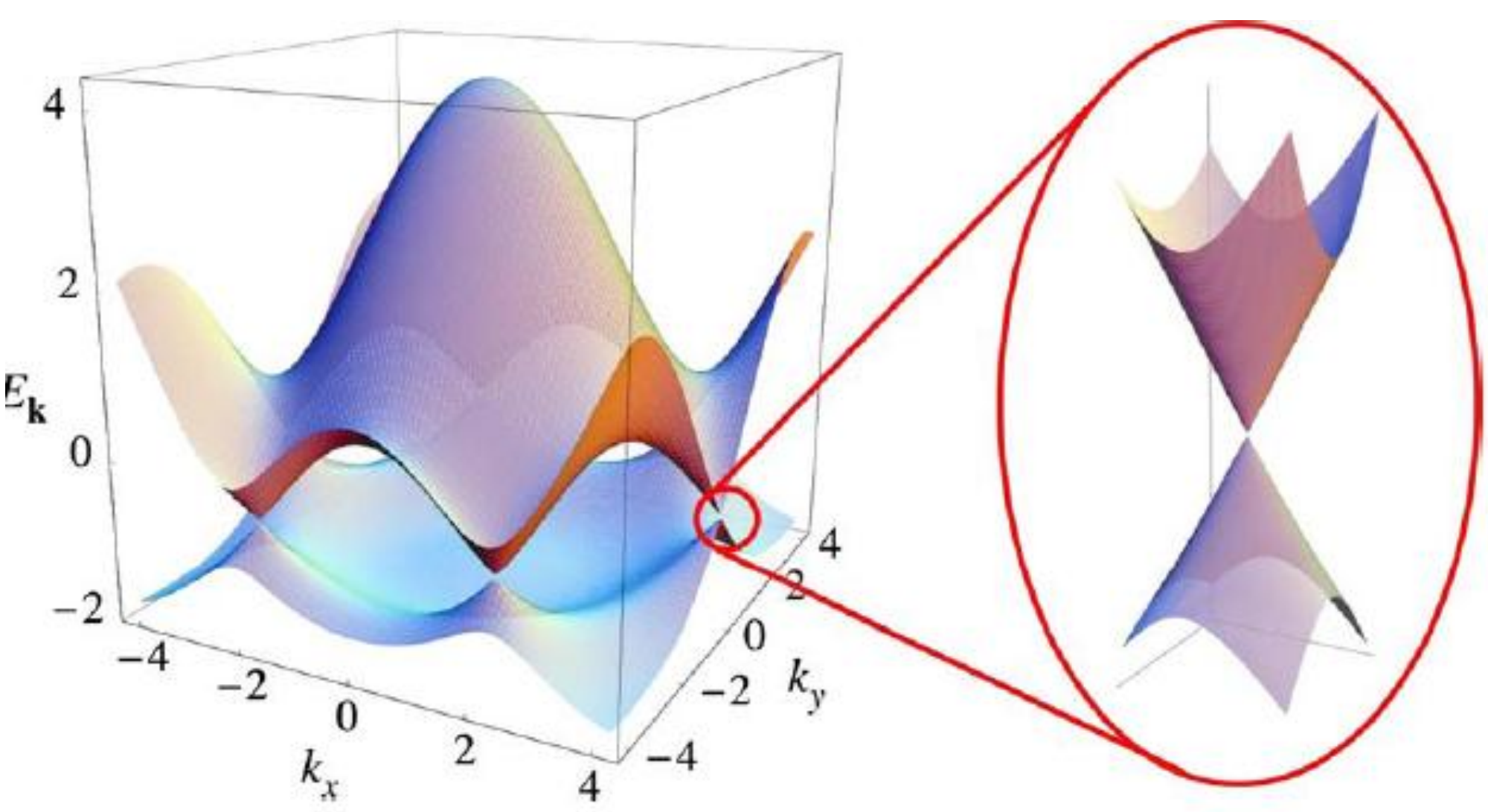

**Рис. 1.8.** Электронная плотность состояний в «сотах» кристаллографической решётки графита. Энергетический спектр представлен в относительных единицах, справа с увеличением показаны соприкасающиеся валентная зона и зона проводимости вблизи одной из точек Дирака [29].

Исследования осцилляционных явлений (эффекты де Газа - Ван-Альфена и Шубникова – де Гааза) позволили уточнить вид поверхности Ферми и эффективные массы носителей заряда для монокристалла графита. В первом приближении поверхность Ферми имеет форму эллипсоида вращения с соотношением осей 12,6:1 для дырок и 11:1 для электронов, причём большие оси совпадают с направлением кристаллографической оси ***c*** графита. Эффективные массы электронов и дырок соответственно равны 0,06 $m_0$ и 0,04 $m_0$ для движения носителей вдоль слоя и 14 $m_0$ и 5,7 $m_0$ для движения перпендикулярно слою, поэтому электронные свойства монокристалла графита имеют значительную анизотропию [30].

Анизотропия графита существенно влияет также на теплопроводность графита, где проявляются [31] полупроводниковые свойства графита, характеризующиеся соотношением тепла, передаваемого фононами и носителями заряда. Графитовые материалы имеют характерную для фононной теплопроводности температурную зависимость с максимумом в районе комнатных температур или чуть выше.

Согласно [32, стр.167] теплопроводность, связанная только с фононами, определяется выражением:

$$\boldsymbol{\lambda} = 1/3\ C \times \upsilon \times \Lambda \qquad (1.6),$$

где: *C* **-** полная теплоёмкость фононного газа;

υ – средеквадратичная скорость фононного газа;

Λ – длина свободного пробега фононов.

Теплоёмкость, связанная с колебаниями решётки (фононы) согласно закону Дебая, определяется в виде:

$$\boldsymbol{C} = \begin{cases} aT^3 \ (T < \theta_D), \\ 3Nk_B \ (T \geq \theta_D). \end{cases} \qquad (1.7)$$

Поэтому при низких температурах определяемая фононами теплопроводность растет пропорционально $\boldsymbol{T^3}$, а при высоких температурах, когда теплоёмкость достигает своего предельного значения, любая зависимость теплопроводности от температуры связана преимущественно с изменением длины свободного пробега фононов.

Хотя столкновение фононов с неоднородностями и границей кристаллитов также могут влиять на теплопроводность **λ**, за температурную зависимость теплопроводности при высоких температурах практически полностью отвечает фонон-фононное взаимодействие, т.е. ангармонизм [33, стр. 564]. Динамика решётки кристалла, в котором учитывается фонон - фононное взаимодействие очень сложна, но в конечном результате оказывается, что длина свободного пробега фононов обратно пропорциональна абсолютной температуре:

$$\Lambda \sim \boldsymbol{T^{-1}} \qquad (1.8)$$

Следовательно, при высоких температурах теплопроводность, обусловленная фононами, обратно пропорциональна температуре.

При высоких температурах в графитах участие в теплопроводности принимают также носители заряда, особенно для графитов с малой степенью упорядоченности решётки [31].

### 1.1.5. Технологические аспекты получения высокопрочных искусственных графитов

Физико-механические свойства искусственных графитов определяются особенностями кристаллической структуры на микро - и макроуровне, которая, в свою очередь, зависит от природы и качества исходного сырья и особенностей технологии изготовления графитов [34]. Графит не является самоспекающимся материалом, так как имеет низкие значения коэффициента самодиффузии даже при температурах обработки 2000-3000°С. В основу современной классификации искусственных графитов положен размер зерна.

Классическая технологическая схема производства искусственного графита включает в себя технологические операции подготовки наполнителя и связующего из сырьевых материалов, смешивание композиции, формование заготовок, их обжиг и графитацию [35, 36]. Размеры зерна графита, как правило, определяются размерами частиц наполнителя. В качестве наполнителей для производства искусственных графитов используются кокс различной структуры [34]. Наибольшую ценность для производства представляют малозольные коксы, с зольностью не выше 1%.

При низкотемпературной карбонизации пекообразной массы нефтяных остатков формируются основные структурные особенности кокса. Нагрев этих остатков приводит к появлению сферических частиц - сферул, имеющих сходство с жидкими кристаллами. Температура начала превращения составляет от 400 до 520$^0$C и зависит от вида карбонизируемого вещества [37]. С повышением температуры сферулы растут, взаимодействуют друг с другом и в результате коалисценции возникает «мозаика». При дальнейшем нагревании кокса мозаичность структуры и её слоистость сохраняются, образуя жёсткий коксовый каркас, что сопровождается нарушением сплошности, и возникают поры.

Малозольные коксы бывают двух видов – нефтяные и пековые. Первые получают коксованием нефтяных остатков, вторые – переработкой на кокс каменноугольного пека. Свойства нефтяных коксов зависят главным образом от вида нефтяных остатков, из которых они получаются, и в меньшей степени от условий коксования. Из малоокисленных, богатых водородом материалов (к которым относятся нефтяные и пековые коксы) получают, как правило, легкографитируемые материалы. Следует отметить, что частицы коксов зачастую имеют удлинённую форму (анизометричны), что приводит к анизотропии свойств конечных графитовых композитов.

С конца XIX века в качестве связующего применяется каменноугольный пек, который представляет собой остаток после разгонки каменноугольной смолы по фракциям, твёрдость его может быть различной, как и температура размягчения. В технологическом процессе приготовления искусственных графитов различают такие стадии, как предварительное дробление углеродного сырья, прокаливание, составление шихты, смешение, формование и отжиг. Процесс термического превращения, или графитации исходных углеродных материалов производят в специальных электрических печах, заканчивают графитацию обычно при температуре 2400-2800$^0$ С.

На рис.1.9 схематично показан процесс превращения исходного аморфного углеродного материала в высокоупорядоченный искусственный графит. При температуре выше 1600 - 1700$^o$С структура углеродного материала начинает перестраиваться: базисные плоскости упорядочиваются, а межплоскостное расстояние **с** несколько уменьшается. Вследствие деструкции боковых радикалов возрастает число свободных атомов углерода. Выше ~2000$^o$С происходит образование трёхмерно-упорядоченной структуры кристаллитов, сопровождаемое резким ростом их высоты $L_c$ и диаметра $L_a$**.**

Различия в структурных характеристиках позволяют объяснить иногда разницу в прочности графитовых композитов. К таким структурным характеристикам относятся, прежде всего, плотность и непосредственно связанная с ней мезо - и микропористость материала, текстурированность, анизотропия, а также размеры кристаллитов и степень совершенства кристаллической решётки.

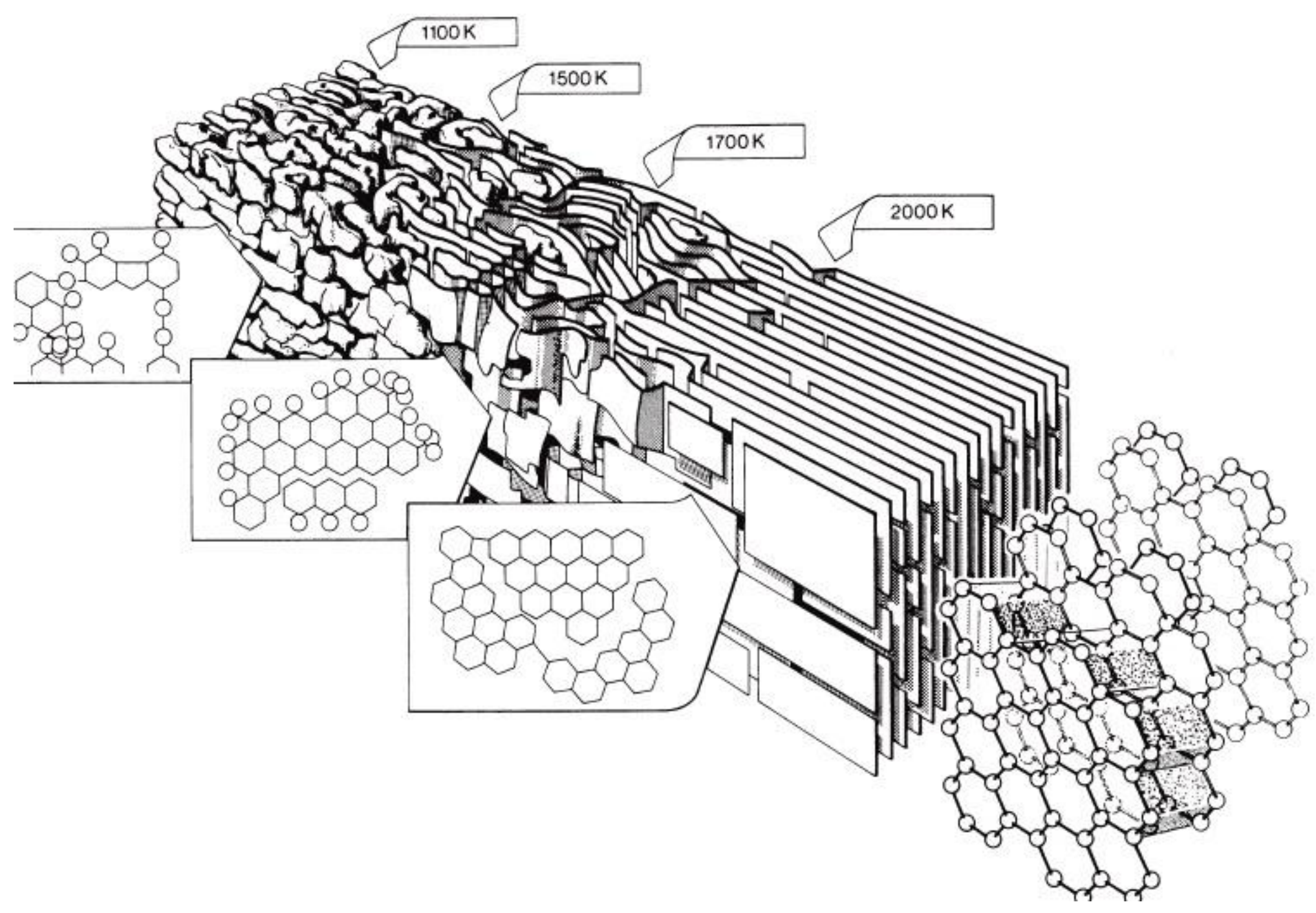

**Рис. 1.9.** Схема изменений мезоструктуры графита в процессе термической обработки [38].

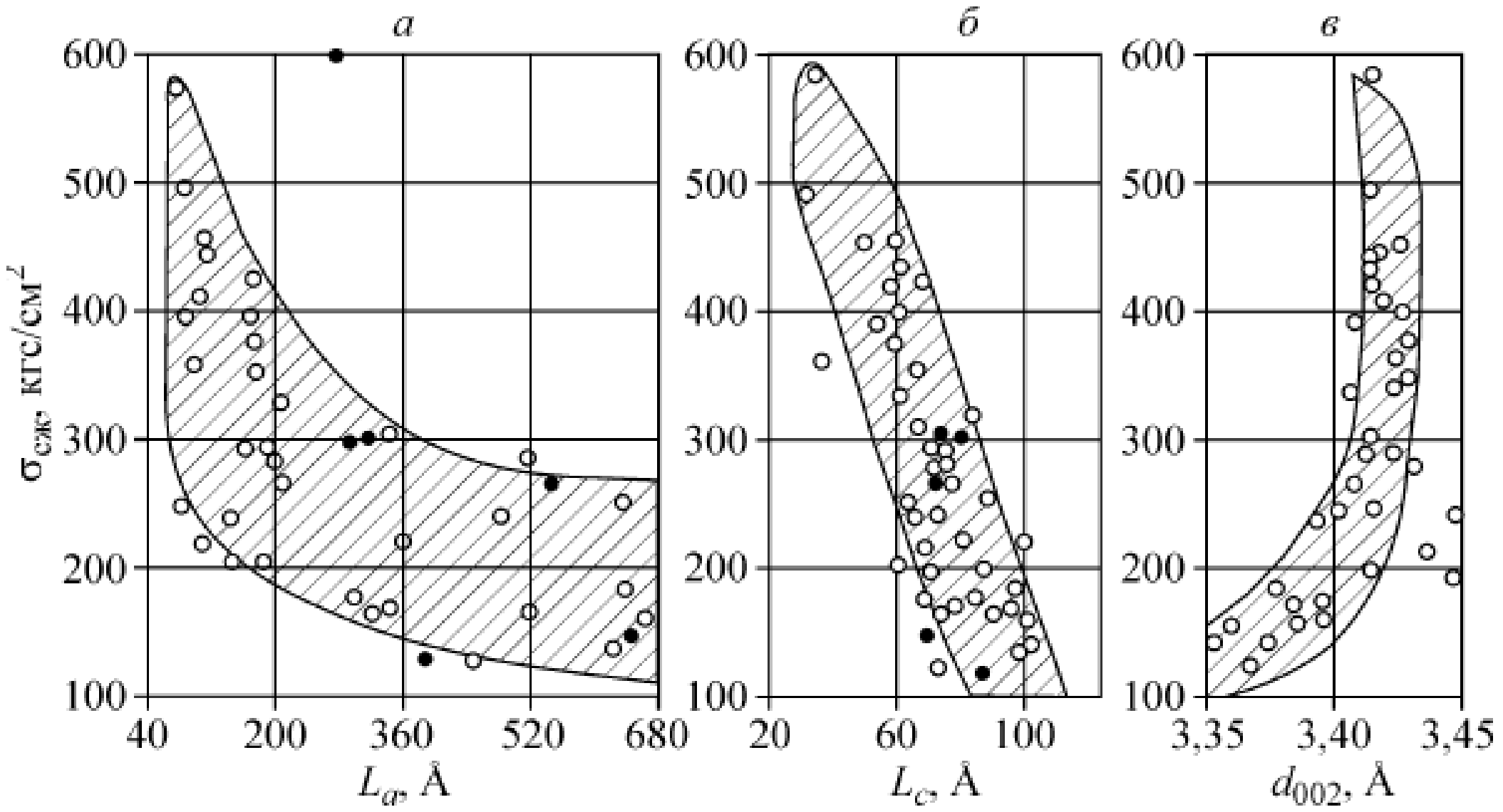

**Рис. 1.10.** Зависимость предела прочности графитов на сжатие от кристаллографических параметров [39]: *а* - от размера кристаллитов $L_a$ ,*б*- $L_c$ и межслоевого расстояния $d_{002;}$ светлые кружки – опытные материалы; тёмные кружки – промышленные материалы.

Так, в частности, при рассмотрении связи трех структурных характеристик: размеров кристаллитов по осям ***a*** и ***c*** ($L_a$ и $L_c$) и межслоевого расстояния $d_{002}$ с пределом прочности на сжатие, в работе Виргильева Ю.С. [39] получена эмпирическая зависимость вида:

$$\sigma_{сж} = const/ L_a, \qquad (1.9)$$

Такая зависимость характерна для материалов, не прошедших до конца графитацию. На (рис. 1.10) приведены данные работы [39], связывающие предел прочности на сжатие с кристаллографическими параметрами различных углеродных композитов.

## 1.2. Основные представления о динамике разрушения

Структурно-аналитическая теория прочности является одним из подразделов нелинейной динамики, которая используется во всех современных концепциях физического материаловедения [40, 41]. Среди теорий нелинейной динамики, получивших экспериментальное подтверждение, кроме структурно-аналитической теории прочности следует назвать также физическую мезомеханику, концепцию элементарного сдвига и фрактальную механику материалов.

В структурно-аналитической теории прочности [41] предпринята попытка объединения основных достижений физики и механики разрушения с целью построения таких уравнений, которые бы правильно учитывали физический аспект явлений и одновременно позволяли производить расчёты инженерного характера. В основе теории заложены аналитические соотношения, которые исходят из того, что движущей силой в структурированном континууме является эффективное поле напряжений $\sigma^*_{ik}$, определяемое через приложенные $\sigma_{ik}$, ориентированные $\rho_{ik}$ и неориентированные $\lambda_{ik}$ напряжения:

$$\sigma^*_{ik} = \sigma_{ik} - \rho_{ik} + \lambda_{ik} \qquad (1.10)$$

где индексы *i, k* в общем случае характеризуют выделенные мезообъёмы анизотропной в целом среды. Это базовое уравнение теории, записанное в тензорной форме для микроуровня, по форме аналогично и для макроуровня. Следует сказать, что в данном случае выбрана двухуровневая модель процесса формирования свойств: приложенные напряжения порождают микронапряжения, которые вызывают физические акты массопереноса и микротекучести, переходящие в деформации макроскопического масштаба. Принципиальным для теории являются понятия представительного объёма, элементарного акта деформации и законов деформационного поведения. При этом свойства представительного объёма должны выражаться через средние значения переменных, характеризующих его как сплошную и относительно однородную среду.

Представительный объём допустимо рассматривать и как математическую точку в сплошном континууме, благодаря чему физические аспекты теории усредняются на нижнем микроструктурном уровне, а механические – на верхнем макроструктурном.

В дополнение к уравнению (1.10) в математическую модель деформируемого твёрдого тела вводятся уравнения для расчёта кинетических коэффициентов структурной податливости, текучести, неоднородности, релаксации, повреждаемости, и снижения концентраторов повреждений. Все эти коэффициенты представляются в виде функционалов как тензорные объекты четвёртого ранга, наряду с коэффициентами упругой податливости и теплового расширения.

В настоящее время проведённые расчёты и эксперименты подтвердили, что структурно-аналитическая теория прочности в состоянии прогнозировать сложные свойства активной пластичности, ползучести и разрушения, а также применима для описания нетривиальных механических свойств с учётом изменений на различных структурных уровнях.

Физическая мезомеханика деформируемого твёрдого тела представляет собой новое направление в физическом материаловедении, в основу которой положена концепция структурных уровней деформации [42, 43]. В основе данной концепции лежат следующие принципиально новые положения:

- деформируемое твёрдое тело есть многоуровневая система, в которой пластическое течение самосогласованно развивается как последовательная эволюция потери сдвиговой устойчивости на различных масштабных уровнях: микро, мезо и макро;

- элементарными носителями пластического течения на мезоуровне являются трёхмерные структурные элементы (зёрна, конгломераты зёрен, субзёрна, ячейки дислокационной субструктуры, деформационные домены, частицы второй фазы, поры и др.), движение которых характеризуется схемой «сдвиг+поворот»;

- в данной концепции органически взаимосвязаны трансляционные и поворотные моды движения трёхмерных структурных элементов. Поворотные моды пластической деформации приводят в самосогласованное движение всю иерархию структурных уровней среды и обусловливают появление в ней новых диссипативных структур;

- основные закономерности пластического течения на мезоуровне связаны с образованием диссипативных мезоструктур и фрагментацией деформируемого материала;

- разрушение есть завершающая стадия фрагментации твёрдого тела, когда она локализуется на макромасштабном уровне;

- механизмы пластического течения, их носители и соответствующие стадии кривой «напряжение-деформация» подчиняются закону подобия (принцип масштабной инвариантности).

Самосогласованная деформация во всём объёме деформируемого твёрдого тела описывается законом структурных уровней деформации, согласно которому при пластическом формоизменении кристалла без нарушения сплошности сумма роторов потоков деформационных дефектов во всей иерархии структурных уровней среды должна быть равна нулю. Разрушение является заключительным этапом, когда локализованные трансляционно-ротационные вихри достигают размеров, соизмеримых с поперечным сечением образца, а ротор первичного скольжения в вихре не компенсируется суммарным ротором всех аккомодационных процессов.

Следует сказать, что введение в физическую мезомеханику понятия структурного уровня в деформируемых твёрдых средах позволило связать механику сплошной среды и теорию дислокаций. Универсальность физической мезомеханики основана на том, что эта теория построена на принципе калибровочной симметрии [44] (или масштабной инвариантности).

С этой точки зрения можно сказать, что физическая мезомеханика является аналогом других калибровочных теорий, таких, например, как электродинамика Максвелла. Ярким примером может служить аналогия между волной пластической деформации и электромагнитной волной, несущими энергию поля или аналогия между электрическим пробоем в газовых средах и разрушением твёрдых тел как итоговыми стадиями диссипации энергии.

Основные уравнения мезомеханики для скоростей сдвигов ν и поворотов ω в поле смещений кристалла по виду действительно похожи на уравнения Максвелла для электромагнитного поля:

$$\mathrm{div}\,\nu = g^{ij}\,\eta^{\alpha}{}_{i}\,\acute{\eta}_{j\alpha};$$

$$\mathrm{rot}\,\nu = \frac{d\omega}{dt};$$

$$(\mathrm{rot}\,\omega)_{\mu} = \frac{1}{C_t^{\,2}}\left[\frac{dv}{dt}\right]_{\mu} + g^{ij}\eta^{\alpha}{}_{i}(D\mu\,\eta^{\alpha}{}_{i}); \qquad (1.11)$$

$$\mathrm{div}\,\omega = 0.$$

Здесь $g^{ij}\,\eta^{\alpha}{}_{i}\,\acute{\eta}_{j\alpha}$ - функция, характеризующая источники вихрей; $C_t$ – предельная скорость распространения калибровочного поля в структурно-неоднородной среде; $g^{ij}\eta^{\alpha}{}_{i}(D\mu\ \ \eta^{\alpha}{}_{i})$ – полевые потоки, обусловленные изменением репера в пространстве; индексы $ij$ нумеруют взаимодействующие мезобъёмы, а индексы α и μ характеризуют размерность гетерогенной среды и могут принимать значения, равные 1, 2, 3 [41].

В настоящее время физическая мезомеханика стала базовой методологией компьютерного моделирования и компьютерного конструирования материалов со сложной внутренней структурой.

<u>Нелинейная динамика теории прочности</u> оперирует с самопроизвольно возникающими специфически-упорядоченными формами, которые называются диссипативными [40, 45]. Для любой среды, где есть взаимодействие компонент, используется уравнение:

$$\dot{g}(i) = \alpha g(i) + \beta g(i)g(j) + f(t), \qquad (1.12)$$

где: $\dot{g}(i)$ – скорость изменения основного параметра, например, обобщенной координаты для выделенного элемента; α, β – смысл этих констант определяется конкретной задачей, а управляющий параметр β характеризует взаимодействия между элементами системы; $f(t)$ – эта функция характеризует вероятность возникновения флуктуаций.

Диссипативными структуры образуются в неравновесных условиях при подводе внешней энергии к материалу. Спонтанное образование диссипативных состояний предопределяет нарушение фрактальной симметрии.

Отрицательные обратные связи в деформируемом материале определяют организацию структуры на квазиравновесной стадии, а положительные – самоорганизацию диссипативных структур в точках неустойчивости системы (точках бифуркаций). В точках бифуркаций снижается степень неравновесности системы в результате действия положительных обратных связей. Со временем степень неравновесности снова увеличивается, и циклы повторяются вплоть до разрушения системы.

При анализе неравновесных систем рассматривают, как правило, не временную эволюцию, а последовательности стационарных неравновесных состояний. Например, при деформировании металлов и сплавов в процессе эволюции системы в зависимости от исходной структуры реализуется спектр точек-бифуркаций, отвечающих смене лидеров-дефектов, ответственных за диссипацию энергии на разных уровнях квазиравновесности системы.

Переход от одного неравновесного состояния к другому осуществляется тогда, когда система достигает некоторого порогового уровня энергии, диссипированной лидер-дефектом [40]. На основе такого подхода можно разбить конструкционные материалы на несколько классов, отличающихся механизмом диссипации энергии и доминирующим типом лидер-дефекта. Физическое материаловедение трактует элементарные акты пластической деформации и разрушения как разные, но взаимообусловленные события. Они реализуются в твёрдых телах, представляющих собой статистические системы из большого числа атомов и взаимодействующих дефектов кристаллического строения.

Кинетическая концепция прочности показала, что разрушение твёрдых тел начинается с момента их нагружения и является кинетическим, термофлуктуационным процессом [46]. Предложенная в данной работе концепция не только показала, что разрушение является термофлуктуационным процессом, но и позволила обосновать принципиальную возможность прогноза времени разрушения. В зависимости от приложенного напряжения **σ** и абсолютной температуры $T$ макроскопическая кинетика разрушения описывается уравнением долговечности:

$$\boldsymbol{t}(\sigma) = \tau_0 \exp[(U_0 - \gamma\sigma)/kT], \qquad (1.13)$$

где: $k$ – постоянная Больцмана; $\tau_0$ – время, по порядку величины близкое к периоду тепловых колебаний атомов ~ $10^{-13}$ сек; $U_0$ – начальная энергия активации процесса разрушения, снижаемая приложенным напряжением σ; $\gamma = qVa$, где $Va$ – есть активационный объём в элементарном акте диссоциации; $q$ – коэффициент локальных перенапряжений. Этот коэффициент достигает в реальных телах значения 10 ÷ 100 и выше, что и характеризует отличие экспериментальной величины прочности от теоретической.

Макроскопическая деформация также описывается кинетической зависимостью вида:

$$d\varepsilon/dt = d\varepsilon_0(\sigma, T)/dt \exp[(Q_0 - \alpha\sigma)/kT]. \qquad (1.14)$$

Здесь $d\varepsilon/dt$ есть скорость деформации, а величина $Q_0$ есть энергия активации ползучести.

Приведённые выше соотношения (1.13) и (1.14) имеют похожий вид, но написаны для разных процессов, и в общем случае $Q_0 \neq U_0$ и $\alpha \neq \gamma$.

Если элементарный акт пластической деформации предполагает временное ослабление или разрыв межатомных связей, то разрушение связано с образованием несплошностей, в пределах которых действием межатомных сил можно пренебречь. Вид формулы (1.13) и обширные экспериментальные данные свидетельствуют о том, что приводящие к разрушению атомные перестройки осуществляются термоактивационно за счёт тепловых флуктуаций. Таким образом, макроскопическое разрушение тела представляет собой процесс последовательных элементарных актов разрыва напряженных межатомных связей флуктуациями тепловой энергии атомов.

**Двухстадийная модель разрушения твёрдых тел**

Макроскопическое время до разрушения материала под нагрузкой $t(\sigma)$ можно представить в виде двустадийного процесса [47, 48]:

$$\tau(\sigma) = \tau_1 + \tau_2\,, \qquad (1.15)$$

где $\tau_1$ – время накопления стабильных микротрещин и образования макротрещины в ансамбле микротрещин, $\tau_2$ – время роста трещины до критического размера потери стабильности, и $\tau_1 \gg \tau_2$

Времена $\tau_1$ и $\tau_2$ можно выразить через среднее время образования микротрещины $\tau_m$ следующим образом:

$$\tau_1 = k_1\,\tau_m \qquad (1.16)$$

$$\tau_2 = k_2\,\tau_m \qquad (1.17)$$

где время образования микротрещины согласно [1.44] равно:

$$\tau_m = \tau_0 \times \exp[(U(\sigma)/kT]. \qquad (1.18)$$

Предполагается, что в данном случае $k_1$ и $k_2$ – известные структурные параметры, при этом для мелкозенистых и наноструктурированных материалов условия образования микротрещин в объёме и на границах зерна становятся сопоставимыми, что приводит к существенной зависимости долговечности мелкозернистых материалов от размеров зерна [47]. В этом случае материал можно представить себе как двухфазный, одной из фаз которого являются границы зёрен, другой – внутризёренный объём. В общем случае, вероятности образования микротрещин на границах зерна и в объёме материала можно считать независимыми процессами, и, в случае, если одна из вероятностей образования микротрещины заметно превышает другую, мы снова приходим к исходной формуле (1.13).

Интересно отметить, что скорость накопления микротрещин $dC/dt$ описывается выражением, аналогичным по форме уравнению (1.13) для долговечности:

$$dC/dt = \acute{C}_0 \exp[(U_0 - \gamma\sigma)/kT] \qquad (1.19).$$

Это говорит о том, что именно кинетика накопления микротрещин в конечном счёте определяет долговечность нагруженного материала.

Согласно [48], предсказание предразрушающего состояния должно было бы заключаться в контроле тем или иным методом за процессом накопления повреждений с тем, чтобы предупредить переход во вторую стадию процесса разрушения. Таким методом мог бы служить, например, использованный авторами данной работы метод акустической эмиссии (АЭ). На рис. 1.11 видно, что измеренная методом АЭ скорость накопления микротрещин сначала нарастает, затем становится пологой, и при появлении очага разрушения снова быстро возрастает вплоть до разрушения.

Необходимость анализировать большие ансамбли микротрещин для понимания закономерностей перехода от микро- к макроразрушению и использовать статистические закономерности для количественного анализа взаимодействия и развития микротрещин привела к представлению о двухстадийной модели разрушения [49, 50], проиллюстрированной на рис. 1.12.

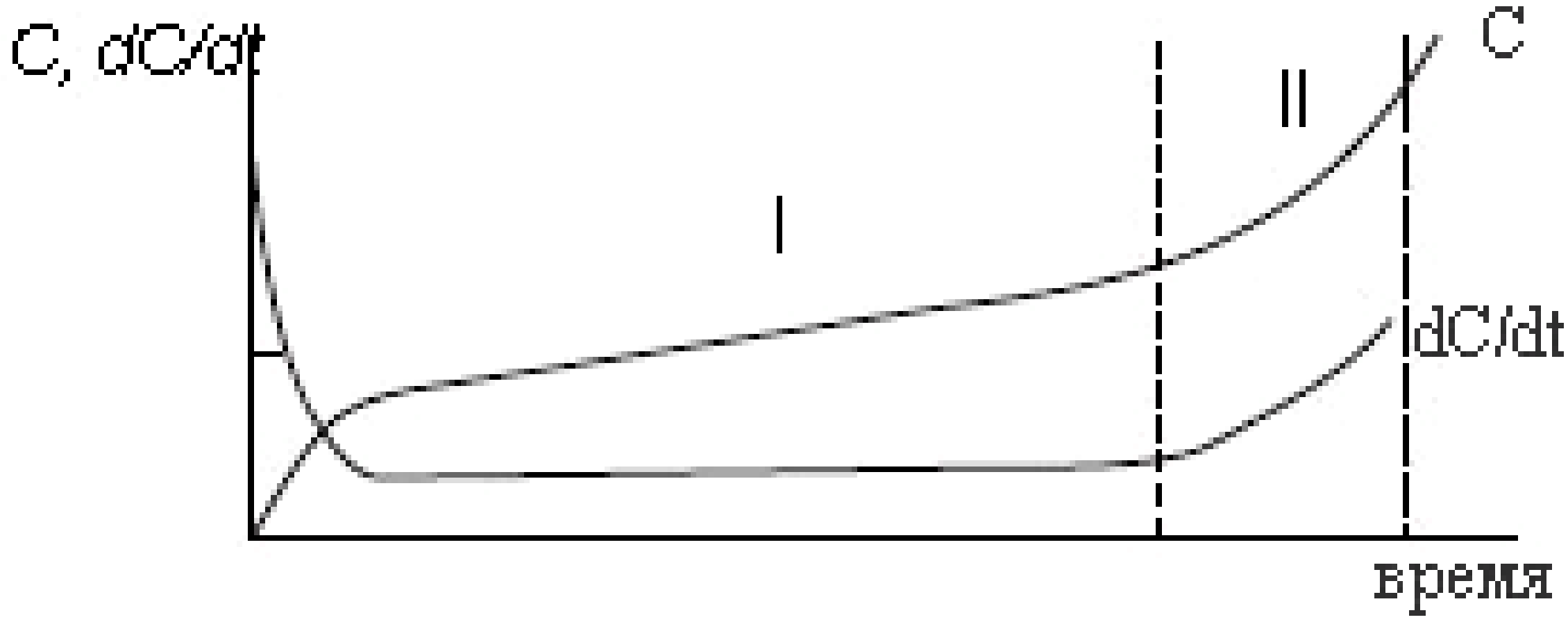

**Рис. 1.11.** Накопление микротрещин (*C*) и скорость их накопления (*dC/dt*) [1.44].

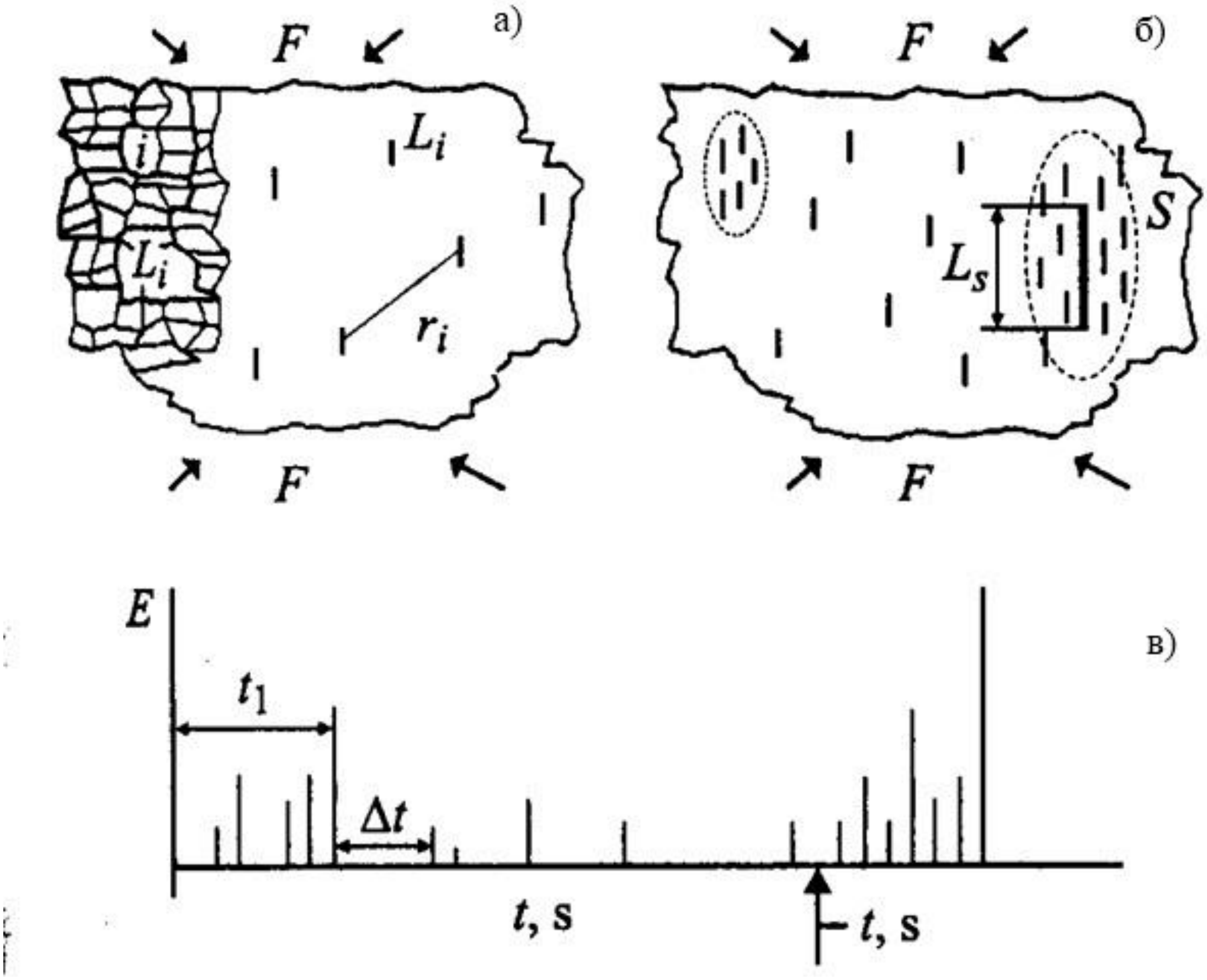

**Рис. 1.12.** Две стадии разрушения: стационарная (а), очаговая (б), и схема излучения акустических сигналов при образовании трещин (в); $E$ — энергия излучения, $t$ — время, $F$ — нагрузка, $Li$ — размер начальных трещин, $Ls$ — размер второго ранга, $S$ — область очага разрушения [1.44].

На первой стадии (рис.1.12, а) происходит делокализованное накопление одиночных стабильных микротрещин в объеме тела, приводящее вследствие флуктуаций концентрации трещин к образованию ансамблей близкорасположенных трещин, способных к взаимодействию и слиянию, и, в конечном итоге, к формированию очага разрушения. Вторая стадия (рис.1.12, б) определяется локализованным развитием очага разрушения, заканчивающимся появлением магистральной трещины с разрушением образца. Количественно переход от первой стадии накопления трещин к появлению ансамблей или кластеров проанализирован в [48]. Определяющим параметром такого перехода является концентрационный параметр

$$K = r_i/L, \qquad (1.20)$$

где $L$ – размер образующихся трещин, $r_i$ – средние расстояния между трещинами. Активное образование кластеров начинается при значениях $K \approx 3$.

Уравнение (1.13) позволяет с логарифмической точностью оценить время до разрушения нагруженного тела, если известны входящие в него параметры. Двустадийная модель, в свою очередь, позволяет сформулировать физические принципы прогнозирования макроскопического разрушения и существенно повысить точность оценки момента разрушения. Если есть возможность контролировать тем или иным методом процесс накопления повреждений и заметить переход из первой стадии накопления повреждений во вторую, это и было бы предсказание предразрушающего состояния. Кроме того, процесс накопления повреждений является стохастическим, что позволяет использовать целый ряд статистических параметров для описания процесса накопления повреждений, которые закономерно должны меняться при переходе из квазистатической стадии накопления повреждений в стадию развития локализованного очага разрушения.

Так, согласно рис.1.11 кривая накопления дефектов (микротрещин) $C$ сначала быстро нарастает, затем становится пологой. При появлении очага разрушения интенсивность зарождения дефектов вновь возрастает до окончательного разрушения. Средние расстояния между дефектами при их хаотическом делокализованном зарождении на первой стадии остаются постоянными и резко уменьшаются при появлении локализованного очага разрушения. Особенно информативным является параметр, связанный с временными интервалами между хронологически последовательными актами образования дефектов. Во-первых, он связан со скоростью зарождения дефектов и может описывать кривую их накопления. На стационарном участке, где скорость накопления дефектов постоянна, этот параметр тоже остается постоянным и уменьшается при активном развитии очага разрушения (рис.1.12, в).

Двухстадийная модель разрушения характерна для весьма широкого класса материалов, в частности, для металлов, полимеров, различных композитов, и т.д. При этом структура материала, гомогенная либо гетерогенная, формируя локальные перенапряжения, обуславливает начальный размер микротрещин.

Анализ особенностей разрушения гомогенных и гетерогенных пористых систем [51] говорит о том, что при “гетерогенном” разрушении суммарная приложенная нагрузка рассредоточена в большом числе вершин растущих микротрещин, а при “гомогенном” разрушении – всего на нескольких вершинах. В результате гетерогенные системы способны выдерживать более высокие предельные нагрузки, т.е. обладают более высокой прочностью.

Исследования, проведенные на различных материалах: полимерах, металлах и композитных материалах, показали [48, 51], что размеры начальных микротрещин обусловлены структурой материала, которая как формирует локальные перенапряжения, так и ограничивает их рост на границах гетерогенности.

## 1.3. Изменение свойств конструкционного графита при облучении

### 1.3.1. Оценка работоспособности графита в кладках уран-графитовых реакторов

Особенности поведения графита для применения в кладках графитовых реакторов исследовались в целом ряде работ, в частности, в [19, 37, 52-57]. Графит, и мелкозернистый графит, в частности, предназначенный для ядерных реакторов, обладает повышенной прочностью при высоких температурах. Это свойство графитов объясняют, как правило, «залечиванием» микротрещин в результате самодиффузии атомов углерода при повышенных температурах [19, 31]. Заметное возрастание прочности графитовых композитов происходит вплоть до 2400÷2500°С, затем прочность начинает резко падать.

Столь уникальный характер прочностных свойств графитовых композитов при повышенных температурах позволяет сохранять работоспособность графитовых кладок уран-графитовых реакторов под воздействием нейтронного облучения в течение многих десятков лет, с почти двукратным превышением ресурсного срока. Например, графитовая кладка реактора АМ 1 Обнинской АЭС остается работоспособной в течение 50 лет [57].

Динамика процесса разрушения графита под воздействием высоких температур и нейтронного облучения происходит, как можно полагать, всё так же в две стадии, и это можно обнаружить, например, по изменениям электрической проводимости, связанной с ней теплопроводности и предела прочности графитовых композитов от флюенса нейтронов (рис. 1.13).

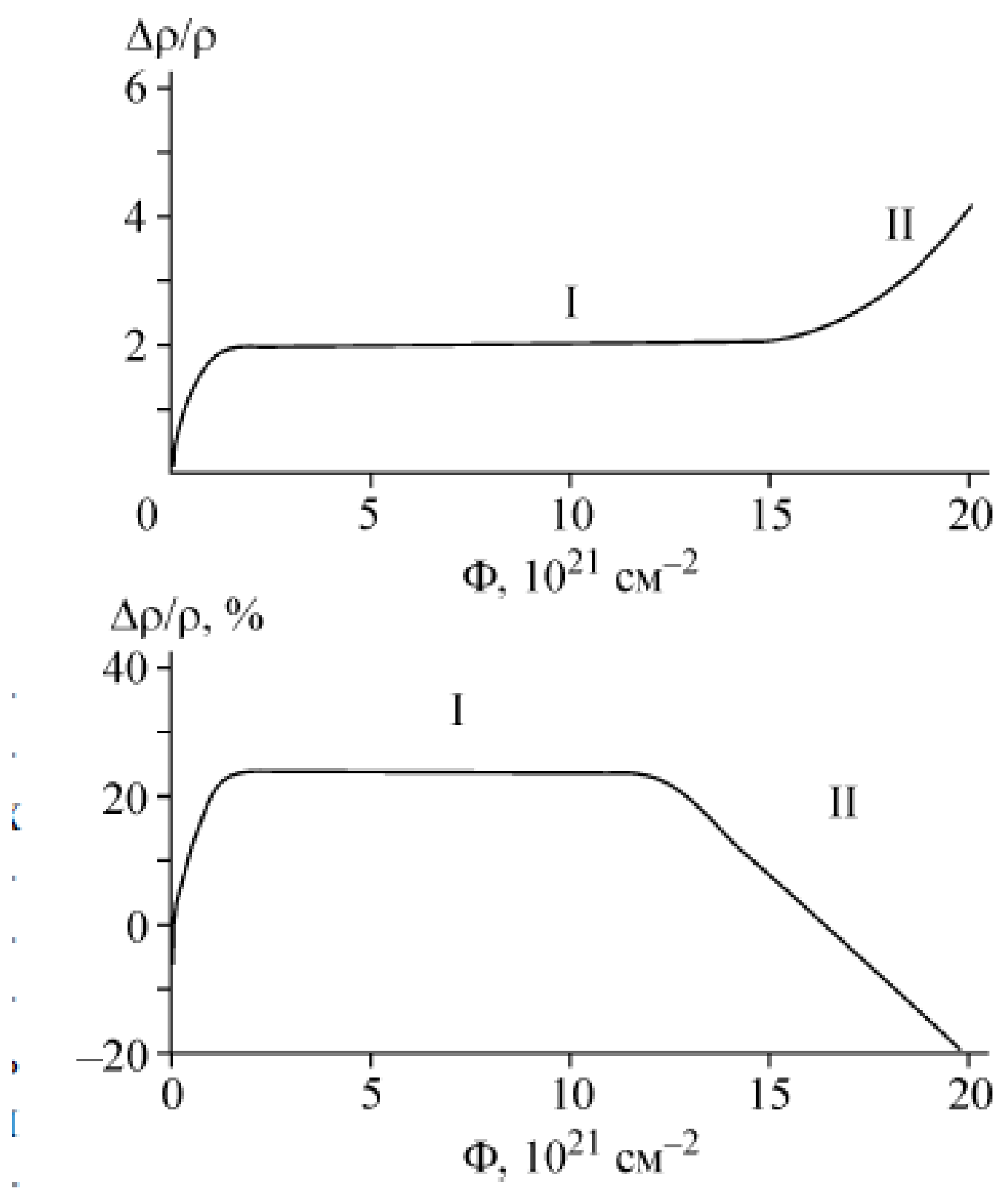

**Рис. 1.13.** Характерная кривая изменения зависимости относительного изменения удельного электросопротивления образцов графитовой кладки от флюенса нейтронов (вверху). Характерная кривая зависимости относительного изменения предела прочности при сжатии образцов от флюенса нейтронов (внизу). Использованы графические данные работы [57].

В работах [53-57] в качестве критерия работоспособности графита из уран-графитовой кладки выбрано отношение пределов прочности при сжатии и изгибе, поскольку это отношение практически не меняется при окислении или облучении дозами ниже критического флюенса нейтронов. Этот эмпирически найденный показатель весьма чувствителен к радиационному распуханию образцов и начинает быстро расти при закритических флюенсах нейтронов.

В работах Виргильева Ю.С. с сотрудниками установлена также взаимосвязь между относительными изменениями размеров графитовых образцов реакторных графитов и физическими свойствами графита (электросопротивление, модуль Юнга, предел прочности) в зависимости от флюенса нейтронов [3]. В работе [58] для обоснования работоспособности графитовой кладки ядерного реактора был предложен комплекс математических моделей и программ для расчёта деформаций графита в процессе эксплуатации с учётом неоднородности распределения температурного и нейтронного поля, а также анизотропной усадки и распухания графита вследствие ползучести.

### 1.3.2. Теория радиационно-индуцированного формоизменения графита

Двустадийная модель разрушения твёрдых тел, несмотря на свою простоту и наглядность, как правило, не применима для сложных гетерогенных систем, находящихся под комплексным воздействием облучения, повышенной температуры и механических напряжений. В этой ситуации требуется иной подход, и такой подход был разработан в ряде работ, в частности *Kelly B.T.* [59], а также в ряде работ Панюкова С.В. с соавторами [60-61].

Согласно [60, стр.268] «…при описании радиационно-индуцированных эффектов графит рассматривается как поликристалл, состоящий из определённым образом упорядоченных кристаллитов» (рис.1.14). Дальний текстурный порядок в графите либо значительно ослаблен, либо отсутствует в зависимости от степени разупорядоченности в ориентации кристаллитов. Различие в ориентации соседних кристаллитов при наличии анизотропии их свойств ведёт к появлению микротрещин на границах кристаллитов, которые вносят определяющий вклад в упругость структурно-разупорядоченных твёрдых тел.

Морфология графита согласно [61] определяется двумя главными составляющими – наполнителем и связующим при наличии ансамбля микротрещин и технологических пор. Связующее имеет мелкокристаллическую однородную структуру, наполнитель - иерархическую, в основе которой лежит микрокристаллит с более или менее совершенной кристаллической структурой. Микрокристаллиты образуют различного рода образования – зёрна с разной степенью выраженности текстуры.

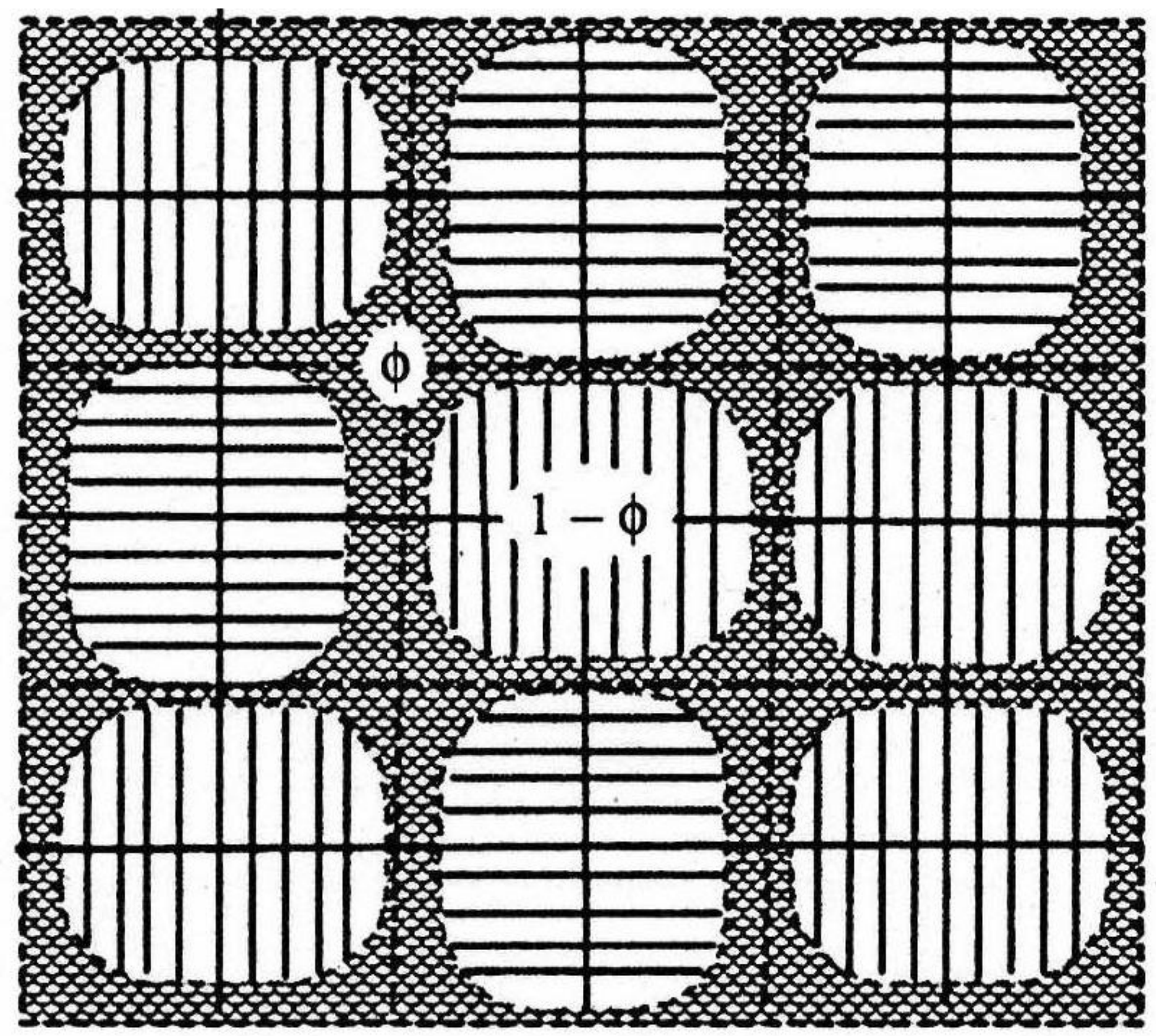

**Рис. 1.14**. Решёточная модель графита с кристаллитами объёмной долей (1-φ) и направленной вдоль главных осей решётки осью анизотропии в узлах матрицы, заполненной однородным связующим объёмной долей φ [60].

Ансамбль микротрещин изначально является следствием сильной анизотропии свойств на уровне микрокристаллита, в то время как анизотропия заметно ослаблена на уровне зерна. Под действием облучения ансамбль микротрещин эволюционирует, что в значительной степени и является причиной наблюдаемых радиационных эффектов. На уровне поликристалла в целом радиационные эффекты либо изотропны, либо ортотропны в зависимости от способа получения графита.

Движущей силой всех вышеперечисленных эффектов является радиационно-индуцированное изменение формы микрокристаллита, вызываемое развитием в нём микроструктуры, преимущественно в базисной плоскости. Хотя перечисленные эффекты описаны эмпирически и поняты на качественном уровне, теория, позволяющая описывать радиационные эффекты в графите на макроуровне отсутствует.

В работе [61] предложена аналитическая модель, в которой на основе рассмотрения ортотропного континуума со случайно расположенными формоизменяющимися включениями (кристаллитами) удалось связать развитие внутренних напряжений с эволюцией ансамбля трещин.

Рассматриваемая модель учитывает два основных эффекта, определяющих упругие свойства графита и ранее не принимавшихся во внимание, - рост внутренних напряжений в графите вследствие формоизменения кристаллитов под действием облучения и изменение упругих свойств графита из-за эволюции в нём ансамбля микротрещин. Для описания упругих свойств графита рассматривается решёточная модель поликристалла (рис.1.14), в которой кристаллиты находятся в узлах решётки с координатами $x_j$. Модель предполагает наличие беспорядка в ориентации и размерах кристаллитов, а их формоизменение происходит под действием случайных сил со стороны матрицы. В гипотетическом случае регулярной решётки кристаллитов эти силы также регулярны и соответствуют сжатию вдоль оси кристалла и более слабому растяжению вдоль других осей.

В реальных неупорядоченных графитах распределение кристаллитов по размеру существенно бимодально, поскольку наряду с кристаллитами больших размеров (наполнитель) имеется множество мелких, входящих в состав связующего решётки. Если кристаллиты больших размеров могут быть учтены в решёточной модели явно, то в отношений мелких кристаллитов связующего такой анализ затруднён по ряду причин. Предполагается, что упругость структурно-разупорядоченных тел в существенной степени определяется наличием микротрещин между соседними кристаллитами. При деформации графита как в результате эволюции внутренних напряжений, так и под действием внешних сил происходит изменение формы и объёма микротрещин.

В модели предполагается, что формоизменение кристаллитов и силы внутренних напряжений $p_i$ пропорциональны флюенсу нейтронов Ф, так что $p_i = p_0$ Ф (1- $c_i$). Доля разорванных связей между кристаллитами $c_i$ из-за наличия микротрещин поначалу уменьшается с ростом флюенса, поскольку сохраняются растущие микротрещины и могут исчезать уменьшающиеся. В конечном счёте оказывается, что зависимость средней доли микротрещин от флюенса можно апроксимировать линейной функцией:

$$c_i = c_{i0}\,(1\text{-}\beta_i \cdot \Phi) \qquad (1.21)$$

где $c_{i0}$ – доля микротрещин в отсутствие облучения. Значения параметров $p_0$, $c_{i0}$ и $\beta_i$ зависят от микроскопических механизмов образования микротрещин и в дальнейшем в рамках данной модели рассматриваются как исходные параметры, в зависимости от состава наполнителя, связующего, температуры графитации и т.д. Расчеты, проведенные в рамках данной решёточной модели, показывают хорошее согласие теории с экспериментом (рис.1.15).

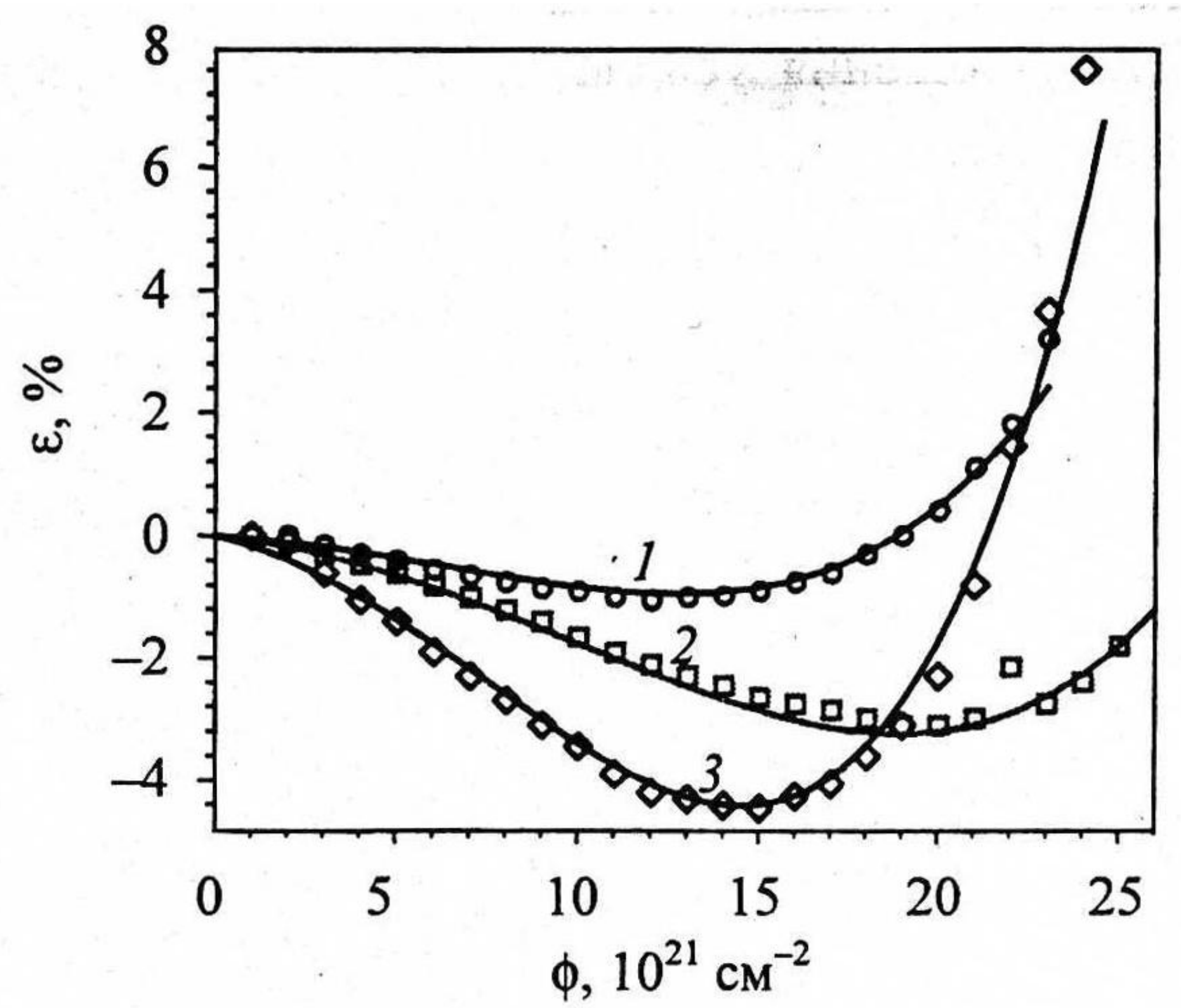

**Рис.1.15.** Зависимость относительного изменения поперечных (1) и продольных (2) размеров, а также объёма (3) от дозы облучения для графита ГР-280 . Расчётные кривые показаны сплошными линиями [61].

В случае большого радиационного распухания развивается новый ансамбль микротрещин, который следует принимать во внимание при нахождении зависимости сил внутренних напряжений $p_{\mathrm{i}}$ от дозы облучения. Параметр $\beta_i$ в данном случае выбран из условия воспроизведения экспериментальной кривой деформации графита и, по сути, является подгоночным.

***

Таким образом, резюмируя первую главу, можно сказать, что прочность и долговечность графитов напрямую связана с микро - и мезоструктурой материала углеродного композита, а также с такими характеристиками, как плотность и анизотропия, размеры кристаллитов, размеры и характер пор, т.е**.** *со структурой* материала в широком смысле слова структура. Изменение прочности материала происходит тогда, когда происходят существенные изменения его фазового состава, микроструктуры и/или пористой структуры, иначе говоря, когда идут процессы фазовых превращений, спекания, рекристаллизации, образования или залечивания дефектов и т.д., все эти процессы вместе или хотя бы один из них. Поэтому так важно контролировать физическими методами такие характеристики как размер кристаллитов, величину микроискажений структуры, соотношение кристаллической и аморфной составляющих, размер и суммарный объем пор и т.д. На основании этих данных не всегда удается предсказать механические характеристики, в частности, прочностные, но можно со значительной долей уверенности говорить о стабильности структуры материала, а значит о сохранении, по крайней мере, исходной прочности.

Вышеизложенное предполагает необходимость исследования особенностей внутренней структуры и дефектности графитовых материалов с целью прогноза долговечности их работы на основе наиболее общих соображений структурно-аналитической теории прочности. Так, в частности, более чем важным представляется вопрос о взаимосвязанных явлениях прочности, ползучести и самодиффузии для случая графитовых композитов, где, как и в металлах, представляется предопределяющей роль межзёренной границы раздела. Так, в частности, для металлов уже было показано [62], что именно ползучесть с энергией активации, близкой к энергии активации самодиффузии, при определённых условиях может стать контролирующим механизмом диссипации энергии.

Можно предположить также, что в графитах, также как и в металлах, основным механизмом диссипации избыточной энергии, подводимой за счёт протонного облучения, является пластическая деформация, а в качестве доминирующего лидер-дефекта можно рассматривать как различного типа дислокации и вакансии, так и другие дефекты кристаллической структуры. Краевые дислокации, междоузельные атомы и атомы примесей, нарушения упаковки слоёв, а также особенности мезоструктуры графитовых композитов могут играть определяющую роль в стабильности и долговечности графитового материала.

ГЛАВА 2

# ПРОГНОЗ ВРЕМЕНИ ЖИЗНИ ГРАФИТОВЫХ КОМПОЗИТОВ НЕЙТРОННОЙ МИШЕНИ МЕТОДОМ ВЫСОКОТЕМПЕРАТУРНЫХ ИСПЫТАНИЙ

Работа по созданию интенсивного источника высокоэнергетичных нейтронов на основе протонного/дейтронного ускорителя проводилась в рамках проекта МНТЦ №3682 [63-67]. Основным узлом такого источника является нейтронопроизводящая мишень, в которой под воздействием мощного протонного либо дейтронного пучка возникают нейтроны. В рамках данного проекта были проведены расчёты полного выхода нейтронов в реакциях типа $^{12}C(p, n)^{13}N$, а также выхода вперед (полярный угол меньше 30 градусов) нейтронов из углеродных мишеней с содержанием изотопа $^{12}C$ - 99 % и $^{13}C$ – 1%.

Расчёты показали, в частности, что при торможении протонного пучка в графитовой мишени имеет место ярко выраженный брэгговский пик выделенной мощности на глубине до шести миллиметров. Кроме того, при работе мишень должна принимать из пучка и рассеивать в непрерывном и импульсном режиме до 150 – 300 кВт в пятне размером около 1 см$^2$, что также предполагает резкую анизотропию тепловой нагрузки. Материалы, из которых может быть изготовлен конвертор, по ряду причин, ограничены узким набором легких элементов – Li, Be, B, а также углеродные материалы; и в данном наборе только углеродные материалы способны выдерживать высокую температуру в сочетании с резко неоднородной по глубине и объёму тепловой нагрузкой. По этой причине изучение свойств и структуры мелко- и тонкозернистых углеродных материалов в данной работе предполагало своей целью прогноз долговечности и стабильности графитовой мишени, работающей в условиях как высоких (до 2000$^0$С) температур, так и радиационного облучения.

В настоящее время представляется, что основными параметрами, влияющим на ресурс конвертора нейтронной мишени из графитового материала является температура и время прогрева [67]. Для исследования закономерностей разрушения графитовой мишени были проведены испытания, моделирующие нагрев под воздействием протонного пучка с помощью пропускания через образец импульсного либо переменного тока [63].

## 2.1. Образцы

Для испытаний использовались мелкозернистые графитовые композиты немецкого производства от SGL Carbon Group [68], тонкозернистые французской фирмы Le Carbone-Lorraine (LeCL) [69] и мелкозернистые композиты марки МПГ производства Новочеркасского электродного завода [70]. В таблице 2.1 представлены паспортные характеристики графитов марки SGL, LeCL и МПГ. Образцы размером 65×5×1 мм прогревались переменным током в вакууме $2\times10^{-4}$ торр, температура образцов измерялась пирометром IS12 производства «Impac electronics». Для устранения механических напряжений на образец во время испытаний использовалась специальная конструкция держателя (см. Приложение I).

**Таблица 2.1.** Характеристики образцов графитовых композитов.

| | Ед. изм | SGL CARBON GROUP | LE CARBONE LORRAINE | МПГ-6 |
|---|---|---|---|---|
| Объёмная плотность | г/см$^3$ | 1,73 – 1,82 | 1,86 | 1.76-1.88 |
| Удельное электро сопротивление | МкОм·м | 9,4 – 10,2 | 16 | 11-16 |
| Прочность на изгиб | МПа | 40-85 | – | 50 – 70 |
| Прочность на сжатие | МПа | 90-170 | 76 | 100 – 120 |
| Пористость | % | 9,5-15 | 6 | 9 |
| Зольность | % | < 0,03 | – | 0,25 – 0,1 |
| Размер зерна | мкм | <3-20 | 5 | 30 – 150 |
| Модуль Юнга | ГПа | 10-13,5 | – | 10 – 12 |

| Теплопроводность | Вт/м*К | 65-130 | 80 | 180 – 190 |
|---|---|---|---|---|
| КТР | $10^{-6}$ К$^{-1}$ | 3,5-5,8 | 5,7 | 8 – 8,8 |

## 2.2. Рентгенографические измерения

Рентгенографические измерения проводили в ИК СО РАН на приборе URD-6 с использованием монохроматизированного Cu$K\alpha$-излучения. Регистрация дифрактограмм проводилась в пошаговом режиме (шаг 0.05°, время накопления 10 сек) в диапазоне углов 2θ от 10 до 100°. Эти измерения показали, что на рентгенограмме образцов SGL, LeCL и МПГ присутствуют практически все рефлексы, характерные для 2H политипа графита (рис. 2.1), однако степень их уширения очень различна. Узкие рефлексы типа 00l и hk0 свидетельствуют об относительно больших размерах областей когерентного рассеяния (областей, в пределах которых сохраняется периодическая структура) как в направлении, перпендикулярном к графитовым слоям, так и в плоскости слоев. В то же время значительное уширение пиков с индексами общего типа hkl означает, что в структуре графита сохраняется большое число дефектов упаковки (ошибок в чередовании слоев). Из рентгенодифрактограмм можно оценить как межплоскостные расстояния, так и величину области когерентного рассеяния (ОКР) методом Вильямсона-Холла. [71].

Параметры элементарной ячейки были уточнены методом наименьших квадратов (МНК) с помощью программы [72] с использованием положений дифракционных максимумов. Оценка размеров ОКР и величины микроискажений проведена по полуширинам дифракционных пиков методом аппроксимации в приближении лоренцевской формы пиков [71, стр.21]. Для разделения эффектов уширения за счет размеров ОКР и микроискажений использовались полуширины дифракционных пиков 002 и 004.

Более точные расчеты показали, что есть как сходство, так и некоторое различие в этих образцах (табл. 2.2). Параметры решетки всех трёх образцов совпадают в пределах оцениваемых погрешностей. Размеры ОКР в плоскости графенового слоя также достаточно близки. Некоторые различия имеются в микроструктуре образцов в направлении 00l. Размеры ОКР в образце *SGL* существенно меньше, хотя меньше и величина микроискажений (вариаций межслоевых расстояний).

Это свидетельствует о наличии в данном образце более тонких, но лучше упорядоченных пакетов графитовых сеток по сравнению с МПГ-6. Под лучшим упорядочением здесь имеется в виду только меньшая величина вариаций межслоевых расстояний, но не концентрация дефектов наложения слоев, которая, судя по анизотропии уширения пиков, одинакова в обоих образцах.

Расчеты ОКР в направлении 001 для образцов марки SGL хорошо коррелируют с данными электронной микроскопии, поскольку видно, что высота упорядоченных пакетов может быть заметно меньше 30 нм. Характерной особенностью графита марки LeCL можно считать несколько увеличенное межслоевое расстояние при стандартной величине микроискажений.

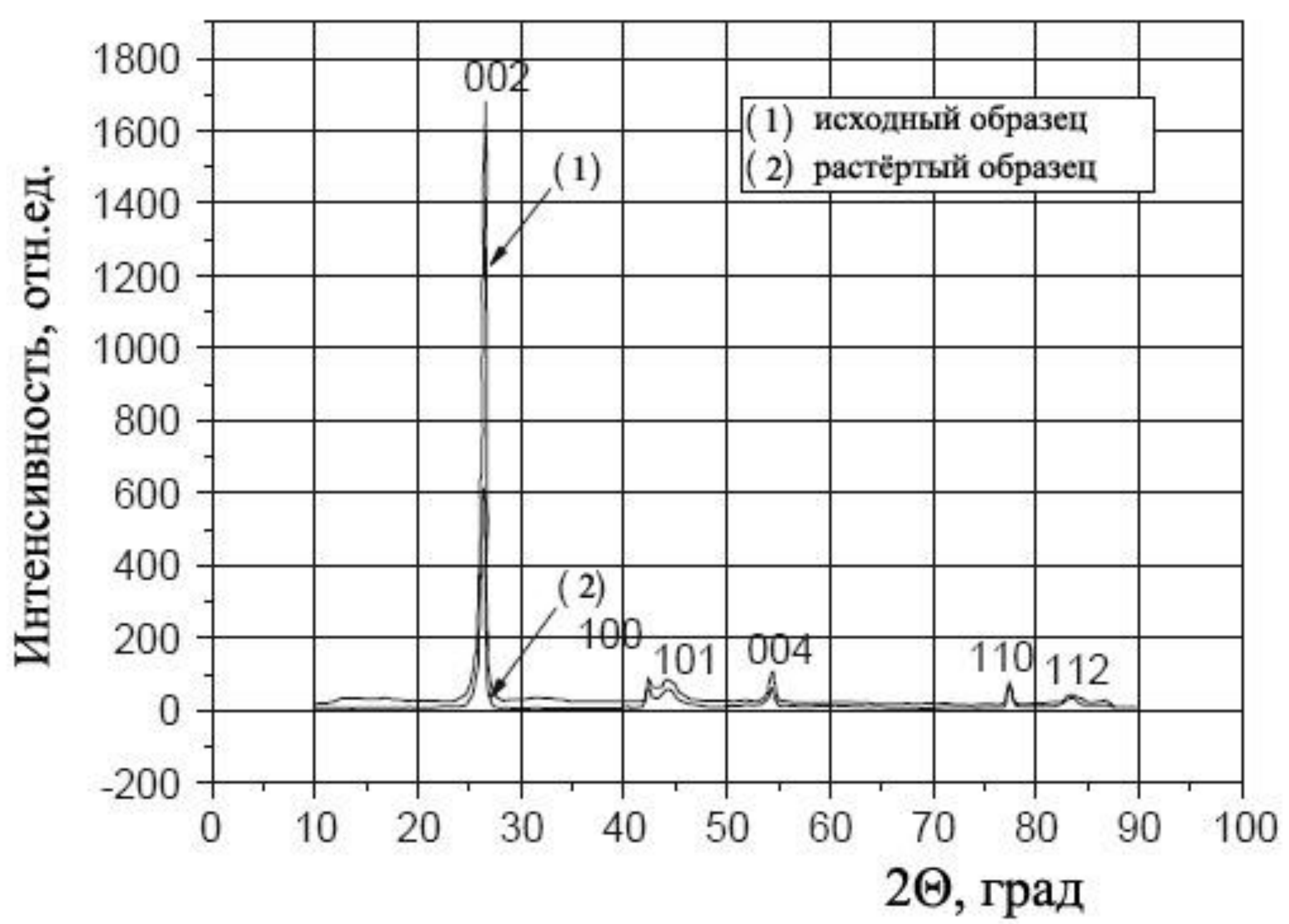

**Рис. 2.1.** Рентгенограмма образцов мелкодисперсного плотного графита SGL Кривая (1) относится к неразрушенной пластине, кривая (2) к растертому образцу. Высота пика 002 заметно больше в первом случае, поскольку для пластины имеет место текстура – преимущественная ориентация в направлении 00l. Для образцов графита марки LeCL и МПГ-6 рентгенограммы практически совпадают с нижней кривой [66].

**Таблица 2.2.** Кристаллографические параметры графитовых композитов.

| Образец | Параметры решетки | | Размеры ОКР, Å | | Величина микроискажений $\varepsilon_{001}$ |
|---|---|---|---|---|---|
| | *a*, Å | *c*, Å | 00l | hk0 | |
| МПГ-6 | 2.464(1) | 6.766(3) | >1500 | 250 | 0.0065 |
| SGL | 2.465(1) | 6.764(4) | 300 | 250 | 0.0030 |
| LeCL | 2.463(1) | 6,792(5) | >1000 | 220 | 0.0060 |

## 2.3. Электронно-микроскопические съемки

Электронно-микроскопические съёмки были выполнены в ИК СО РАН на просвечивающем электронном микроскопе (TEM) JEM-100C (Япония) при ускоряющем напряжении 100 кэВ и разрешении 0,5 нм. Образцы для измерений приготавливались с использованием стандартной процедуры. Углеродный материал измельчался в агатовой ступке, диспергировался ультразвуком в растворе этилового спирта, и затем полученная суспензия наносилась на медную сетку, покрытую углеродной пленкой с отверстиями микронных размеров.

На микрофотографии высокого разрешения (ВРЭМ) у мелкодисперсного углеродного композита марки LeCL вблизи изгибов графенового слоя можно наблюдать многочисленные дислокации (рис.2.2). Природа этих дефектов объясняется разрывом графеновых плоскостей и формированием дислокационных петель вследствие накопления междоузлий.

Такие агломераты образуются, когда отдельные подвижные атомы междоузлий диффундируют между двумя базовыми плоскостями, и объединяются, образуя менее мобильную группу [73]. Таким образом, междоузлия накапливаются в некоторое исходное ядро, которое впоследствии образуют дислокацию.

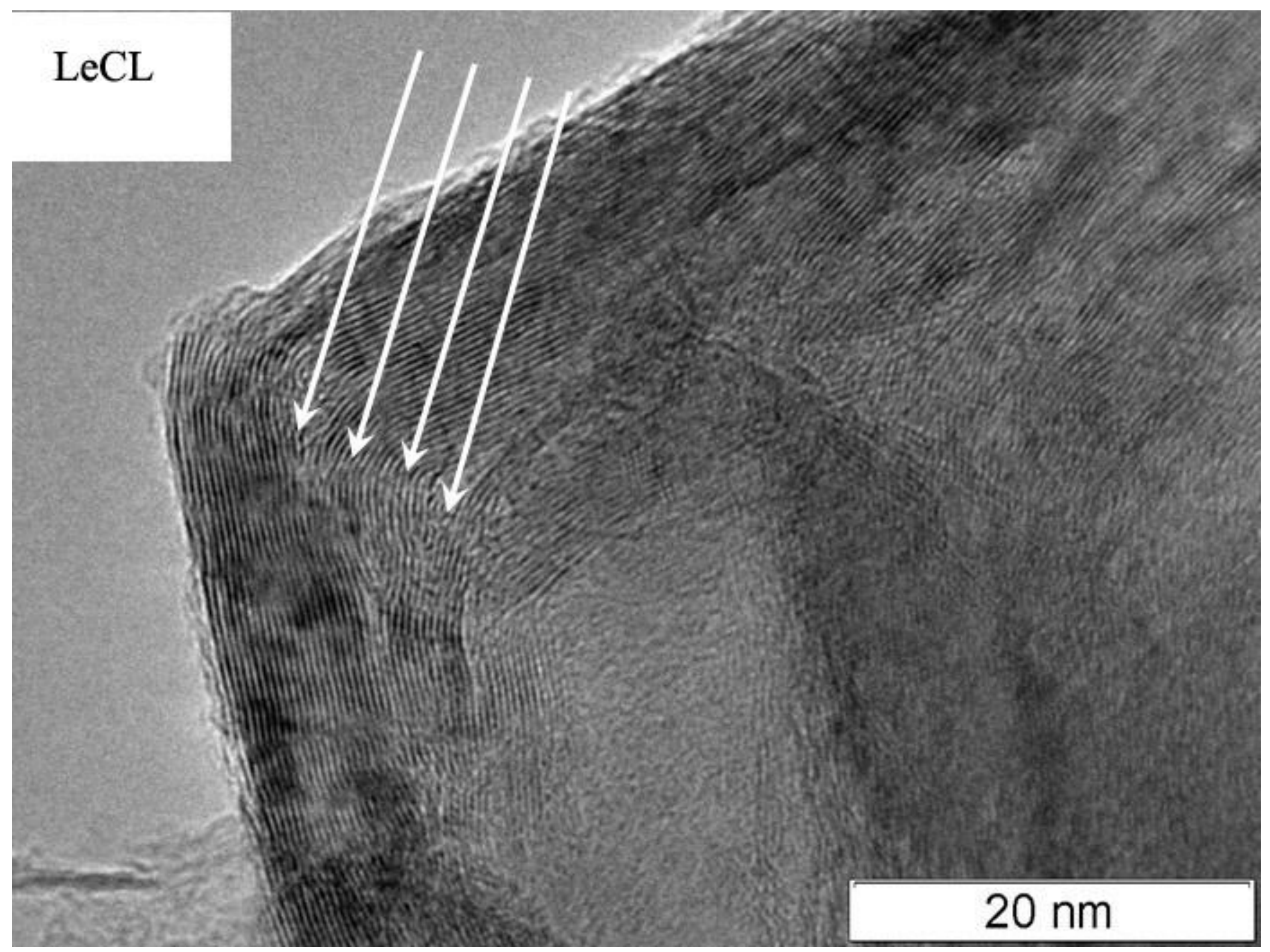

**Рис. 2.2.** ВРЭМ микрофотография графитового композита марки LeCL. Стрелками показаны дислокационные дефекты. [78].

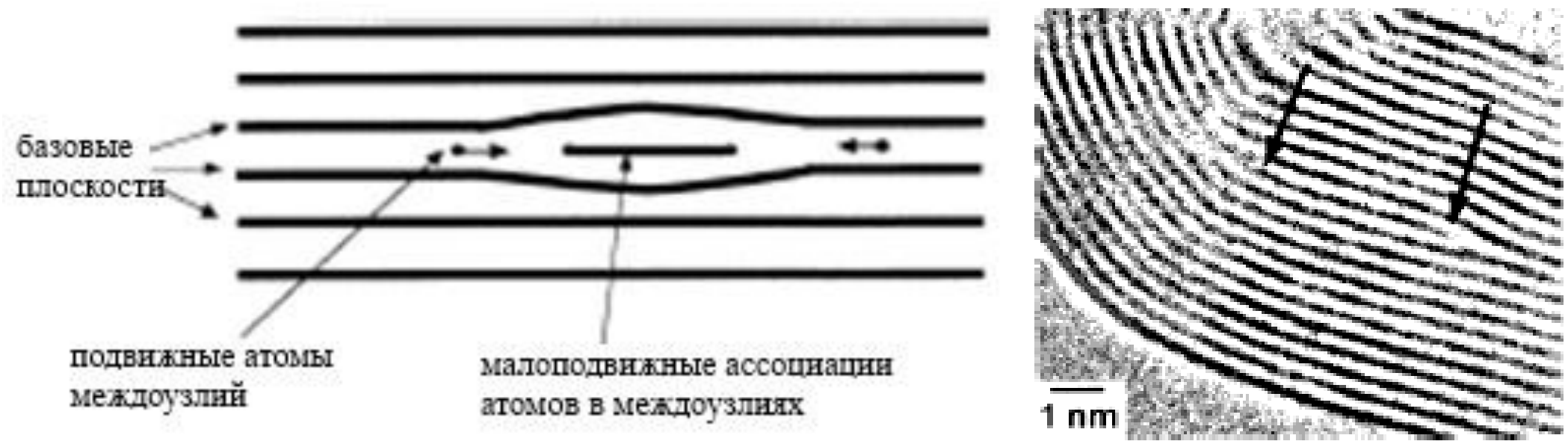

**Рис. 2.3.**) Схематическое изображение подвижных междоузлий (слева); ВРЭМ микрофотография дислокационных петель в пространстве между графеновыми плоскостями (справа). Стрелками указаны концы графеновой плоскости внедрения. [73].

Эта дислокация, расталкивая соседние базовые плоскости, приводит к формированию новой субатомной решётки (рис.2.3). В атомных слоях углеродного композита марки SGL можно наблюдать так называемую надмолекулярную структуру [21], состоящую из участков с параллельной ориентацией углеродных слоев (рис. 2.4). Хорошо виден участок внебазисного двойникования (наклонная межкристаллитная граница с углом, достаточно близким к 48°). Этот участок характеризуется многочисленными разрывами связей и достаточно развитой системой краевых дислокаций.

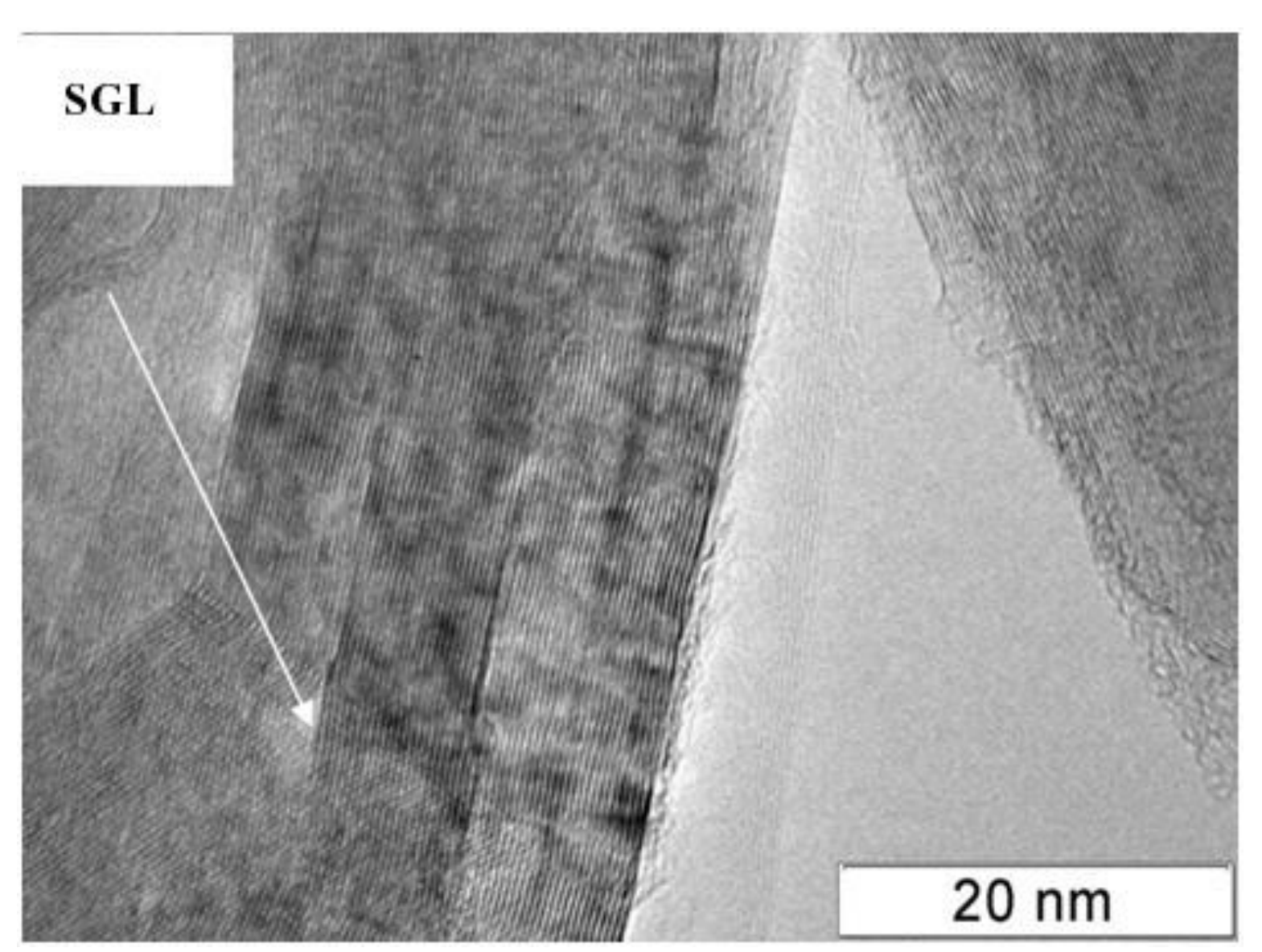

**Рис.2.4** ВРЭМ-микрофотография образца углеродного композита марки SGL [78]. Стрелкой указан дефект двойникования [30].

На микрофотографии (рис. 2.5) хорошо видны «дефекты расщепления», которые приводят к появлению мезопор с характерными размерами до 10 нм. На рис. 2.6 можно наблюдать дефект, связанный с появлением гетерогенно-графитированых областей. Такие дефекты сажевого типа предположительно возникают вокруг металлических частиц примеси, являющихся катализатором процесса графитации.

То, что такие частицы примеси существуют в данном типе графита, показывают результаты измерений методом ускорительной масс-спектроскопии (AMS) [74]. Хорошо видно наличие примесей калия, например, а также серы и кислорода (рис. 2.7). Следует отметить, что медь не является характерной примесью для данного графита, пик меди в данном случае связан с держателем образца.

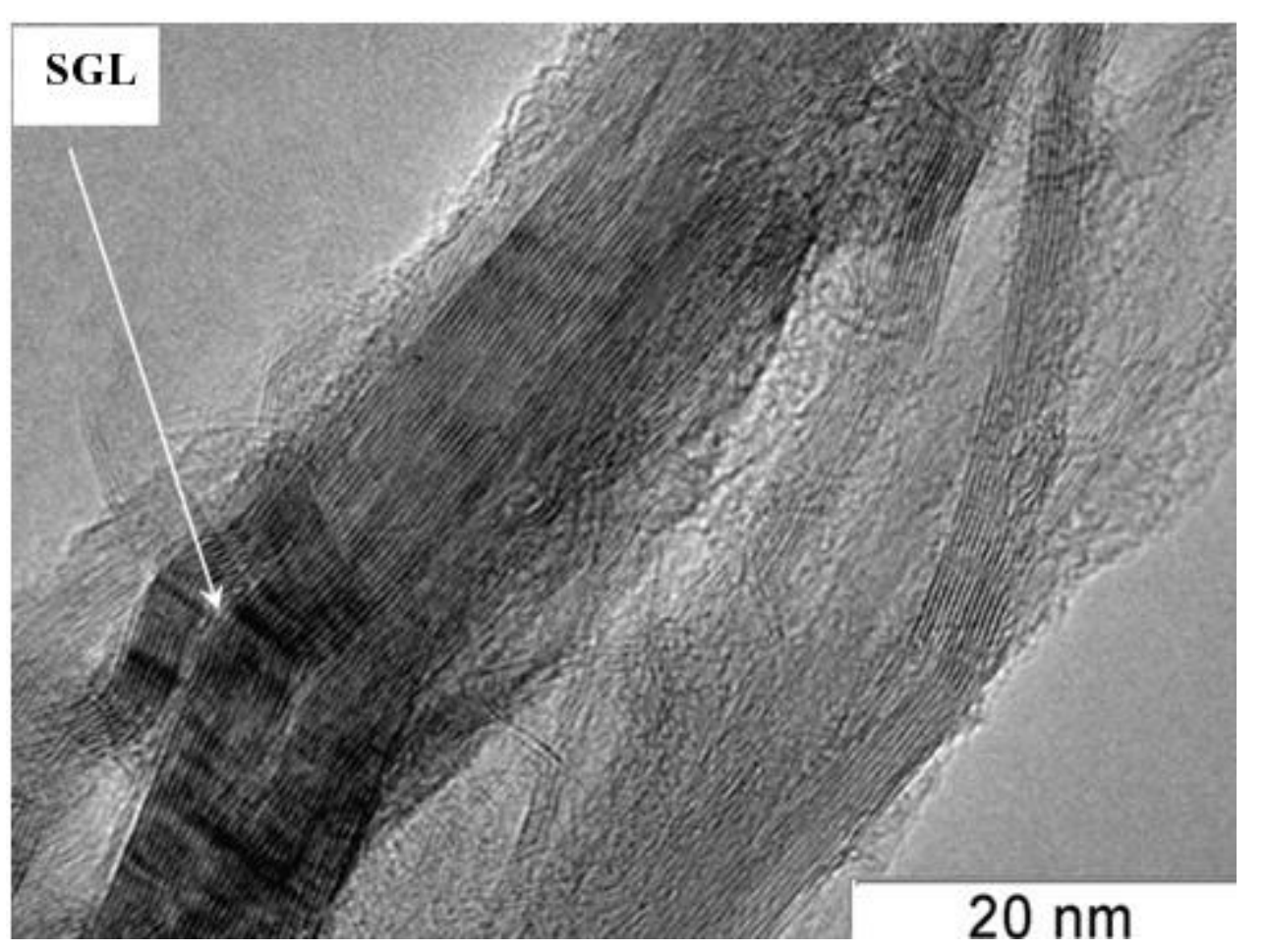

**Рис.2.5.** ВРЭМ - микрофотография углеродного композита марки SGL. Стрелкой указан так называемый «дефект расщепления» [78].

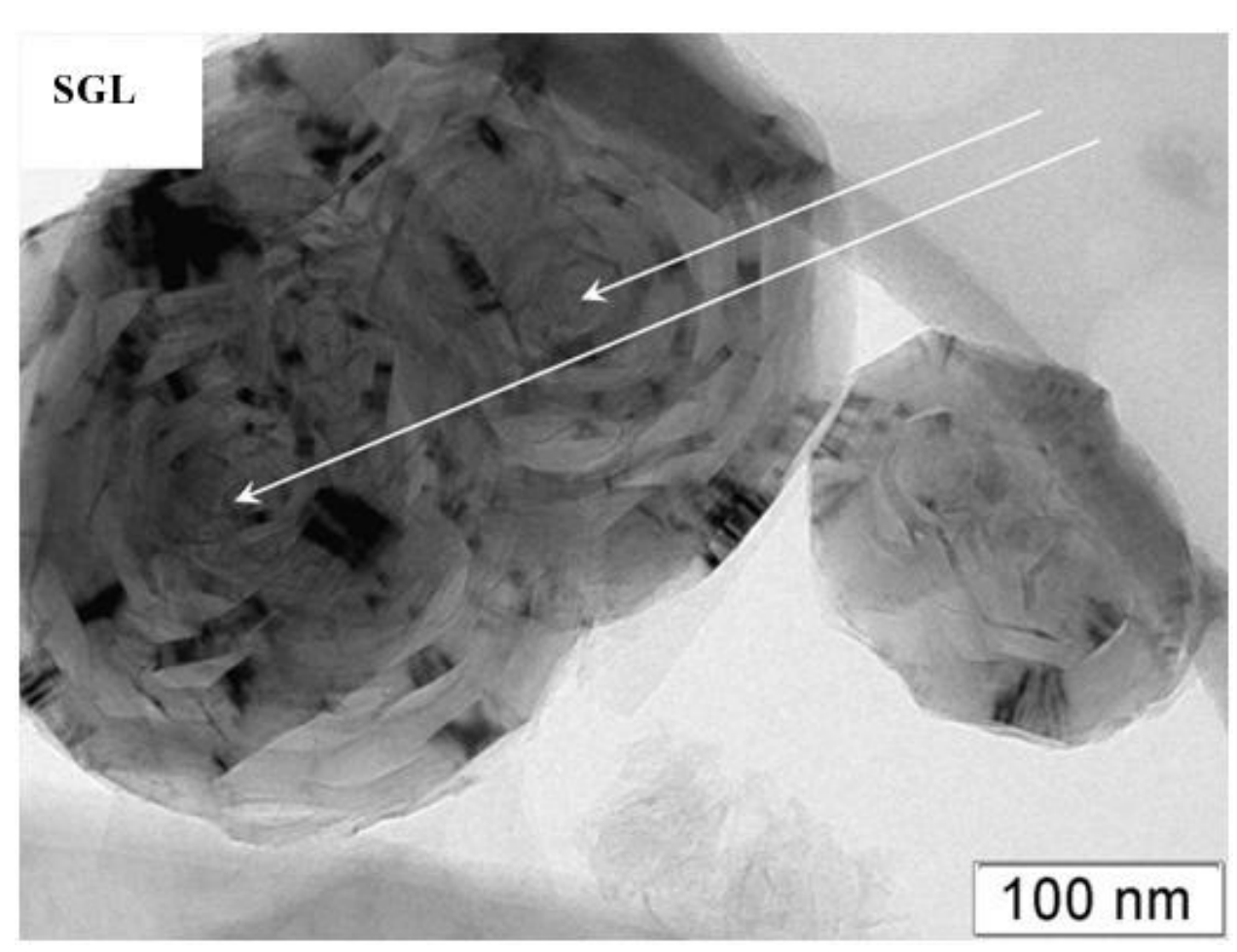

**Рис.2.6.** ВРЭМ-микрофотография образца углеродного композита марки SGL. Стрелками указаны дефекты, связанный с появлением гетерогенно-графитированых областей [78].

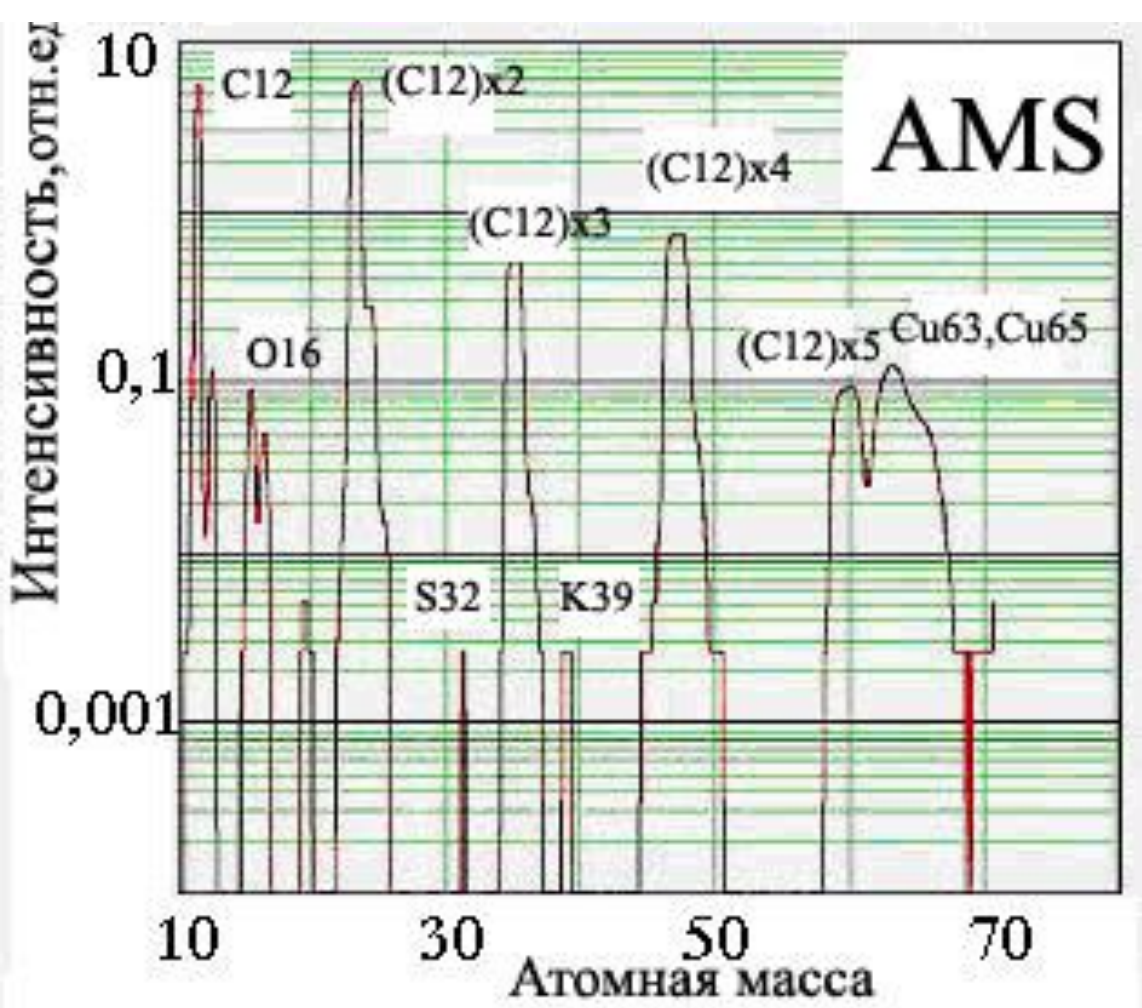

**Рис.2.7.** Результаты измерения состава примесей в графите марки SGL методом ускорительной масс-спектрометрии [74].

**Графит марки LeCL** представляет собой плотный, хорошо окристаллизованный графит без видимых примесей. Агрегаты частиц имеют размеры от 500 нм до нескольких микрон. Толщина отдельных графитовых слоёв варьируется в достаточно широких пределах: наблюдаются как слои состоящие из нескольких графеновых листов, так и слои графита толщиной 100нм и более. При этом на снимках высокого разрешения проявляется большое количество дефектов различного типа.

Так, например, на поверхности графитовых слоев наблюдаются фуллерено-подобные образования с размерами от 1 до 3 нм (рис. 2.8, слева). Слои углерода с краёв агрегатов и в местах изгиба имеют ступенчатую структуру (справа), причем края ступеней образованы замкнутыми дугообразными графеновыми слоями (рис.2.9, слева). Такие образования можно обозначить как закрученные террасы (*curled terraces)*. Также наблюдаются скрученные слои (*twisted layers*) графита толщиной 3–5 нм (рис.2.9 справа).

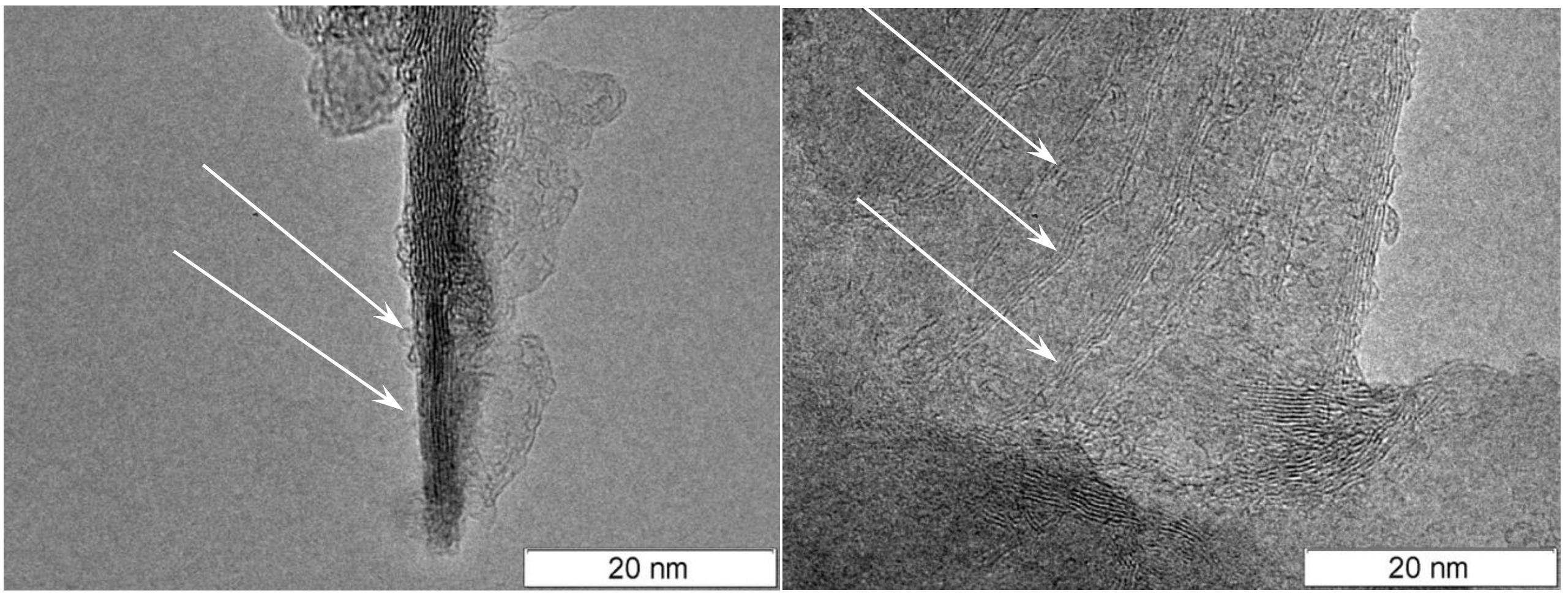

**Рис.2.8.** ВРЭМ-микрофотография графитового композита марки *LeCL*: *слева* - фуллерено-подобные образования с размерами от 1 до 3 нм; *справа* – ступенчатая структура слоёв с краёв агрегатов и в местах изгиба (показано стрелками) [78].

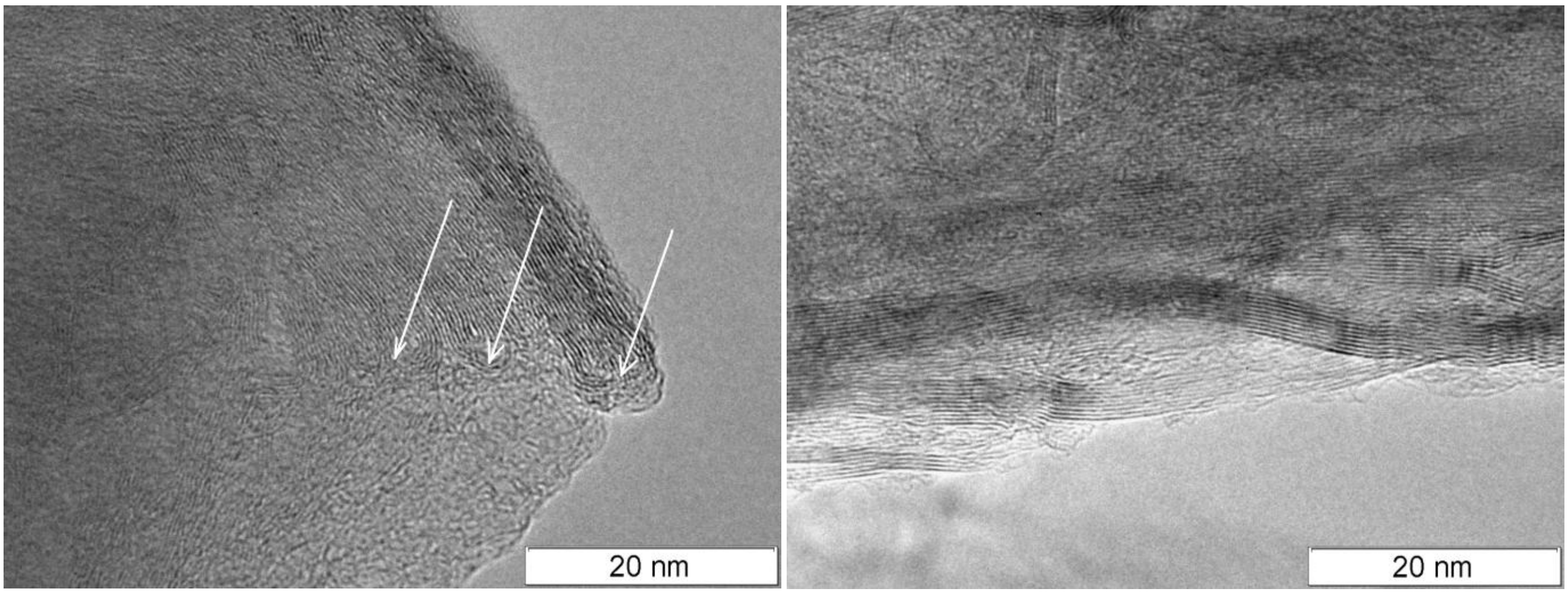

**Рис.2.9.** ВРЭМ-микрофотография графита марки *LeCL*: *слева* – террасы (указаны стрелками); *справа* - скрученные слои [78].

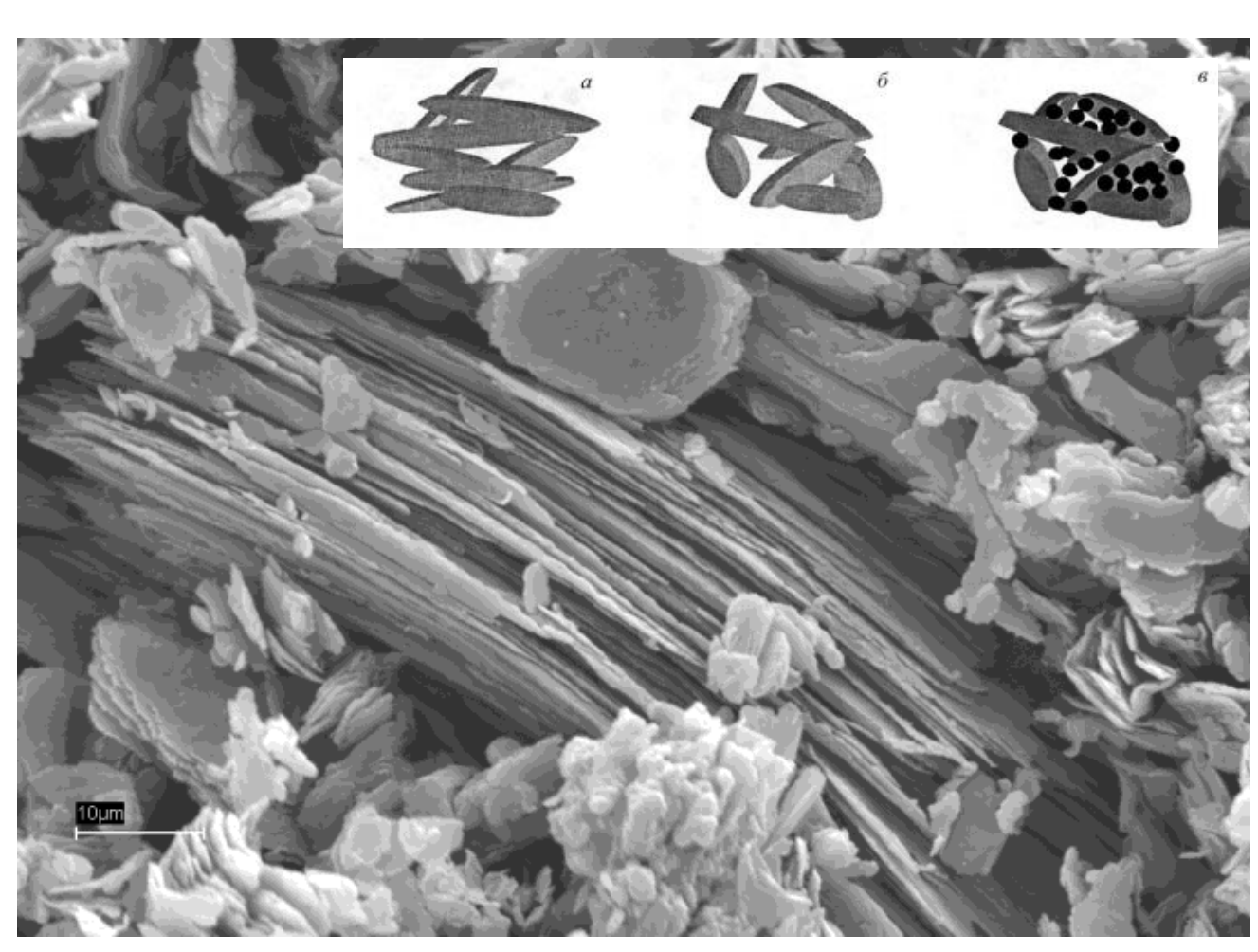

**Рис. 2.10.** Электронные растровые микрофотографии поверхности образца марки SGL. Снимок выполнен в зоне разрушения образца в режиме «вторичных электронов». На поверхность образца для улучшения разрешения напылялось золото толщиной до 100 Å. Прогрев осуществлялся переменным током до разрушения [78]. На вставке: упаковка частиц анизометрического наполнителя при разных способах прессования: *а* – одноосное прессование в матрицу; *б* – изостатическое прессование; *в* – изостатическое прессование комбинированного наполнителя [34].

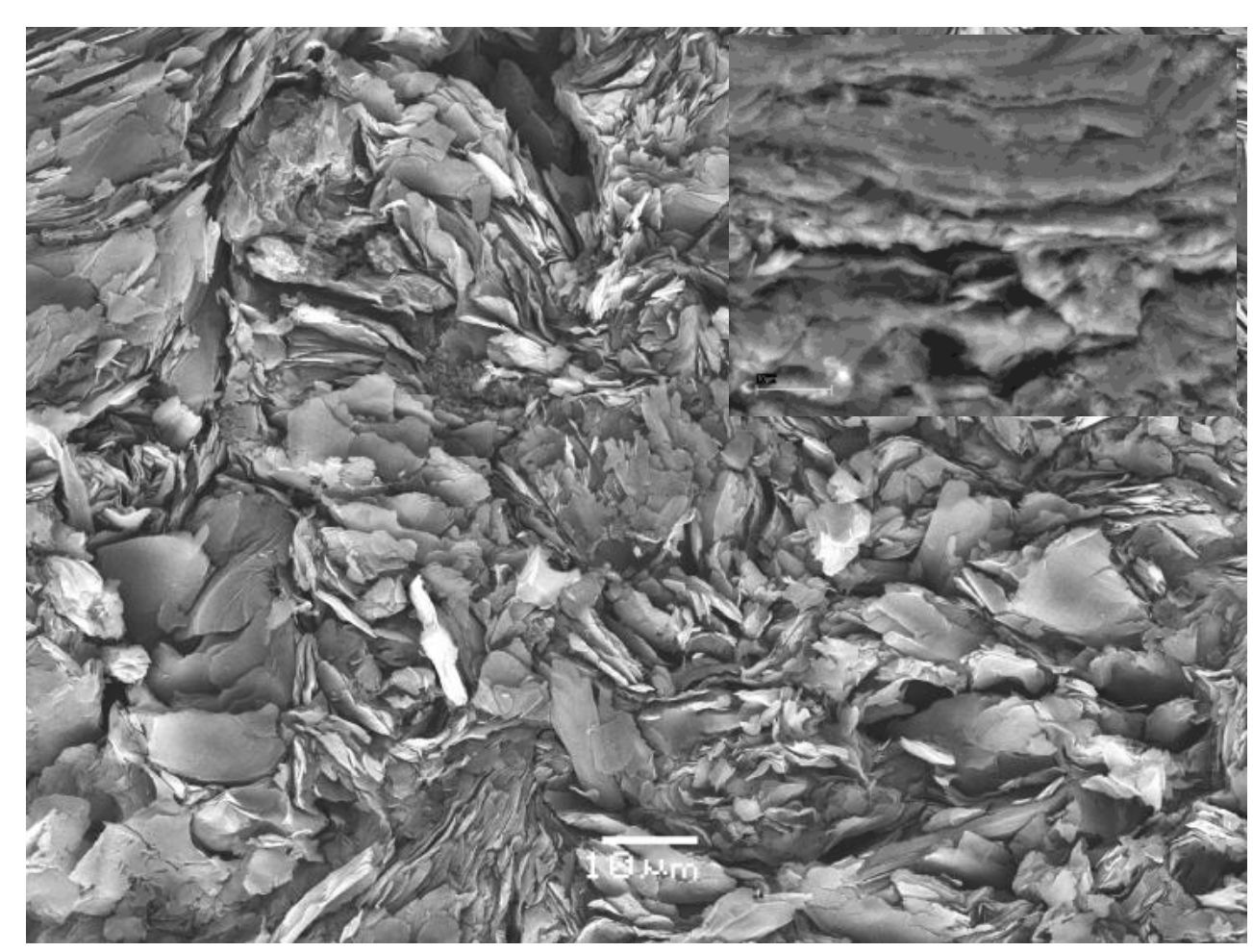

**Рис. 2.11.** Электронные растровые микрофотографии поверхности образца марки SGL Снимок выполнен немного в стороне от зоны разрушения образца в режиме «вторичных электронов» Вставка вверху слева – исходная поверхность образца, снимок выполнен в *backscattering* моде [78].

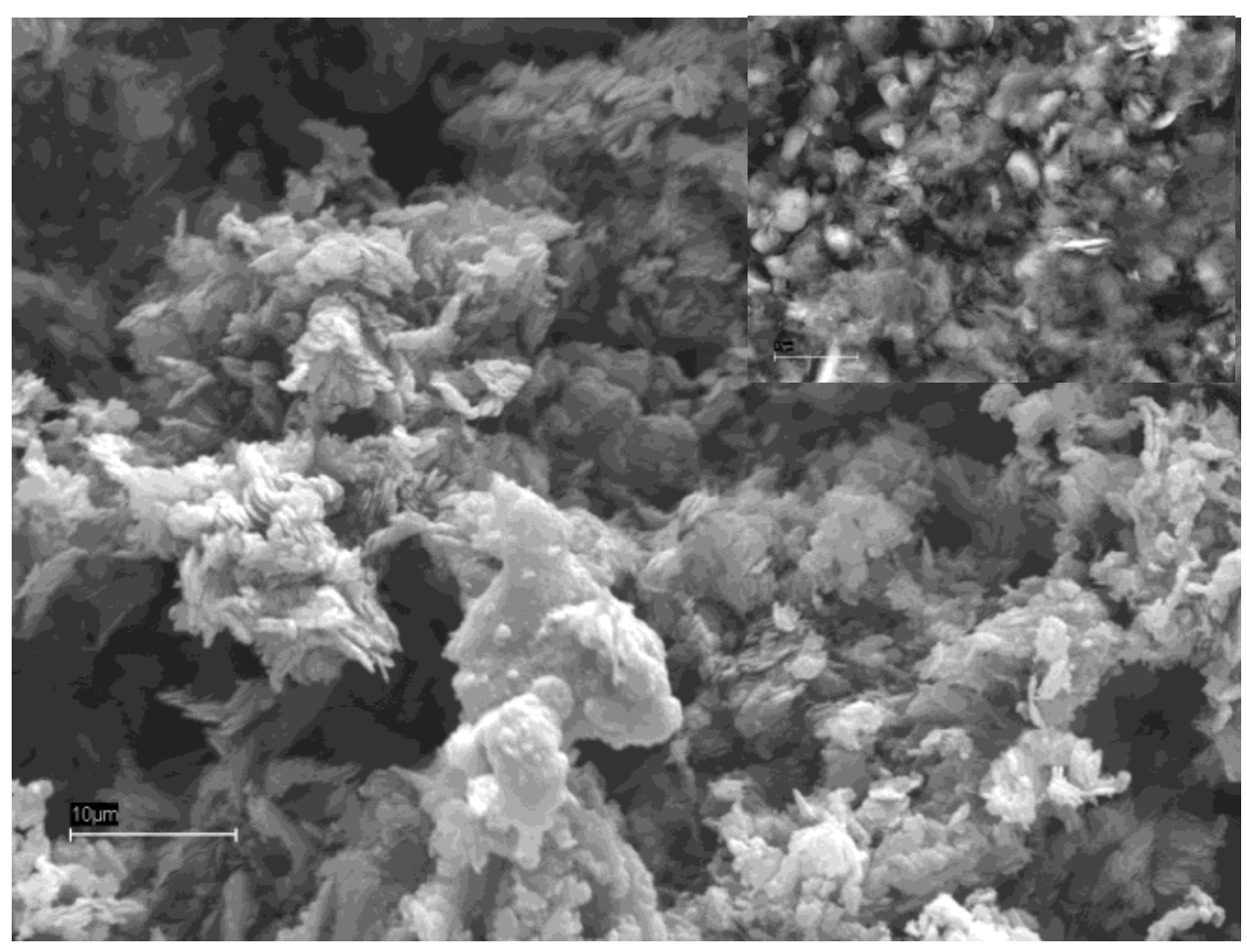

**Рис. 2.12.** Электронные растровые микрофотографии поверхности образца марки МПГ-6. Снимок выполнен в зоне разрушения образца в режиме «вторичных электронов», использовалось напыление золотом до 100Å. На вставке – исходная поверхность образца, режим *backscattering* [78].

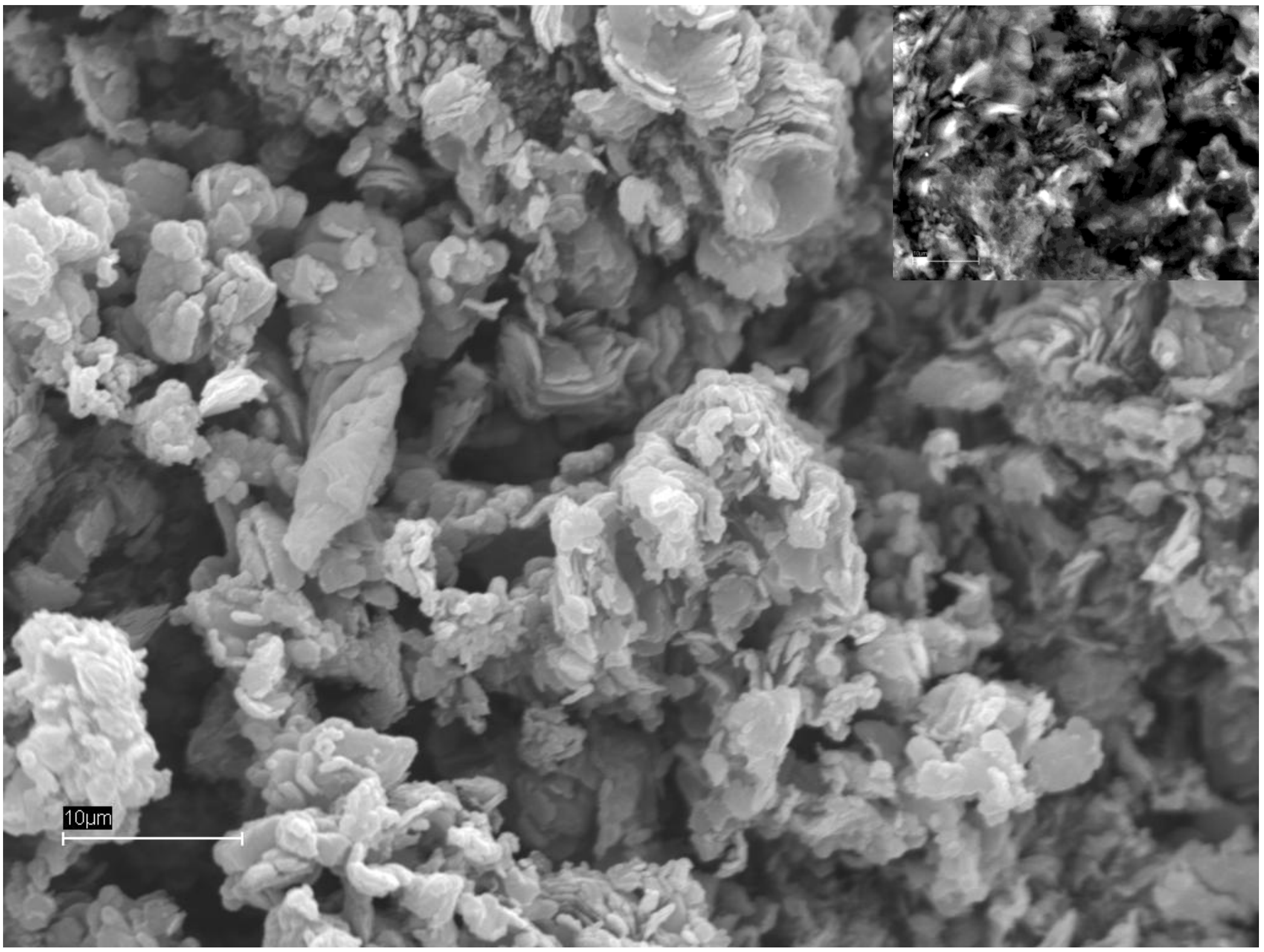

**Рис. 2.13.** Электронные растровые микрофотографии поверхности образца марки LeCL. Снимок выполнен в зоне разрушения образца в режиме вторичных электронов, использовалось напыление золотом до 100Å. На вставке – исходная поверхность образца, снимок выполнен в режиме *backscattering* [78].

**Графит марки SGL** по данным высокоразрешающей электронной растровой микроскопии отличается заметной анизотропией структуры (рис. 2.10) и (рис. 2.11). Эта анизотропия предположительно связана с использованием игольчатого кокса с размерами зерна наполнителя 20–30 мкм в качестве базового материала графитового композита. При измельчении этого кокса образуются частицы с высокой анизометричностью, и при прессовании порошка в матрицу можно получить искусственный графит с высокой плотностью, но в сочетании с очень высокой и нежелательной анизотропией физико-механических свойств [34]. Прямым подтверждением такой анизотропии является рентгенофазная диаграмма для графитового композита марки SGL (рис. 2.1). На этой рентгенограмме высота рефлекса 002 заметно больше для не растёртого в порошок образца. Текстура для пластины образца *SGL* связана с преимущественной ориентацией кристаллитов в направлении 00l. Для графитов марки LeCL и МПГ-6 рентгенограмма говорит о практически изотропном характере мезоструктуры этих образцов, что подтверждается измерениями, выполненными на высокоразрешающем растровом микроскопе (рис. 2. 12) и (2. 13).

Для образцов LeCL и МПГ-6 в отличие от графита марки SGL хорошо просматривается развитая, пористая структура материала, достаточно заметны также поворотные моды деформации.

## 2.4. Влияние мезоструктуры на прочность и долговечность графитового композита

Предположение о связи мезоструктуры образца с энергией активации разрушения подтверждается данными измерений времени разрушения образцов при различных температурах (рис. 2.14). Согласно [2] высокотемпературная область (2500÷3000°С) ползучести поликристаллического графита, объясняемая большинством авторов самодиффузией углерода в графите, зависит от явления сублимации графита. Тем более, что энергия активации процесса ползучести для этих температур, находящаяся, по Мартенсу (H.E. Martens), в пределах 720÷1130 кДж/моль, близка к энергии сублимации графита $\Delta H_{субл}$ ~716,7 кДж/моль, при этом энергия активации самодиффузии может быть оценена как $680^{\pm 50}$ кДж/моль. Очевидно, что именно сублимация является источником вакансий в кристаллитах, т.е. первопричиной вакансионной диффузии, а, следовательно, и диффузионного механизма ползучести при высоких температурах.

Следует отметить также, что явление ползучести в графите может быть связано с движением краевых и винтовых дислокаций. В то же время, в области температур до 2500°С вполне приемлемое объяснение ползучести поликристаллического графита, сделано S. Mrozowski и J.E.Hove [2, стр.112].

Они объясняют механизм ползучести разрывом периферийных связей С-С и скольжением кристаллитов друг относительно друга.

Высокоупорядоченный графит характеризуется энтальпией плавления $\Delta$*Нпл* равной ~104 кДж/Моль, энтальпией сгорания $\Delta$*Нсгор* равной ~395 кДж/Моль [5]. Последняя примерно соответствует энергии σ-связи для графита, согласно [75] равной 418,7÷460,6 кДж/Моль. Начальная энергия активации разрушения, которая может быть получена из данных работы [63] по времени разрушения графита *МПГ-6* составляет величину около 890 кДж/Моль.

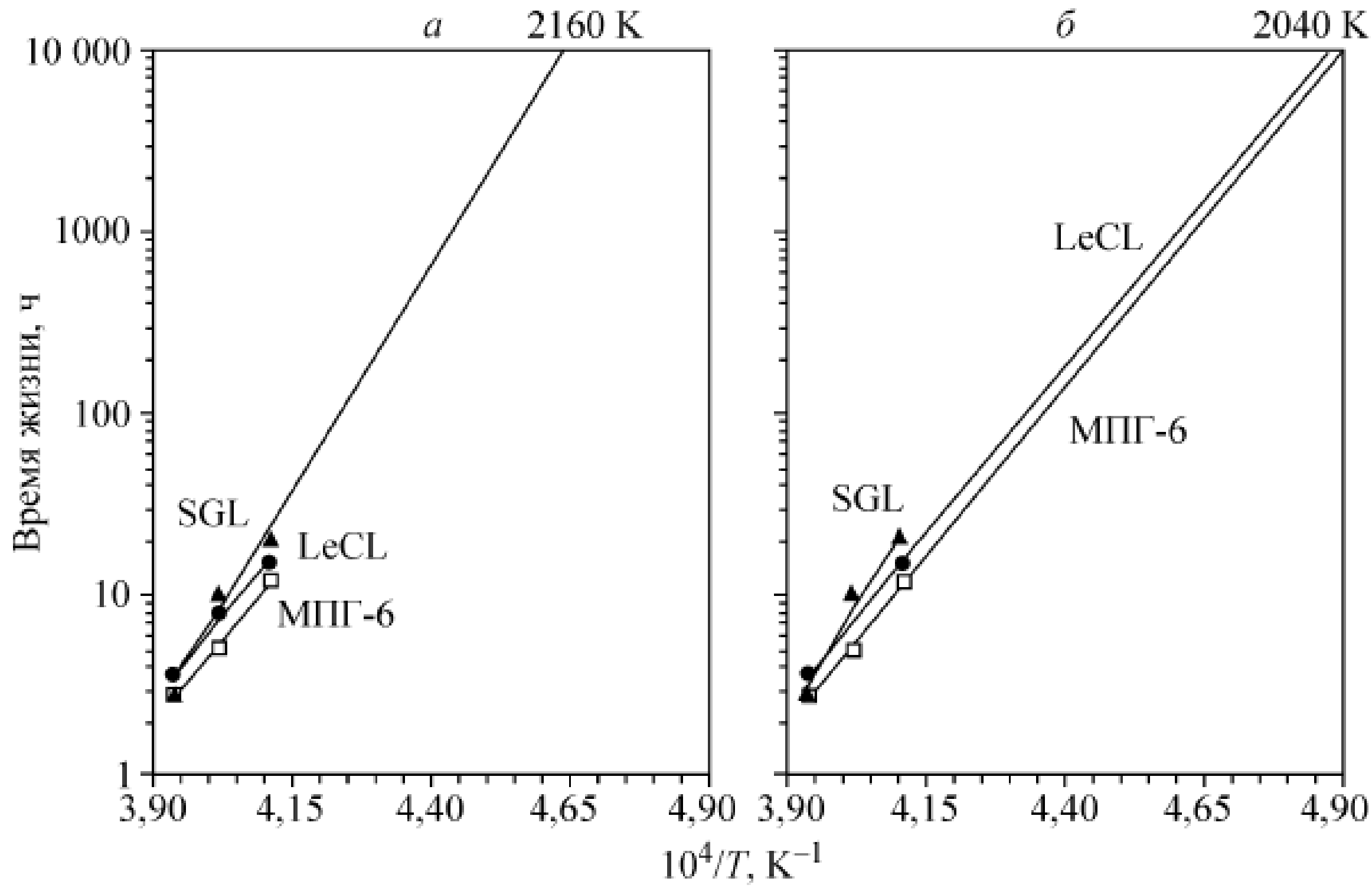

**Рис. 2.14.** Зависимость времени жизни образцов *SGL*, *МПГ-6* и *LeCL* от обратной температуры: *слева* – SGL, *справа* – LeCL и МПГ [66].

Эта энергия заметно превышает величину энтальпии возгонки (сублимации), и может соответствовать, например, энергии ползучести, которая по некоторым данным [2, стр.128] может достигать величины ~1200 кДж/Моль. Но более аккуратная постановка эксперимента с применением пирометра IMPAC IS-12 взамен устаревшего ОПИР-09 показывает заметно меньшую энергию активации процесса разрушения для графитового композита МПГ-6 (а также композита LeCL) равную ~$690^{\pm 60}$ кДж/Моль. Эта величина весьма близка к энергии сублимации графита 716,7кДж/моль [5] либо, что более вероятно, к энергии активации самодиффузии углерода в графите (рис. 2.14).

Для графитового композита марки *SGL* начальная энергия активации разрушения оказывается более высокой и равной $\Delta H$~1000÷1100кДж/Моль. Это заметно больше, чем для образцов марки LeCL или МПГ-6. Последнее обстоятельство позволяет предположить несколько иной механизм разрушения, связанный с ярко выраженной анизотропией данного материала, а также с некоторыми особенностями проявления ползучести графита при повышенных температурах. Так, в частности, в [2, стр.137] утверждается, что «изучение прочностных свойств и модуля упругости для различных марок графита совместно с предварительными измерениями на нитриде бора наводит на мысль о справедливости для условий высоких температур гипотезы S. Mrozowski - J.E.Hove, касающейся влияния кристаллической анизотропии на температурную зависимость прочности и упругости».

Возрастание прочности графита с ростом температуры до ~ $2500^{o}$ С и его модуля упругости согласно гипотезе S. Mrozowski можно объяснить влиянием двух факторов:

- анизотропией отдельных кристаллитов графита

- полимерной природой межкристаллитной валентной связи.

Согласно [2, стр.112] можно предположить, что значительное расширение кристаллитов в направление оси ***c*** приводит к заполнению межкристаллитных пустот и это сжатие зёрен делает структуру жёсткой. Очевидный недостаток такого объяснения связан с тем, что для такого, близкого к графиту по структурным свойствам материалу, как нитрид бора, возрастания предела прочности с температурой не происходит. В противоположность прочностной гипотезе, H.E. Martens и др. предположили, что возрастание пластичности графита при высоких температурах уменьшает концентрации напряжений в результате пластических деформаций. Последнее хорошо согласуется с идеологией физической мезомеханики, где аккомодационные процессы происходят в результате сдвиговых и поворотных мод деформации [44].

Так или иначе, несомненным представляется влияние поликристаллической, полимерной структуры графита на его прочностные свойства. В частности, согласно [2, стр.115]: «…Кристаллические полимеры состоят из чередующихся кристаллических и аморфных областей с отдельными полимерными цепочками, проникающими через следующие друг за другом кристаллические области (кристаллиты) и аморфные области. В некоторых случаях кристаллиты ориентированы вдоль определённых направлений или плоскостей; в других случаях ориентация кристаллитов случайная».

Схематическое изображение структуры поликристаллического полимера представлено на рис. 2.15 а. Для более совершенных кристаллических полимеров можно полагать [2, стр.115], что края соседних кристаллитов могут соединяться, чтобы «…сцепиться вторичными валентными силами той же природы, что и силы, удерживающие вместе цепочки в параллельных плоскостях кристаллитов.». Речь, как можно понять, может идти только **σ**-связях, благодаря которым образуются графеновые плоскости.

Итак, «…углеродный материал представляет собой по существу пространственную сетку из кристаллитов, жёстко скреплённых сложной системой пересекающихся валентных связей, которые стабилизируют структуру на такое расстояние, что параллельные перемещения плоскостей в кристаллитах могут происходить только под действием сил, способных разорвать краевые С-С связи».

Схематическое изображение структуры графита по S. Mrozowski и J.G. Castle показано на (рис. 2.15 б, в). Как более отчётливо показано на схеме J.G. Castle, молекулярная цепочка сцепленных слоёв может проходить через различные кристаллиты, будучи закреплена в каждом пересечении.

Присутствие ориентированных кристаллитов в «сшитых» полимерах оказывает влияние на их механические свойства в определённых направлениях, и, как понятно из рис. 2.15, модуль упругости ориентированного кристаллического полимера может быть больше в направлении, параллельном цепочкам.

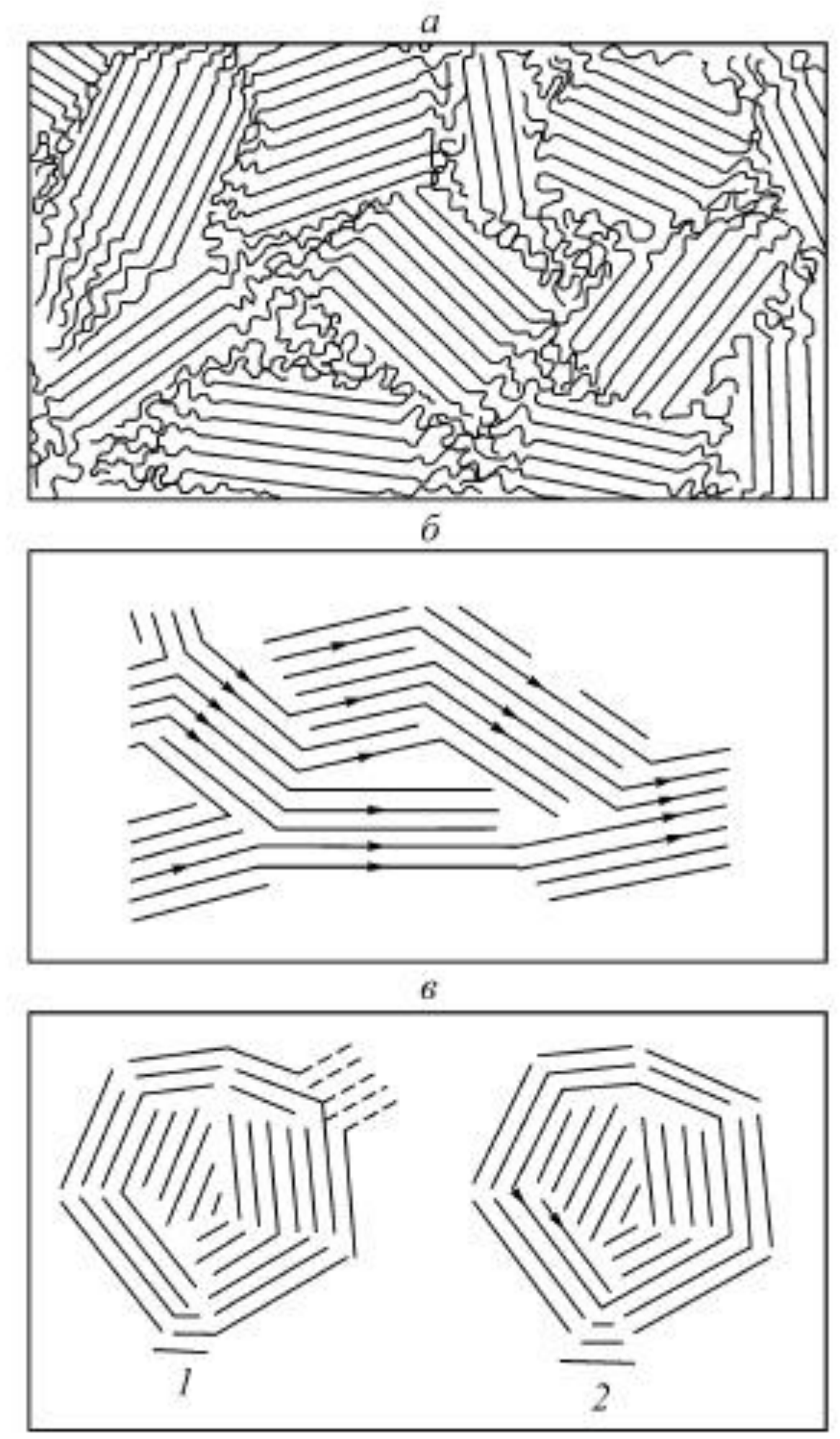

**Рис. 2.15.** Сравнение схематических изображений структуры графита и полимера согласно [2, с.115] а) - морфология полукристаллического полимера по P.J. Flory; б) – схематическое расположение кристаллитов в поликристаллическом графите по J.G. Castle; в) – микрокристаллы углерода, связанные валентными связями: расположение в виде плотно упакованного кольца (1); расположение в виде неупорядоченного кольца (2).

Энергия активации ползучести в направлении, перпендикулярном и параллельном к преимущественной ориентации зёрен, как было показано при высокотемпературных испытаниях на ползучесть, при растяжении и кручении для графита марки H4LM могут отличаться в два раза [2, стр.131]. Эти данные говорят о том, что механизм ползучести в области малых деформаций включает в себя сдвиг параллельно плоскости базиса кристаллитов одной пачки относительно другой с ограниченной скоростью скольжения, при которой периферийные связи С-С разрываются и восстанавливаются. В этих испытаниях было получено, что среднее значение энергии активации ползучести составляет 348 кДж/моль, что очень близко к энергии **σ –**связи углерода в графите, равной 357,4 кДж/моль.

В то же самое время, сравнение средней энергии активации ползучести со значением $683^{\pm 50}$кДж/моль для объёмной самодиффузии графита наводит на мысль, что скорость ползучести при достаточно высоких температурах контролируется диффузионными процессами, которые включают в себя движение атомов углерода на границе зёрен или кристаллитов.

## 2.5. Межкристаллитная фаза графитового композита по данным рентгенофазного анализа и электрофизических измерений

Согласно данным рентгенофазного анализа [76 - 78] размеры кристаллитов (точнее, размеры ОКР) остаются неизменными вплоть до самых высоких температур отжига образцов графита типа МПГ. Порошковый рентгенофазный анализ образцов, доведённых в результате высокотемпературного прогрева или электронного облучения до разрушения, выявляет в точности те же самые кристаллографические параметры, что и в исходном образце.

На рис. 2.16, *а* представлена микрофотография образца МПГ-6, на которой видно, что он состоит из агрегатов большого размера (свыше 1000 нм), образованных, в свою очередь, тонкими ограненными пластинками в виде неправильных многогранников [76]. Отдельная типичная пластинка с размером в поперечном направлении около 500 нм представлена на рис. 2.16, *б*.

Картина микродифракции электронов, приведенная на врезке к рис. 2.16 *а*, является точечной, что свидетельствует о монокристаллическом характере структуры отдельной пластинки, составляющей агрегат, а ее гексагональная симметрия указывает на то, что развитой плоскостью пластины является плоскость (111) графита.

Вполне логично данные электрофизических измерений образцов графитовых композитов типа МПГ после высокотемпературных отжигов связать именно с изменениями параметров межкристаллитной границы (рис. 2.17).

Сдвиг кривых температурной зависимости проводимости (рис. 2.17) можно пояснить, если в соответствие с [79, стр.65] согласиться, что проводимость поликристаллического материала складывается из проводимости кристаллитов и проводимости межкристаллитной границы:

$$\sigma(T) = \sigma_к(T) + \Delta\sigma_0 . \qquad (2.1)$$

Слагаемое $\sigma_к(T)$ представляет собой температурную зависимость проводимости отдельного кристаллита. Для мелкозернистых плотных графитов класса *МПГ* эта зависимость хорошо укладывается в рамки теории, развитой Котосоновым А.С. для квазидвумерных графитов [30].

Определяющим для классификации графитового материала как квазидвумерного графита (КДГ) является увеличенное (>0,342 нм) межслоевое расстояние и нарушенный азимутальный порядок. Увеличенное межслоевое расстояние ***c*** согласно [30] позволяет пренебречь межслоевыми взаимодействиями и использовать в расчётах для проводимости зонную модель двумерного графита. Для *МПГ-6* параметр ***c*** близок к 0,34 нм и расчет температурной зависимости проводимости по формулам [30] дает хорошее согласие с экспериментом [76, 77] - (*частное сообщение профессора Котосонова А.С.*).

Поскольку общая проводимость графита складывается из проводимостей отдельных кристаллитов, то в работе [30] вводится текстурный фактор ***F***, сложным образом зависящий от анизотропии композита, от проводимости отдельных кристаллитов, а также от взаимной пространственной ориентации соседних кристаллитов и макротекстуры образца в целом. Кроме того, в [30] вводится также коэффициент связности ***K***, учитывающий пористость образца.

Как показывает рис. 2.17, в результате отжига сдвиг кривой зависимости сопротивления от температуры происходит в сторону уменьшения сопротивления образца. Такой характер сдвига температурной зависимости характер для перколяционных систем при превышении порога перколяции [80]. В общем случае, выше порога протекания кристаллиты в графитовом композите будут образовывать своеобразную «сетку», по которой осуществляется проводимость. Величина этой проводимости будет определяться ветвистой неупорядоченной структурой кристаллитов (см. рис. 2.18), а формирование токопроводящей сети выше порога перколяции можно проиллюстрировать на примере углеродной плёнки, содержащей графитоподобные наноразмерные кристаллиты [81].

Величина электрического сопротивления межкристаллитной фазы может быть достаточно высокой по той причине, что безотносительно к структуре самой этой фазы контакты между соседними кристаллитами могут либо прерываться, либо быть энергетическими барьерами для носителей заряда [30]. Тогда температурно-независимый сдвиг величины проводимости $\boldsymbol{\Delta\sigma_0}$ возникает естественным образом как следствие изменения высоты и ширины энергетического барьера, связанного с межкристаллитной фазой.

В приближении аппроксимации эффективной среды согласно [82] этот вклад в макроскопическую проводимость образца может быть с точностью до множителя оценён в виде:

$$1/\Delta\sigma = \rho_0\, \boldsymbol{a}\, \boldsymbol{u_0}^{\nu} \exp(\boldsymbol{u}_c), \qquad (2.2)$$

где $\boldsymbol{a}$ – ширина энергетического барьера; а $\boldsymbol{u_0}$, $\boldsymbol{u_c}$ представляют собой предельную и критическую высоту энергетического барьера. Выражение

(2.2) получено в [82] в предположении равномерного распределения высоты барьеров:

$$F(u) = \begin{cases} 1/u_0 & \text{при} \quad u \le u_0 \\ 0 & \text{при} \quad u > u_0 \end{cases} \qquad (2.3)$$

Здесь $\nu$ является так называемым критическим индексом, значение которого зависит только от размерности решетки и в случае трехмерной случайной решетки $\nu = 0.875$.

Можно предположить, что как облучение электронами, так и прогрев образца переменным током может приводить к уменьшению высоты и ширины барьеров и, соответственно, к увеличению общей проводимости образца. Сдвиг $\Delta\sigma_0$ может быть связан с изменениями в межкристаллитной фазе графитового композита, эти изменения могут инициироваться, например, увеличением концентрации носителей, которое, в свою очередь, может быть связано с ростом числа дефектов в межкристаллитной фазе. Последнее обстоятельство прямо подтверждается измерениями относительного магнетосопротивления при гелиевых температурах (рис. 2.19).

В данном случае, мы наблюдаем классическое положительное магнетосопротивление, связанное с увеличением протяжённости траектории движения электрона (или дырки) из-за её закручивания в скрещенных электрических и магнитных полях. Отжиг уменьшает величину этого относительного магнетосопротивления тем больше, чем больше температура или время термообработки.

Необходимо отметить, что механизм возникновения температурно-независимой прибавки к проводимости $\Delta\sigma_0$ в формуле (2.1) может быть заметно более сложным, чем описано выше, поскольку рост дефектности межкристаллитной границы в результате термообработки будет увеличивать число центров для неупругого резонансного туннелирования [80].

## 2.6. Интеркристаллитное разрушение

Согласно современным представлениям, прогноз времени жизни нанокомпозитов основан [47, стр.799] на некоторых соотношениях между прочностью кристаллитов и прочностью межкристаллитной фазы. Согласно [47] фрактографическое исследование поверхностей хрупкого разрушения нанокомпозитов выявили доминирующую роль механизма интеркристаллитного разрушения, показано также, что в случае металлических нанокомпозитов прочность при растяжении выше прочности крупнозернистых материалов в 1,5 – 8 раз. Роль границы раздела в данном случае связана с особенностями разрушения гетерофазных материалов. Поверхность раздела двух фаз, в данном случае, границы зёрен, являются препятствием на пути распространения дислокаций и трещин, что и предопределяет повышение прочности и твёрдости нано- и мелкодисперсных материалов, если нет искажающих факторов. Можно полагать, что это правило вполне применимо и для случая графитовых композитов. Так, в работе [34] показано, что предел прочности на сжатие тонкозернистых графитов из мезофазных пеков может быть выше предела прочности среднезернистых графитов из непрокалённого кокса в два раза, предел прочности на изгиб выше почти в три раза.

В мелкодисперсных и нанокристаллических материалах (НМ) кинетические условия образования трещин становятся сопоставимыми [79], что может приводить к существенной зависимости материала от размера зерна.

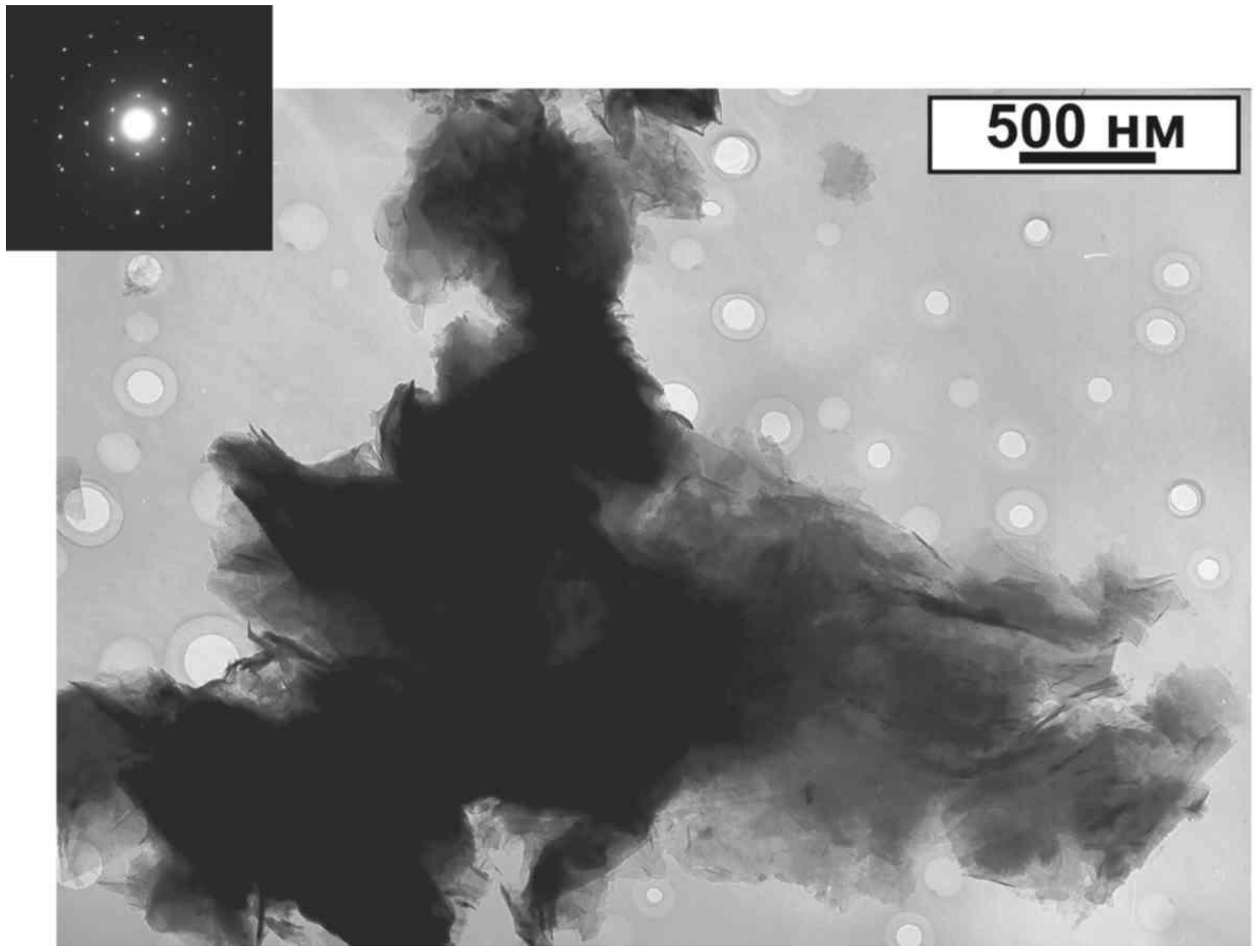

а)

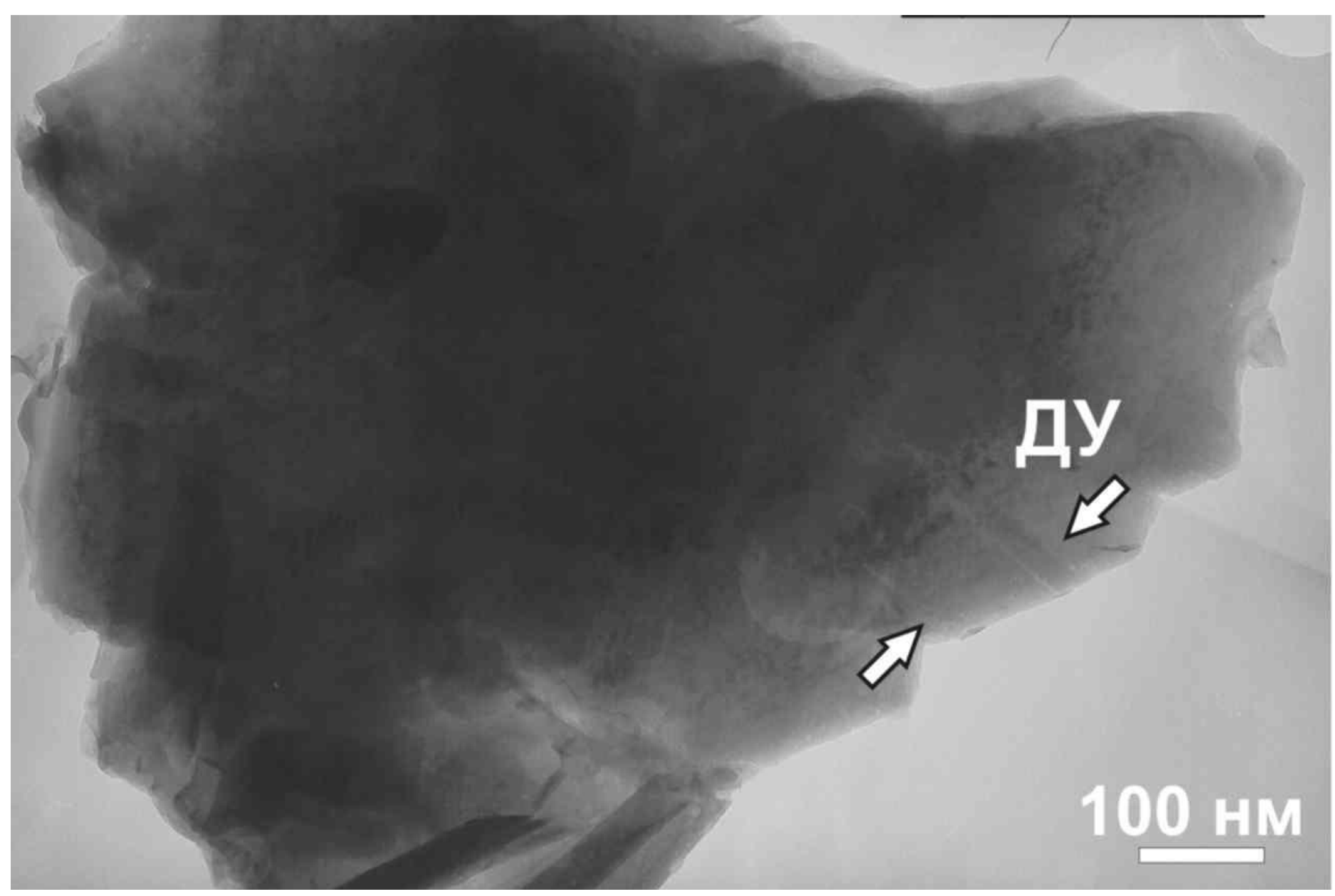

б)

**Рис. 2.16.** Микрофотографии и микродифракционная картина образца *МПГ-6* [76].
На вставке – микродифракционная картина образца МПГ-6. Стрелками показаны дефекты упаковки.

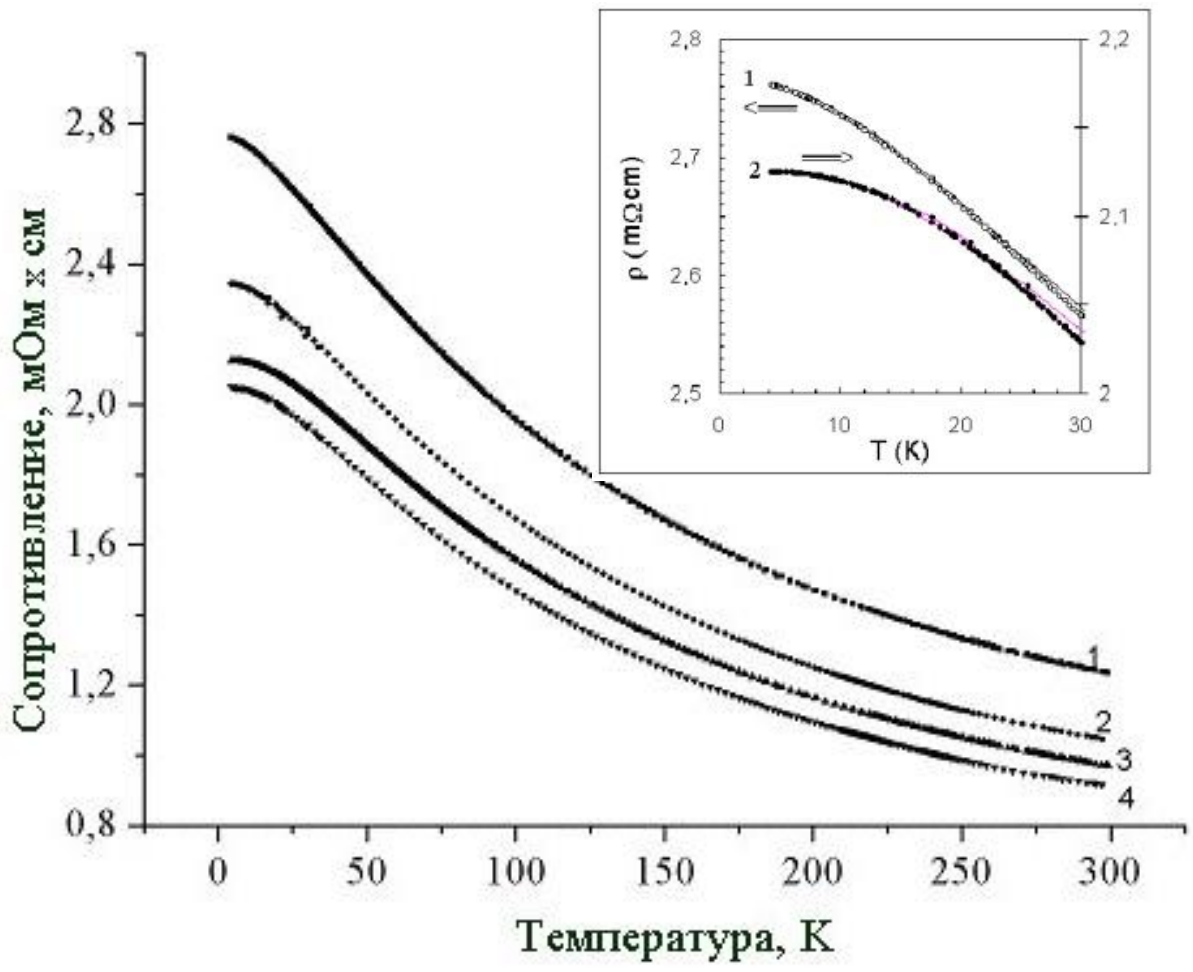

**Рис. 2.17.** Измерения удельного электросопротивления графитовых композитов MPG-7 до и после термообработки [76]:1 - исходный образец; 2 - время прогрева 1 час 30 минут при 2250$^{O}$C; 3 - время прогрева 1 час 20 минут при 2450$^{O}$C; 4 - время прогрева 50 часов при 2250$^{O}$C.

На вставке: температурная зависимость в области низких температур электросопротивления ρ(T) исходного (1) и прогретого при T =2450$^{O}$C в течении 80 мин (2) образцов МПГ-7.

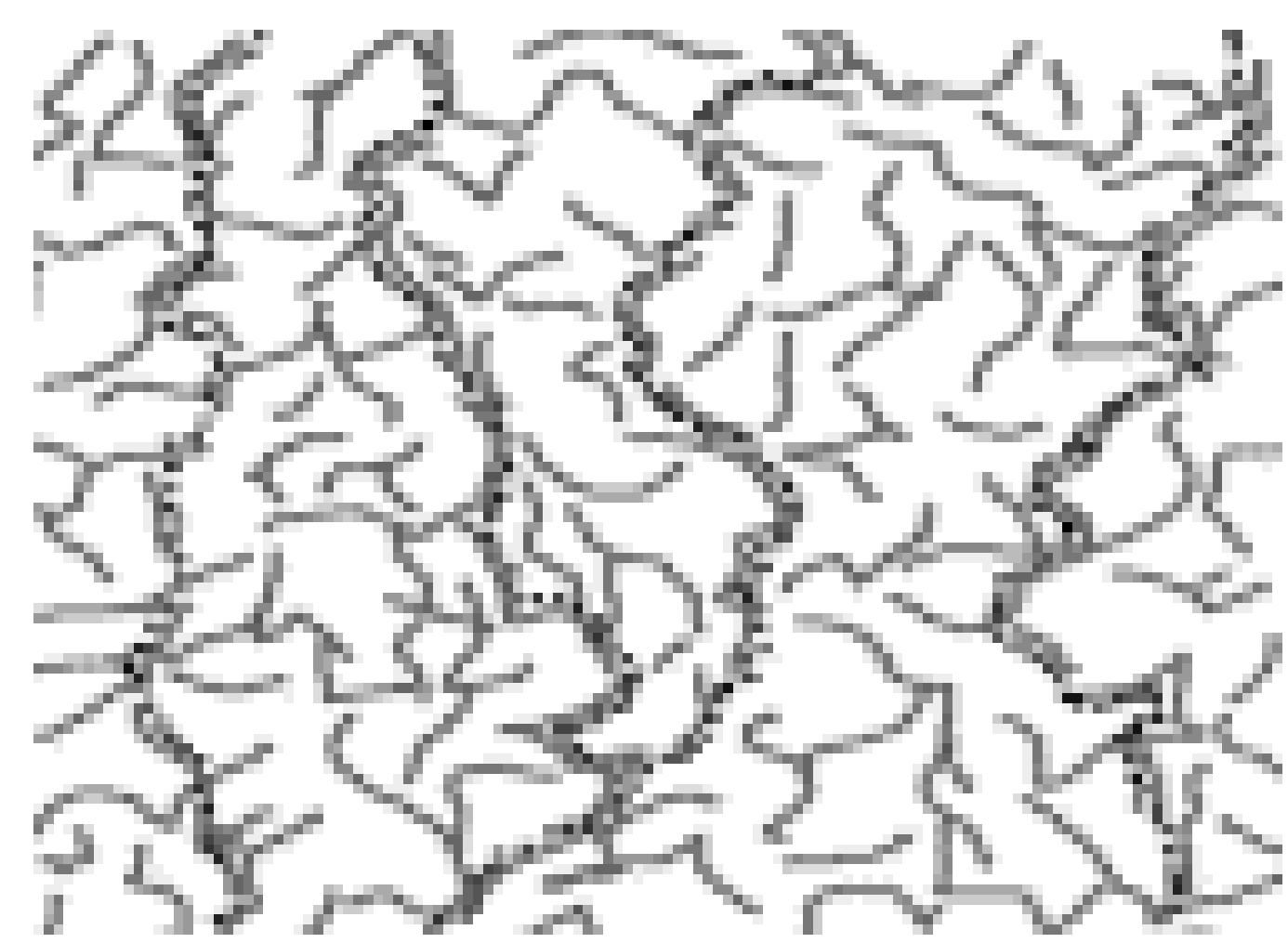

**Рис. 2.18.** Схема формирования двумерных токопроводящих путей (выделены пунктиром) на углеродных плёнках, содержащих графитоподобные кристаллиты [81].

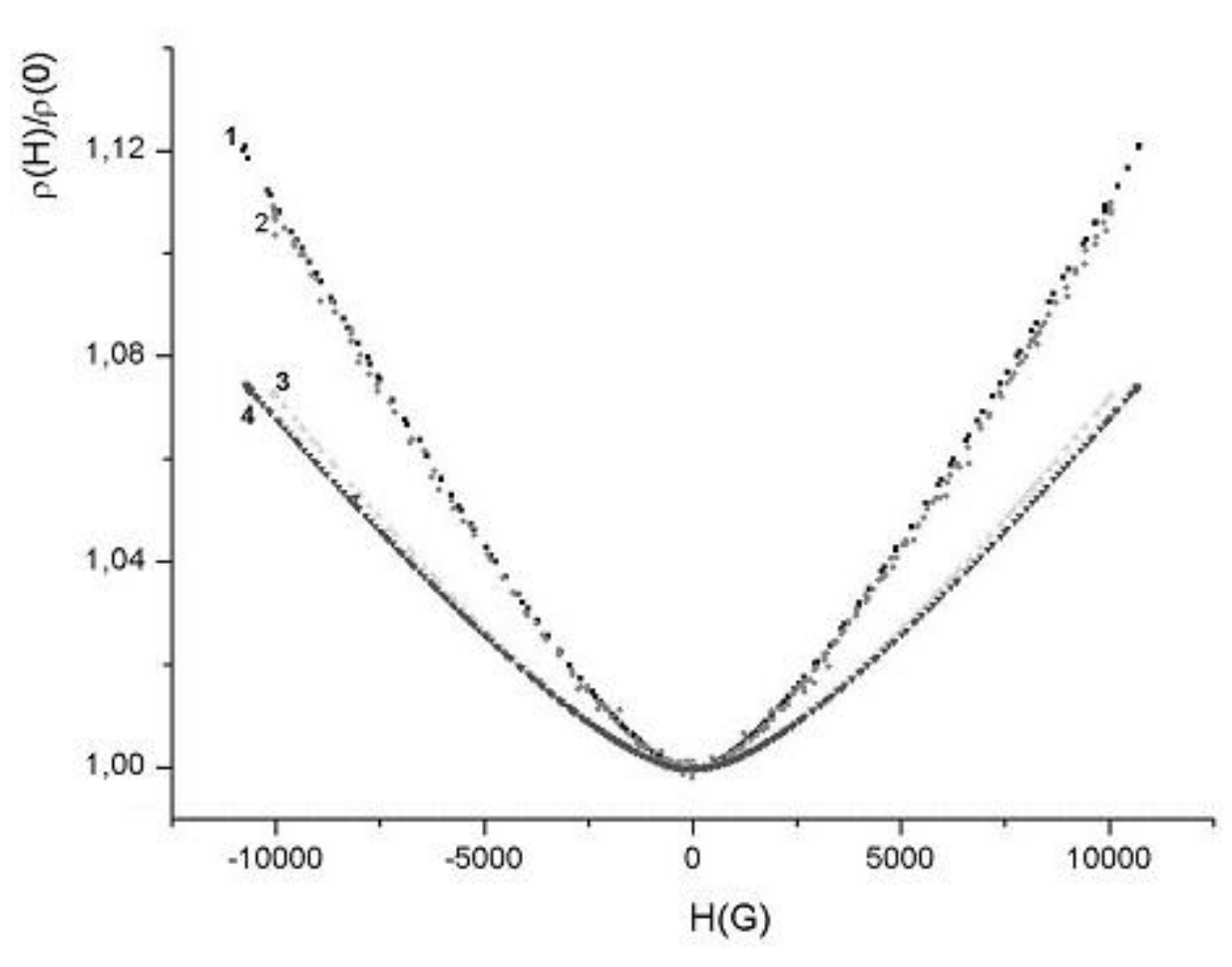

**Рис. 2.19.** Измерения относительного магнетосопротивления графитовых композитов MPG-7 до и после термообработки:1 - исходный образец; 2 - время прогрева 1 час 30 минут при 2250$^{O}$C; 3 - время прогрева 1 час 20 минут при 2450$^{O}$C; 4 - время прогрева 50 часов при 2250$^{O}$C.

Измерения проводились при температуре 4.2 K в атмосфере гелия на оригинальной установке со сверхпроводящим соленоидом в ИНХ СО РАН. Магнетосопротивление измерялось как вдоль поля, так и поперек. Рисунок взят из работы [76].

Для металлических крупнозернистых материалов влияние размера зерна на твёрдость (прочность) описывается эмпирическим соотношением Холла-Петча [79, стр.81]

$$H_v\,(\sigma_T) = H_0\,(\sigma_0) + kL^{-1/2} \qquad (2.4)$$

где $H$ – твёрдость; $\sigma_T$ – предел текучести; $H_0$ – твёрдость объёма зерна; $\sigma_0$ – внутреннее напряжение, препятствующее распространению пластического сдвига в объёме зерна; $k$ – коэффициент пропорциональности.

Считая процессы образования микротрещин в объёме и на границах зёрен независимыми процессами, для вероятности образования микротрещины можно записать согласно [47]:

$$W = W_V\, f_V + W_B\, f_B \qquad (2.5)$$

$$W_V = \tau^{-1}{}_V = \nu_0 \exp[-U_V/kT], \qquad (2.7)$$

$$W_B = \tau^{-1}{}_B = \nu_0 \exp[-U_B/kT] \qquad (2.8)$$

где (2.7) и (2.8) есть вероятности образования микротрещин в объёме и в границах зёрен соответственно, $f_V$ и $f_B$ – объёмные доли материала объёма и границ соответственно, $\nu_0 = 1/\,\tau^{-1}{}_0$ – частота атомных колебаний.

В первом приближении $U_V(\sigma) = U(0) - \gamma_V\,\sigma$, $U_B(\sigma) = U(0) - \gamma_B\,\sigma$

где $U_{(0)}$ – не зависящая от внешнего напряжения $\sigma$ часть энергии образования микротрещины. Структурные параметры $\gamma_i$ учитывают концентрацию напряжений.

Тогда согласно [47] долговечность материала может быть представлена в виде:

$$\tau = (k_1+k_2)\ \frac{\tau_V}{f_V} \frac{}{+ f_B} \frac{}{(\tau_V / \tau_B)_{V\,/}} \qquad (2.9)$$

где предполагается, что $k_1$ и $k_2$ суть известные структурные параметры формулы (1.13).

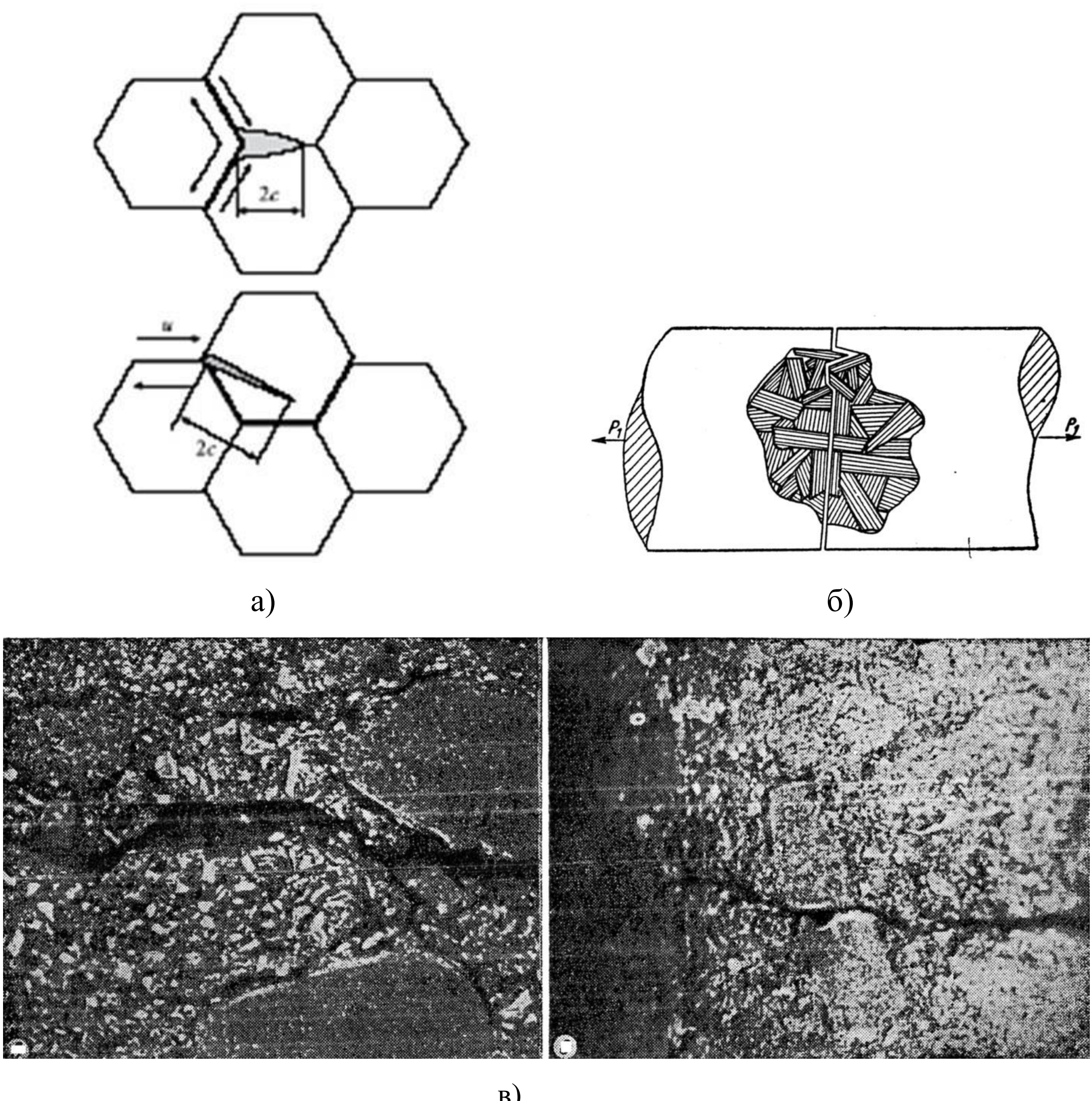

**Рис. 2.20.** Механизм образования микро- и макротрещин: а) - микротрещины в металлических нанокомпозитах; вверху - на границе зёрен; внизу - в области зерна [47]; б) схема продвижения трещины в микрообъёме графита в обход кристаллитов и вдоль базисных плоскостей; в) *слева* – разрушение по коксу от связующего, в обход зёрен для графита типа ВВП; *справа* – пересечение трещиной зерна [83].

В нашем случае, благодаря рентгенографическими и электрофизическим измерениям, можно предположить, что вероятность образования микротрещины на границе зерна заметно больше вероятности образования микротрещины в объёме кристаллита:

$$W_B >> W_V$$

В общем случае, вероятности образования микротрещин на границах зерна и в объёме материала можно считать независимыми процессами, и, в случае, если одна из вероятностей образования микротрещины заметно превышает другую, мы снова приходим к исходной формуле (1.13).

## 2.7. Механизм разрушения графитовых материалов

Характер разрушения углеродных материалов изучался ранее в [83] и согласно фрактологическим данным этой работы носит хрупкий характер, связанный с неоднородностью структуры – анизотропией свойств, трещинами, развитой пористостью, и т.д. Механизм разрушения по логике авторов следует рассматривать прежде всего макроскопически, поскольку поликристаллические графиты структурно состоят из наполнителя и связующего (см. также §1.3.5.). Кокс связующего, как правило, более рыхлый, кроме того, пековый кокс обладает более низкой пикнометрической плотностью по сравнению с плотностью используемого в наполнителе нефтяного кокса. Это обстоятельство уже само по себе предопределяет преимущественное разрушение углеродного материала по связующему.

Проведённый в работе [83] анализ показал, что как в процессе нагружения до разрушения, так и после снятия напряжения на образец без разрушения, процессы деформирования и разрушения графита в макрообъёме в основном проходят по границам зёрен: по коксу связующего путём развития и объединения уже имевшихся трещин и пор (рис. 2.20 в, слева).

Однако если в исходных зёрнах имеются определённым образом ориентированные трещины, то магистральная трещина может эти зёрна беспрепятственно пересекать. В закритической стадии разрушения, когда скорость распространения магистральной трещины в условиях растяжения велика, магистральная трещина также может пересекать отдельные зёрна (рис. 2.20 в, справа).

Магистральная трещина, обходя макроскопические зёрна наполнителя, распространяется в микрообъёме обычно по границам кристаллитов и параллельно базисным плоскостям в кристаллите, расщепляя слабые связи, так что разрушение происходит в основном по границам малоразориентированных кристаллитов (рис. 2.20 б). Чем меньше диаметр кристаллитов, тем более затруднено распространение магистральной трещины, отсюда и большая прочность мелкокристаллических материалов. Хотя макротрещина может пересекать как отдельные кристаллиты, так и макроскопические зёрна, однако такой характер разрушения не является основным.

Важнейшими факторами, влияющими на прочность графитов, является степень совершенства кристаллической структуры, а также наличие макродефектов, таких, как поры и трещины. Ранее [39] было показано, как зависит предел прочности от размеров кристаллитов и межслоевого расстояния в кристаллитах. В целом, изменение предела прочности при сжатии и модуля упругости в зависимости от температуры обработки графитового материала немонотонно. Степень совершенства кристаллической структуры прямо зависит от температуры отжига, поэтому в зависимости предела прочности от температуры наблюдается экстремум в диапазоне температур 2100-2300$^0$C. Было показано также [83], что для материалов, обработанных выше температуры 2300$^0$C, усилие разрушения при сжатии обратно пропорционально диаметру кристаллитов в степени ½. Иначе говоря, разрушение достаточно совершенного графита объяснялось, в соответствии с теорией Гриффитса-Орована, спонтанным распространением трещин по кристаллиту. Для такого характера разрушения материала будет справедливым соотношение:

$$\sigma = \left(2/\pi \times Ep/L\right)^{1/2} \qquad (2.10)$$

где σ – усилие разрушения при сжатии;        *E* – модуль упругости;

*p* – удельная поверхностная энергия скола.

Для двумерно-упорядоченных материалов, температура обработки которых ниже $2000^0$C, справедлив механизм, описываемый уравнением Петча, что свидетельствует в пользу межкристаллитного характера разрушения:

$$\sigma = \sigma_0 + kL^{-1/2}$$

где $\sigma_0$ – напряжение трения в плоскости скольжения;

*k* – эмпирическая постоянная.

В итоге, как следует из работы [83], прочность полученных по электродной технологии конструкционных углеродных материалов определяется в основном диаметром кристаллитов и общей пористостью материала.

## 2.8. Теплофизические свойства графита МПГ - 6

Экспериментальные исследования теплофизических свойств графита МПГ 6 были выполнены методом лазерной вспышки на автоматизированном экспериментальном стенде LFA 427 фирмы Netzsch (Германия) [84]. Образец устанавливался в держатель на игольчатых подставках. Его нижняя поверхность нагревалась лазерным импульсом (1.064 мкм) длительностью 0.8 мс и энергией до 10 Дж. Изменение температуры верхней поверхности образца регистрировалось ИК_детектором (InSb), который охлаждался жидким азотом. Измерения проводились после длительного термостатирования образцов при постоянной температуре в серии из трех “выстрелов”. Интервал времени между “выстрелами” составлял 5 мин.

Теплоемкость МПГ-6 измерялась тремя различными методами. До 1650 K теплоемкость определялась относительным методом на установке LFA 427 с использованием в качестве эталонов POCO графита и молибдена, на поверхность которых напылялся слой графита для обеспечения одинаковой черноты поверхностей эталонов и исследуемых образцов. (Для молибденового эталона учитывались только данные, полученные до 1100 К, так как при более высоких температурах молибден начинал реагировать с графитовым покрытием). До 1100 К теплоемкость также измерялась на дифференциальном сканирующем калориметре DSC - 404 F1 (скорость нагрева–охлаждения – 10 К/мин) с использованием платиновых тиглей с корундовыми вкладышами и сапфира для тарировки, а в интервале 298–385 К – динамическим методом на *C*, λ калориметре, который подробно описан в [85].

Перед нагревом установки LFA -427 и DSC - 404 F1 вакуумировались до ($10^{-2}$ Тор) и заполнялись аргоном чистотой 99.998 об. % ($O_2$ – 0.0001%, $N_2$ – 0.0005%, $H_2O$ – 0.0004%, $CO_2$ – 0.00002%, $CH_4$ –0.0001%, $H_2$ – 0.0001%), который дополнительно очищался системой BI - GAScleaner (ИК СО РАН). Измерения на *C*, λ - калориметре проводились на воздухе.

Плотность образцов ρ при комнатной температуре определялась путем прямых измерений геометрических размеров и массы. Масса образцов измерялась на аналитических весах AND GH 300 с погрешностью менее 0.3 мг, диаметр – электронным штангенциркулем Kraftool с погрешностью 0.03 мм, толщина – электронным длинномером Tesa Digico 10, который поверялся по образцовым мерам непосредственно перед измерениями, с погрешностью порядка 2 мкм.

Погрешности измерений были оценены по измерениям свойств эталонных материалов и составляли для температуропроводности 2–4%, теплоемкости – 3–5%, плотности – 0.5%, интегрального среднего температурного коэффициента линейного расширения – $1 \times 10^{-7}$ $К^{-1}$.

Образцы для LFA_427 имели форму цилиндров диаметром 12.6 мм и толщиной около 2.5 мм с плоскопараллельными шлифованными торцами, а для *C*, λ-калориметра – диаметром 19 мм и толщиной около 6 мм. Масса цилиндрической навески для DSC-404 F1 составляла 35 мг.

**Таблица 2.3.** Свойства образцов МПГ-6 при 293 К

| Номер образца | Плотность, г/см$^3$ | Температуропроводность, мм$^2$/с | Теплопроводность, Вт/(м К) |
|---|---|---|---|
| № 1 | 1,783 | 97.0 | 120.4 |
| № 2 | 1,825 | 113.2 | 143.8 |
| № 3 | 1,776 | 96.1 | 118.7 |
| № 4 | 1,664 | 80.5 | 93.3 |

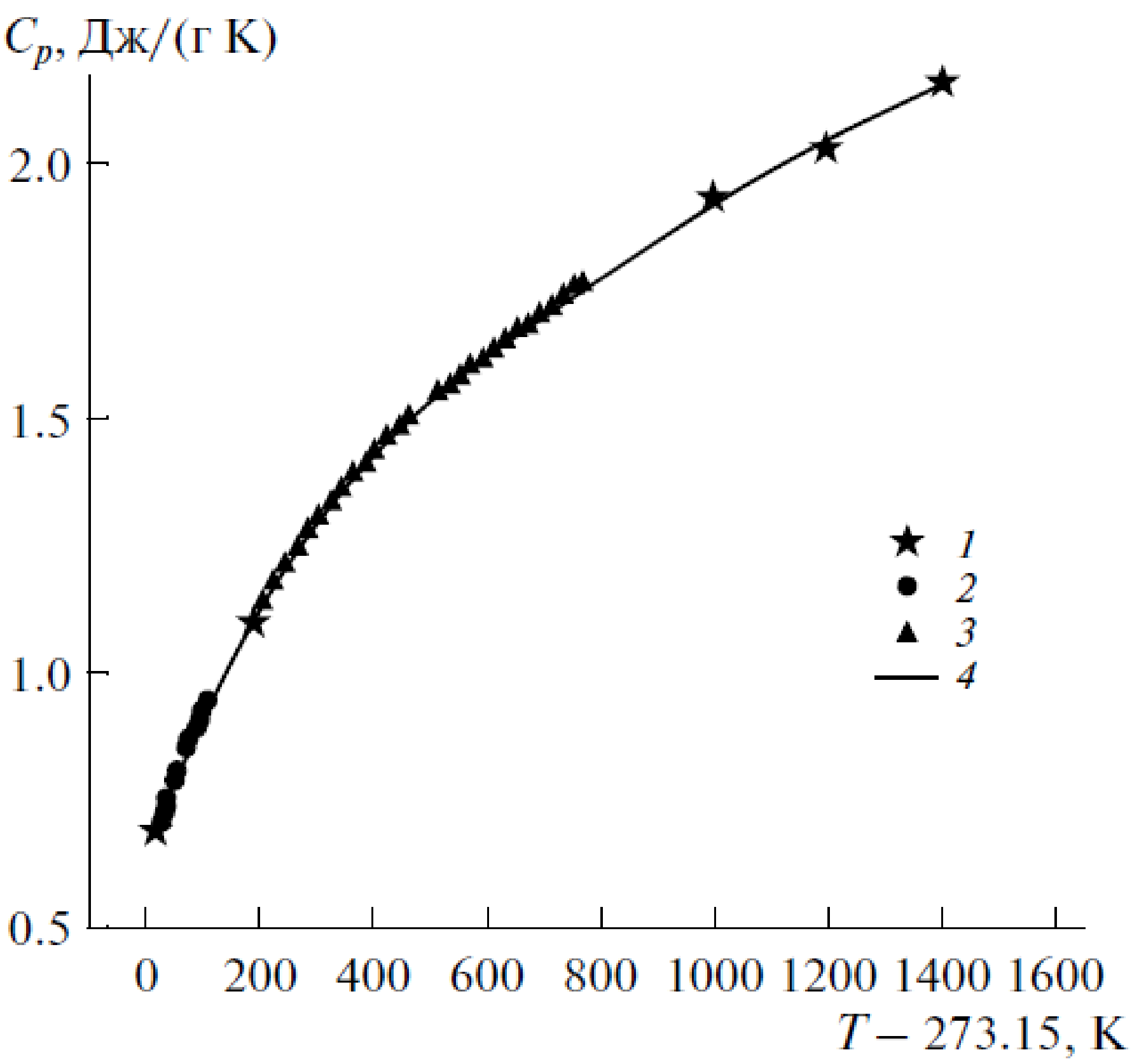

**Рис. 2.21.** Теплоемкость графита марки МПГ_6: *1* – результаты измерений на установке LFA - 427, *2* – *C*, λ - калориметр, *3* – DSC 404 F1, *4* – уравнение (1).

На рис. 2.21 приведены результаты измерений теплоемкости, полученные всеми тремя методами. Видно, что данные хорошо согласуются между собой. Среднеквадратичное отклонение экспериментальных точек от аппроксимирующей зависимости

$$C_P(T)=0.636+3.09\times10^{-3}\,t - 3.724\times10^{-6}\,t^2 + 2.609\times10^{-9}\,t^3 - 6.92\times10^{-13}\,t^4 \qquad (2.11)$$

составляет 0.7%. Здесь $t = T - 273.15$, размерность $C_P(T)$ – Дж/(г К), $T$ – К. В пределах оцениваемых погрешностей измеренные значения теплоемкости МПГ-6 согласуются со справочными данными для POCO графита [86]. Это подтверждает тот факт, что теплоемкость графитов практически не зависит от технологии их получения и пористости [87].

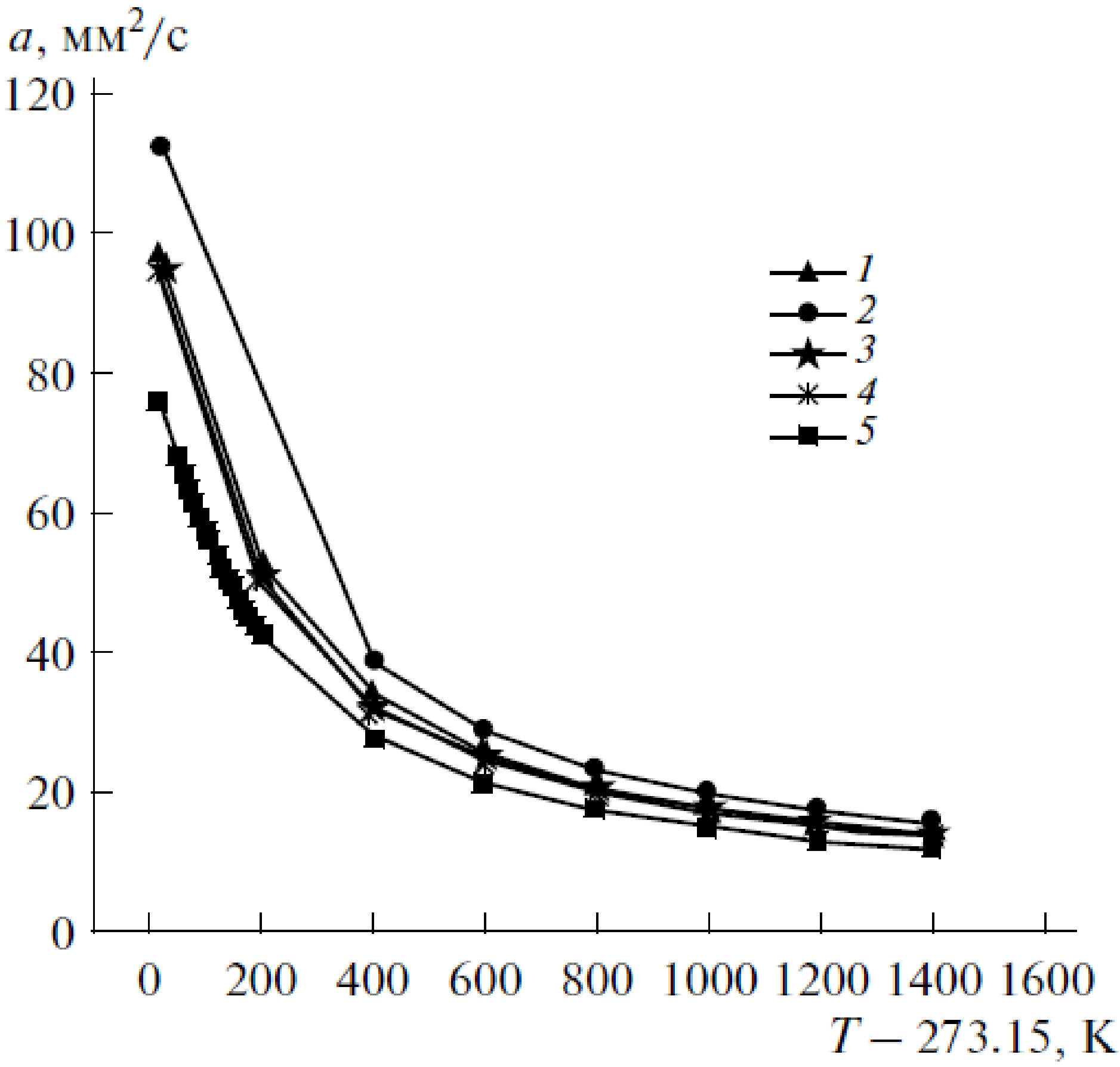

**Рис. 2.22.** Результаты измерений температуропроводности образцов МПГ-6: *1* – образец № 1; *2* – № 2; *3*, *4* –№ 3 (две серии экспериментов); *5* – № 4.

На рис. 2.22 приведены результаты измерений температурной зависимости коэффициента температуропроводности четырех образцов графита марки МПГ-6, полученных как при нагреве, так и при охлаждении для постоянной толщины образца. Можно заметить сильное изменение ***a*** в интервале температур 293–800 К и значительное расслоение результатов для разных образцов (30–50%). Воспроизводимость данных проверялась в экспериментах с третьим образцом МПГ-6. Отличие результатов двух серий не превышало 2.7%,что меньше, чем оцениваемые погрешности измерений температуропроводности.

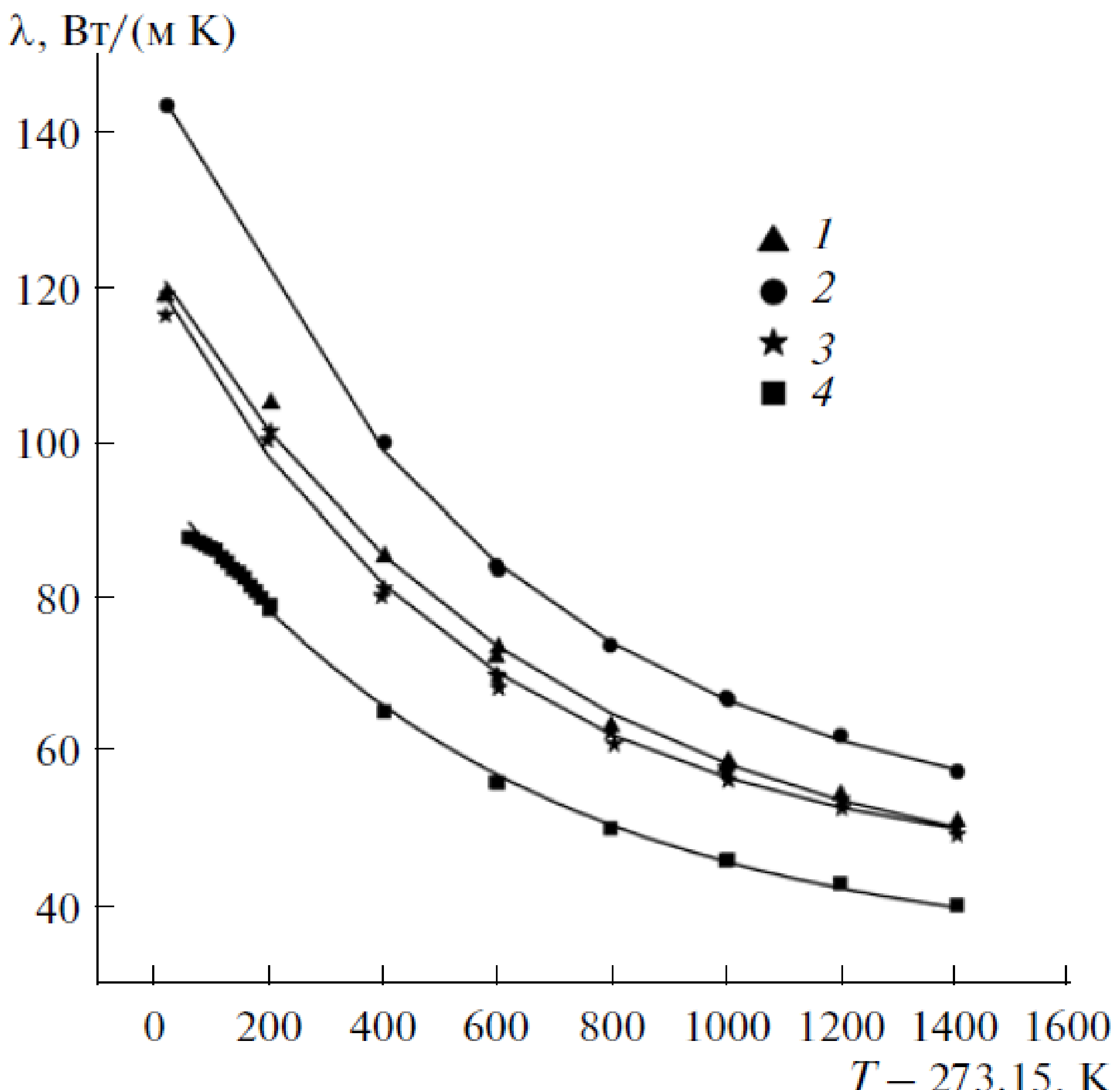

**Рис. 2.23.** Теплопроводность образцов МПГ-6: *1* – образец № 1, *2* – № 2, *3* – № 3, *4* – № 4.

Коэффициенты теплопроводности и температуропроводности связаны известным соотношением:

$$\lambda = a\rho C_P \qquad (2.12)$$

Используя результаты наших измерений по температуропроводности $a$, данные по теплоемкости (2.11), плотности (табл. 2.3) и тепловому расширению МПГ-6 можно рассчитать теплопроводность графита (рис. 2.23).

Из рисунка видно, что коэффициент теплопроводности изменяется менее сильно, чем температуропроводность. Практически эквидистантное расположение температурных зависимостей теплопроводности образцов МПГ-6 (рис. 2.23) позволяет надеяться на получение достаточно простой обобщающей зависимости.

Ранее [78] было показано, что теплопроводность поликристаллических графитов в диапазоне температур от комнатной до 3300 K может быть описана эмпирической зависимостью вида $\lambda(T) \sim 1/T$, которая, однако, не учитывает пористость образца. Согласно общим представлениям теории фонон-фононного взаимодействия [89] изменение теплопроводности в области промежуточных температур (выше комнатной, но ниже температуры Дебая) имеет вид $\lambda(T) \sim T^n \exp(b/T)$. Однако использование этого выражения для аппроксимации первичных экспериментальных данных по МПГ-6 приводит к систематическим отклонениям сглаженных значений. Хороших результатов удалось добиться с помощью эмпирической зависимости:

$$\lambda(T) = A + B\exp\left(-\frac{t}{C}\right) \qquad (2.13)$$

Используя (2.13) определялась теплопроводность образца при комнатной температуре $\lambda_{293}$ и строились безразмерные зависимости $\lambda(T)/\lambda_{293}$. Из рис. 2.24 видно, что приведенная теплопроводность для всех образцов практически совпадает между собой. Совместная обработка результатов измерений дала уравнение:

$$\lambda(T)/\lambda_{293} = 0.351 + 0.675\exp\left(-\frac{t}{597}\right) \qquad (2.14)$$

Среднеквадратичное отклонение первичных данных от (4) составило 2%.

Как указывалось ранее в главе 1, теплопроводность графитов в значительной степени определяется пористостью [90]. Для графитов одной марки пористость однозначно связана с макроскопической плотностью образцов. Это подтверждают результаты наших измерений при комнатной температуре (табл. 2.3, рис. 2.25). В интервале плотностей 1,66-1,825 г/см$^3$ теплопроводность образцов МПГ-6, со среднеквадратичным отклонением 0.6%, описывается зависимостью:

$$\lambda_{293} = 5246.2 - 6.2051\,\rho + 1.86803\times10^{-3}\,\rho^2, \qquad (2.15)$$

где $\lambda$ измеряется в Вт/(м·К), ρ в г/см$^3$.

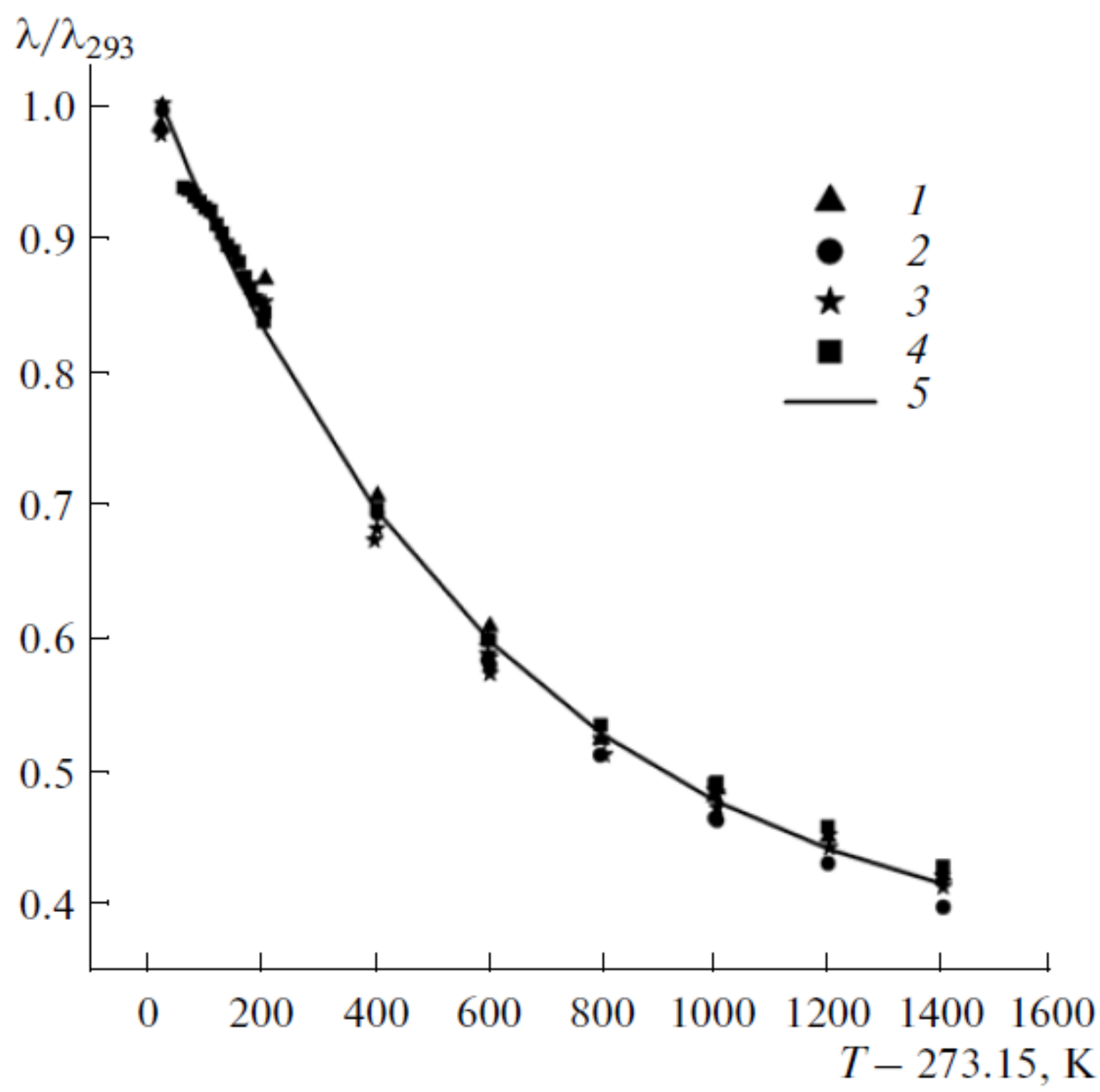

**Рис. 2.24.** Нормированная на λ293 теплопроводность образцов МПГ_6: *1* – образец № 1, *2* – № 2, *3* – № 3, *4* –№ 4, *5* – уравнение (3).

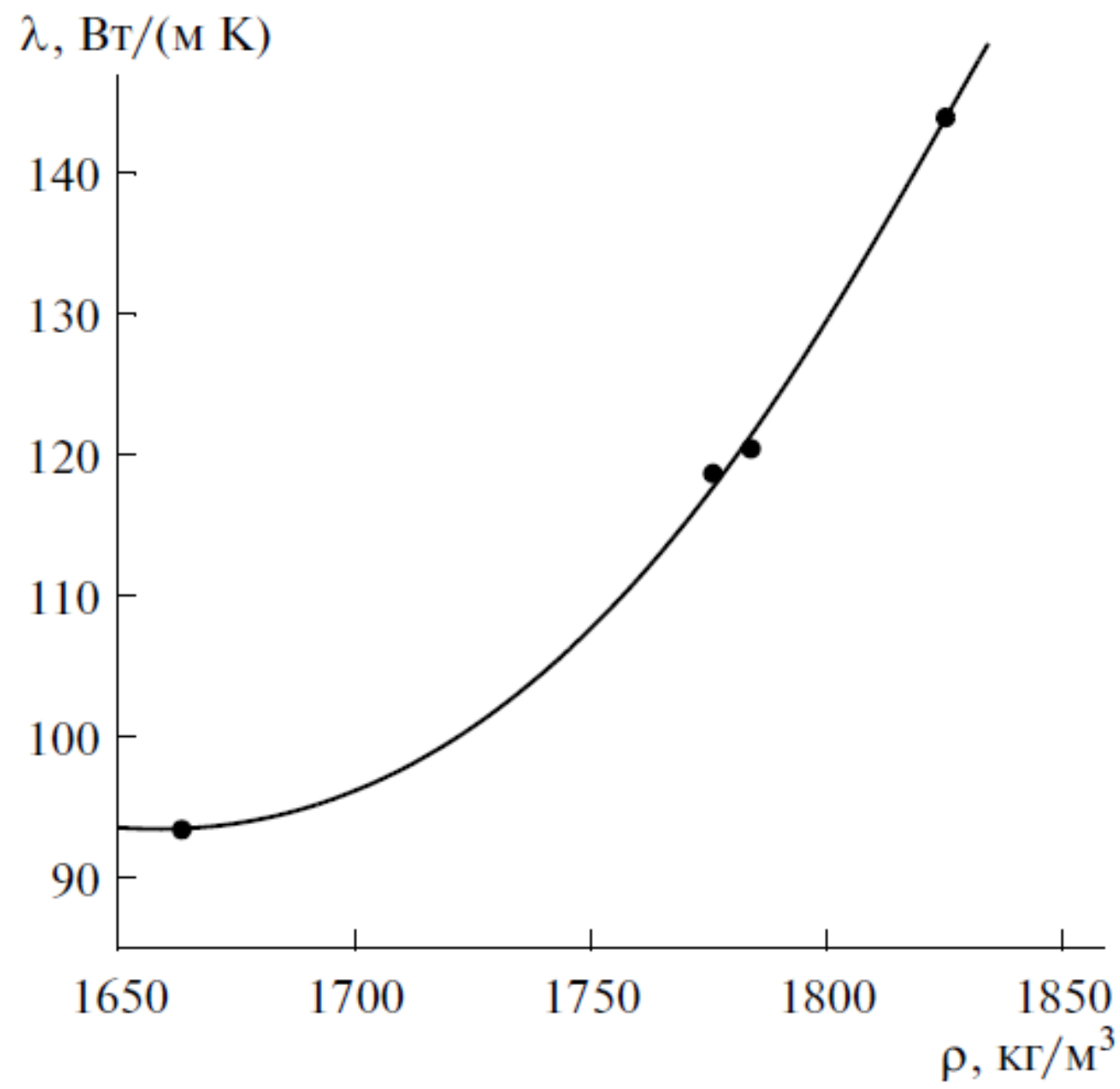

**Рис. 2.25.** Зависимость теплопроводности образцов МПГ-6 от макроскопической плотности при 293.15 К. Точки – экспериментальные значения, линия – уравнение (5).

**Таблица 2.4** Зависимость теплопроводности графита марки МПГ-6 от плотности и температуры.

| $T$, К | Плотность, г/см$^3$ | | | | | | | |
|---|---|---|---|---|---|---|---|---|
| | 1,650 | 1,675 | 1,700 | 1,725 | 1,750 | 1,775 | 1,800 | 1,825 |
| 293 | 93.9 | 94.0 | 96.5 | 101.4 | 108.5 | 118.1 | 129.9 | 144.2 |
| 400 | 83.8 | 84.0 | 86.2 | 90.5 | 97.0 | 105.5 | 116.1 | 128.8 |
| 500 | 76.0 | 76.1 | 78.1 | 82.0 | 87.9 | 95.6 | 105.2 | 116.7 |
| 600 | 69.3 | 69.4 | 71.3 | 74.9 | 80.2 | 87.2 | 96.0 | 106.5 |
| 700 | 63.7 | 63.8 | 65.5 | 68.8 | 73.7 | 80.1 | 88.2 | 97.8 |
| 800 | 58.9 | 59.0 | 60.6 | 63.6 | 68.1 | 74.1 | 81.6 | 90.5 |
| 900 | 54.9 | 55.0 | 56.5 | 59.3 | 63.5 | 69.1 | 76.0 | 84.3 |
| 1000 | 51.5 | 51.6 | 53.0 | 55.6 | 59.5 | 64.8 | 71.3 | 79.1 |
| 1200 | 46.2 | 46.3 | 47.5 | 49.9 | 53.4 | 58.1 | 63.9 | 70.9 |
| 1400 | 42.4 | 42.4 | 43.6 | 45.8 | 49.0 | 53.3 | 58.7 | 65.1 |
| 1600 | 39.7 | 39.7 | 40.8 | 42.8 | 45.9 | 49.9 | 54.9 | 60.9 |
| 1800 | 37.7 | 37.8 | 38.8 | 40.7 | 43.6 | 47.4 | 52.2 | 57.9 |

Уравнения (2.14) и (2.15) дают возможность рассчитать теплопроводность графитов марки МПГ-6 разной пористости практически с погрешностью экспериментальных результатов, привлекая лишь данные по макроскопической плотности образцов (табл. 2.4). При этом максимальная температура оценки $\lambda$, по-видимому, значительно превышает 1650 К. Во всяком случае сопоставление с данными [91], где измерения теплопроводности мелкозернистого графита с плотностью 1,7 г/см$^3$ при 3010 К согласуются с расчетом по уравнениям (2.14), (2.15) в пределах 2%, что существенно меньше оцениваемых погрешностей.

***

В этой главе проведён анализ физических свойств и дефектности различных углеродных композитов и показано, что кинетика разрушения графитовых композитов под воздействием высоких температур в целом хорошо согласуется с представлениями термофлуктуационной концепции и укладывается в двустадийную модель разрушения твёрдых тел. Рентгенографическими, электрофизическими и другими измерениями показано, что кристаллическая структура графитовых композитов не меняется или даже совершенствуется под воздействием высокотемпературного прогрева, как электронным пучком, так и переменным током. Обнаружено, что энергия активации разрушения может быть связана с такими явлениями, как ползучесть или самодиффузия углерода, где важную роль играет особенности мезоструктуры графитового композита, в частности, анизотропия материала.

Кроме того, в этой главе приведены новые данные по температуропроводности, теплопроводности и теплоемкости графита марки МПГ-6. Подтверждено, что теплоемкость графитов практически не зависит от пористости и способа получения. Установлена связь между теплопроводностью графита и его макроскопической плотностью, а также предложен способ оценки температурной зависимости теплопроводности графитов марки МПГ-6 в широком интервале температур.

ГЛАВА 3

# МЕТОДЫ МОЛЕКУЛЯРНОЙ ДИНАМИКИ ДЛЯ РАСЧЁТА И МОДЕЛИРОВАНИЯ МОЩНОСТИ ДОЗЫ И ЧИСЛА СМЕЩЕНИЙ АТОМОВ В ГРАФИТОВОЙ МИШЕНИ

В настоящее время не существует надежных способов экспериментального определения числа смещенных атомов, вызываемых каким либо видом излучения. Все существующие методы определения числа смещенных атомов основаны на косвенных измерениях тех или иных характеристик вещества [92]. С другой стороны, число смещенных атомов принято считать одной из основных характеристик, определяющих степень воздействия радиационных излучений на характеристики материала (прочность, коэффициент теплового расширения, сопротивление для проводников и др.). В связи с этим возникает необходимость оценки числа смещений на атом (СНА) или величин определяющих эту характеристику, таких как скорость генерации дефектов и скорость отжига численными методами.

В случае нейтронопроизводящей мишени, облучаемой протонами или дейтронами (см. гл. 5), основную радиационную нагрузку будет нести графитовая мишень, предназначенная для получения высокоэнергетических нейтронов. В этой мишени будет нарабатываться наибольшее количество дефектов, образующихся в результате атомных каскадов инициируемых дейтронами с начальной энергией 40 МэВ. Влияние нейтронов на образование дефектов в мишени будет незначительно, поскольку коэффициент преобразования $D \rightarrow n$ составляет величину ~4%, к тому же вероятность вторичного нейтрона вступить в реакцию ядрами углерода в мишени весьма невелика. В то же самое время простая оценка на примере кремния показывает, что дейтерий в процессе замедления и каскадообразования может вызывать до 200 смещений атомов. Кроме этого, термализованные дейтроны будут образовывать в графитовой мишени области с повышенным содержанием газа D, $D_2$, (возможно метана и других C-H(D) структур) [93, 94].

## 3.1. Общие представления

### 3.1.1. Взаимодействие протонов с веществом.

Для протонов широкого диапазона энергий от единиц МэВ до $10^3$МэВ согласно [95, 96] характерны следующие основные процессы взаимодействия с веществом:

1) электромагнитное взаимодействие с электронами среды;

2) электромагнитное взаимодействие с ядрами среды;

3) упругое рассеяние на ядрах среды;

4) неупругое взаимодействие с ядрами среды.

Первый процесс дает значительный вклад при прохождении протона на больших расстояниях от атома. В этом случае энергия протона расходуется в основном на ионизацию материала, т.е. передается электронам среды. При каждом таком взаимодействии протон теряет энергию малыми порциями и практически не меняет направления своего движения. Однако благодаря большому количеству электронов, с которыми протон взаимодействует, общие потери энергии могут быть значительными. Указанные потери энергии называют ионизационными и количественно выражают величиной $\left(\frac{dE}{dx}\right)_i$ которая зависит от энергии протона и облучаемого материала. Зависимость ионизационных потерь от энергии протонов для случая кремния и германия приведена на рис.3.1.

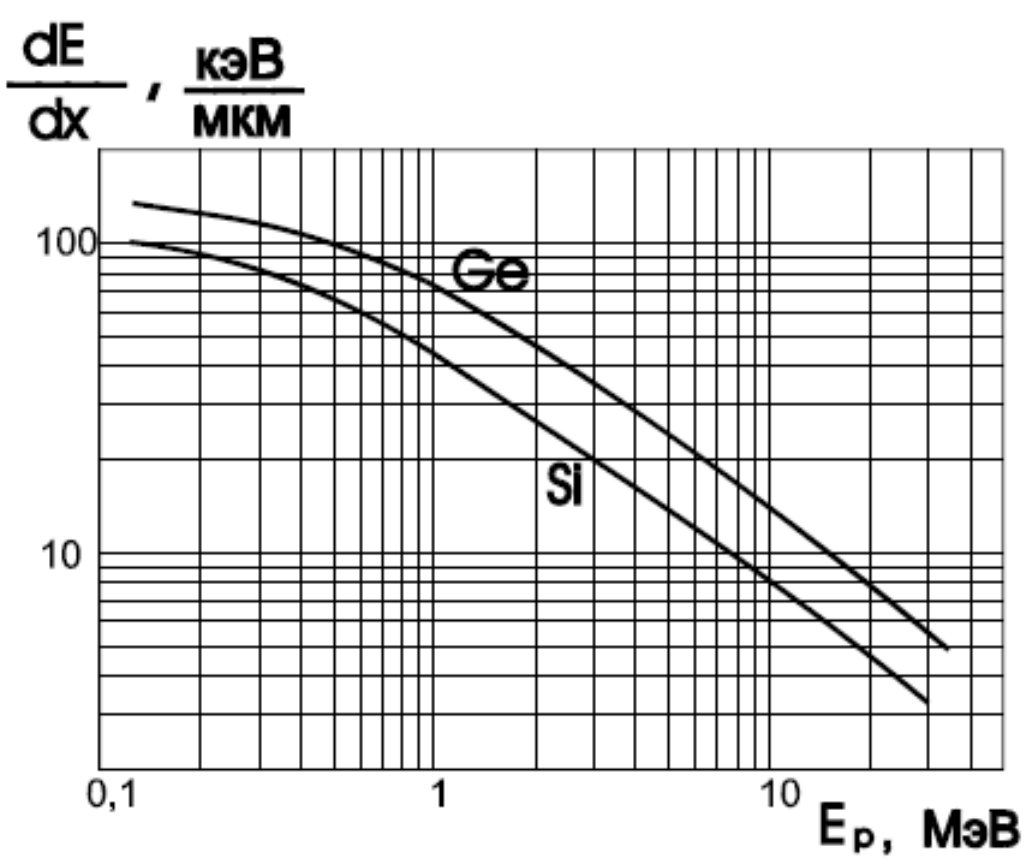

**Рис. 3.1.** Зависимость ионизационных потерь энергии протонов в германии и кремнии от их энергии [95].

Если протон проходит от ядра на расстоянии, меньшим размера атома, то наблюдается упругое кулоновское рассеяние протона в поле ядра, т.е. упругое электромагнитное рассеяние на кулоновском потенциале. Отклонение протонов в поле ядра происходит, как правило, на малые углы, т.е. атом отдачи в среднем получает энергию, значительно меньшую, чем при упругом рассеянии типа твердых шаров. Но поперечное сечение упругого кулоновского рассеяния значительно больше, так как для такого взаимодействия необязательно попадать в ядро.

В случае попадания протона в ядро возможно ядерное упругое взаимодействие, которое аналогично упругому рассеянию нейтрона с ядрами, и ядерное неупругое взаимодействие, которое приводит к протеканию следующих внутриядерных процессов:

а) неупругое рассеяние типа $(p, p')$;

б) ядерные реакции типов $(p, n)$, $(p, d)$, $(p, \alpha)$ и др;

в) реакция перезарядки протона $(p, n)$;

г) реакции расщепления, сопровождающиеся испусканием большего числа частиц, типов $(p, pn)$, $(p, 2n)$, $(p, p\ 2n)$, и т.д.:

д) ядерные реакции, приводящие к образованию ядерных фрагментов с атомным номером 3 и больше $(Z \geq 3)$;

е) ядерные реакции деления тяжелых ядер.

Указанные процессы характерны для протонов высоких энергий (более 50÷100 МэВ). Полные поперечные сечения неупругого взаимодействия протонов с веществом близки к геометрическим размерам ядер. Протоны в этом случае взаимодействуют не с ядром в целом, а с его нуклонами, которые, в свою очередь, могут взаимодействовать с другими нуклонами ядра и т.д., что приводит к развитию лавинообразного каскада нуклон-нуклонных соударений, сопровождающихся выбиванием частиц из ядра. Направление вылета этих частиц, называемых каскадными, анизотропно, преимущественно - по направлению движения протона. Их энергетический спектр находится в широком диапазоне - от 5 МэВ до энергии падающего протона, которая может достигать десятки и сотни МэВ. Ядро после окончания каскада находится в возбужденном состоянии, так как часть поглощенной энергии была передана нуклонам отдачи, не вылетевшим из ядра. Эта энергия возбуждения перераспределяется между нуклонами ядра и вследствие флюктуаций может передаваться одному или нескольким нуклонам, которые могут покинуть ядро. Такой процесс называется испарительной стадией, а испускаемые частицы - испарительными частицами. Они состоят в основном из нейтронов (~ 50%) и протонов (~25%). Кроме того, среди них имеются дейтроны, тритоны, α - частицы. Могут испускаться и гамма-кванты. Вылет испарительных частиц происходит изотропно.

Пробег протонов в веществах при энергиях до 50 - 100 МэВ определяется главным образом ионизационными потерями энергии. При больших энергиях значительный вклад в торможение вносит неупругое ядерное взаимодействие.

Для тонких образцов (размеры меньше длины пробега протона) велика роль электромагнитного взаимодействия, для толстых – при больших энергиях протонов необходимо учитывать вклад ядерного взаимодействия. В конце пробега протон может захватить электрон и начать упруго сталкиваться с ядрами по механизму твердых шаров. Зависимости пробега протонов от их энергии для случая кремния и германия приведены на рис. 3.2.

#### 3.1.2. Пороговая энергия смещения атома из узла кристаллической решетки

Пороговая энергия смещения атома из узла представляет собой минимальную кинетическую энергию, которую нужно сообщить атому для выхода его в междоузельное положение, и обозначается $E_d$. Один из наиболее ранних подходов к определению величины $E_d$ принадлежит Зейтцу (F.Seitz, 1956), который связал $E_d$ с энергией сублимации атома $E_S$ [97]. Он считал, что для смещения атома в объеме затрачивается энергия в 2 раза большая, чем для отрыва атома с поверхности, так как требуется разорвать в два раза больше связей.

Но такое соотношение характерно для относительно медленного перемещения атома из узла, когда соседние атомы успевают вернуться в исходное положение. При резком ударе этого не происходит, поэтому пороговая энергия смещения должна быть еще больше, а именно равна (4÷5) $E_S$, т.е. (20-25) эВ. В более поздних подходах величину пороговой энергии $E_d$ определяли, исходя из энергии отдельной связи $D$ как $E_d=4D$. Энергия связи $D$, в свою очередь, обусловлена энергией сцепления и энергией перехода из состояния $2s^2\ 2p^2$ в состояние $2sp^3$. При этом для полупроводников IV группы (германия, кремния) получалась величина, близкая по значению $E_d$, полученному Зейтцем.

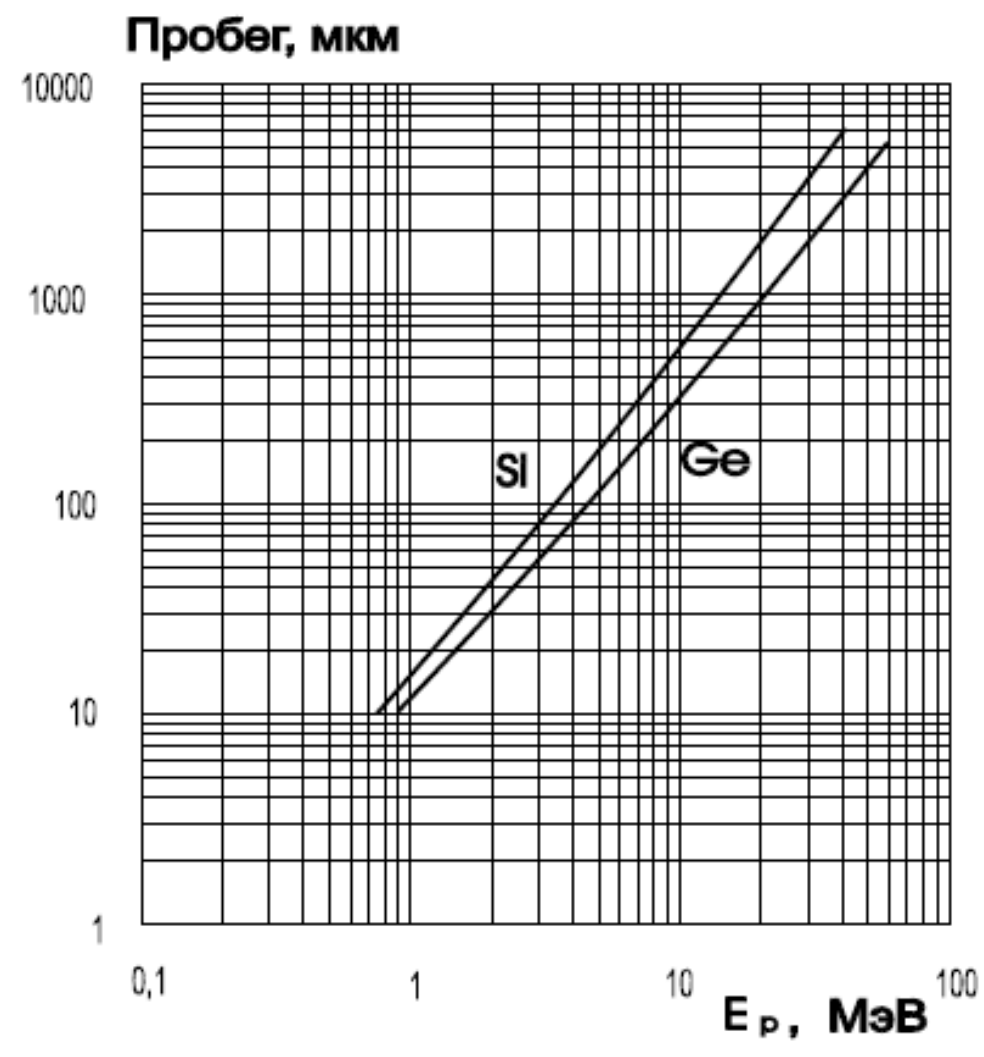

**Рис.3.2.** Зависимость длины пробега протонов в германии и кремнии от их энергии [95].

Экспериментальные данные величины $E_d$ для германия и кремния хорошо согласуются с указанными подходами. Так, для кремния $E_d$≈21эВ, а для германия ~ 27,5 эВ. Однако следует отметить, что величина $E_d$ может зависеть от кристаллографического направления, по которому происходит движение выбитого атома, и весьма чувствительна к температуре при облучении (с ростом температуры $E_d$ падает).

Для графита в [92] приводятся несколько значений энергии смещения, полученные при разных температурах в диапазоне от 15 ($E_d$=60 эВ) до 290 К ($E_d$=24.7 - 42 эВ). Принимая во внимание высокую рабочую температуру графитовой мишени нейтронного конвертора в сборке [98], чувствительность энергии смещения к температуре, а также довольно сильный разброс экспериментальных значений, нами был произведен независимый теоретический расчет пороговой энергии методом молекулярной динамики (МД).

При МД моделировании углерода наибольшую распространенность получили потенциалы класса REBO [99-100].

Короткодействующая часть потенциала REBO $E_{REBO}$ имеет вид

$$E_{REBO}=\frac{1}{2}\sum_{i=1}^{N}\varphi_i=\frac{1}{2}\sum_{i=1}^{N}\sum_{\substack{j=1\\(j\neq i)}}^{N} S(r_{ij})\left[V_R(r_{ij})-b_{ij}V_A(r_{ij})\right]\ , \qquad (3.1)$$

где:

$$V_R(r_{ij})=\frac{D_e}{S-1}e^{-\alpha\sqrt{2S}\ r_{ij}-r_e}\ , \qquad (3.2)$$

$$V_A(r_{ij})=\frac{SD_e}{S-1}e^{-\alpha\sqrt{\frac{2}{S}}\ r_{ij}-r_e}\ . \qquad (3.3).$$

Здесь $V_R(r)$ и $V_A(r)$ – отталкивающий и притягивающий члены, соответственно, а $S(t)$ – обрезающая функция, обеспечивающая непрерывное и гладкое зануление взаимодействия на выбранном отрезке, которая имеет вид:

$$S(r)=\begin{cases}1, & r\le r_{\min}\\ \left(1-\left[\dfrac{r^2-r_{\min}^2}{r_{\max}^2-r_{\min}^2}\right]^2\right)^2, & r_{\min}<r<r_{\max}\\ 0, & r\ge r_{\max}\end{cases}\ . \qquad (3.4)$$

Множитель $b_{ij}$, учитывающий влияние окружения на связь $i$-го и $j$-го атомов, имеет вид:

$$b_{ij}=\frac{1}{\sqrt{1+G_{ij}\sum_{k(\neq)}S(r_{ik})e^{\,m_{ij}\ r_{ij}-r_{ik}}}}\ . \qquad (3.5)$$

В выражении (3.5) слагаемое $G_{ij}\sum_{k(\neq)} S(r_{ik})e^{m_{ij}\ r_{ij}-r_{ik}}$ имеет смысл влияния окружения на связь $i-j$. Множитель $G_{ij}$ – весовой множитель, характеризующий восприимчивость связи между атомами $i$ и $j$ к окружению. В сумме под знаком корня функции от длин связей $i-j$ и $i-k$ также гладко обращаются в 0 на отрезке $r_{\min}, r_{\max}$ .

Короткодействующая часть потенциальной энергетической функции имеет шесть параметров: $\alpha$ , $S$ , $D_e$ , $r_e$ , $m$ , $G$ , а также расчетные параметры $r_{\min}$ и $r_{\max}$ . Путем варьирования этих параметров необходимо стабилизировать требуемую кристаллическую структуру.

Короткодействующая часть потенциала описывает взаимодействие атомов в графеновом слое. Взаимодействие между слоями описывается дальнодействующей частью потенциала, в качестве которой может быть выбран простой парный потенциал типа Леннарда-Джонса. Параметры потенциала REBO для углерода приведены в [100].

Расчеты проводились при рабочей температуре мишени. Для определения энергии смещения в МД расчетах заранее выбранному атому (в графите все атомы эквивалентны) придавалась скорость в некотором направлении, определяемом углами $(\theta,\varphi)$ (см. рис. 3.3). Для построения функции $E_d(\theta,\varphi)$ полярный и азимутальный угол менялись с шагом $10^{o}$ и $5^{o}$ соответственно. Диапазон изменения полярного угла [0, π/2], азимутального - [0, 2π].

В МД расчетах движение частиц в ближайшем окружении выбранной частицы является скоррелированным и при смещении выбранной частицы происходит релаксация ближайшего окружения с учетом ее изменяющегося положения.

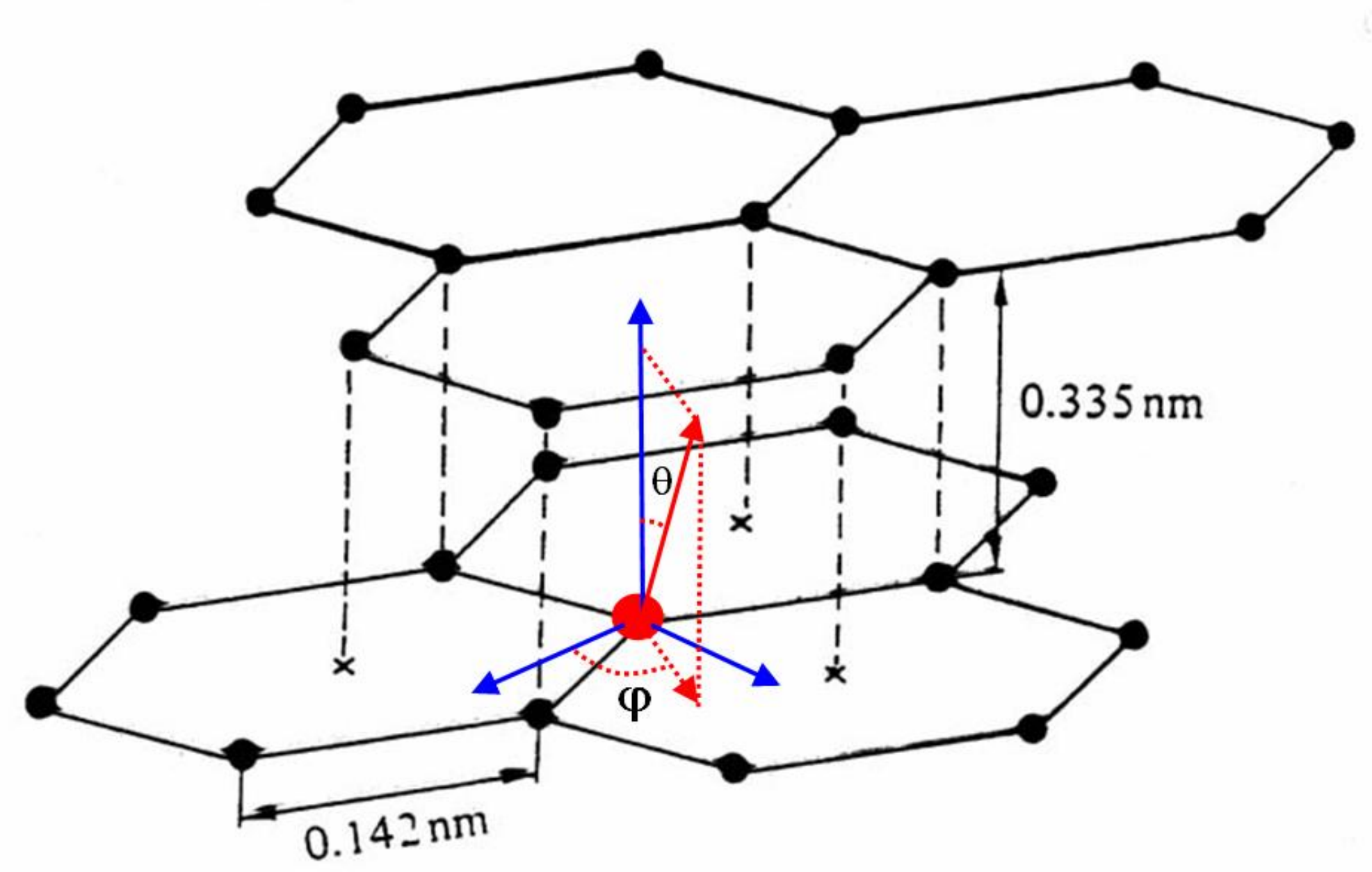

**Рис. 3.3.** Кристаллическая структура графита и схема постановки расчетов. Красной стрелкой показано направление вылета выбранного атома из узла.

При смещении атома из положения равновесия потенциальная энергия начинает расти, затем, при некотором смещении $r^m$ проходит максимум и начинает снижаться. Разность энергий в точке $r^m$ и в положении равновесия называется энергией смещения. При некоторых направлениях смещения (например, в направлении ближайшего соседа) целесообразно применить другой критерий для определения энергии смещения, а именно, энергией смещения называется разность энергий в точке пересечения смещаемой частицей границ ячейки Вигнера-Зейца и в положении равновесия. Таким образом, энергия смещения $E_d$ определяется по двум критериям, в зависимости от того какой из них выполнится первым.

На рис. 3.4 энергия смещения графита представлена как функция полярного и азимутального углов. Из рисунка видна сильная зависимость энергии смещения от направления. Усреднение по углам дает значение $\langle E_d \rangle = 17\ \text{eV}$, что является вполне разумным, принимая во внимание чувствительность энергии смещения к температуре.

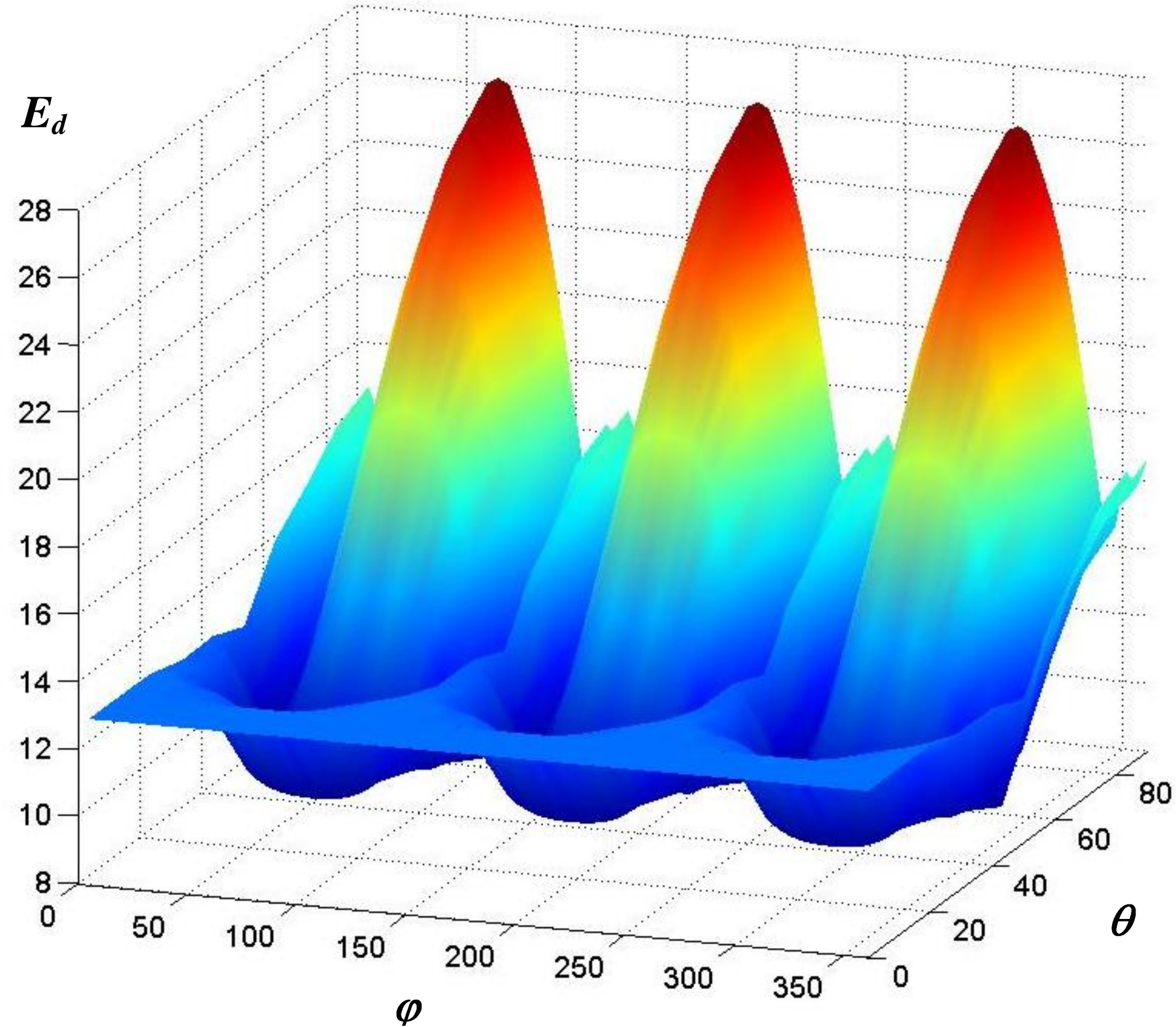

**Рис. 3.4.** Зависимость энергии смещения $E_d$ в графите от полярного и азимутального углов.

3.1.3. Пороговая энергия ионизации

Из рассмотрения взаимодействия тяжелых заряженных частиц с веществом следует, что в некотором энергетическом диапазоне их потери энергии определяются кулоновским взаимодействием с электронами среды, а при снижении энергии начинают превалировать упругие потери энергии за счет аналогичного взаимодействия с ядрами атомов. Это демонстрирует рис.3.5, где кривая справа отражает ионизационные потери энергии, а кривая слева - упругие потери энергии, которые составляют величину $\sigma_d N\langle E_A\rangle$, т.е. пропорциональны полному поперечному сечению образования первичных смещений ($E_A$ - средняя энергия, полученная атомом отдачи при образовании смещений). В общем случае $\langle E_A\rangle$ подсчитывается следующим образом:

$$\langle E_A\rangle = \frac{1}{\sigma_d}\int_{E_d}^{E_{A\max}} E_A \left|\frac{d\sigma}{dE_A}\right| dE_A \qquad (3.6)$$

Согласно [95], соотношение (3.6) может быть преобразовано к виду:

$$< E_A > = \frac{E_d \ \ln \dfrac{E_{A\max}}{E_d}}{1 - \dfrac{E_d}{E_{A\max}}} \qquad (3.7)$$

В целом упругие потери энергии тяжелых заряженных частиц растут с уменьшением их энергии. Точка на оси абсцисс, соответствующая равенству ионизационных и упругих потерь энергии, получила название пороговой энергии ионизации $E_i$. Выше этой энергии преобладают ионизационные потери, ниже - упругие.

Зейтц для количественного определения $E_i$ предложил следующий подход. Ионизационные потери энергии должны преобладать, когда скорость налетающей заряженной частицы превышает скорость валентных электронов, а равенство указанных скоростей наблюдается при $E_{част}= E_i$. Исходя из этого Зейтц предложил следующие соотношения для $E_i$:

а) для полупроводников и диэлектриков:

$$E_i = \frac{1}{8}\frac{M_{част}}{m_e}E_g \qquad (3.8);$$

б) для металлов:

$$E_i = \frac{1}{16}\frac{M_{част}}{m_e}E_F, \qquad (3.9)$$

где $E_g$ - ширина запрещенной зоны, $E_F$ - энергия Ферми соответственно.

Понятие пороговой энергии ионизации позволит упростить оценку концентрации смещенных атомов при воздействии излучений. Следует также отметить, что атомы отдачи, получив достаточно большую энергию, выходят из узлов в виде ионов, так как скорости их валентных электронов оказываются меньше скорости ядра. Поэтому к таким атомам применимы положения для тяжелых заряженных частиц.

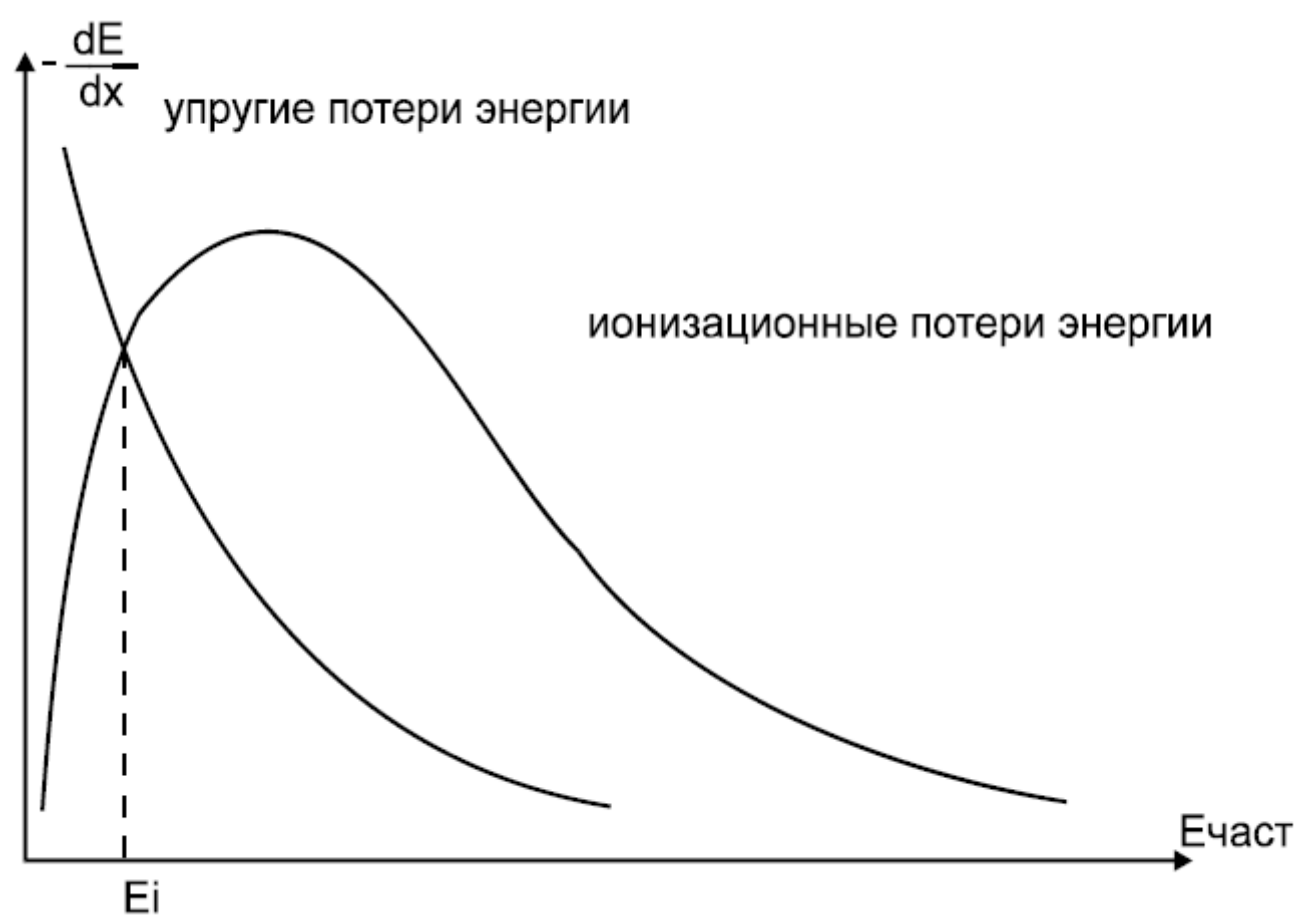

**Рис.3.5.** Зависимость упругих и ионизационных потерь энергии заряженной частицы от ее энергии [95].

3.1.4. Каскадная функция

Так как общее количество выбитых атомов может существенно превышать количество первичных смещений за счет взаимодействия с ядрами вещества атомов отдачи, для удобства расчета вводится понятие каскадной функции **ν**, которая показывает, во сколько раз общее количество смещений превышает количество первично смещенных атомов. Определим каскадную функцию для случая упругих взаимодействий атомов отдачи с другими атомами среды. Расчет был продемонстрирован в [95] при следующих допущениях:

1). Общее количество смещений растет, когда налетающий атом имеет энергию, превышающую $2E_d$. Это допущение связано с тем, что при столкновении с аналогичным по массе атомом налетающий атом в среднем будет отдавать половину своей энергии (полагаем, что взаимодействие происходит по закону твердых шаров). В противном случае один из участвующих во взаимодействии атомов будет иметь энергию, меньшую $E_d$. Это приведет к тому, что либо сидящий в узле атом не выйдет из узла, либо налетающий атом займет пустой узел, т.е. не произойдет увеличения количества смещений.

2). При $E_A>E_i$ атом теряет энергию только на ионизацию вещества, при $E_A<E_i$ - только на упругие столкновения по механизму твердых шаров (в этом случае атом движется по решетке в виде нейтральной частицы, так как скорость валентных электронов больше скорости атома).

Схема создания последующих смещений от первично выбитого атома показана на рис. 3.6.

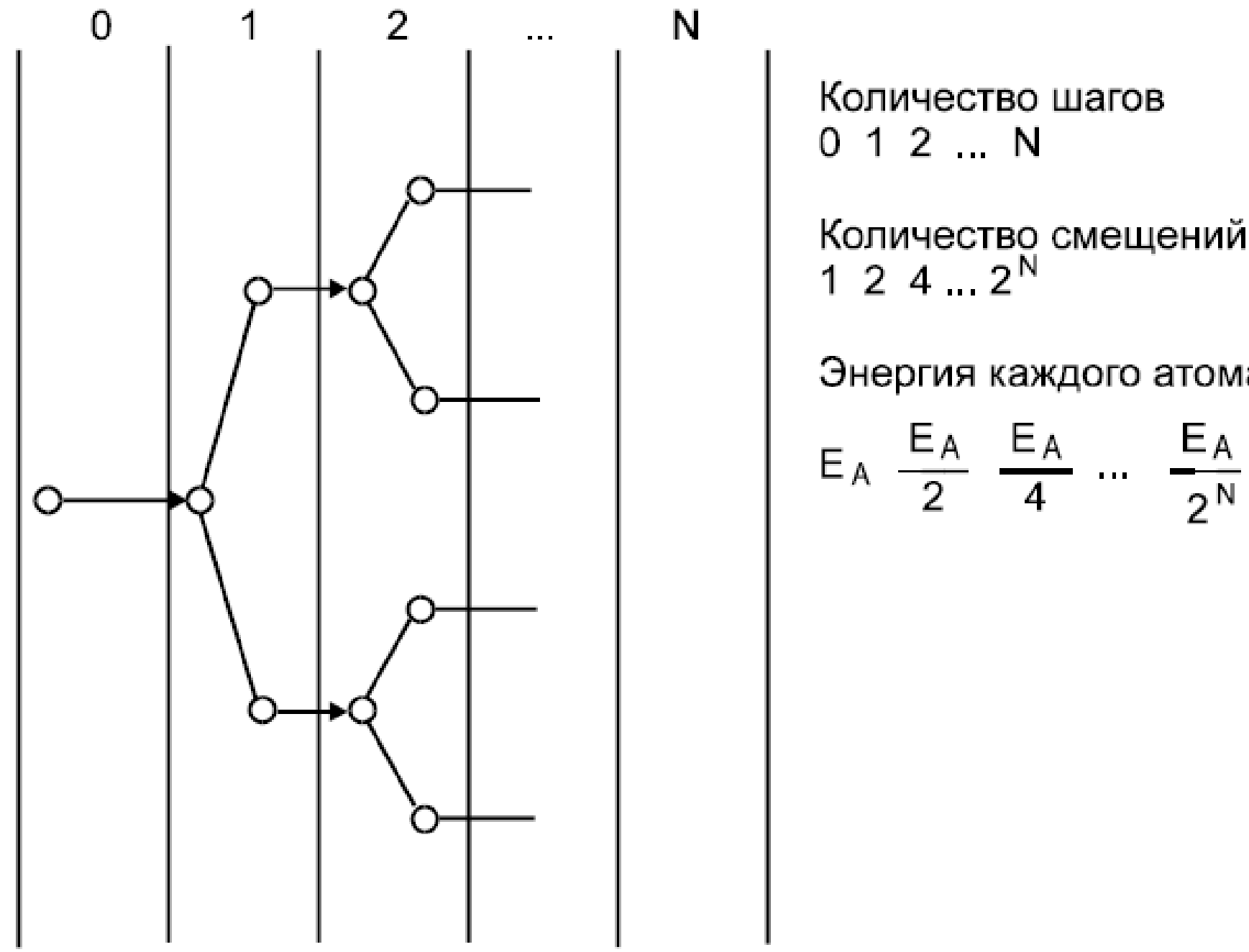

**Рис.3.6**. Схема образования каскада смещений согласно модели *Kinchin-Pease* [101].

Пусть первичный атом имеет энергию $E_A<E_i$. При равенстве масс сталкивающихся атомов он будет отдавать в среднем половину своей энергии. Так будет при каждом столкновении. Смещения будут происходить, пока энергия у последних атомов отдачи не станет равной $E_d$ (первое допущение). Это условие можно записать так:

$$\frac{E_A}{2^N} = 2\,E_d \tag{3.10}$$

Но $2^N$ есть как раз каскадная функция ν. Таким образом,

$$\nu = \frac{E_A}{2E_d} \tag{3.11}$$

Если принять во внимание допущение 2) и понятие пороговой энергии $E_d$, то для ν можно записать:

$\nu=0$ при $E_A < E_d$

$\nu$=1 при $E_d < E_A < 2E_d$ (3.12)

$\nu = E_A / 2E_d$ при $E_d < E_A < 2E_d$

$\nu = E_i / 2E_d$ при $E_A > E_i$

График функции $\nu(E_A)$ показан на рис.3.7. сплошной линией.

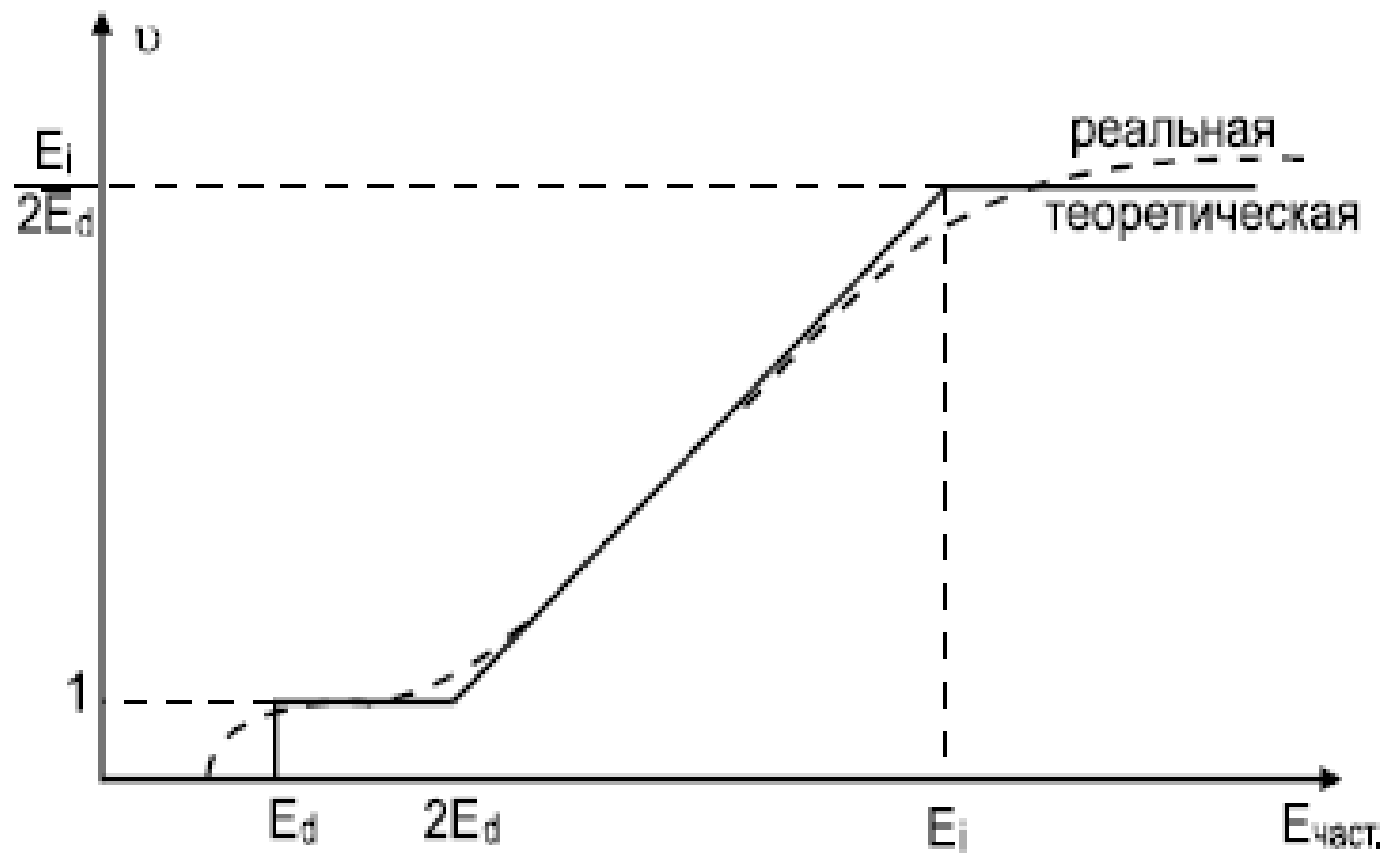

**Рис. 3.7.** Каскадная функция

На самом деле, если не использовать допущения 1) и 2), каскадная функция должна иметь вид, отмеченный пунктирной линией. Использование же полученной теоретической зависимости $\nu(E_A)$ существенно упрощает расчеты, не приводя к значительным погрешностям в оценках.

3.1.5. Оценка полного количества смещений при облучении протонами на примере кремния

Количество актов взаимодействия ($A$), происходящих в единицу времени в единице объема в веществе, определяется соотношением:

$$A = \sigma \cdot \phi \cdot N, \qquad (3.13)$$

где $\phi$ - плотность потока налетающих частиц, $N$ - концентрация объектов, с которыми происходит взаимодействие, $\sigma$ – коэффициент пропорциональности, называемый полным поперечным сечением взаимодействия.

С точки зрения классических представлений полное поперечное сечение взаимодействия $\sigma$ представляет собой площадь вокруг объекта взаимодействия, при попадании в которую налетающей частицы происходит акт взаимодействия. Произведение $\sigma \cdot N \cong \Sigma$ называется макроскопическим сечением взаимодействия. Величина $1/\Sigma$ является средним расстоянием, на котором происходит один акт взаимодействия.

В ряде случаев недостаточно знать только полное поперечное сечение $\sigma$, а необходимо пользоваться понятием дифференциального поперечного сечения $d\sigma$. Специальной единицей измерения величины поперечного сечения является барн (1 барн $=10^{-24}$см$^2$).

Оценку концентрации смещенных атомов $N_{\text{см}}$ при упругом взаимодействии налетающей частицы с веществом проводят по формуле:

$$N_{\text{см}} = \sigma_d N \nu \Phi \qquad (3.14)$$

При этом величину $\boldsymbol{\nu}$ берут в соответствии с соотношениями (3.12) или из графика на рис.3.7, принимая за $E_A$ среднюю энергию атомов отдачи.

Для случая облучения кремния протонами с энергией 30МэВ оценка согласно [95] проведена, исходя из $\sigma_d = 7\times10^{-22}$ см$^2$. Формула (3.6), исходя из того, что масса $M_{si} >> M_p$ может быть преобразована к виду:

$$< E_A > = E_d \ln \frac{4 \frac{M_P}{M_{Si}} E_P}{E_d} \qquad (3.15)$$

т.е. $< E_A > \sim 25{,}6$ эВ. Поскольку $<E_A>$ □ $<E_i>$ то $v=6$, а количество смещённых атомов составляет примерно $N_{\text{см}} = 210\ \Phi$.

## 3.2. Расчёт каскада смещений в графите

### 3.2.1. Постановка задачи

Известные методики расчета первичных радиационных дефектов, основанные на методе Монте-Карло MARLOW [102, 103], TRIM-91 [104], а также методики, разработанные в РФЯЦ ВНИИТФ: ATOCAS, TRCR2 [105], предназначены для оценки наработки дефектов от частиц относительно невысоких энергий, когда не столь важно учитывать флуктуации энергии в процессе ионизационного торможения инициирующей частицы. В рассматриваемой задаче, по предварительным оценкам [106], учет флуктуаций потерь энергий, для дейтронов 40 МэВ, падающих на графитовую мишень может менять пиковое значение плотности дефектов на два порядка. На этом основании можно полагать, что в задачах, где принципиально важную роль играют ионизационные потери энергии, необходимо учитывать флуктуации потерь энергии на ионизацию и возбуждение атомов среды.

К особенностям радиационного воздействия ионов дейтерия на материалы можно отнести следующее: большая часть энергии расходуется на ионизацию и возбуждение атомов среды $\Delta E_i$, - процесс, который имеет статистический характер и приводит к разбросу не только термализованных внедренных ионов и энерговклада, но и формирует пространственную зависимость плотности смещенных атомов. Пространственная зависимость плотности дейтронов от глубины определяется не только средними потерями энергии на ионизацию и возбуждения атомов среды $\left(\frac{dE}{dx}\right)_i$, но и угловым разбросом, который обусловлен упругими столкновениями, а также флуктуациями потерь энергии, вследствие статистического характера процесса торможения. Меньшая доля энергии $\Delta E_d$ дейтерия тратится на образование дефектов. Потери энергии ионов в результате образования дефектов также носят статистический характер, и формируют пространственную зависимость плотности дейтерия, а, следовательно, и пространственную зависимость плотности дефектов. Однако, в случае с достаточно высокими начальными энергиями дейтронов (в нашем случае 40 МэВ) разброс плотности, обусловленный процессами дефектообразования (упругого взаимодействия дейтерия с ядрами среды) незначителен, а учет флуктуаций энергии может оказаться весьма важным для корректной оценки пространственной зависимости плотности дефектов и внедренных ядер дейтерия.

При проведении оценок роли учета флуктуации ионизационных потерь энергии на распределение плотности дефектов и внедренных атомов, в программу ATOCAS были введены возможности моделирования флуктуаций энергии по модели Вавилова и гауссовского распределения, а также введен учет малоуглового рассеяния при высоких энергиях ионов в гауссовом приближении. Кроме описанных выше изменений, моделирование траекторий дейтерия и выбитых ядер углерода производилось в соответствии со схемой, описанной в работе [105]. При этом для описания упругого рассеяния использовался потенциал Мольера, а для описания ионизационной тормозной способности использовалась модель Биерсакка [107] с корректирующим множителем при формуле Линхарда-Шарфа $k/k_L$=1,25.

3.2.2. Оценка полного количества смещений в графитовой мишени при облучении протонами

Графитовая мишень ($\rho$=1.85 г/см$^3$) представляет собой колесо, вращающееся с частотой 1 Гц, радиусом 65 см и центр пучка находится на расстоянии 60 см от оси вращения мишени (рис. 3.8). Для расчёта были использованы следующие параметры: ток пучка дейтронов составлял 5 mA, распределение гауссовское, с плотностью:

$$f_{xy} = \frac{1}{2\pi \cdot \sigma^2} e^{-\frac{(x-x_0)^2 + (y-y_0)^2}{2\cdot\sigma^2}} \qquad (3.16)$$

где ($x_0$, $y_0$) – координаты центра пучка, $\sigma$=1см; энергия дейтронов 40МэВ; время работы пучка 10 000ч. Энергия смещения атомов бралась из расчетов молекулярной динамики (МД) согласно [108] и составляла $E_d$=17 эВ.

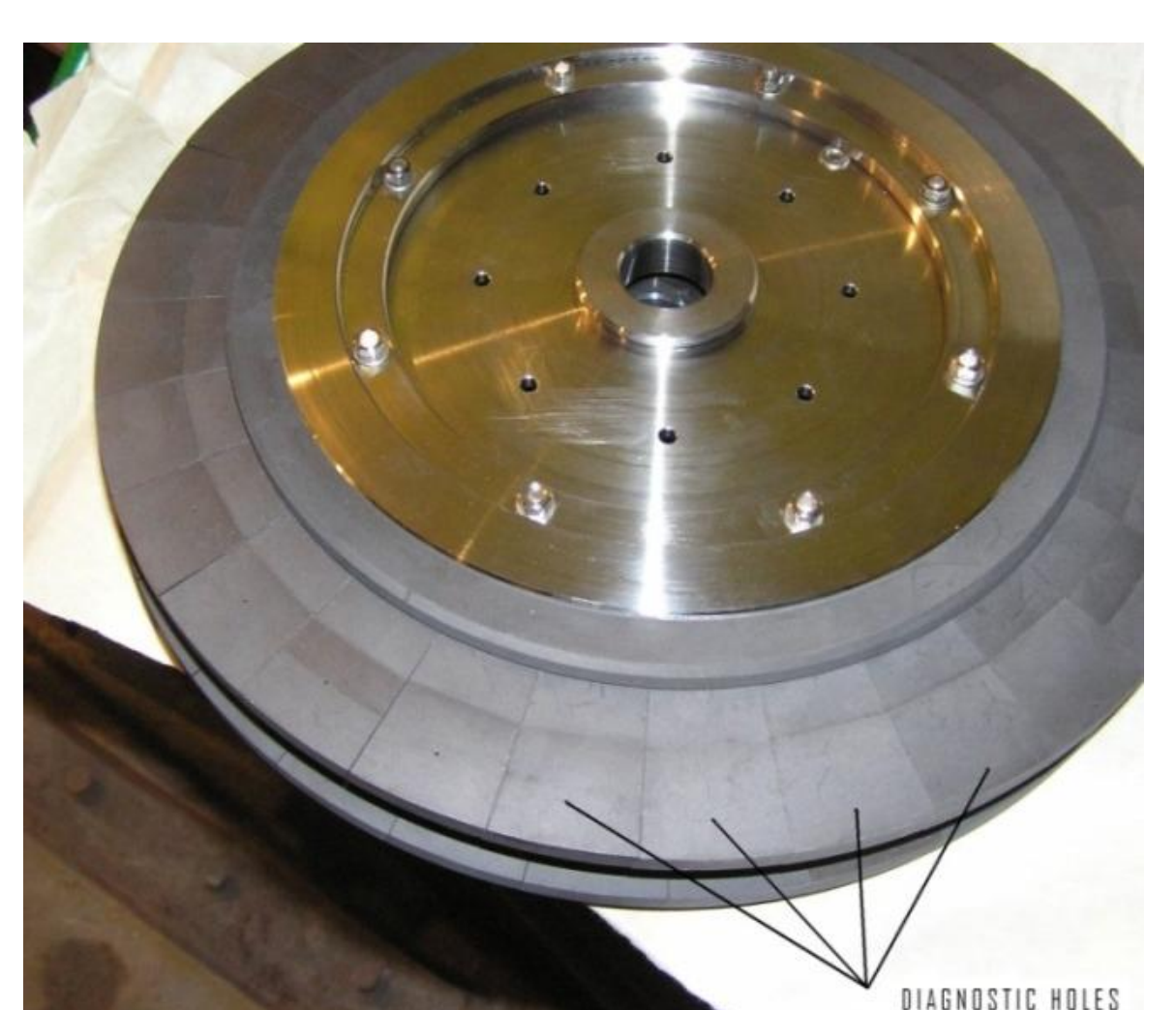

**Рис. 3.8.** Конструкция графитовой мишени нейтронного конвертора в сборке [98].

Используя программу ATOCAS и данные параметры, было рассчитано пространственное распределение скорости наработки смещенных атомов (среднее по вращениям), скорость внедрения ядер дейтерия и распределение мощности поглощенной дозы в углеродной мишени (программа EDVOXEL).

Средний проективный пробег дейтерия в углеродной мишени составил: $R_p$=0.525см стандартное отклонение от среднего σ=0.006 см (рис.3.9)

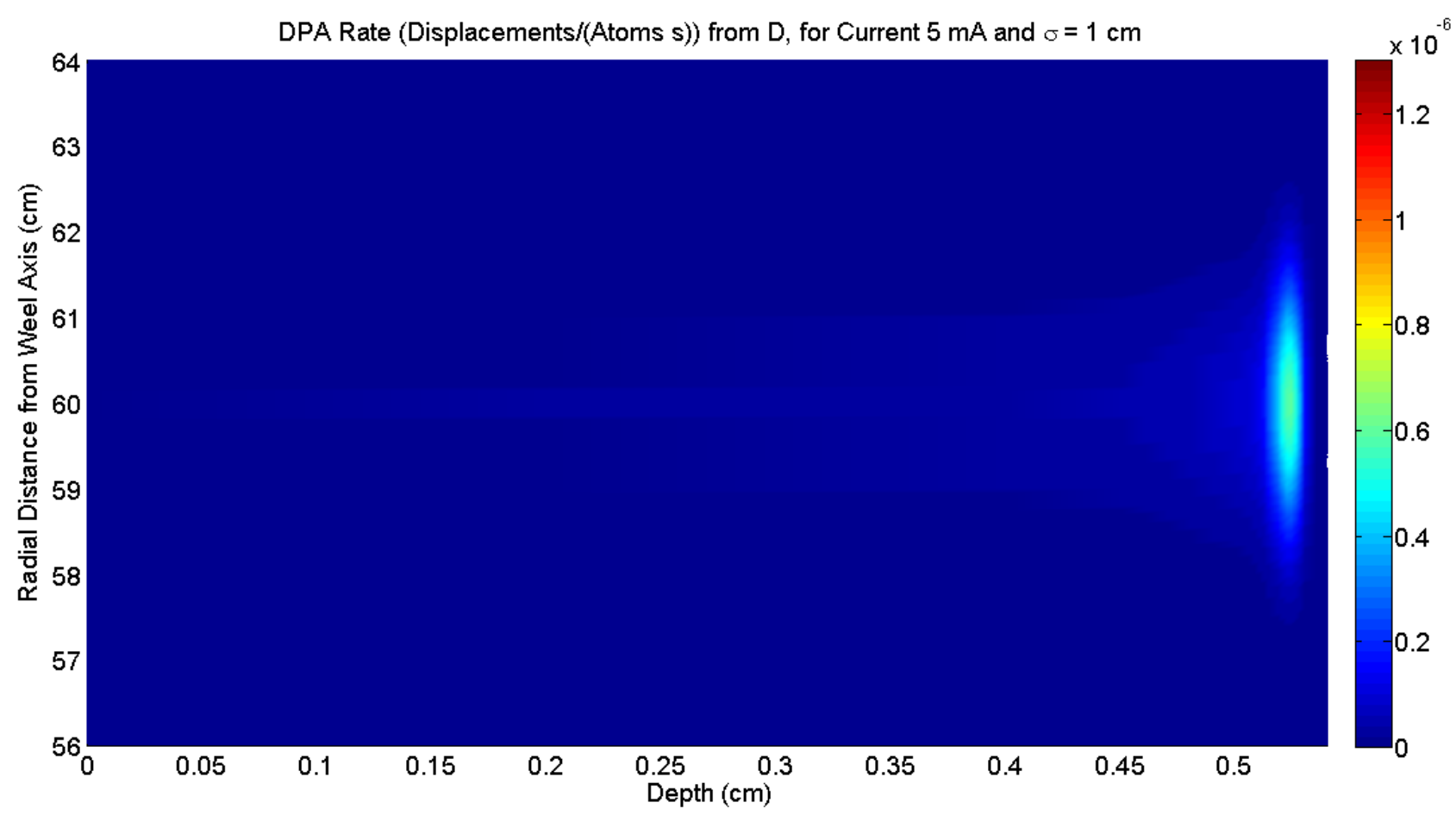

**Рис. 3.9.** Пространственное распределение скорости наработки смещенных атомов от ядер дейтерия. Среднее число смещенных ядер в расчете на одно ядро дейтерия 74 [98].

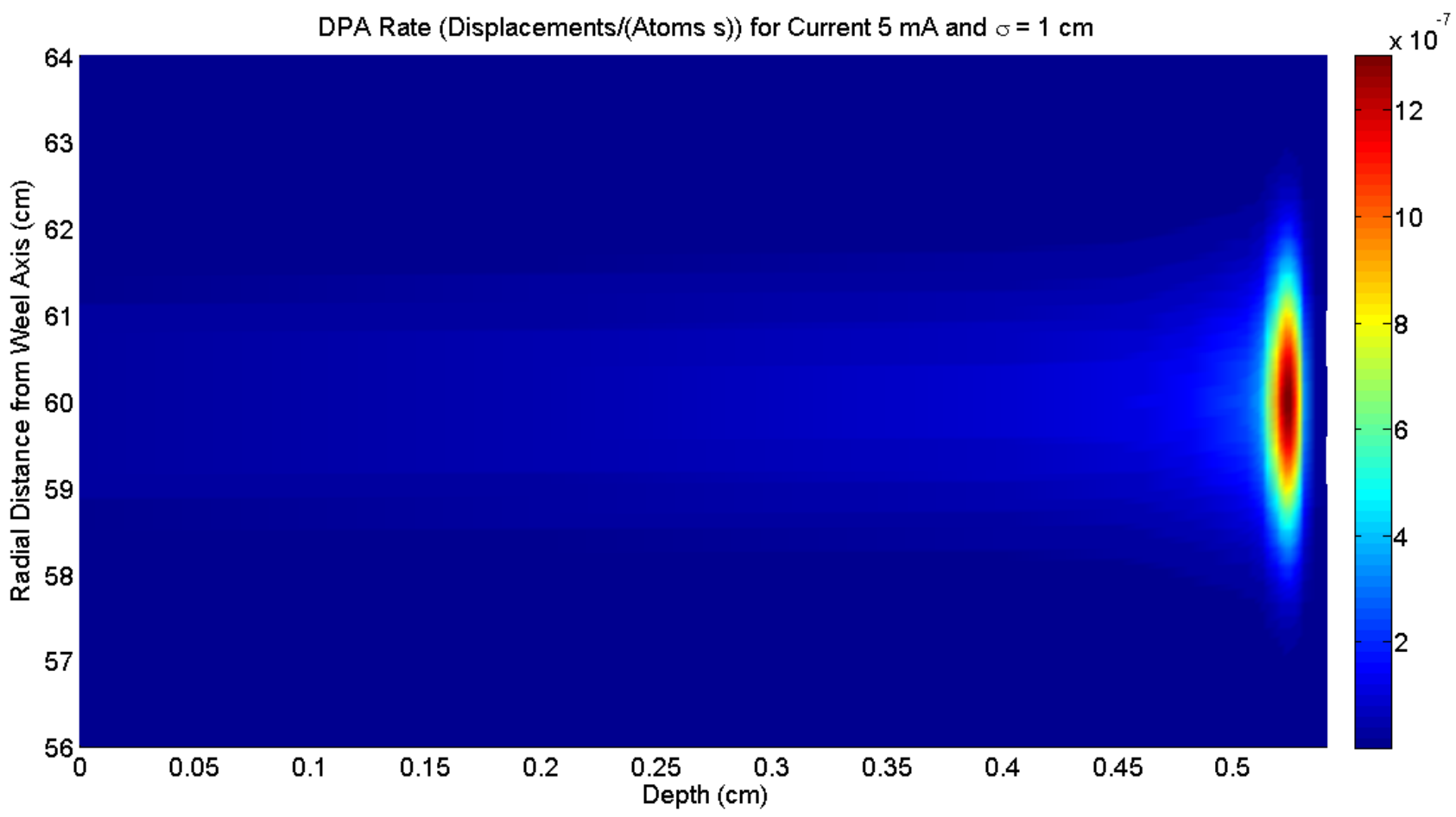

**Рис. 3.10.** Пространственное распределение скорости наработки смещенных ядер. Среднее число смещенных ядер в расчете на одно ядро дейтерия 188 (замещений 52) [98].

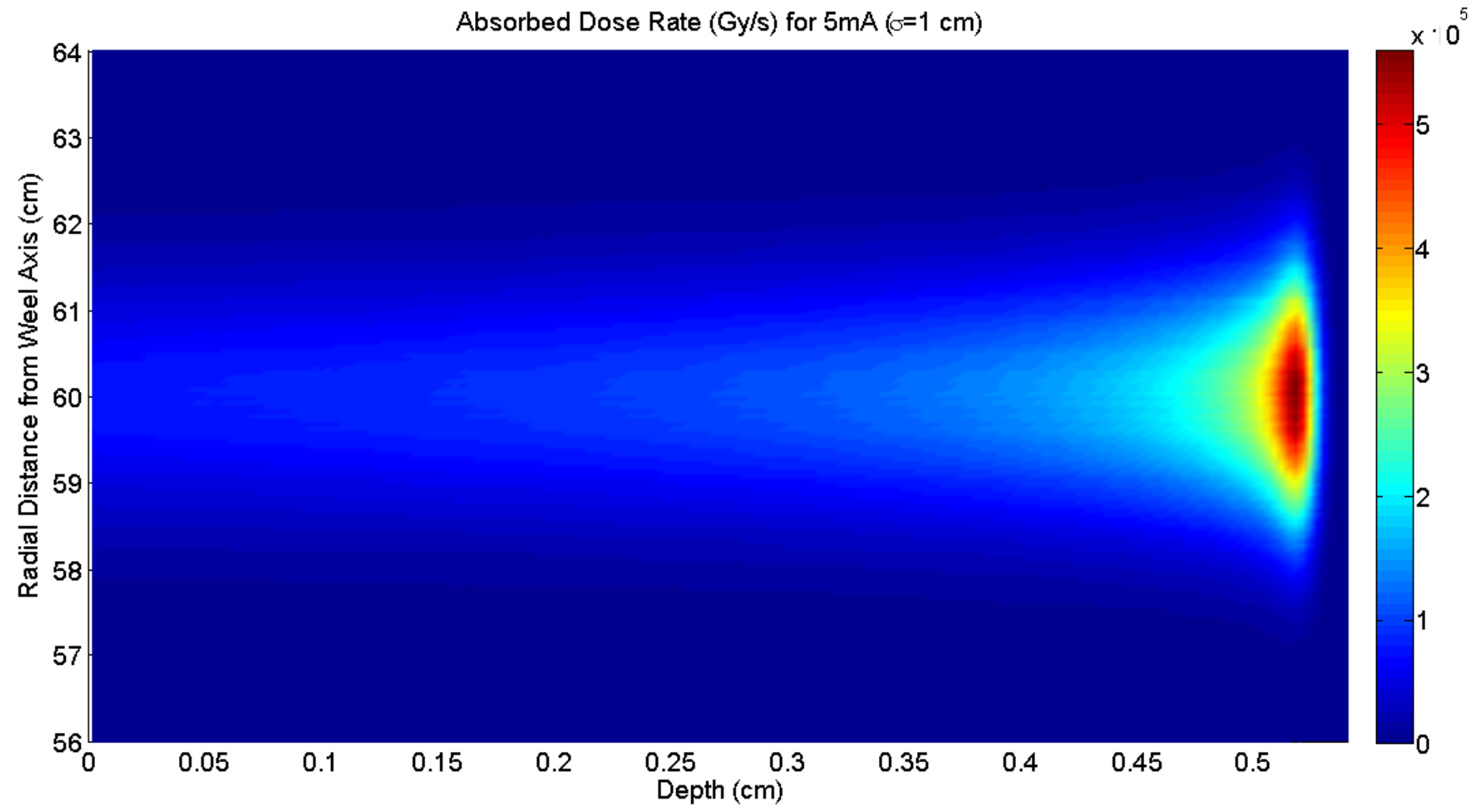

**Рис. 3.11.** Пространственное распределение мощности дозы в графитовой мишени [98].

Из результатов моделирования (рис. 3.10 и 3.11) видно, что пик мощности дозы находится несколько раньше пика наработки смещенных ядер, который, в свою очередь находится несколько раньше пика имплантированных ядер дейтерия. Максимальная концентрация ядер дейтерия находится очень близко к области максимального радиационного воздействия на материал (графит) и составляет ~2.3 $10^{-8}\ \dfrac{\textit{дейтронов}}{\textit{атом}\cdot\textit{сек}}$.

Максимальная скорость наработки смещенных атомов в графите составляет ~1.3 $\times 10^{-6}\ \dfrac{\textit{смещений}}{\textit{атом}\cdot\textit{сек}}$, при этом максимальная мощность дозы составляет 0.55 МГр/сек.

## 3.3. Подвижность водорода (дейтерия) в графите и аморфном углероде

Исходя из конструкции мишени, легко оценить, что за время работы ~10 000 часов в углеродной пластине в области накопления дейтерия его концентрация в три раза превысит концентрацию атомов углерода, если не учитывать подвижность дейтерия. Поэтому, с целью оценки выхода (диффузии) дейтерия из области его накопления при торможении в материале мишени были проведены расчеты подвижности в графите и аморфном углероде при рабочей температуре мишени ~1800$^0$С. Расчеты для аморфного углерода проводились исходя из предположения о возможной аморфизации графита в области подвергающейся наиболее сильным радиационным повреждениям.

В МД расчетах подвижности использовался потенциал REBO-2002 [100] параметризованный для расчетов систем C+H, включая возможность образования углеводородов.

3.3.1. Постановка задачи

В образец графита, содержащий ~30 000 атомов помещались 100 атомов водорода (дейтерия). Шаг интегрирования 0.1 фс. Поведение системы атомов водорода отслеживалось в течении 0.2 нс. Из анализа треков атомов водорода была вычислена их подвижность в заданных условиях.

На рисунке 3.12. показан срез образца графита с внедренными атомами водорода (красные атомы). Из рисунка видно, что из-за слабой связи между слоями графита и высокой температуры слои подверглись искажению.

Треки атомов водорода за время моделирования показаны на рис. 3.13.

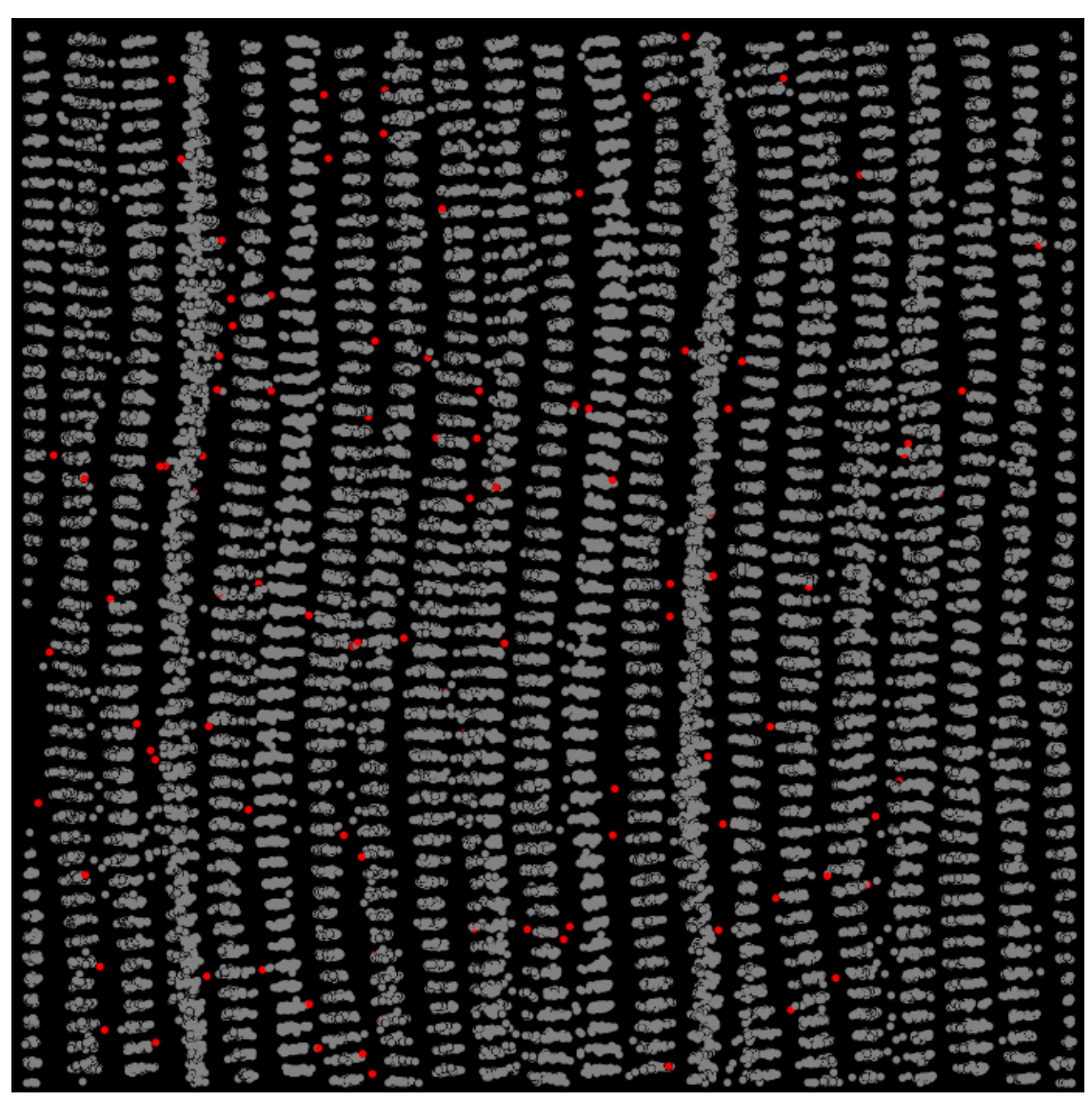

**Рис. 3.12.** Образец графита с внедренными атомами водорода (показаны красным цветом).

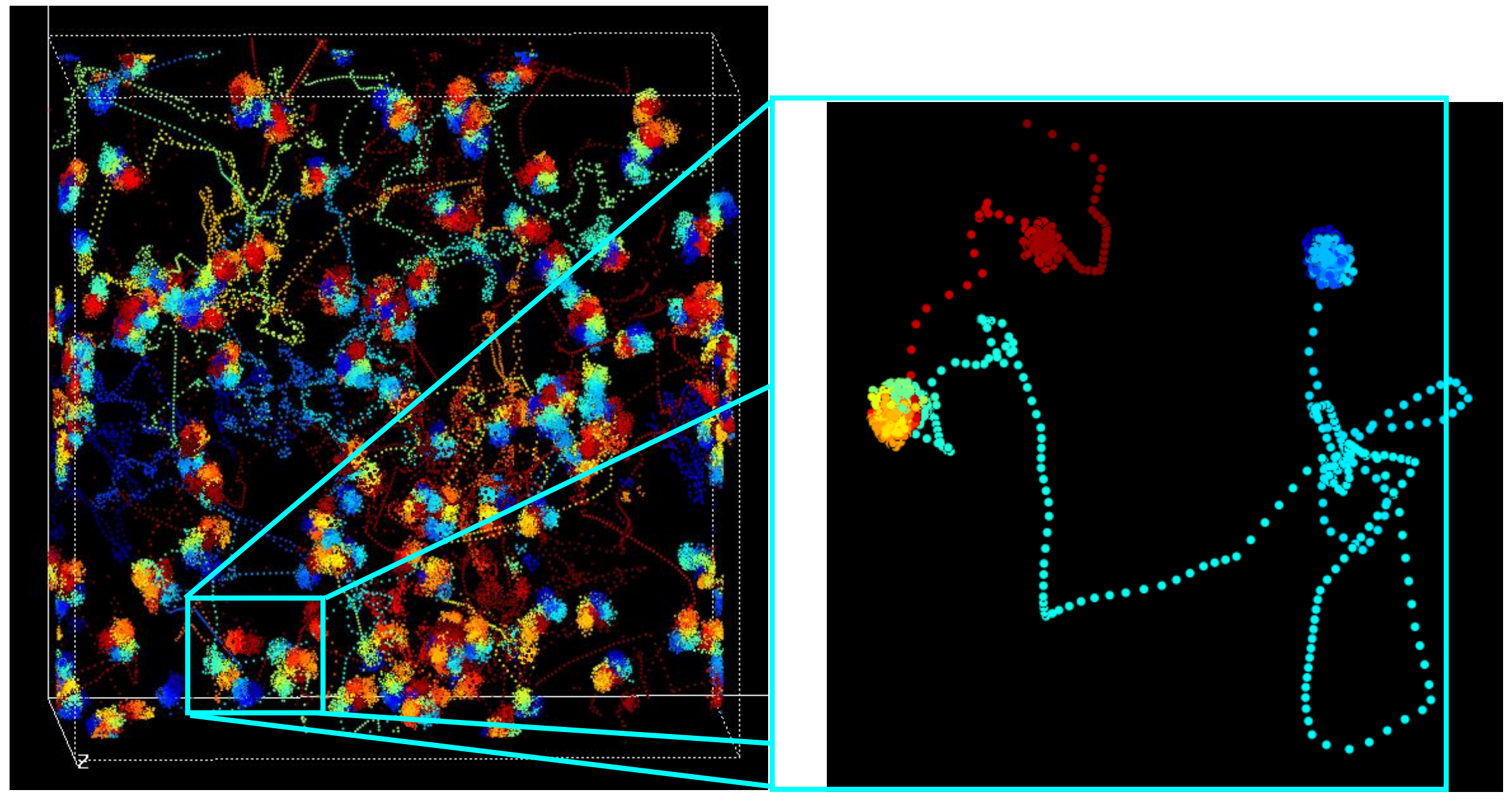

**Рис.3.13.** Треки атомов водорода в графите.

На рис. 3.13. справа показан участок трека одного атома водорода. Цветом отмечены времени – от красного (ранние моменты) к синему (поздние моменты). Характер диффузии следующий: атом довольно быстро движется в кристаллическом поле между соседними графеновыми слоями (участки траектории, отмеченные точками), в некоторые моменты образуются химические связи с атомами углерода, в связи с чем водород проводит относительно продолжительное время, фактически не перемещаясь относительно решетки («клубки»). Усредненное по всем атомам водорода расстояние между связанными состояниями ~1 нм. Время оседлой жизни ~15 пс. Время свободного перемещения между связанными состояниями пренебрежимо мало по сравнению с временем нахождения в связанном состоянии. Смещения между связанными состояниями происходят в плоскости графеновых слоев.

### 3.3.2. Подвижность водорода в аморфном углероде

Подвижность водорода в аморфном углероде значительно ниже, чем в графите. Фактически все время водород проводит в связанном состоянии, перемещения между связанными состояниями происходят на расстояние соответствующее среднему межатомному состоянию в аморфном углероде. Перемещения происходят в случайном направлении. Время оседлой жизни ~80 -100 пс.

На рисунках 3.14, 3.15 показаны положения атомов водорода в решетке аморфного углерода и треки в всех атомов водорода за время 200 пс. Полученная информация о подвижности водорода при рабочей температуре мишени будет использована в дальнейшем для оценки скорости выхода водорода и материала мишени.

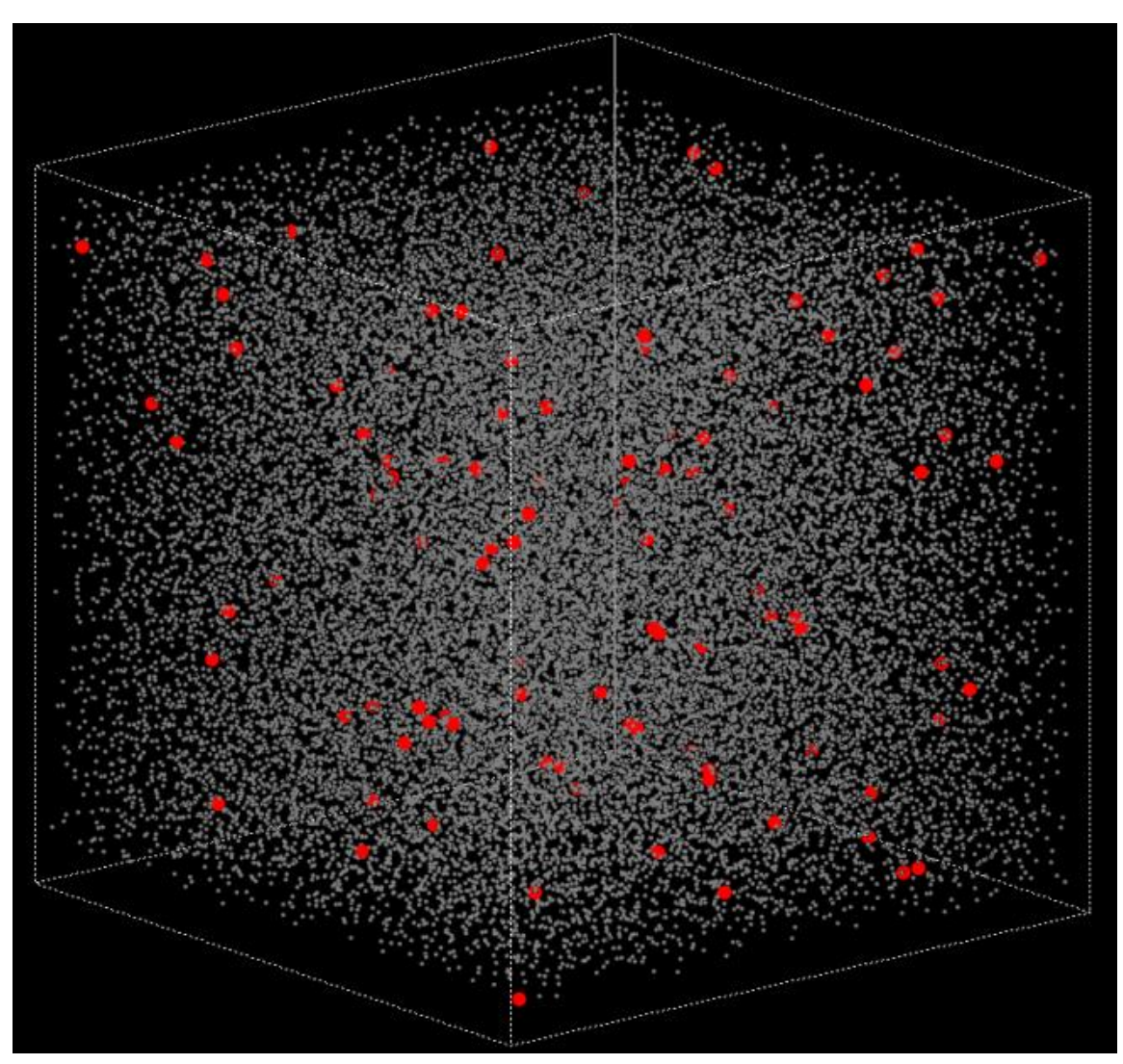

**Рис. 3.14.** Образец аморфного углерода с внедренными атомами водорода (показаны красным цветом).

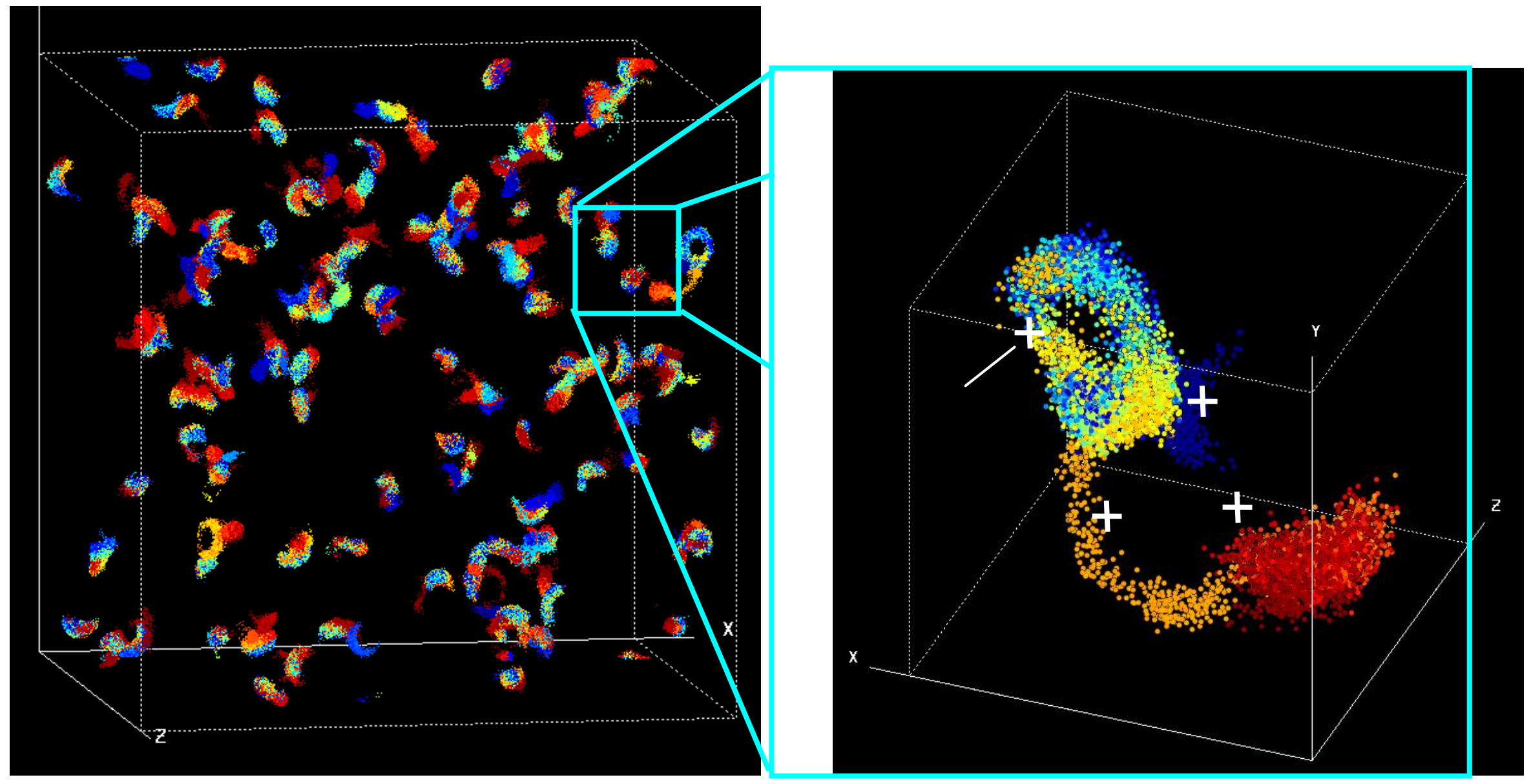

**Рис. 3.15.** Треки атомов водорода в аморфном углероде. Белыми крестами отмечены положения атомов углерода с которыми последовательно связан атом водорода.

Из полученных результатов можно оценить коэффициент диффузии дейтерия в монокристаллическом графите и аморфном углероде, а так же в поликристаллическом графите, предполагая, что межзеренные границы являются областями аморфного углерода:

$D_{Graphite} = 10^{-8}$ m²/c   $D_{Amorphous} = 10^{-10}$ m²/c   $D_{poly} \sim 10^{-10}$ m²/c

На рисунке 3.16 расчетная точка показана в сравнении с имеющимися экспериментальными данными.

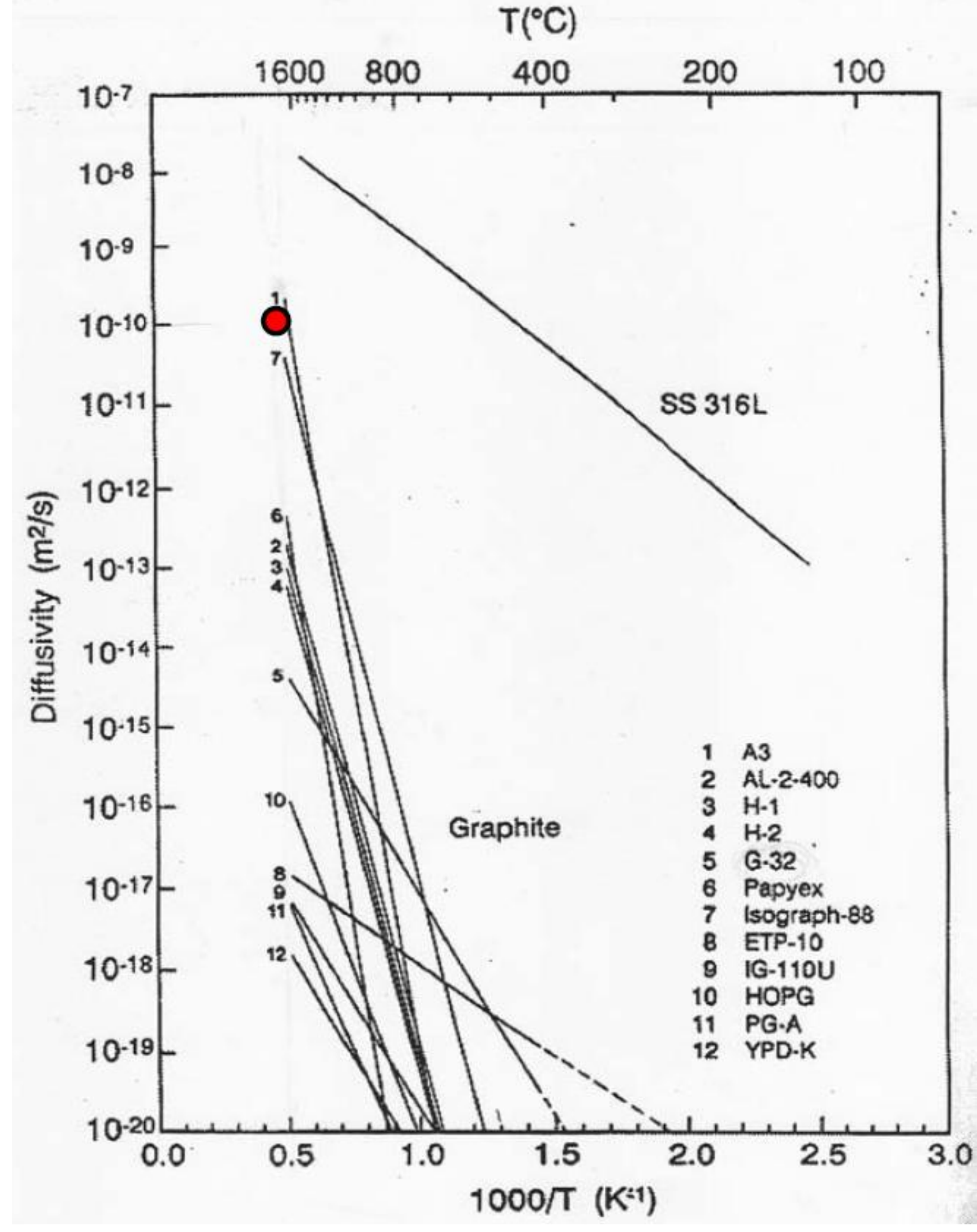

**Рис. 3.16.** Коэффициент диффузии в зависимости от температуры для различных графитов. Красная точка результат расчетов методом МД.

На приведенных ниже рисунках для различных значений коэффициента диффузии приведены расчетные данные по концентрации дейтерия в плоскости Брегговского пика в зависимости от времени (рис. 3.17) и распределению дейтерия по толщине графитовой пластины на момент времени 10 000 часов (рис 3.18).

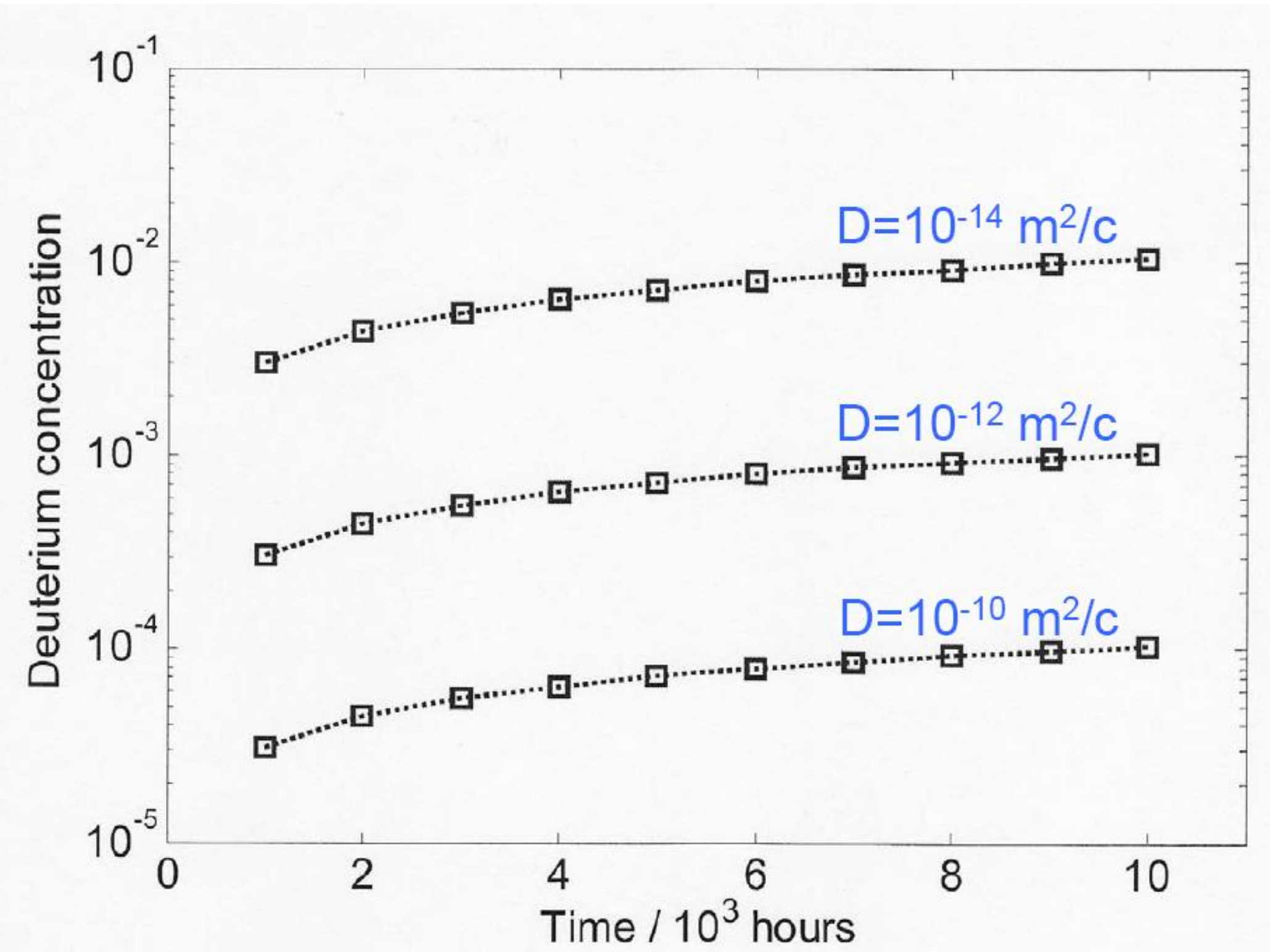

**Рис. 3.17.** Концентрация дейтерия в плоскости Бреговского пика в зависимости от времени для различных значений коэффициента диффузии.

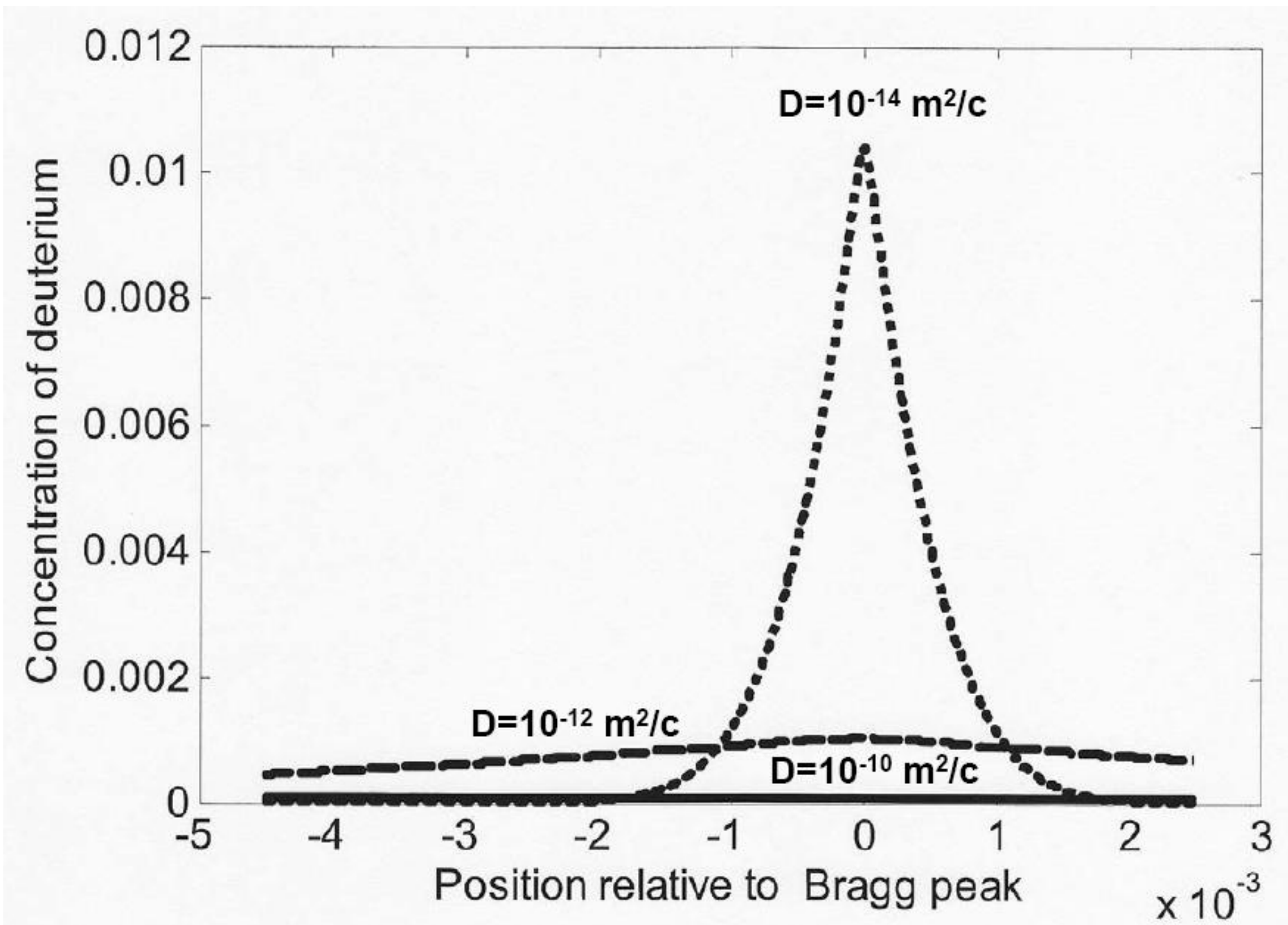

**Рис. 3.18.** Распределение концентрации дейтерия по толщине графитовой пластины на момент времени 10 000 часов для различных значений коэффициента диффузии.

## 3.4. Оценка скорости отжига области аморфизации в графите

При торможении высокоэнергетических частиц в материале мишени в области наибольшего выделения энергии (приблизительно на глубине 5 мм от бомбардируемой поверхности) возможно образование аморфной областей. Так как выше было показано, что скорость диффузии водорода (дейтерия) сильно зависит от структуры материала, встает вопрос о времени существования аморфных областей, вызывных радиационным повреждением. Эта информация также будет включена в модель накопления и выхода водорода из материала мишени.

С целью оценки времени отжига были проведены прямые МД расчеты в следующей постановке. В центре образца графита создавалась аморфная область размером 1/27 от объема образца (см. рис. 3.19)

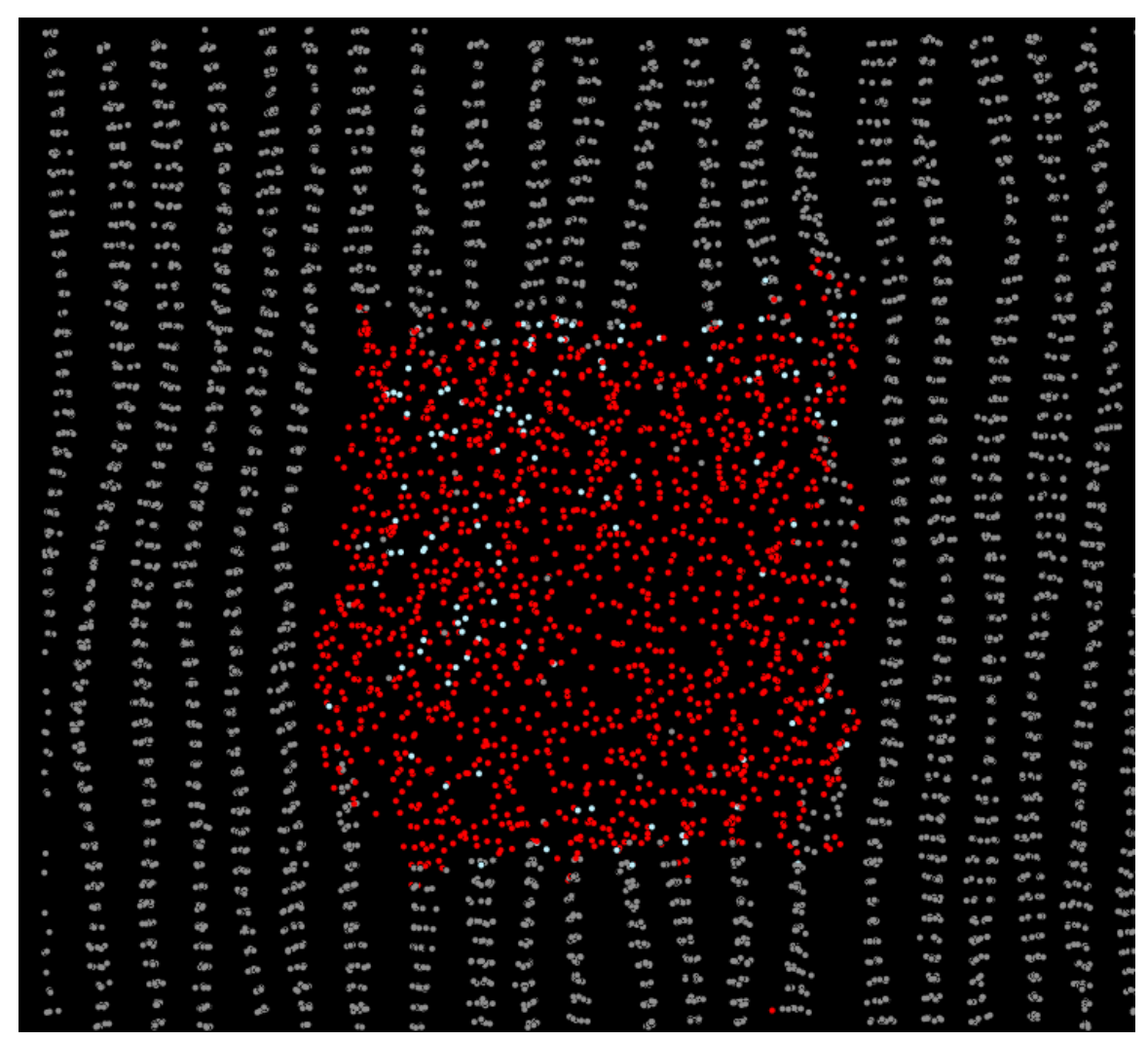

**Рис. 3.19** Аморфная область в центре графитового образца

Образец выдерживался при температуре $1800^0$С в течении 400 пс. Для определения структуры использовался Метод Адаптивного Шаблона [109], позволяющий определить принадлежность атома к кристаллической решетке или к аморфной структуре. Эволюция размера аморфной области во времени показана на рис.3.20. Аппроксимация результатов расчетов дает величину скорости отжига $\upsilon = A\ L/L_0\ {}^{2}$, где *A=1.5 ×10*$^{11}$ *1/c*, *L*$_0$*=4 нм.*

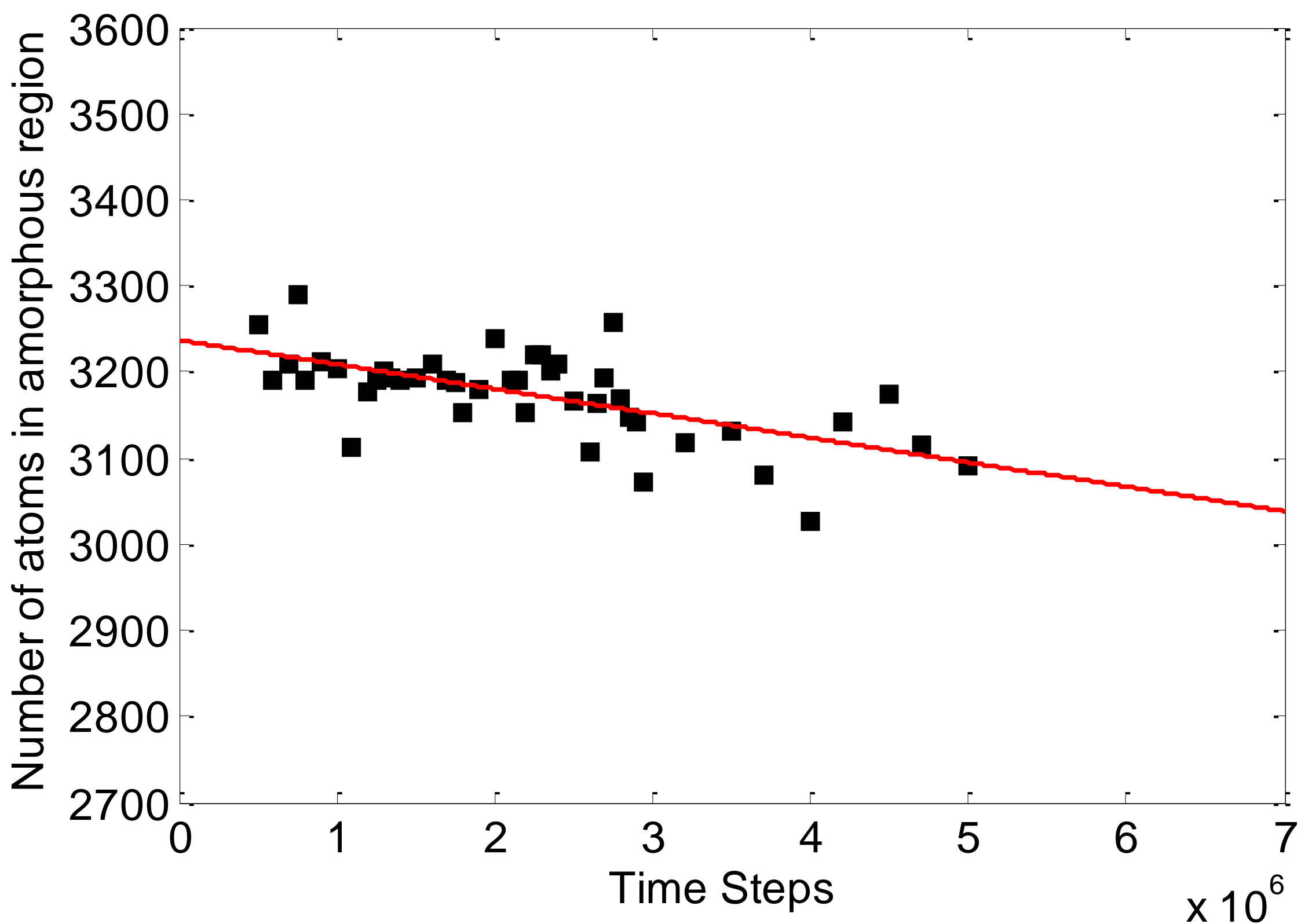

**Рис.3.20.** Эволюция размера аморфной области во времени.

***

Таким образом, используя методы молекулярной динамики и компьютерного моделирования, в этой главе была проведена оценка скорости наработки дефектов в углеродной мишени инициируемых пучком дейтонов с энергией до 40 МэВ. Оценено распределение скорости внедрения ионов дейтерия в мишени, а также распределение мощности дозы. Показано, что наибольшие концентрации дейтерия сосредоточены за областью максимального радиационного воздействия (область максимальной скорости наработки дефектов). Для оценки влияния радиационного воздействия дейтерия на материал мишени необходимо изучение вопросов отжига дефектов при рабочих температурах мишени. Особенно это важно, когда радиационная нагрузка на материал неравномерна из-за вращения мишени (радиационная нагрузка продолжительностью ~0.01 сек, сменяется ~0.99 сек отжига). Другим вопросом является особенности взаимодействия дейтерия с графитом: подвижность дейтерия, его способность образовывать химические соединения $D_2$ , $CD_4$ и способность к десорбции, как в атомарном состоянии, так и в составе молекул [93] при рабочих параметрах мишени.

ГЛАВА 4

# ОСОБЕННОСТИ ГРАФИТАЦИИ УГЛЕРОДНОГО МАТЕРИАЛА НА ОСНОВЕ ИЗОТОПА УГЛЕРОДА $^{13}C$

Для получения образцов нейтронных мишеней с повышенным содержанием изотопа $^{13}C$ для работы с протонным пучком необходимо исследование графитируемости изотопа $^{13}C$, который используется в качестве наполнителя углеродного композиционного материала. Способность к графитации изотопа $^{13}C$, в значительной степени, будет определять термопрочность материала в условиях воздействия пучков частиц с высокой энергией, а также технологические параметры его получения. При бомбардировке мишени протонами она нагревается до 2000°С [63-67]. В данной главе будут рассмотрены влияние температуры обработки порошка изотопа $^{13}C$ на электрические, рентгеноструктурные и текстурные параметры исходного наполнителя с высоким, до 98% содержанием изотопа углерода $^{13}C$. В дальнейшем, для краткости, этот материал будет обозначаться как исходный порошок $^{13}C$.

## 4.1. Свойства исходного порошка изотопа углерода $^{13}C$

Классификация углеродных структур по структурным составляющим была впервые предложена В.Г. Нагорным в работе [110]. В дальнейшем эта классификация понадобится при анализе исходного порошка изотопа углерода $^{13}C$.

В работе [111], было проведено определение зольности усредненного образца и спектральный анализ примесных элементов исходного порошка, а также оценены поверхностные свойства материала. Зольность усредненного образца определяли методом сжигания при температуре 850°С, спектральный анализ примесных элементов проводили на спектрометре ДФС-8, поверхностные свойства оценивали газохроматографическим методом по низкотемпературной адсорбции аргона при температуре −196°С на установке Газометр ГХ-1. Данные представлены в табл. 4.1.

Порошок изотопа углерода $^{13}C$ содержит до 1 масс. % железа, содержание остальных элементов относительно небольшое (табл. 1), имеет высокую удельную поверхность, соизмеримую с удельной поверхностью активированных углей (800 $м^2/г$). Например, удельная поверхность изотропного пекового кокса, используемого в настоящее время для получения конструкционных графитов составляет 0,2 $м^2/г$ (для более мелкой фракции –250 мкм).

**Таблица 4.1.** Свойства исходного порошка изотопа $^{13}C$ [4.2].

| Свойства изотопа | Величина |
|---|---|
| Общая зольность, масс. % | 1,7 |
| Содержание примесей, масс. % (спектральный анализ): | |
| железо | до 1 |
| алюминий | $1{,}5\cdot10^{-2}$ |
| магний | $5{,}4\cdot10^{-3}$ |
| марганец | $3{,}6\cdot10^{-2}$ |
| кремний | $2{,}2\cdot10^{-2}$ |
| бор, никель, хром | следы |
| Удельная поверхность, $м^2/г$ (фракция −400/+315) | 360 |

## 4.2. Влияние температуры обработки на удельное электросопротивление

Для измерения удельного электросопротивления (четырехзондовым методом) использовали узкую фракцию порошка изотопа −400/+315 мкм. Обнаружена резко экстремальная зависимость величины удельного электросопротивления от температуры обработки (рис. 4.1). Максимум при температуре 2000°С, по всей видимости, обусловлен разупорядочением структуры, поскольку без этого невозможен поворот слоев в азимутальной плоскости для перестройки турбостратной структуры в графитовую. В работе [112] отмечается, что перед началом интенсивного образования трехмерной структуры как при атмосферном, так и при повышенных давлениях происходит незначительное увеличение среднего межслоевого расстояния и одновременно с этим несколько увеличивается дефектность структуры.

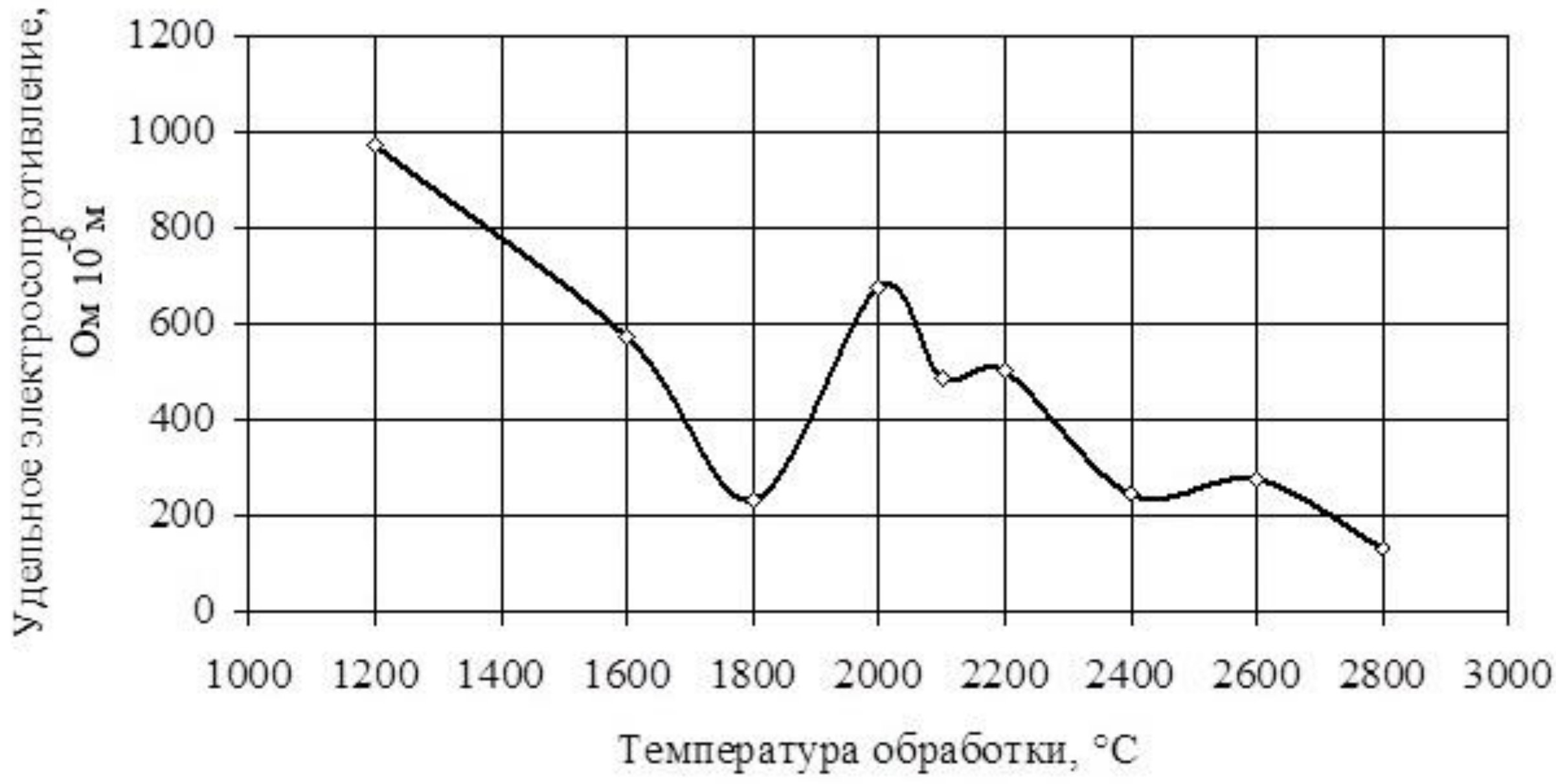

**Рис. 4.1.** Зависимость удельного электросопротивления порошка изотопа $^{13}C$ от температуры обработки [111].

## 4.3. Изменение свойств углеродных материалов в предкристаллизационный период графитации

Известно, что в предкристаллизационном периоде графитации углеродного вещества происходит некоторое разупорядочение его структуры. Одним из первых этот период графитации выделил в самостоятельный период структурных преобразований углеродного вещества В.И.Касаточкин [113,114]. Он показал, что в интервале температур обработки 1600–2000°С энтальпия углеродного материала имеет максимум. При температурах обработки выше 1600°С энтальпия углеродного материала возрастает. На этой стадии в углеродном материале проходят эндотермические процессы, приводящие к поглощению энергии веществом. В этой же области температур остаточная энтропия прессованных образцов крекингового, пиролизного и пекового кокса имеет максимальное значение [35].

Обнаруженное возрастание энтропии на предкристаллизационной стадии свидетельствует о глубокой перестройке структуры углеродного материала, которая обусловлена не только процессами упорядочения и формирования турбостратной структуры, но и, наоборот, процессами частичного разупорядочения.

В этом интервале температур имеют максимумы термо-э.д.с. и коэффициент Холла, несколько возрастает замкнутая пористость [18]. В работе [115] исследованы характеристики стеклоуглерода и различных коксов методами малоуглового рентгеновского рассеяния. Установлено, что предкристаллизационный период характеризуется также ростом степени искажений, уровня внутренних напряжений, причем, чем труднее графитируется данный кокс, тем сильнее выражено его напряженное состояние в предкристаллизационном периоде.

В.И.Касаточкин [114] выдвинул предположение, что гомогенно-графитирующийся углерод на предкристаллизационной стадии термообработки характеризуется процессом деструкции менее прочных боковых связей с сохранением жесткого полимерного каркаса. В целом, образец углерода, подвергнутый термической обработке в указанном интервале температур, претерпевает качественные и количественные преобразования межатомных связей.

Столь глубокие преобразования в углеродном веществе, термообработанном при температурах предкристаллизации, не могут не сказаться на процессах его взаимодействия с химически активными расплавами [116].

В работе [115] было показано, что минимальные уровни внутренних напряжений и замкнутой пористости характерны для игольчатых коксов, максимальные – для стеклоуглерода. Аналогичная зависимость получена и после взаимодействия углерода с эвтектическим сплавом NiMn.

Минимальное количество графитированной фазы получено при температуре обработки 2000°С игольчатого кокса, максимальное – для КРФС

2000 (кокс резольной фенолформальдегидной смолы с температурой обработки 2000°С) [116].

Таким образом, ранее было установлено, что углеродные материалы, сформированные в предкристаллизационном периоде графитации, характеризуются некоторым разупорядочением структуры, появлением ненасыщенных валентных связей, что в свою очередь, оказывает существенное влияние на процессы их взаимодействия с расплавами металлов. Науглероживание расплавов такими аномальными структурами приводит к необычному повышению концентраций насыщения расплава углеродом. Так, концентрация насыщения расплава никеля коксом КРФС 2000 достигает величины 5,0 масс.%, что почти в 2 раза превышает концентрацию насыщения искусственным графитом (2,7 масс. % при 1500°С).

### 4.4. Разупорядочение структуры исходного порошка изотопа углерода $^{13}C$ в предкристаллизационный период, механизм графитации углеродных материалов

Предположение о разупорядочении структуры углеродного материала в предкристаллизационный период достаточно убедительно подтверждают данные по изменению межслоевого расстояния (рис. 4.2), высоты кристаллитов (рис. 4.3), степени графитации (рис. 4.4), рентгеновской плотности (рис. 4.5). Температура начала разупорядочения составляет 1800°С. В экстремальной точке (температура 2000°С) происходит увеличение межслоевого расстояния, уменьшение высоты кристаллитов, степени графитации и рентгеновской плотности. Отмечен значительный рост высоты кристаллитов после температуры обработки 2600°С. Наблюдающиеся на кривой изменения удельного электросопротивления изотопа $^{13}C$ максимумы при температурах 2200 и 2600°С, вероятно, свидетельствуют о дальнейшем совершенствовании кристаллической структуры углерода. Важно подчеркнуть, что данный процесс происходит дискретно, посредством чередования нескольких циклов разупорядочения – упорядочения. В работе [112] приведены расчетные данные методом атом-атомного потенциала, которые свидетельствуют о том, что для каждого размера слоя при их парном ротационном взаимодействии имеется свой определенный угол, отвечающий минимуму потенциальной энергии, который зависит от количества атомов в ансамбле и угла поворота графенового слоя.

В приведенной работе количественно показано, что энергетические минимумы отвечают большеугловым ротационным границам. Поворот одной плоскости относительно другой, выводящий границу из симметричного положения, расстраивает хорошее совпадение атомов в этой границе. Расчет энергии ротационной границы показывает, что в зависимости от угла поворота соседних слоев, кроме основного, глубокого минимума энергии, соответствующего графитовому состоянию, имеют место ряд частных минимумов, которые отвечают устойчивым промежуточным конфигурациям. Подобная зависимость состояния соседних слоев, очевидно, может служить причиной предположения о том, что общая энергия, необходимая для преобразования ансамбля атомов в графит, должна суммироваться изо всех частных состояний пар слоев, соответствующих углам рационального двойникования, то есть достаточно симметричной конфигурации расположения атомов в соседних слоях. Этим, по всей видимости, можно объяснить растянутый по температуре процесс графитации, который для графитируемых веществ начинается при 1600°С и еще далек от завершения при температурах порядка 3000 °С.

Нагорным В.Г. показано [110] , что в искусственных графитах могут существовать четыре типа структур: $G_0$, $G_1$, $G_2$, $G_3$. Расчеты, проведенные на ЭВМ, показали, что межслоевые расстояния, соответствующие различным трансляционным сдвигам при получении этих структур соответственно равны: $d_{G0}$ = 0,365 нм, $d_{G1}$ = 0,343 нм, $d_{G2}$ = 0,3354 нм (графитовая структура), $d_{G3}$ = 0,334 нм. Соответствующие трансляционные сдвиги равны: $a_0$ = 0 нм, $a_1$ = 0,71 нм, $a_2$ = 0,142 нм, $a_3$ = 0,213 нм. В структуре $G_0$ атомы каждого последующего слоя находятся под атомами предыдущего слоя без смещения.

Структура $G_1$ называется «турбостратной» и, как правило, образуется при температурах, близких к температуре начала трехмерного упорядочения.

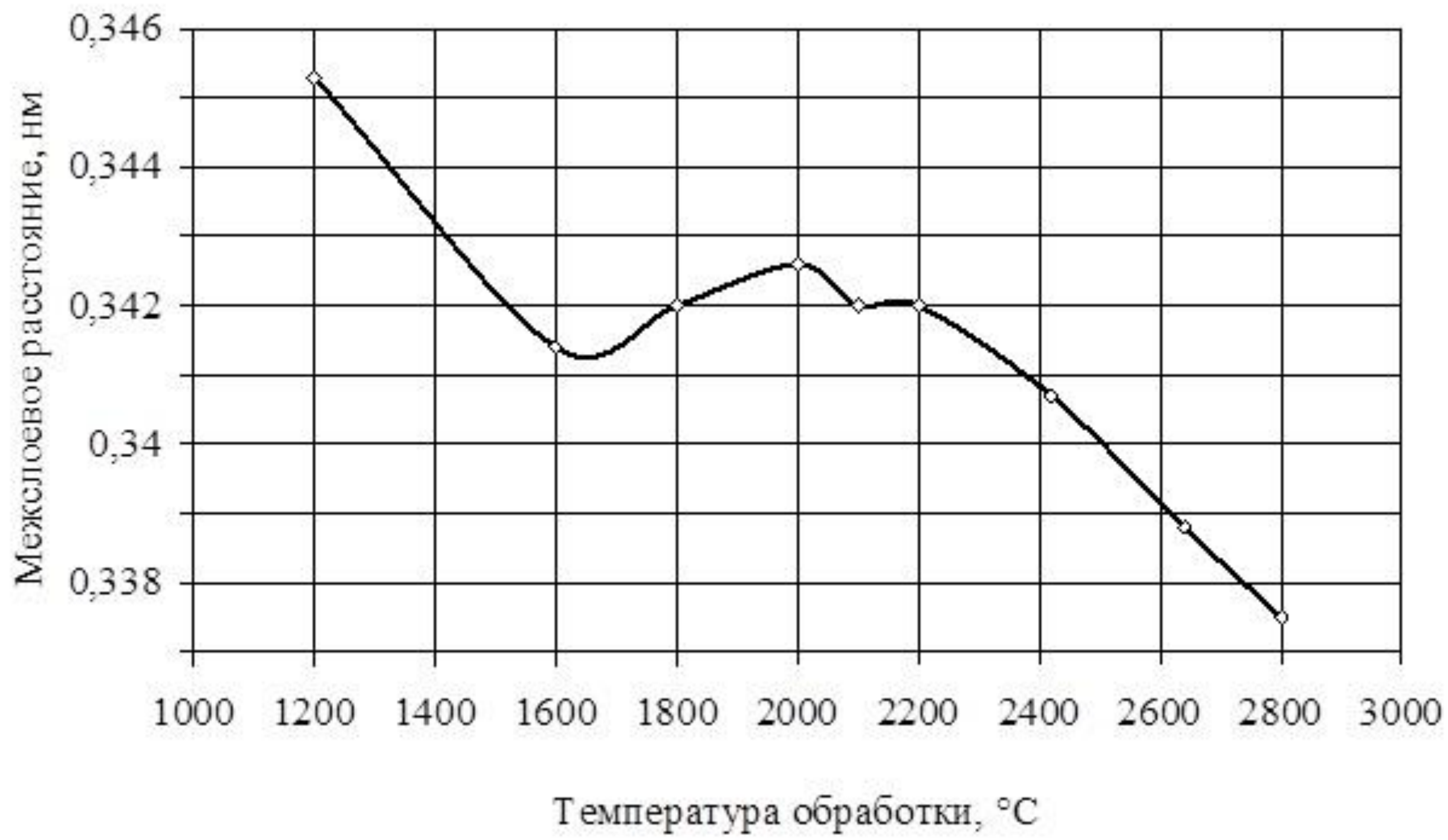

**Рис. 4.2**. Влияние температуры обработки на межслоевое расстояние материала на основе изотопа $^{13}C$ [111].

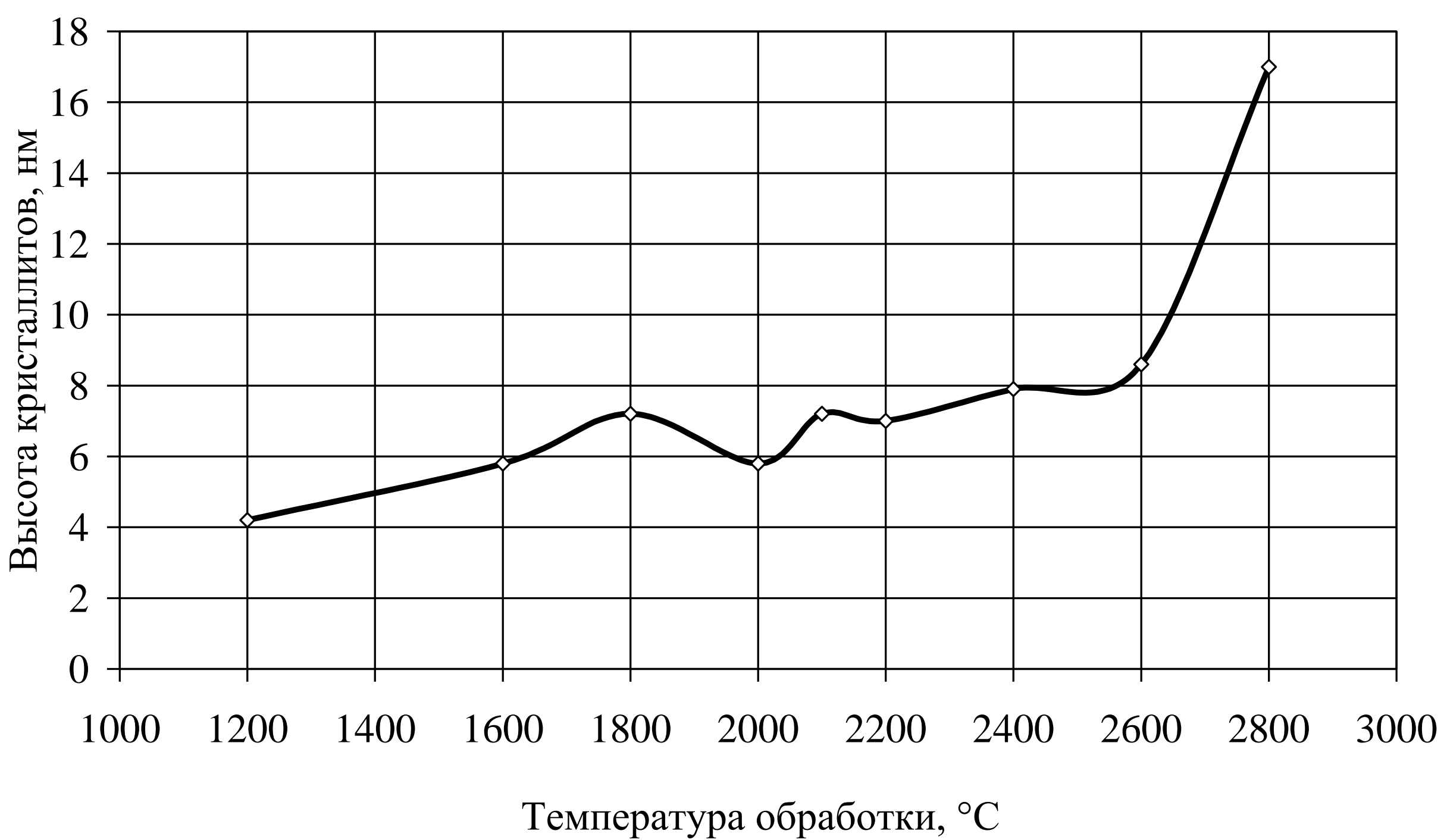

**Рис. 4.3.** Зависимость высоты кристаллитов материала на основе изотопа $^{13}C$ от температуры обработки [111].

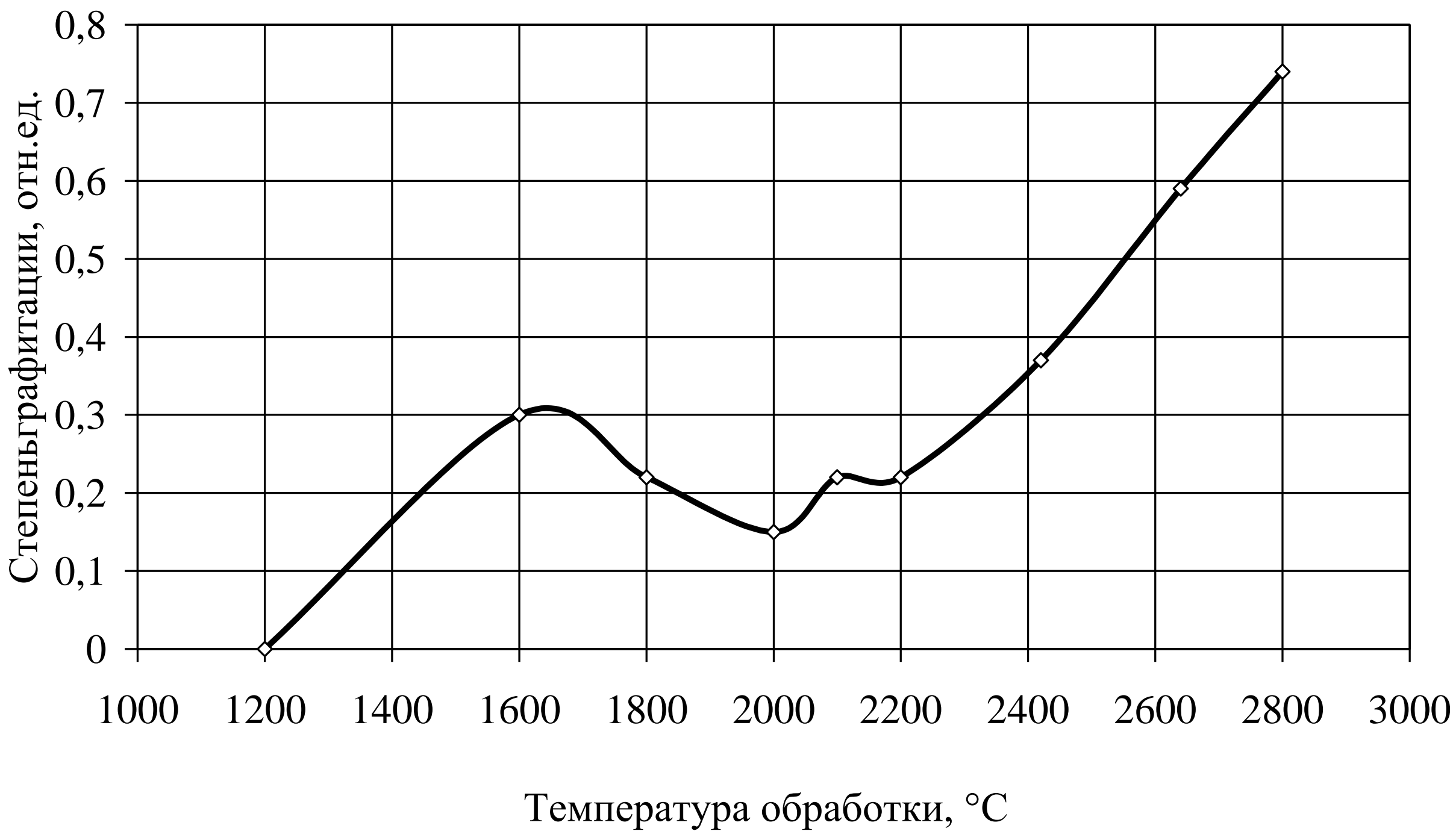

**Рис. 4.4**. Влияние температуры обработки на степень графитации изотопа $^{13}C$ [111].

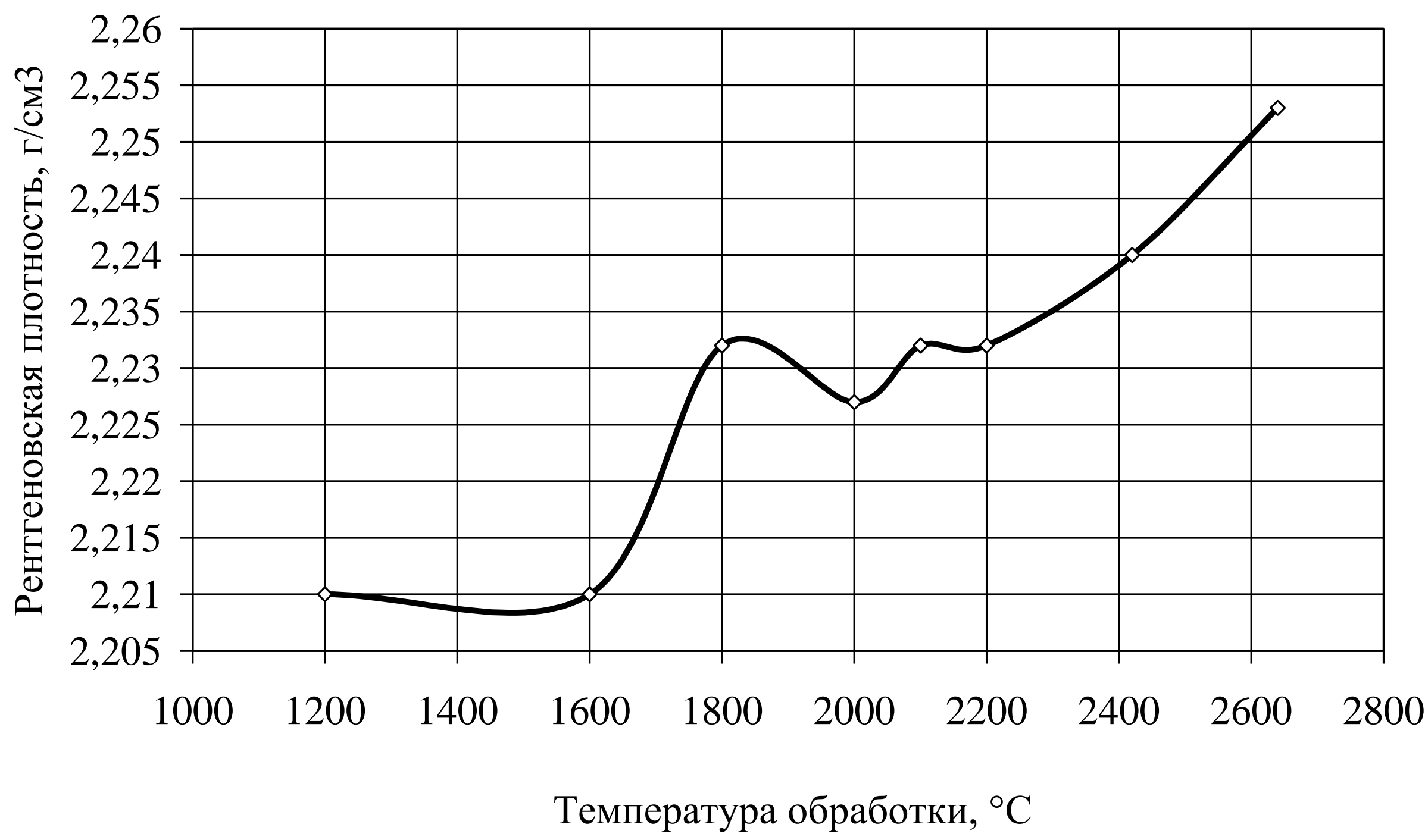

**Рис.4.5.** Влияние температуры обработки на рентгеновскую плотность композита на основе изотопа углерода $^{13}C$ [111].

Установлено, что структура этого углерода является более сложной, чем считалось ранее. Поворот соседних плоскостей, вероятно, составляет не случайный, а определенный рациональный угол для пары плоскостей с набором углов во всей упаковке. Расчеты интерференционных функций показали, что структуру турбостратного углерода можно представить в таком виде, когда центры поворотов соседних плоскостей расположены у соседних пар статистически. В противном случае на интерференционной функции наблюдаются экстрарефлексы, которые не подтверждаются экспериментальными данными. Структура $G_2$ является графитовой. Атомы последующего слоя в ней находятся под центрами шестиугольника, образованного атомами предыдущего слоя.

Несмотря на то, что межслоевое расстояние в структуре $G_3$ меньше, чем в $G_2$, значительное смещение атомов второго слоя не дает возможности считать ее графитовой. Структура $G_3$ была предсказана Китайгородским А.И. [117], кроме того, в данной работе были сделаны выводы о механизме графитации как процессе совершенствования структуры за счет термического сброса двойниковых образований, соответствующего сбросу поверхностной энергии.

Анализ энергетического состояния кристаллической решетки графитоподобных структур установил, что устойчивые межслоевые расстояния для выделенных подструктур (структурных составляющих) зависят от диаметра ароматического слоя. От слоя диаметром 5 нм до почти бесконечного слоя параметр $d_{002}$ изменяется в следующих пределах: для подструктуры $G_0$ : 0,364 – 0,341 нм, для подструктуры $G_1$: 0,344 – 0,340 нм, для подструктуры $G_2$: 0,343 – 0,335 и для подструктуры $G_3$: 0,344 – 0,338 нм.

В работе [118] было экспериментально подтверждено образование в четырех видах коксов с различной надкристаллитной структурой и термообработанных в интервале температур 1200 – 2600 °С структурных политипов углерода $G_0$, $G_1$, $G_2$,, которые были теоретически предсказаны Нагорным В.Г., а так же было высказано предположение, что межплокостное расстояние для каждой из подструктур лежит в определенном интервале значений. На основе анализа температурной зависимости хаотической составляющей микроискажений предложен механизм фазовых переходов в коксах при термообработке, включающий перестройку структуры $G_0$ в структуру $G_1$ в температурном интервале 1400 – 1600°С и перестройку структуры $G_1$ в $G_2$, в температурном интервале 1800 – 2600 °С.

Эксперименты, проведенные в высокотемпературной рентгеновской камере, которая позволяла работать при температурах от комнатной до 2800°С и проводить нагрев с очень большими скоростями (до 1500 °С/сек), позволили обнаружить несколько стадий процесса графитации, одна из которых протекает за очень короткий промежуток времени (секунды) при энергии активации порядка 8 кДж/моль [119].

Быстрое охлаждение анизотропного пекового кокса от температуры 2600°С при малой выдержке до температуры около 100°С показало, что основная масса кокса состоит из турбостратного углерода (эксперименты проведены автором данной главы). Этот факт доказывает, что процесс графитации обязательно проходит через стадию образования турбостратного углерода.

При исследовании процесса графитации на 3-х углеродных материалах с различной надкристаллитной структурой (кокс фенолформальдегидной смолы, сферолитовая составляющая кокса КНПС, красноводский игольчатый кокс) и способностью к совершенствованию структуры установлено, что все материалы при температуре 2000°С по значению межплоскостного расстояния представляют собой одну и ту же кристаллическую структуру – турбостратный углерод (межплоскостное расстояние равно 3,44 Å). Экспериментальные данные представлены на рисунке 4.6.

Таким образом, процесс термического совершенствования кристаллической структуры углеродных объектов (графитация) по мнению Нагорного В.Г. [110] невозможно представлять как обычный процесс рекристаллизации, хотя он и включает некоторые черты последнего.

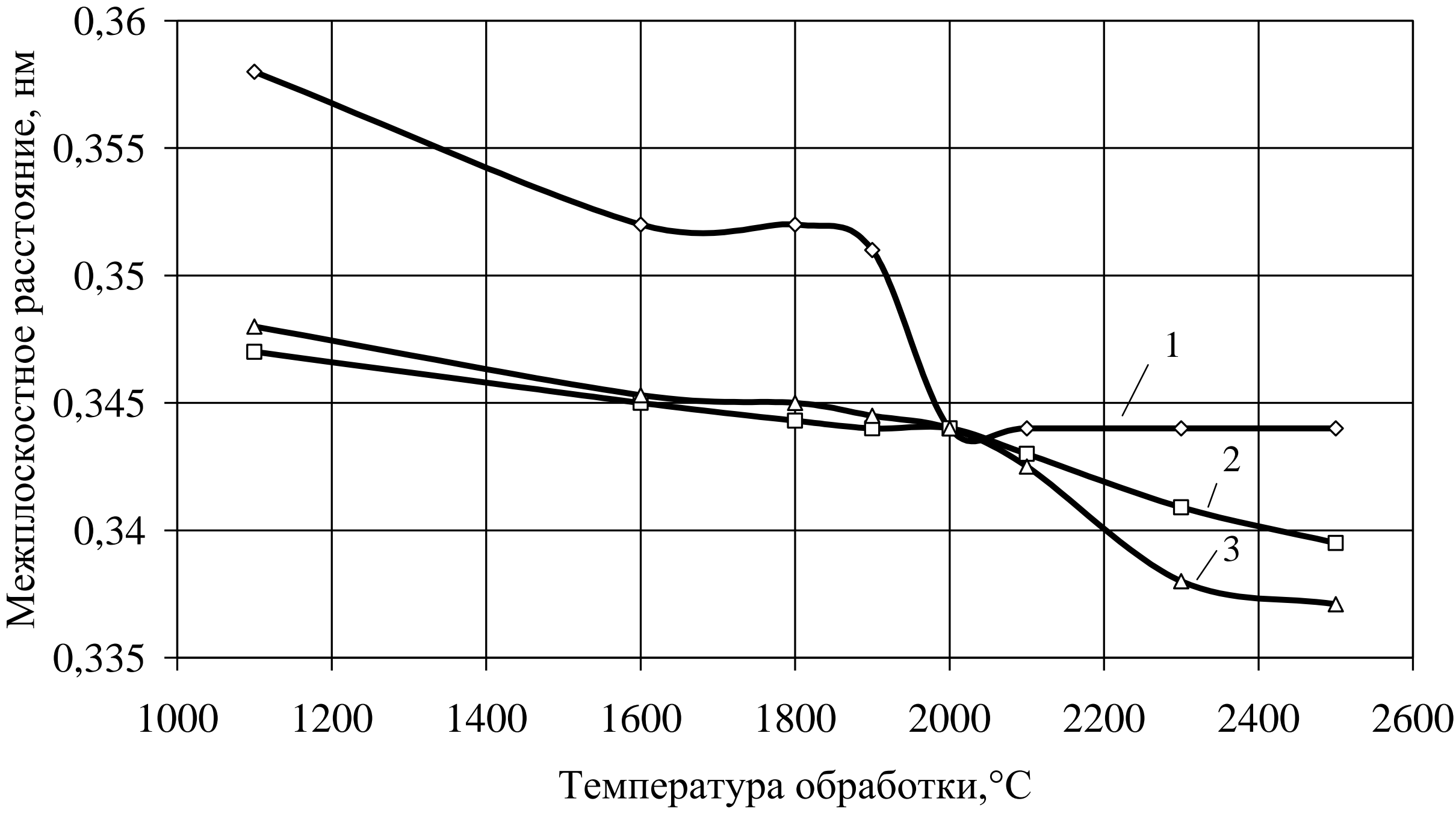

**Рис.4.6.** Зависимость межплоскостного расстояния исходных коксов от температуры обработки: 1 – кокс ФФС;2 – сферолитовая составляющая КНПС;3 – игольчатый кокс [120].

В исходном состоянии углеродные графитируемые материалы представляют собой образования из достаточно крупных по величине углеродных гексагональных слоев с большим количеством изгибов, вакансий, групп вакансий и существенного количества гетероатомов. Отмечаются и плоские участки нескольких слоев, расположенных примерно параллельно, образующие области когерентного рассеяния, которые на этом этапе часто неправильно называют кристаллитами. Группы плоскостей могут располагаться либо в виде линейно-протяженных систем (струйчатые составляющие), либо в виде сферических образований (сферолитовая структура). В начале термического воздействия происходит постепенное выпрямление отдельных слоев, миграция вакансий и удаление гетероатомов, что приводит к увеличению параллельности в группах слоев и определенной стабилизации по местоположению краевых и винтовых дислокаций. Причем краевые дислокации в системах упаковок располагаются статистически. Предполагается, что для графитруемых углеродов это происходит до температур порядка 1300 – 1500°C.

Следующий этап включает процесс перемещения полигонизационной стенки дислокаций с образованием мало- и большеугловых границ и протекает, вероятно, в предкристаллизационный период (~ 2100°C), когда дальнодействующие поля дислокаций полигонизационной сетки приводят к значительному увеличению внутренних напряжений, что отчетливо выявляется в эксперименте по графитациии изотопного материала. Полигонизация* дислокаций, вызываемых термическими напряжениями, приводит к образованию внебазисных двойников. Стабилизация двойниковых образований на этом этапе происходит путем скольжения вследствие аккомодации, приводящей к снятию напряжений на краях двойника.

Дальнейшим этапом совершенствования структуры является термический сброс двойниковых образований, отвечающий сбросу поверхностной энергии. При этом вначале происходит уменьшение углов внебазисного двойникования (выпрямление систем плоскостей), затем – процесс взаимодействия винтовых дислокаций с вакансиями в системе пар плоскостей, что приводит к перераспределению по количеству вышеуказанных последовательностей ($G_i$ ) в соседних парах плоскостей.

Сброс двойниковых образований носит в основном характер мартенситных превращений. Следует помнить, что выпрямление двойникового образования может происходить ступенчато, проходя через дискретные углы, соответствующие рациональным углам, которым отвечают максимумы на зависимости энергии границы от угла наклона. Очевидно, этим можно объяснить растянутость по температуре и времени процесса структурного совершенствования, хотя каждый элементарный акт сброса двойника в данном материале может протекать с большими скоростями.

Влияние выпрямления двойникового образования экспериментально доказано в работе [120] при исследовании процесса графитации при высоких давлениях при температуре ниже плавления металла или сплава – катализатора, используемого для синтеза алмазов. Процесс твердофазной графитации в присутствии атомов переходных металлов состоит в распрямлении областей когерентного рассеяния, причем в этом случае высота области когерентного рассеяния (ОКР) от времени процесса практически не зависит, поскольку совершенствование структуры протекает очень быстро. Происходит диффузия атомов металла по границе внебазисного двойникования, где формируется полигонизационная стенка дислокаций, обладающих повышенной энергией. Далее происходит одновременный разрыв углеродных связей и уменьшение поверхностной энергии границы областей когерентного рассеяния с одновременным уменьшением угла внебазисного двойникования (предположение автора настоящей главы).

Разделение процесса совершенствования структуры на этапы условно, так же как и привязка их к конкретным температурам термического воздействия. Необходимо учесть, что некоторые рассмотренные процессы могут протекать и параллельно.

В работе [110] Нагорный В.Г., представляя механизм графитации, указал, что настоящая работа в определенной степени носит описательный характер и не претендует на окончательное постулирование выдвинутых положений графитации.

В то же время экспериментальные данные, накопленные за последние 20 лет по различным аспектам графитации, полностью подтверждают теоретические выводы, в частности, о существовании структурных политипов углерода, о роли турбостратной структуры в данном процессе, об уменьшении угла внебазисного двойникования и свидетельствуют о том, что многие положения этой теории правильные.

Исследование диамагнитной восприимчивости изотопа $^{13}C$ с различной температурой обработки показали, что данная величина зависит от температуры обработки и существенно меньше, чем для графитов (табл. 4.2).

**Таблица 4.2.** Диамагнитная восприимчивость порошка изотопа $^{13}C$ с различной температурой обработки, а также искусственных графитов.

| Образец | Диамагнитная восприимчивость, $10^{-6}$, СГМС/г |
|---|---|
| Изотоп $^{13}C$, 1200 °С | 0,9 |
| Изотоп $^{13}C$, 1600 °С | 2,1 |
| Изотоп $^{13}C$, 2800 °С | 3,8 |
| Графит с температурой обработки 2000 – 2800 °С | 6,3 |

Экспериментальные данные свидетельствуют о том, что в массе изотопа углерода $^{13}C$ основную долю составляют кристаллиты малых размеров с неупорядоченной структурой, в то же время имеются области с высокой степенью упорядоченности, которые фиксируются при рентгеноструктурном анализе. Таким образом, порошок изотопа $^{13}C$ является гетерогенно-графитирующимся материалом, несмотря на то, что степень графитации, рассчитанная по межслоевому расстоянию, равна 0,74. В массе графитированного изотопа обнаружены кристаллы монокристаллического графита (рис. 4.7), что подтверждается лауэграммой (рис. 4.8).

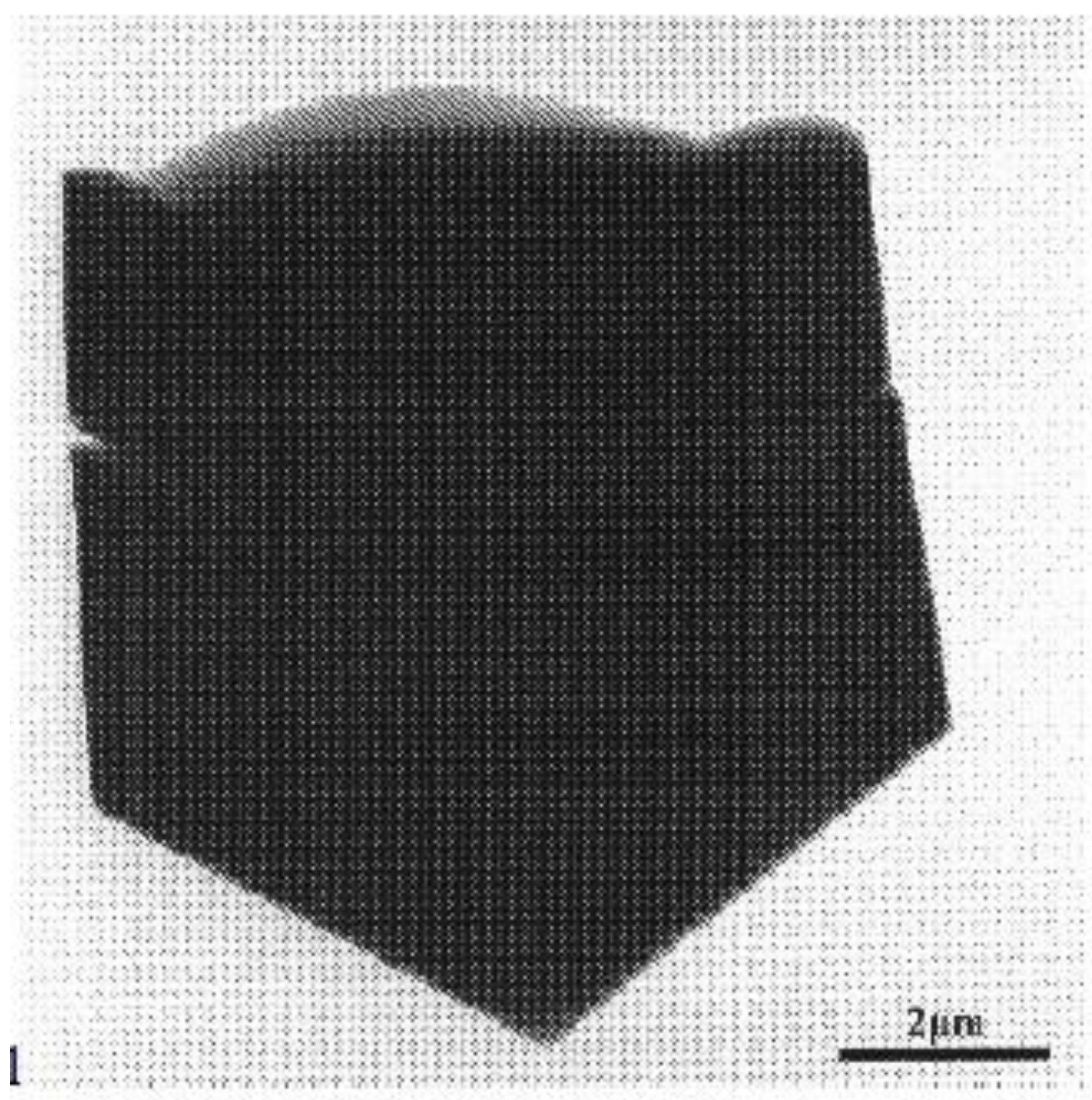

**Рис. 4.7**. Монокристаллическое образование в изотопе углерода $^{13}$C [111].

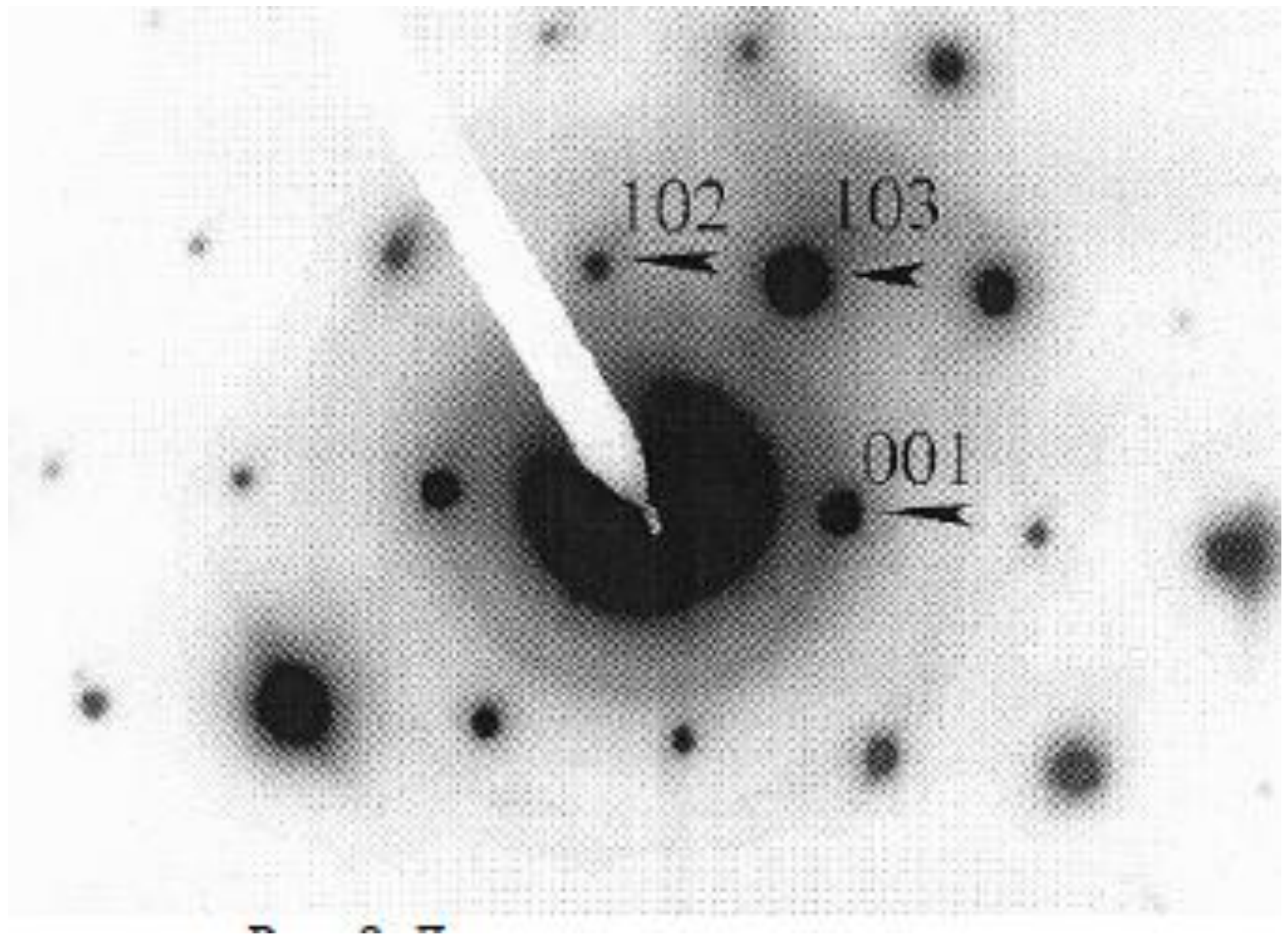

**Рис.4.8.** Лауэграмма монокристалла в изотопе углерода $^{13}$C [4.2].

## 4.5. Технологические особенности получения углеродного композиционного материала на основе изотопа $^{13}C$

Нейтронная мишень должна выдерживать высокие тепловые нагрузки, то есть обладать достаточной термопрочностью, а также в минимальной степени изменять структуру под действием частиц с высокой энергией [63-67]. В соответствии с критерием Кинджери максимальной термопрочностью обладают материалы с высокими прочностными характеристиками, высокой теплопроводностью и минимальным тепловым коэффициентом линейного расширения и динамическим модулем Юнга.

Анализ свойств исходного изотопа $^{13}C$ показал, что данный углерод имеет невысокую структурную прочность частиц, высокую удельную поверхность и низкую графитируемость. Как правило, труднографитирующиеся материалы имеют низкую теплопроводность, поэтому, для получения максимально возможной термопрочности требовалось создание композиционного материала с максимальными прочностными характеристиками и низким тепловым коэффициентом линейного расширения. Прочность увеличивается по мере приближения прочности карбонизованного связующего к прочности дисперсного наполнителя и возрастания адгезии между ними, а также при повышении гомогенности материала. Эта взаимосвязь определяет и другое свойство материалов – радиационную стойкость [121]. Кроме того, максимальные прочностные характеристики имеют углеродные материалы с малыми размерами областей когерентного рассеяния (см. главу 1).

Наиболее подходящим углеродным материалом для модификации малопрочного и пористого наполнителя на основе изотопа $^{13}C$ является стеклоуглерод, получаемый отверждением термореактивных фенолформальдегидных смол с последующей высокотемпературной обработкой. Его отличительными особенностями являются низкая газопроницаемость, высокая термическая стойкость, большая твердость, высокие прочностные характеристики, инертность к агрессивным средам, невысокий термический коэффициент линейного расширения.

Таким образом, в качестве связующего и импрегната была выбрана фенолформальдегидная смола. Общая технологическая схема получения композиционного углеродного материала на основе изотопа $^{13}C$ включает: дробление и рассев наполнителя, смешивание с жидким связующим, низкотемпературную обработку, размол и рассев прессмассы (получение пресспорошка), прессование, полимеризацию, обжиг предварительных заготовок, размол, рассев, повторное смешивание со связующим композиционного наполнителя, низкотемпературную обработку, размол и рассев прессмассы (получение пресспорошка), прессование, полимеризацию, обжиг, графитацию. Такой тип технологической схемы называется «нудель» процесс. Использование данной технологии позволяет увеличить плотность и прочность частиц наполнителя, увеличить адгезию между наполнителем и связующим, получить максимально гомогенный материал.

Исследование формоизменений образцов композиционного углеродного материала на основе изотопа $^{13}C$ после полимеризации показало наличие очень интенсивного расширение образца начиная с температуры 200°С, максимум приходится на температуру 400°С, затем происходит усадка, но после обжига происходит необратимое увеличение объема (рис. 4.9). В соответствии с полученными формоизменениями был разработан специальный режим обжига заготовок.

Температура графитации конечных изделий была определена на основании изменения высоты кристаллитов наполнителя от температуры обработки (рис. 4.3). Ранее было отмечено, что максимальные прочностные характеристики возможно получить при малых размерах ОКР. Начиная с температуры более 2600°С наблюдается резкий рост высоты кристаллитов, поэтому температура графитации составила 2600°С.

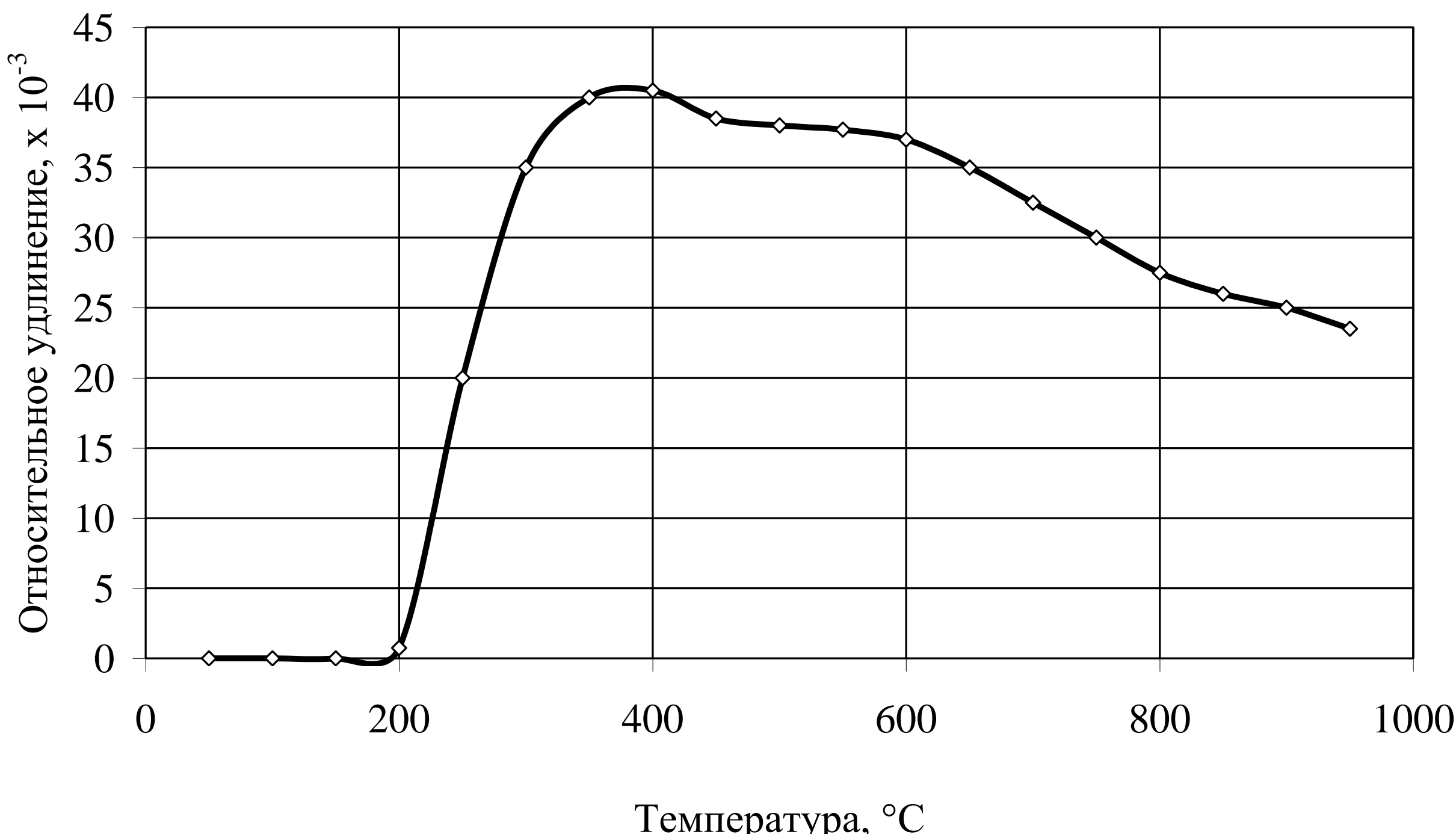

**Рис. 4.9.** Зависимость относительного удлинения образца композиционного углеродного материала на основе изотопа $^{13}$С от температуры.

Увеличение плотности композита позволяет значительно улучшить характеристики, в то же время существует оптимальный диапазон значений плотности, при котором отсутствуют трещины при обжиге и графитации. Трещинообразование при графитации обусловлено значительным изменением кристаллической структуры изотопа углерода $^{13}$С в предкристаллизационный период.

В таблице 4.3 приведены свойства композиционного материала с различной плотностью.

**Таблица 4.3.** Влияние плотности на свойства углеродного композиционного материала на основе изотопа $^{13}$С (температура графитации 2400°С)

| Свойства | Плотность 1,23 г/см$^3$ | Плотность 1,45 г/см$^3$ |
|---|---|---|
| Прочность на сжатие, $\sigma_{сж.}$, МПа | 29,2 | 62,6 |
| Коэффициент теплопроводности при 25 | 8,5 | 15,2 |

| °C, λ, Вт/м· К | | |
|---|---|---|
| Коэффициент линейного термического расширения, α, 1/град | 3,8 | 4,3 |
| Удельное электросопротивление, ρ, мкОм·м | 78,3 | 51,5 |

Установлено, что с ростом температуры графитации с 2400 до 2600°С также наблюдается некоторое увеличение прочностных характеристик (табл.4.4). Следует заметить, что данная зависимость не характерна для углеродных материалов на основе углерода $^{12}C$.

**Таблица 4.4.** Влияние температуры обработки (графитации) на прочность на сжатие и статический модуль упругости (плотность композита 1,23 г/см$^3$).

| Температура обработки, °С | Прочность на сжатие, σ, МПа | Модуль упругости, Е, ГПа |
|---|---|---|
| 2400 | 29,6 | 1,87 |
| 2600 | 33,9 | 2,97 |

***

Таким образом, подводя итог этой главы можно считать доказанным, что порошок изотопа углерода $^{13}C$ является гетерогенно-графитирующимся материалом. На основе данных по диамагнитной восприимчивости показано, что в массе изотопа углерода $^{13}C$ основную долю составляют кристаллиты малых размеров с неупорядоченной структурой. С помощью структурно-чувствительного метода измерения удельного электросопротивления порошков и рентгеноструктурного анализа впервые экспериментально установлено, что в предкристаллизационный период (1800–2100°С) графитации изотопа $^{13}C$ происходит разупорядочение его структуры. Разработана технология получения углеродного композиционного материала на основе изотопа $^{13}C$.

# ГЛАВА 5

# ИССЛЕДОВАНИЕ ЭЛЕКТРОННОЙ СТРУКТУРЫ И СВОЙСТВ КОМПОЗИТОВ НА ОСНОВЕ ИЗОТОПА УГЛЕРОДА $^{13}$C

Новые возможности для исследований в области физики ядра открывают установки, производящие высокоинтенсивные радиоактивные пучки (РИП) [122 - 125]. Установка проекта *EURISOL* (EURopean Isotope Separation On-Line), введение в эксплуатацию которой планируется в 2015 г., предназначена как для расширения изотопного ряда, так и на получение намного, на порядок более высоких, по сравнению с существующими, интенсивностей радиоактивных пучков. Завершение предварительных проектных работ по *EURISOL* включает в себя реализацию в ближайшее время установок второго поколения: *SPIRAL-II* (*GANIL*, Франция) и *SPES* (*LNL*, Италия). Обе установки – *SPES* и *SPIRAL-II* – обещают получение широкого диапазона интенсивных стабильных и радиоактивных пучков. В них используется двухступенчатая схема получения РИП. Первичный пучок (протоны в SPES и дейтроны в SPIRAL-2) с энергией до 50-100 МэВ и средней мощностью до 200 кВт направляется в конвертор нейтронный конвертор, где производит интенсивный (до $3 \cdot 10^{14}$см$^{-2}$ сек$^{-1}$) поток быстрых нейтронов. Произведенный поток нейтронов попадает на горячую толстую мишень деления, изготовленную из соединений $^{238}$U. Продукты деления диффундируют из мишени при высокой температуре, ионизируются и, после разделения по массам, направляются в экспериментальную зону.

Вариант конвертора с мишенью, изготовленной из углерод-углеродного композита с повышенным содержанием изотопа $^{13}$C разрабатывался для протонного первичного пучка, поскольку для изотопа $^{13}$C порог ядерной реакции с выходом нейтронов $^{13}$C(p, n)$^{14}$N составляет величину 3,24 МэВ, в то время как порог реакции $^{12}$C(p, n)$^{13}$N для углерода существенно выше и равен 20,1 МэВ [126, стр.895]. Соответственно, при энергиях протонного пучка меньших 20 МэВ выход нейтронов из мишени, изготовленной из чистого изотопа $^{13}C$, может быть заметно выше. При более высоких энергиях разница не столь заметна [127], тем не менее, такая охлаждаемая излучением мишень может позволить получить в 3-10 раз больший выход нейтронов, чем мишень на основе натурального углерода $^{12}C$ [127]. Дополнительным аргументом в пользу такой мишени является то, что применение дейтронного пучка, равно как и увеличение его энергии, приводит к существенному усложнению и удорожанию всего проекта *EURISOL*.

Кроме того, углеродный композит на основе изотопа $^{13}C$ находит своё применение в резонансной гамма-спектрометрии [128, 129]. Этот метод основан на том, что ядра многих химических элементов обладают свойством резонансного поглощения γ излучения. В частности, таким свойством обладают ядра атомов азота, которые способны резонансно поглощать γ-кванты с энергией 9,1724 МэВ, причём ширина пика поглощения составляет всего 125эВ. Метод обнаружения азотосодержащих веществ основан на сравнении величин резонансного и нерезонансного поглощения при прохождении γ излучения через вещество. Для генерации γ-квантов с резонансной энергией используется обратная поглощению ядерная реакция $^{13}C$ (p, γ) $^{14}N$, когда ускоренным до 1,75 МэВ пучком протонов бомбардируется графитовая мишень, обогащенная изотопом углерода $^{13}C$, при этом образуются возбуждённые ядра $^{14}N$, излучающие γ-кванты с энергией 9,17 МэВ. В данном случае необходимой для задач спектрометрии резонансной энергией обладают γ-кванты, угол вылета которых составляет $80{,}7 \pm 0{,}1^0$ к оси пучка протонов [129].

В данной главе обобщены результаты исследований электронной структуры и свойств углерод-углеродного композита на основе изотопа углерода **$^{13}C$,** поскольку именно понимание свойств может позволить в дальнейшем прогнозировать его время жизни и его возможности в качестве конструкционного материала. Образцы такого композита с повышенным содержанием изотопа $^{13}C$ (далее – графит $^{13}C$) были изготовлены в НИЦ

ФГУП «НИИграфит» из порошка $^{13}C$, полученного методом газофазного осаждения, особенности графитации и технологической схемы приготовления данного композита описаны ранее в главе 4 и приведены в [111, 130 - 131].

## 5.1. Рентгенография и высокоразрешающая микроскопия углеродного композита на основе $^{13}C$

Рентгенодифракционные профили порошка изотопа $^{13}C$ и композита на его основе были измерены в ИК СО РАН в группе Цыбули С.В. согласно п.2.2 предыдущего раздела и опубликованы в [132]. На рис. 5.1 приведена рентгенодифрактограмма высокоупорядоченного поликристаллического мелкозернистого графита марки МПГ-6 (рис. 5.1, линия 1). В рентгенодифракционном профиле углеродного композита на основе $^{13}C$ (рис. 5.1, линия 2) проявляются 00l и hk0 отражения от графитовых плоскостей, причем hk0 рефлексы имеют характерную асимметричную форму с более значительным размытием в сторону больших углов, чем это имеет место в упорядоченном поликристаллическом графите. Такая дифракционная картина соответствует турбостратной структуре графита, в которой отсутствует упорядочение графеновых слоев вдоль кристаллографического направления ***c*** [133].

Дифрактограмма чистого порошка изотопа $^{13}C$ (рис. 5.1., линия 3) показывает, по существу, только один хорошо выраженный уширенный дифракционный пик, соответствующий по своему положению рефлексу 002 для структуры графита. Увеличенное изображение этой дифрактограммы позволяет идентифицировать широкий пик 100, расположенный при $2\theta \approx 44°$ (рис. 5.2.). Размеры области когерентного рассеяния ОКР (Å) для структурных составляющих порошка изотопа $^{13}C$ приведены в таблице 5.2.

**Таблица 5.2.** Размеры области когерентного рассеяния (Å) для структурных составляющих исходного порошка изотопа $^{13}C$

| Структурная составляющая | $d_{002}$ (Å) | $00l$ (Å) |
|---|---|---|
| $G_1$ | 3.43 – 3.44 | 40 |
| $G_0$ | 3.6 – 3.65 | 20 |

Более детальная структура рефлекса 002 порошка изотопа $^{13}C$ представлена на вставке к рис. 5.2. Наличие перегиба указывает на то, что порошок состоит из двух структурных составляющих, различающихся по размеру. Действительно, 002 пик является суперпозицией двух компонент, относящихся к графитовым частицам с областью когерентного рассеяния 20Å и 40Å и с различным межплоскостным расстоянием $d_{002}$ (таблица 5.2.). Близкие значения интегральных интенсивностей компонент указывают на равное соотношение этих частиц в образце. Низкая интенсивность 100 пика не позволяет отнести его к той или иной структурной составляющей.

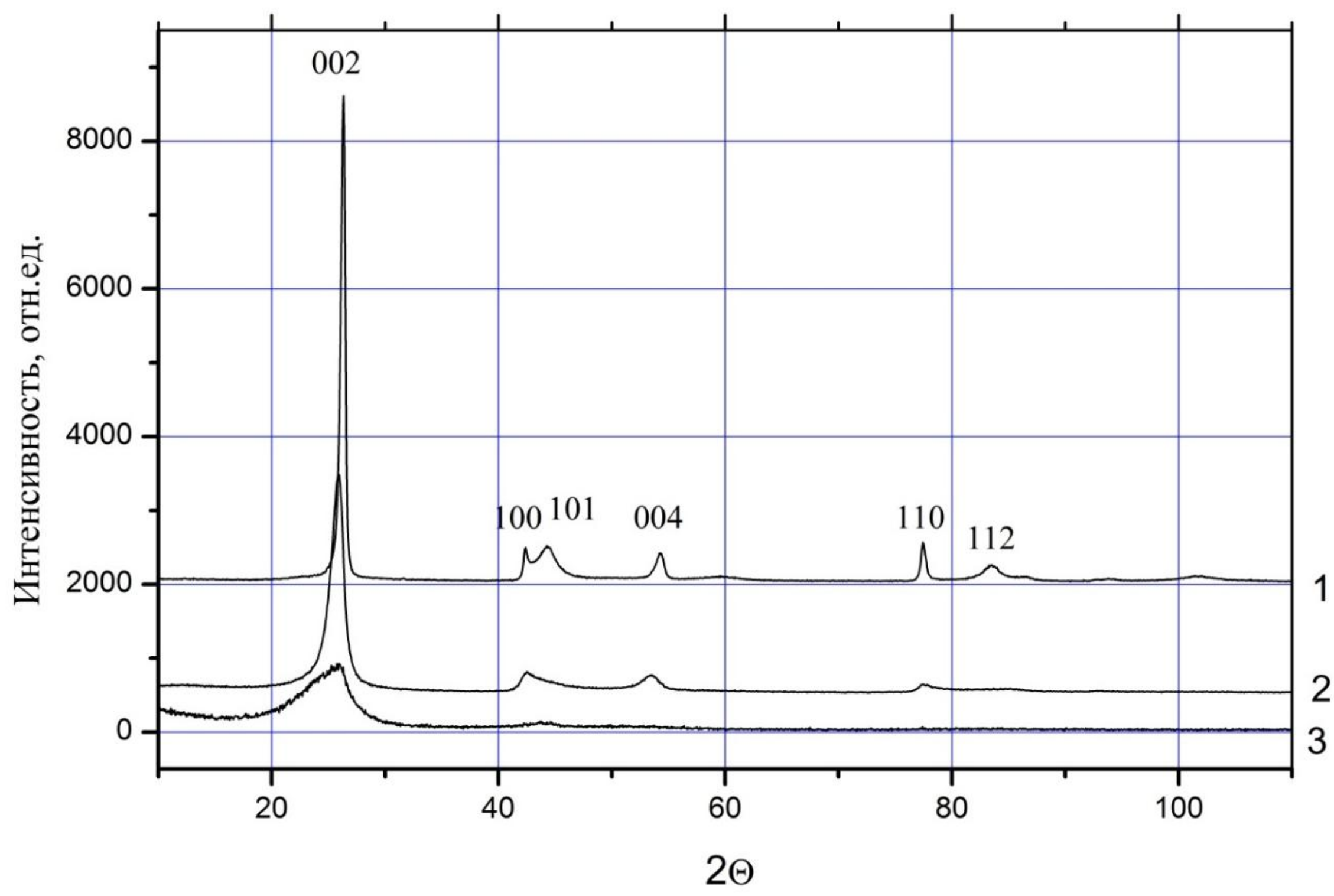

**Рис 5.1.** Рентгенофазная диаграмма: 1) *МПГ-6*; 2) исходная таблетка углерод-углеродного композита на основе изотопа $^{13}C$; 3) исходный порошок чистого изотопа $^{13}C$ [151].

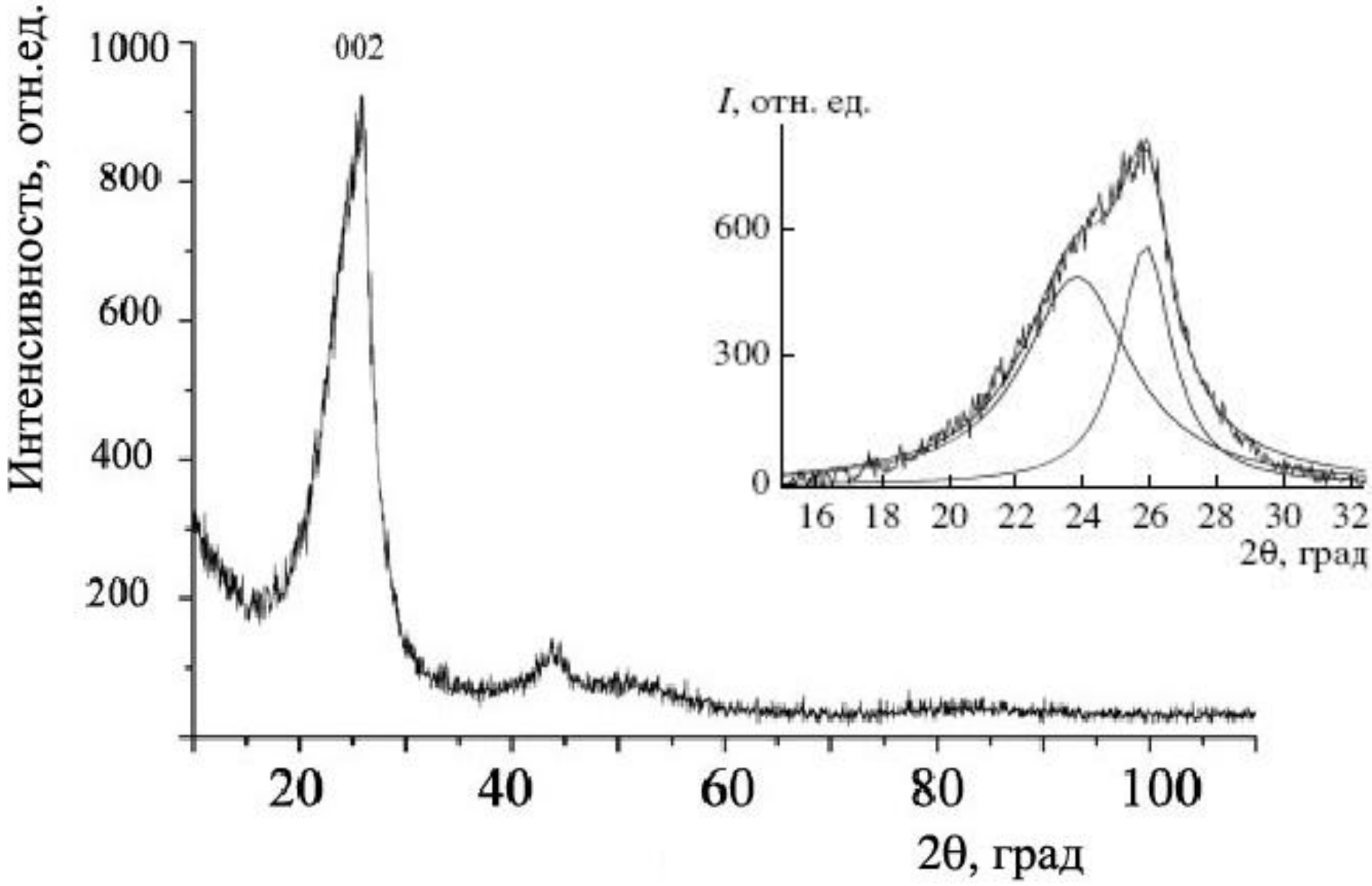

**Рис 5.2.** Рентгенофазная диаграмма порошкового образца чистого изотопа $^{13}C$. Форма дифракционного пика ***002*** (вставка вверху справа) указывает на наличие в образце двух структурных составляющих с различной дисперсностью и, по-видимому, с различным межплоскостным расстоянием $d_{002}$. [151].

**Высокоразрешающая электронная микроскопия** (ВРЭМ) показала, что образец углерод-углеродного композита на основе изотопа $^{13}C$ с плотностью менее 0,8 г/см$^3$ состоит из агрегатов частиц до 1000 нм (рис. 5.3, а). Однако, морфология частиц, составляющих агрегат, резко отличается от образца углеродного композита класса МПГ-6, поскольку каждая тонкая пластинка устроена так, что похожа на лист бумаги, смятый в центре, а затем немного расправленный [132].

Структура такой пластинки не является монокристаллической, а соответствует поликристаллическому состоянию – набору случайно ориентированных между собой блоков, дающих кольцевую микродифракцию, представленную на врезке к рис. 5.3 а. Интересно, что загибающиеся края таких пластинок похожи на волокнистый углерод, что хорошо видно на микрофотографии более крупного масштаба (рис. 5.3 б).

Конструкционные композиты на основе изотопа $^{13}C$ с повышенной плотностью (ρ~1,55 г/см$^3$) и содержанием изотопа около 75% состоят из трех типов углеродных частиц. Основную массу образца составляют смятые и изломанные графитовые пластины толщиной от 1 до 50 нм, которые имеют тенденцию к агломерации. Кроме того, в образце содержатся графитовые глобулы, которые, как правило, хорошо огранены и имеют размеры от 50 до 150 нм; некоторые глобулы частично разрушены. Каждая грань глобулы представляет собой графитовую пластину толщиной 15-20 нм. На рис. 5.4 приведено высокоразрешающее изображение образований, названных «кружевами». Эта форма углеродных структур представляет собой смятые графитовые слои толщиной около 2 нм, перемежающиеся графитовыми пластинами толщиной в 10 нм и длиной до 0.1 мкм. Увеличенное изображение таких образований приведено на рис. 5.5.

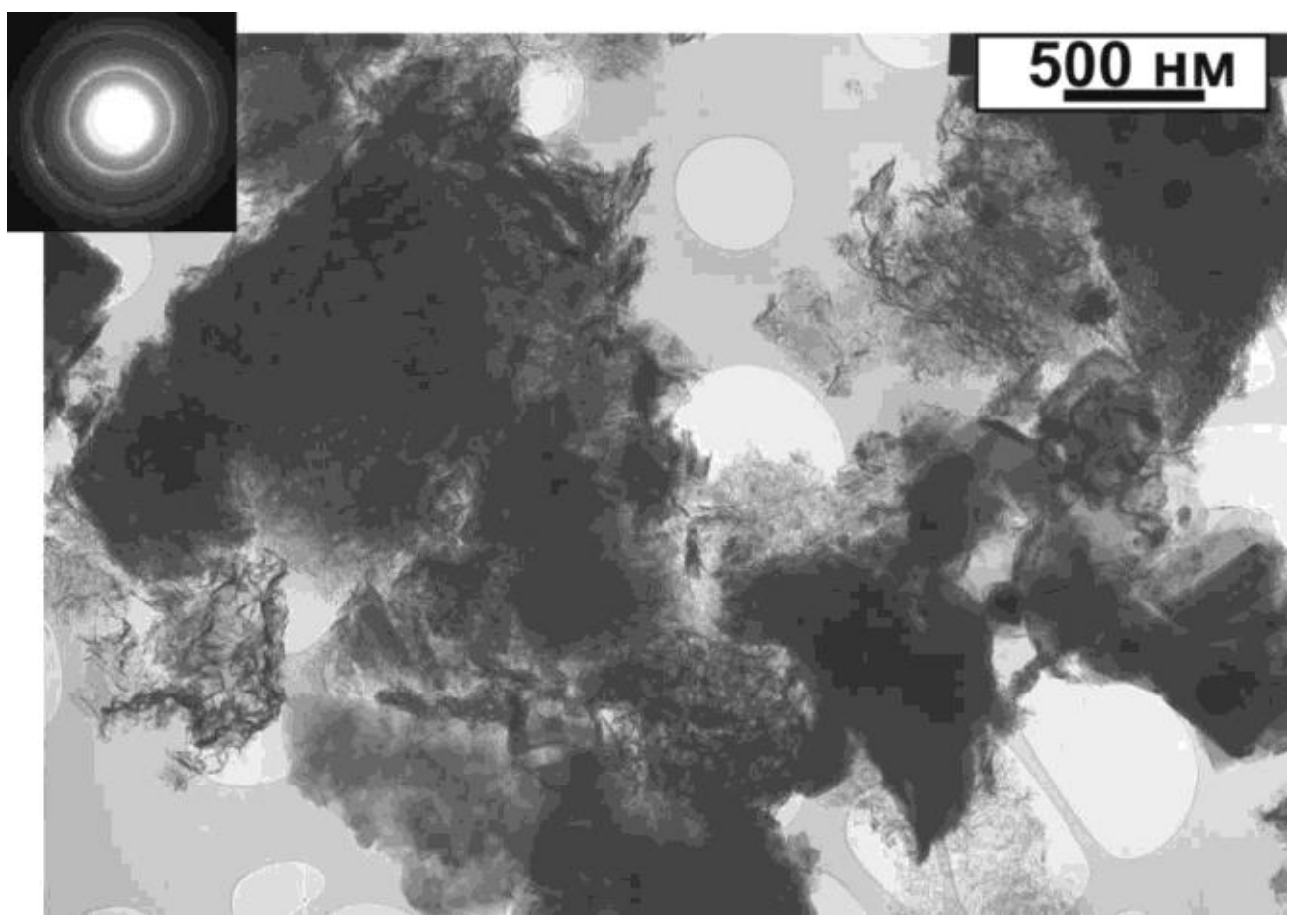

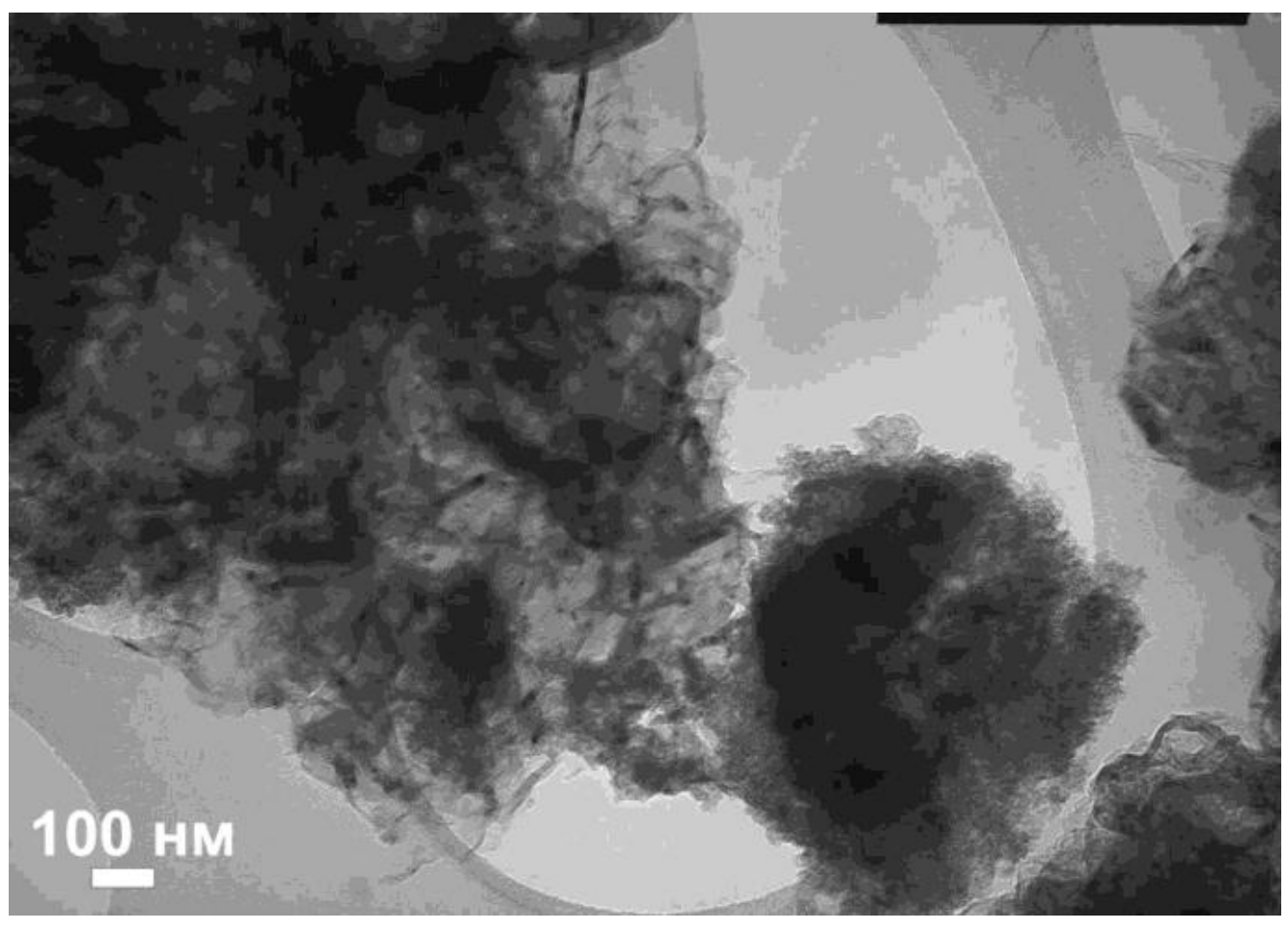

**Рис. 5.3.** Микрофотографии и микродифракция образца углерод-углеродного композита плотностью (ρ≤0,8 г/см$^3$) [132].

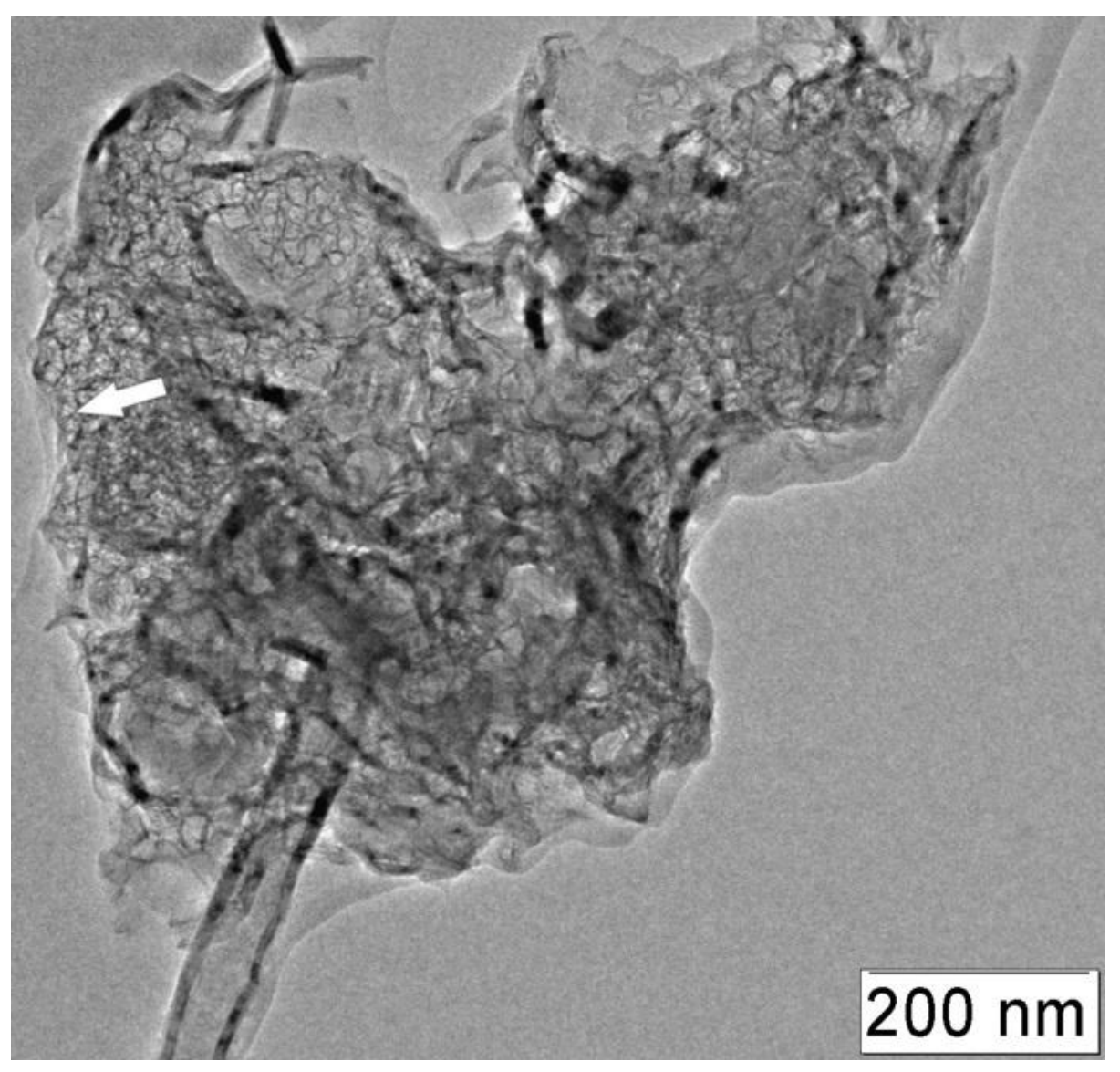

**Рис. 5.4.** Изображение структуры углеродного композита на основе изотопа $^{13}C$ морфологического типа «кружева». «Кружева» состоят из смятых листов графита (толщиной около 2 нм), перемежающихся графитовыми пластинами толщиной в 10 нм и длиной до 0.1 мкм. Стрелкой показано место, с которого получен снимок высокого разрешения (см. рис. 5.5.) [151]

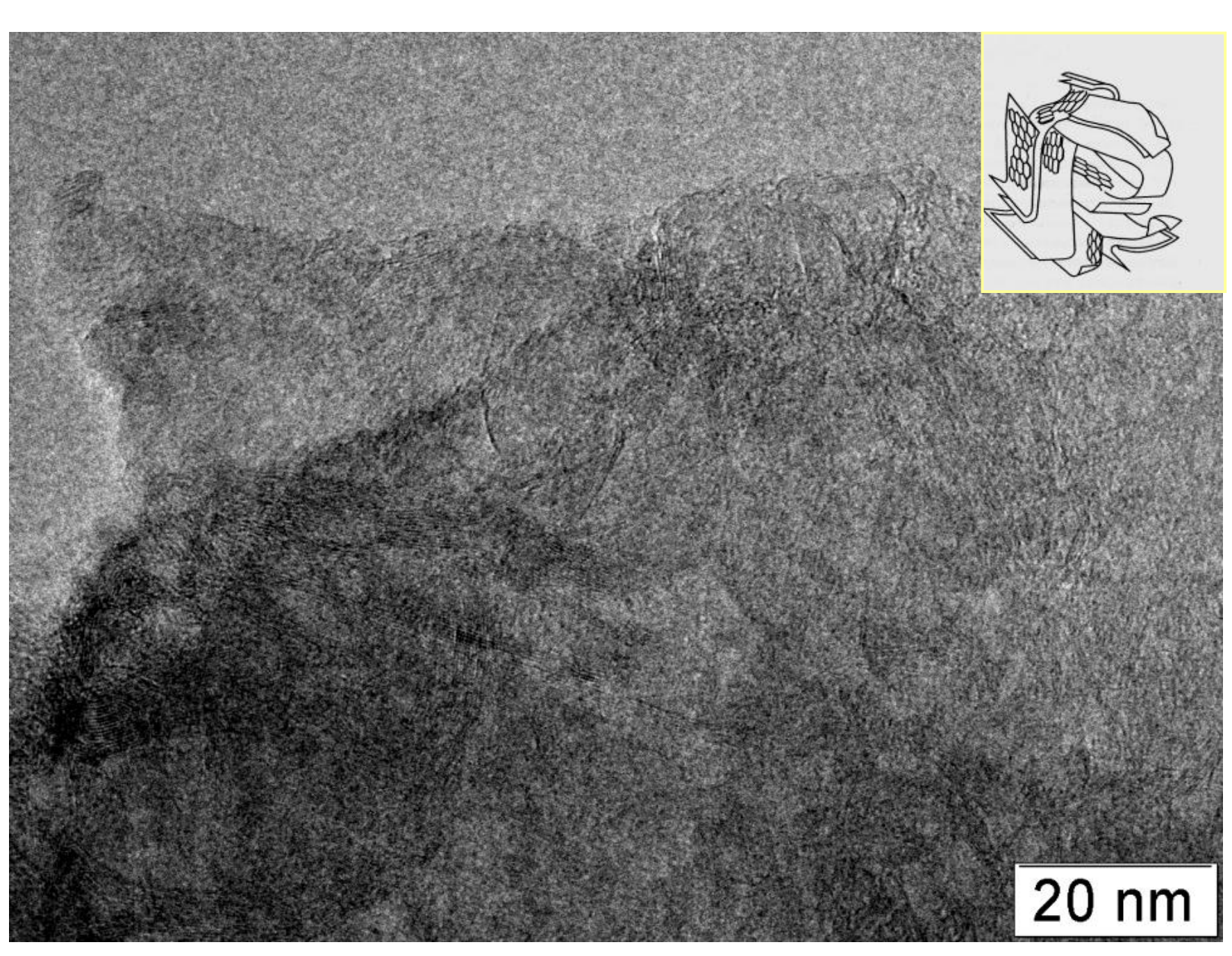

**Рис. 5.5.** Изображение высокого разрешения структуры морфологического типа «кружева». Вещество состоит из хаотически расположенных искривленных и изломанных листов графита, состоящих из 10-20 слоев. На вставке вверху справа - модель турбостратной структуры может быть представлена как «ком мятой бумаги» [151].

## 5.2. Исследования электронной структуры углеродного композита на основе $^{13}C$ методами рентгеновской флуоресцентной спектроскопии и квантово-химического моделирования

Это исследование было выполнено в ИНХ СО РАН, в группе Окотруба А.В. Рентгеновские флуоресцентные С***К***α-спектры поликристаллического графита, порошка изотопа $\mathbf{C^{13}}$ и композита на его основе были измерены на лабораторном спектрометре. В качестве кристалла-анализатора использовался монокристалл бифталата аммония; для учета нелинейной эффективности отражения кристалла применялась математическая процедура, описанная в [134]. Образцы наносились на медную подложку и охлаждались в вакуумной камере рентгеновской трубки до температуры жидкого азота. Режим работы рентгеновской трубки составлял U=4 кВ, J=0.8 А. Регистрация рентгеновского излучения осуществлялась газовым пропорциональным счетчиком, заполненным метаном при давлении ~ 0.1 атм. Разрешающая способность спектрометра 0.4 эВ. Энергия рентгеновских полос определена с точностью ± 0.15эВ.

Рентгеновский флуоресцентный спектр эмиссии возникает в результате заполнения предварительно созданных вакансий на внутренних уровнях соединения валентными электронами. Благодаря дипольным правилам отбора рентгеновский переход в углеродном соединении осуществляется между 2*p*- и 1*s*-орбиталями атома углерода и, таким образом, С$K\alpha$-спектр содержит информацию о плотности С*2p*-состояний в валентной полосе соединения. С$K\alpha$-спектры графита, порошка $^{13}$С и композита представлены на (рис. 5.6,а). В спектрах можно выделить три основных особенности, отмеченные буквами *A*, *B*, *C*. Детальная интерпретация С***К***α-спектра графита проведена в [26]. Интенсивный максимум *C* (E = 276.5эВ) соответствует **σ-** электронной подсистеме графита, высокоэнергетический максимум ***A*** (E = 282 эВ) относится к $\pi$-системе, в формировании максимума *B* участвуют электроны обоих типов. С$K\alpha$-спектр композита на основе изотопа $^{13}$С очень похож на спектр графита по положению и относительной интенсивности основных особенностей (рис. 5.6, а, линия 2), что согласуется с данными рентгеновской дифракции. В С$\boldsymbol{K}\alpha$-спектре порошка изотопа $^{13}$С (рис. 5.6, а, линия 3) по сравнению со спектром графита наблюдается увеличение относительной интенсивности максимумов *B* и *A*. Повышенная плотность занятых высокоэнергетических состояний в спектре может быть связана с наличием значительного числа дефектов, нарушающих однородность гексагональной углеродной сетки [135, 136].

Ренгенодифракционный анализ порошка $^{13}$C (рис.5.1) показал наличие в образце структурных составляющих $G_0$ и $G_1$ с размером упаковки порядка 20Å и 40Å, однако такие размеры должны быть характерны и для графеновых фрагментов, составляющих эти частицы. Было предположено, что CKα-спектр крупных частиц (~40 Å) имеет форму, близкую к спектру графита. Для выделения электронного состояния мелких частиц (~20 Å) из C***K***α-спектра порошка $^{13}$C было вычтено ~60% интенсивности спектра CKα-графита. Нормированный результат вычитания представлен на рис.5.6, б. При таком подходе особенности результирующего спектра оказываются более четко выделены по сравнению с исходными спектрами графита, порошка $^{13}$C и композита. Такой нормированный спектр показывает увеличение интенсивности линии *A* и смещение максимума этой линии в область высоких энергий.

Для подтверждения того, что результирующий спектр относится к малым по размеру графитовым частицам, был рассчитан теоретический C*Kα*-спектр графеновой структуры, состоящей из 150 атомов углерода и имеющей размер ~20 Å. Геометрия фрагмента $C_{150}$ была оптимизирована в приближении теории функционала плотности с использованием трехпараметрического гибридного функционала Беке [137] и корреляционного функционала Ли, Янга и Парра [138] (B3LYP метод) в рамках пакета квантово-химических программ *jaguar* [139]. Атомные орбитали описывались 6-31G** базисным набором. Энергия рентгеновского перехода определялась как разница энергий между одноэлектронными ***i*** и ***j*** уровнями, если уровень ***j*** является внутренним, а уровень ***i*** лежит в валентной полосе молекулы. Интенсивность линии, соответствующей рентгеновскому переходу, вычисляется по формуле:

$$I_{ij} = \sum_{A}\sum_{n}\sum_{m}\left|C_{jm}^{A} C_{in}^{A}\right|^2 \qquad (5.1),$$

где А обозначает углеродные атомы молекулы, $C_{jm}^{A}$ и $C_{in}^{A}$ - коэффициенты, с которыми **1s**-АО и **2p**-АО входят в состав i-ой и j-ой МО. Каждая линия в спектре, была уширена 0,6 эВ линиями Лоренца и нормирована на максимальное значение.

Было проведено сравнение между С***K****α*-спектром, полученным в результате вычитания из спектра порошка $^{13}$C спектра графита, с теоретическим спектром, рассчитанным для всех атомов фрагмента $C_{150}$ (рис. 5.6, б). Спектры показывают два максимума ***C*** и ***A***, расстояние между которыми и относительная интенсивность близки для обоих спектров.

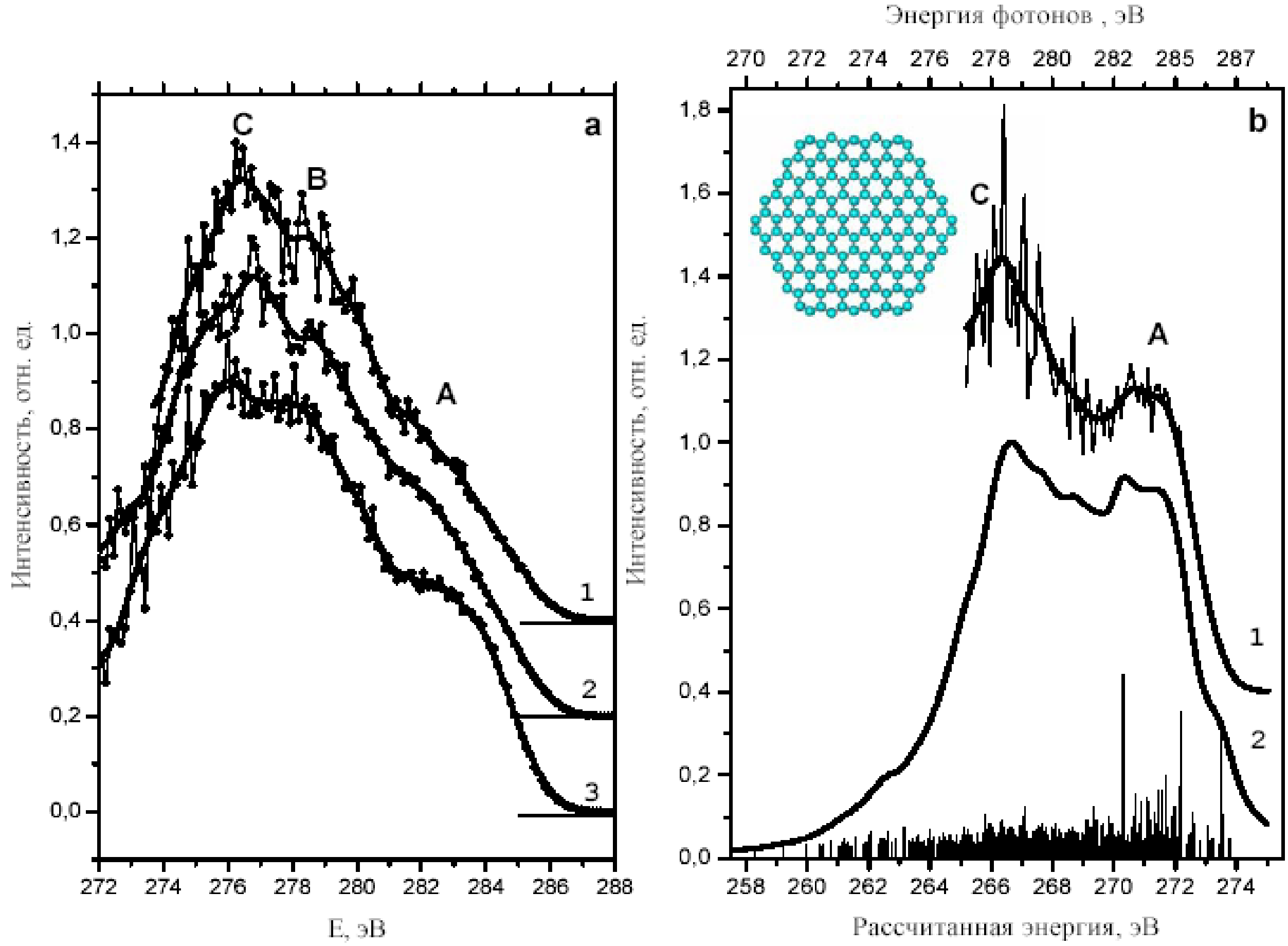

**Рис. 5.6.** а). Сравнение СKα-спектров, измеренных для графита (1), порошка $^{13}$C (2) и композита на основе изотопа (3). б). - Сравнение спектра, полученного вычитанием из СKα-спектра порошка $^{13}$C 60% интенсивности спектра графита, (1) и теоретического CKα-спектра графенового фрагмента $C_{150}$ (2), структура которого представлена в левом верхнем углу рисунка [151].

Высокая интенсивность максимума ***A*** связана с вкладом электронов разорванных связей двухкоординированных атомов углерода, составляющих границу фрагмента. Наибольшая локализация электронов характерна для атомов, составляющих зигзагообразный край [140]. Несколько завышенная интенсивность максимума ***A*** в расчетном спектре по сравнению с экспериментом может указывать на немного больший размер графитовых частиц мелкой фракции порошкообразного $^{13}C$, чем рассчитанный кластер $\mathbf{C_{150}}$.

## 5.3. Электрофизические измерения

Электрофизические измерения были выполнены в ИНХ СО РАН, в группе Романенко А.И. Температурные зависимости электропроводности образцов композитов на основе изотопа $^{13}$C измерялись четырехконтактным методом при постоянном токе.

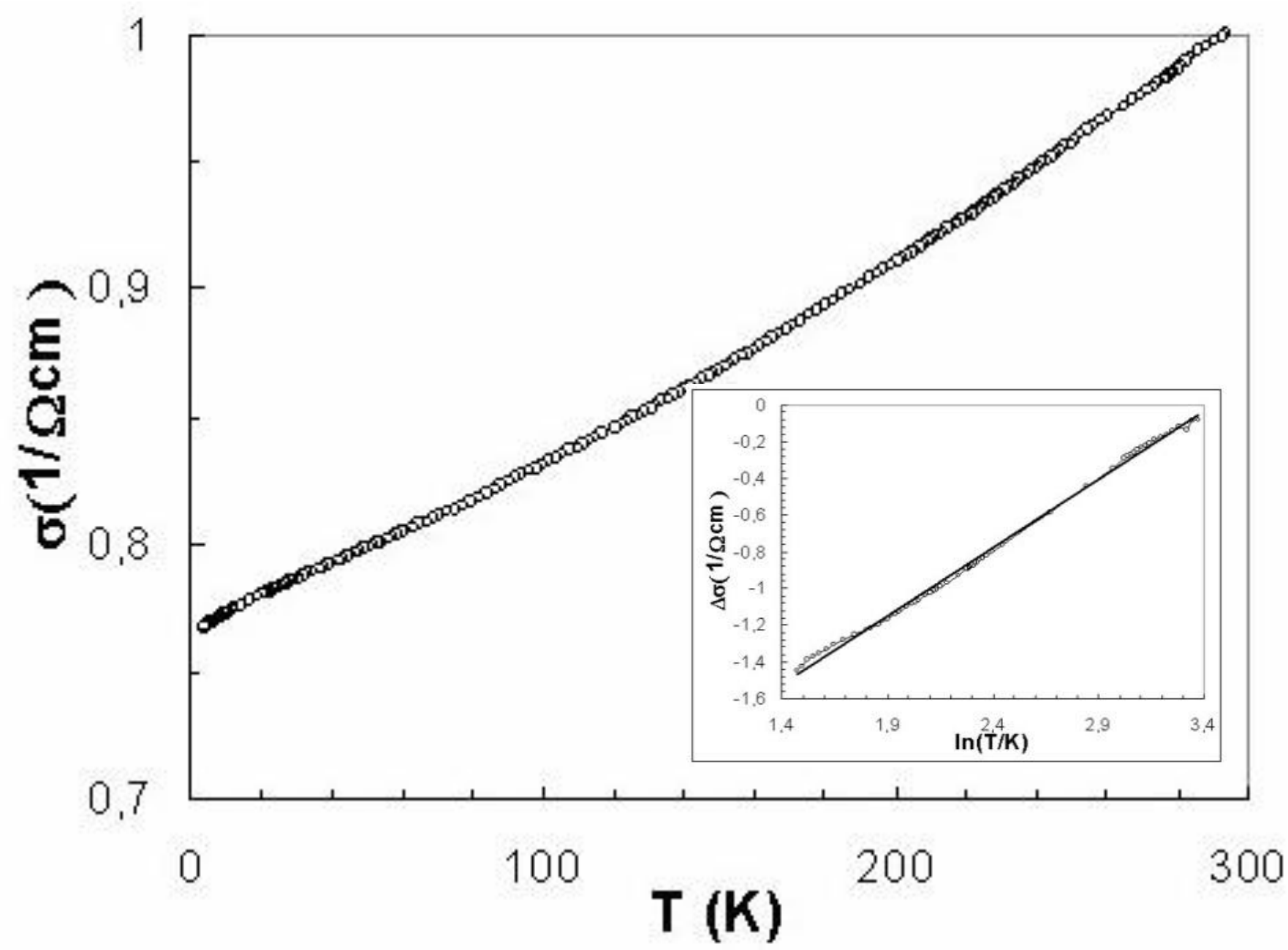

**Рис.5.7.** Температурная зависимость проводимости углерод-углеродного композита на основе изотопа $^{13}$C с повышенной плотностью $\rho \sim 1.55$ г/см$^3$. На вставке: аномальная часть низкотемпературной зависимости электропроводности углеродного композита в координатах $\Delta\sigma(T)$– lnT [151].

Образец для измерений представлял собой блок размером 1×0.5×10 мм$^3$, вырезанный из исходного углеродного материала. Электрические контакты наносились серебряной пастой G3692 Acheson Silver DAG 1415 mit Pinsel (Германия) и имели сопротивление ≅1 Ом.

Измерения проводились в атмосфере гелия и на воздухе в интервале температур 4.2-300 K. Температурная зависимость проводимости близкого к конструкционному плотного композита (ρ ~1,5 г/см$^3$) на основе изотопа $^{13}$C близка к линейной (рис. 5.7) и может быть объяснена доминированием трехмерных квантовых поправок, связанных со слабой локализацией носителей [141, 142], во всем температурном диапазоне. На определяющую роль квантовых поправок указывают измерения магнетосопротивления изотопного композита при гелиевых температурах. Магнетосопротивление является отрицательным во всем диапазоне от 0 до ±1.15 Тл и логарифмическим по асимптотике, что однозначно связано с подавлением интерференционных квантовых поправок магнитным полем.

Используя процедуру, изложенную в [143], в низкотемпературной части зависимости проводимости может быть выделена так называемая аномальная часть, которая представлена вставкой на рис. 5.7 в координатах $\Delta\sigma(T) - \ln T$, где $\Delta\sigma(T) = \sigma(T)_{эксп} - \sigma(T)_{экстр}$. Аномальная часть, полученная вычитанием регулярной части из экспериментальных данных, достаточно хорошо аппроксимируется прямой в координатах $\Delta\sigma(\boldsymbol{T}) - \ln\boldsymbol{T}$. Это также свидетельствует в пользу того, что двумерные квантовые поправки, связанные со слабой локализацией, доминируют при низких температурах.

Возможной причиной отклонения температурной зависимости проводимости от логарифмической при повышении температуры может быть ненулевая вероятность перехода носителей между слоями, что связано с общей высокой дефектностью графеновых слоев и сложной структурой композита (рис. 5.4). Влияние подобного рода переходов на квантовые поправки проанализированы в [144] для случая оксидных сверхпроводников, обладающих перовскитной слоистой структурой. Следуя [144], влияние переходов носителей между слоями на характер проводимости можно качественно пояснить следующим образом. Полная величина интерференционной поправки пропорциональна вероятности возвращения в стартовую точку за время, меньшее времени сбоя фазы $\tau_\varphi$. Пусть, например, структура представляет собой два параллельных слоя. Если переходов между ними нет, то каждый слой дает поправку к проводимости согласно формуле (5.2) в виде:

$$\delta\sigma/\sigma \sim -e^2/\hbar \ln(L\varphi/l) \qquad (5.2).$$

В данном случае длина диффузии связана со временем релаксации его волновой функции соотношением $L_\varphi = (D\ \tau_\varphi)^{1/2}$ (см. Приложение 1). Общая поправка к проводимости для такой двухслойной структуры будет соответственно, в два раза больше. Если окажется, что время перехода между слоями $\tau_{ij}$ сравнимо со временем сбоя его волновой функции $\tau_\varphi$, то вместо того, чтобы вернуться в стартовую точку, электрон может оказаться в соседнем слое с другой координатой $z$. Ясно, что такие траектории перестанут вносить вклад в интерференцию, и величина поправки будет меньше, чем $-2e^2/\hbar \ln(L_\varphi/l)$. Кроме того, должна измениться и форма кривой температурной зависимости проводимости. В том случае, когда время переходов между слоями $\tau_{ij}$ меньше времени энергетической релаксации $\tau_\varepsilon$ (или времени релаксации импульса $\tau_P$ по терминологии авторов) то согласно [144] образуется трехмерная квазиповерхность Ферми и квантовые поправки для такой структуры следует рассматривать так же, как и для трехмерного анизотропного проводника.

## 5.4. Теплофизические измерения

Теплофизические измерения были выполнены в ИТ СО РАН в группе Станкуса С. В. Исследования температуропроводности графитов проводились методом лазерной вспышки на автоматизированном экспериментальном стенде LFA-427 фирмы Netzsch [145] по методике, детально описанной в [146]. К основным достоинствам стенда относятся:

широкий интервал температур, доступный для измерений (25…2000°C); возможность исследовать различные классы твердых материалов; малые размеры образца (толщина – 0.1…6 мм, диаметр – 6…12 мм); возможность работать в вакууме (до $10^{-5}$ Торр), окислительной и защитной (Ar, He) атмосферах; широкий диапазон измерений величины температуропроводности (0.01…10 $см^2/с$); высокая (2…5 %) точность и производительность измерений; наличие автоматизированной системы управления и обработки данных. Оцениваемая погрешность измерений составляет 3 – 6 %. На рис.5.8 представлены результаты по температурной зависимости теплопроводности различных графитов. Видно, что в отличие от температуропроводности, температурная зависимость теплопроводности *МПГ-6* и *SGL* является слабо нелинейной, а теплопроводность композита на основе изотопа $^{13}C$ практически не изменяется с температурой.

За температурную зависимость теплопроводности графитов при высоких температурах практически полностью отвечает фонон-фононное взаимодействие и процессы переброса, а также рассеяние фононов на границах кристаллитов, неоднородностях структуры и дефектах решётки [147-150].

Исследование структуры и электронного строения графитовых композитов на основе изотопа $^{13}C$ методами высокоразрешающей электронной микроскопии на просвет (HRTEM), рентгеновской дифракции, рентгеновской флуоресцентной спектроскопии, комбинационного рассеяния и т.д. показало [132, 151], что достаточно совершенная кристаллическая структура графитов типа *МПГ* радикально отличается от мелкокристаллической, турбостратной структуры графитов на основе изотопа углерода $^{13}C$.

В частности, в случае композита на основе изотопа $^{13}C$ использование фенолформальдегидной резольной смолы в качестве биндера (см. главу 4) приводит к структуре, морфологически схожей со структурой стеклоуглерода [111]. Макромолекулярная, полиэдрическая структура стеклоуглерода детально проанализирована в [36, стр.208].

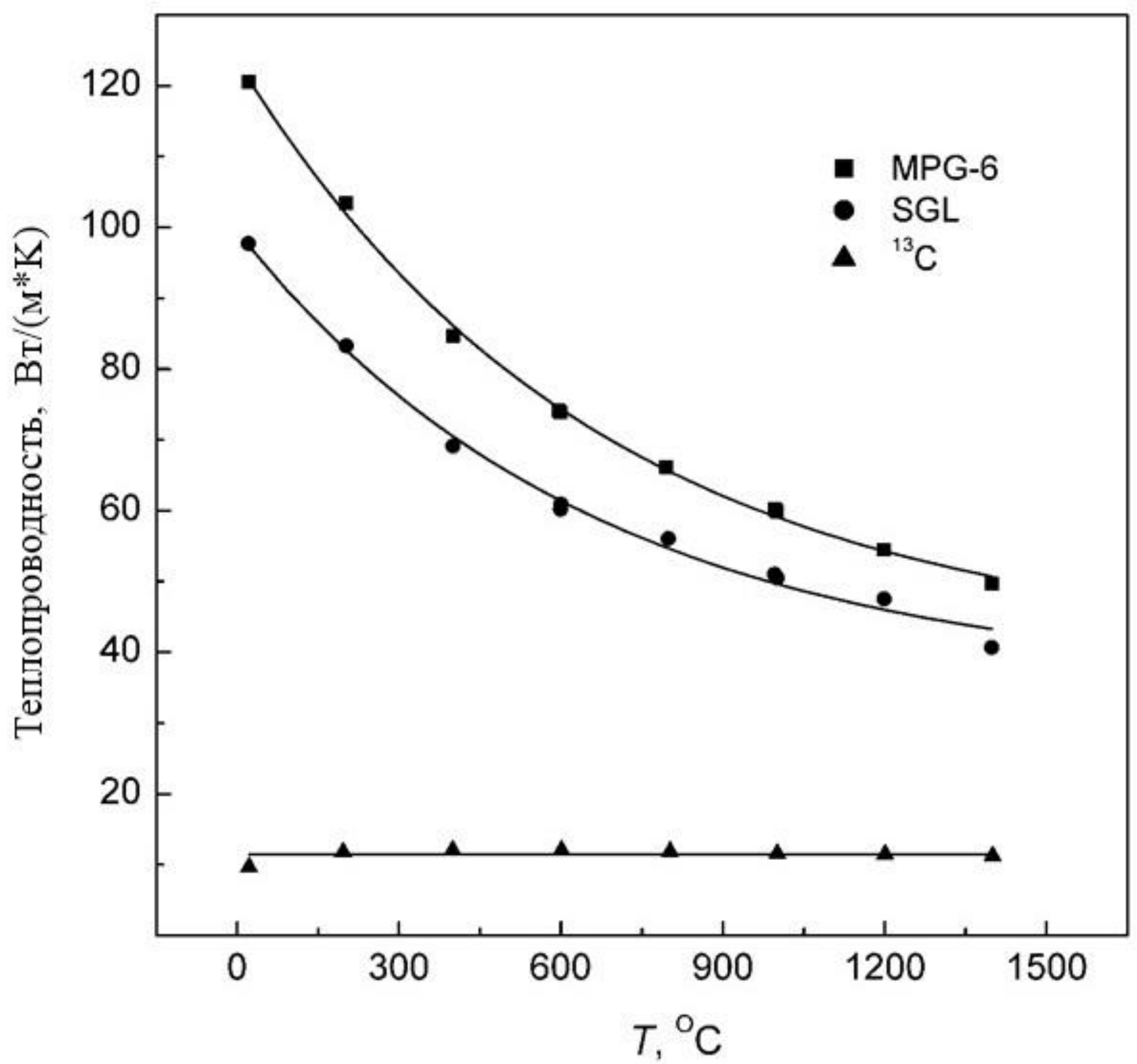

**Рис. 5.8.** Температурная зависимость теплопроводности графитов, рассчитанная из измерений по температуропроводности графитовых композитов [146].

Просвечивающая электронная микроскопия показала [132], в частности, что размер области когерентного рассеяния (ОКР) в композитах на основе $^{13}$C составляет величину порядка ~ 100Å, длина когерентности по данным электрофизических измерений также составляет около 150Å. Таким образом, средний размер микрокристаллитов может быть оценен величиной 100-150 Å. Представляется, что в данном случае так же точно, как и в случае стеклоуглерода, резкое снижение теплопроводности связано с тем, что длина свободного пробега фононов $l_f$ ограничена размером микрокристаллитов и не зависит от температуры (рис.5.8).

**Особенности фононных спектров графита** обсуждались в работе [152]. Явления теплопереноса в твёрдых телах принято рассматривать, используя теорию Дебая, однако эту теорию нельзя считать строгой, в силу приближений, положенных в её основу. Так, ограничиваясь приближением континуума, он использовал только одну акустическую ветвь, предположив, что оптические ветви отсутствуют, а три акустические совпадают. Кроме того, в теории Дебая предполагается линейный закон дисперсии, и, и соответственно, не учитывается фонон-фононное взаимодействие. С физической точки зрения это означает, что волны упругой деформации распространяются по кристаллу, не взаимодействуя между собой, а тепловая энергия переносится фононами со скоростью звука.

Одним из главных упрощений теории Дебая является выбор квадратичной зависимости для спектральной плотности G(ω), которая может радикально отличаться от её истинного вида (исключая область очень малых частот). Наиболее существенные особенности фононных спектров графита, в свою очередь, всецело определяются особенностями спектра графена. Графен имеет гексагональную структуру с двумя атомами углерода в элементарной ячейке, что приводит к появлению в спектре колебаний шести ветвей поляризации, при этом акустические моды LA и TA соответствуют продольным и поперечным колебаниям атомов в графеновой плоскости. Акустическая мода ZA соответствует  колебаниям атомов углерода в направлении, перпендикулярном LA и TA колебательным модам. Акустические моды LA и TA имеют линейную дисперсию и скорость звука для этих мод составляет (LA)=$2.13\times10^6$ см/сек и (TA)=$1.36\times10^6$ см/сек, соответственно [153]. В настоящее время нет полного согласия в отношении вида дисперсионной зависимости моды ZA. В работах [150, 154] предполагается квадратичный характер такой зависимости для моды ZA, в то время как в [155] на основании расчётов предполагается линейная дисперсия моды ZA, где скорость звука предполагается равной $\upsilon$ (ZA) = 0.16 $\times 10^6$ см/сек [156].

Высокочастотные оптические моды LO и TO в графене вырождены в Г-точке зоны Бриллюэна и принадлежат к так называемому $E_{2g}$ двумерному представлению. Эти моды с характерной частотой ~ 1580 $cm^{-1}$ находят своё наилучшее отражение в спектрах комбинационного рассеяния первого и второго порядка [157].

Эти спектры комбинационного рассеяния были измерены на КР-Фурье-спектрометре 100/S BRUKER, в качестве источника возбуждения использовалась линия 1064 нм Nd-YAG-лазера мощностью 100мВт (рис.5.9.). Данные спектры ранее подробно обсуждались в [132], в данном случае следует отметить только рост D - полосы для композита на основе изотопа углерода $^{13}C$ и стеклоуглерода. В данном случае D - полоса связана с неупорядоченным углеродом, и полностью соответствует результатам микроструктурного анализа [132]. Стоит отметить также сдвиг *G* – полосы для композита на основе изотопа углерода, этот сдвиг согласно базовым представлениям физики конденсированного состояния пропорционален: $\Delta\omega(q) \sim (13/12)^{1/2}$ и полностью соответствует наблюдаемому в эксперименте сдвигу главной полосы примерно в 40 $см^{-1}$.

**Механизм теплопереноса**.

Как следует из данных по измерению эффекта Холла [132], концентрация носителей заряда в углеродном композите на основе $^{13}C$ не превышает $4\times10^{19}$ $cm^{-3}$. Это почти на четыре порядка меньше, чем для меди, например. В нашем случае соотношение Видемана – Франца почти в сто раз больше, чем для обычных металлов.

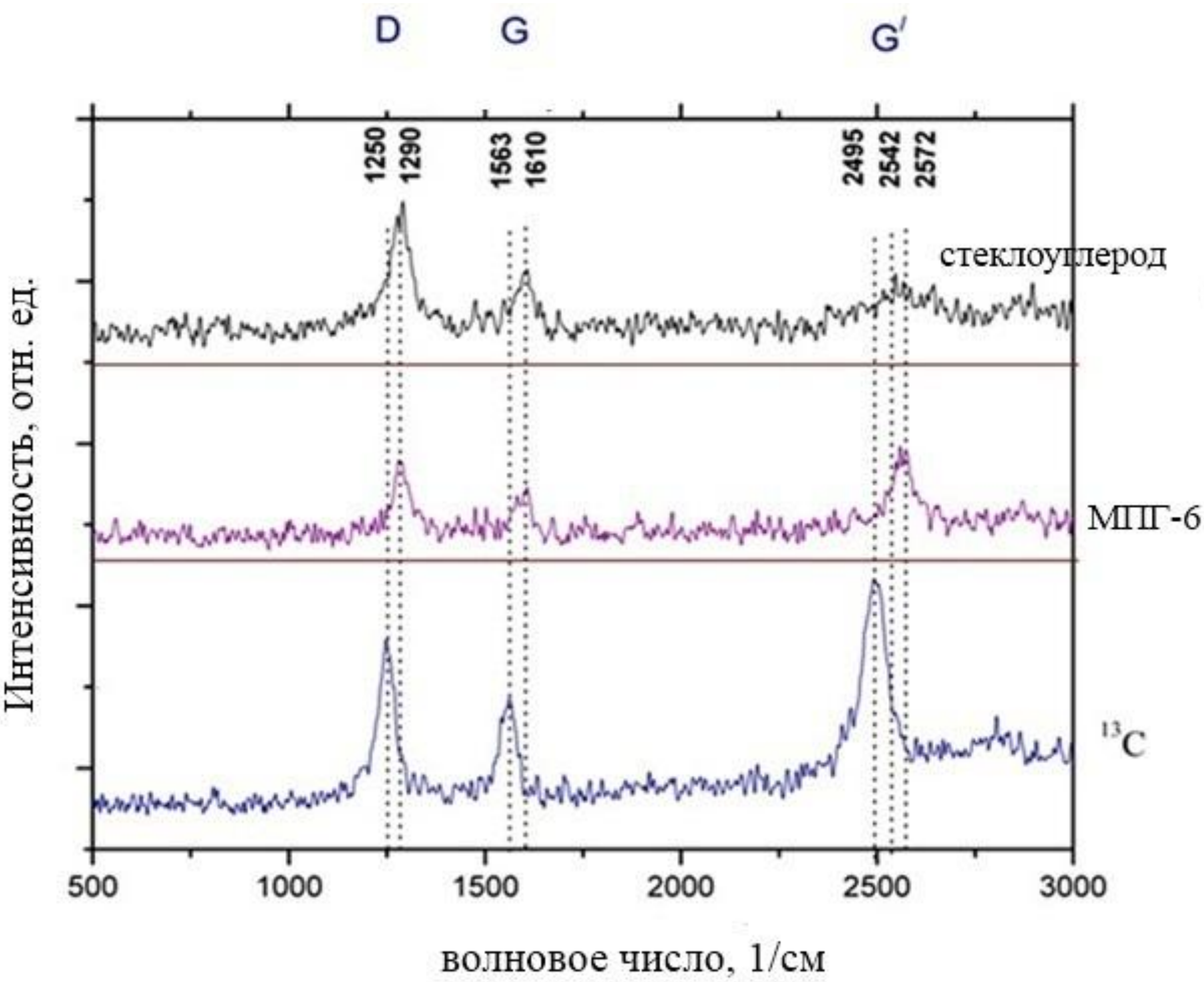

**Рис.5.9.** Спектры комбинационного рассеяния углеродных материалов. Сверху вниз: стеклоуглерод; МПГ-6; графитовый композит на основе $^{13}$C [152].

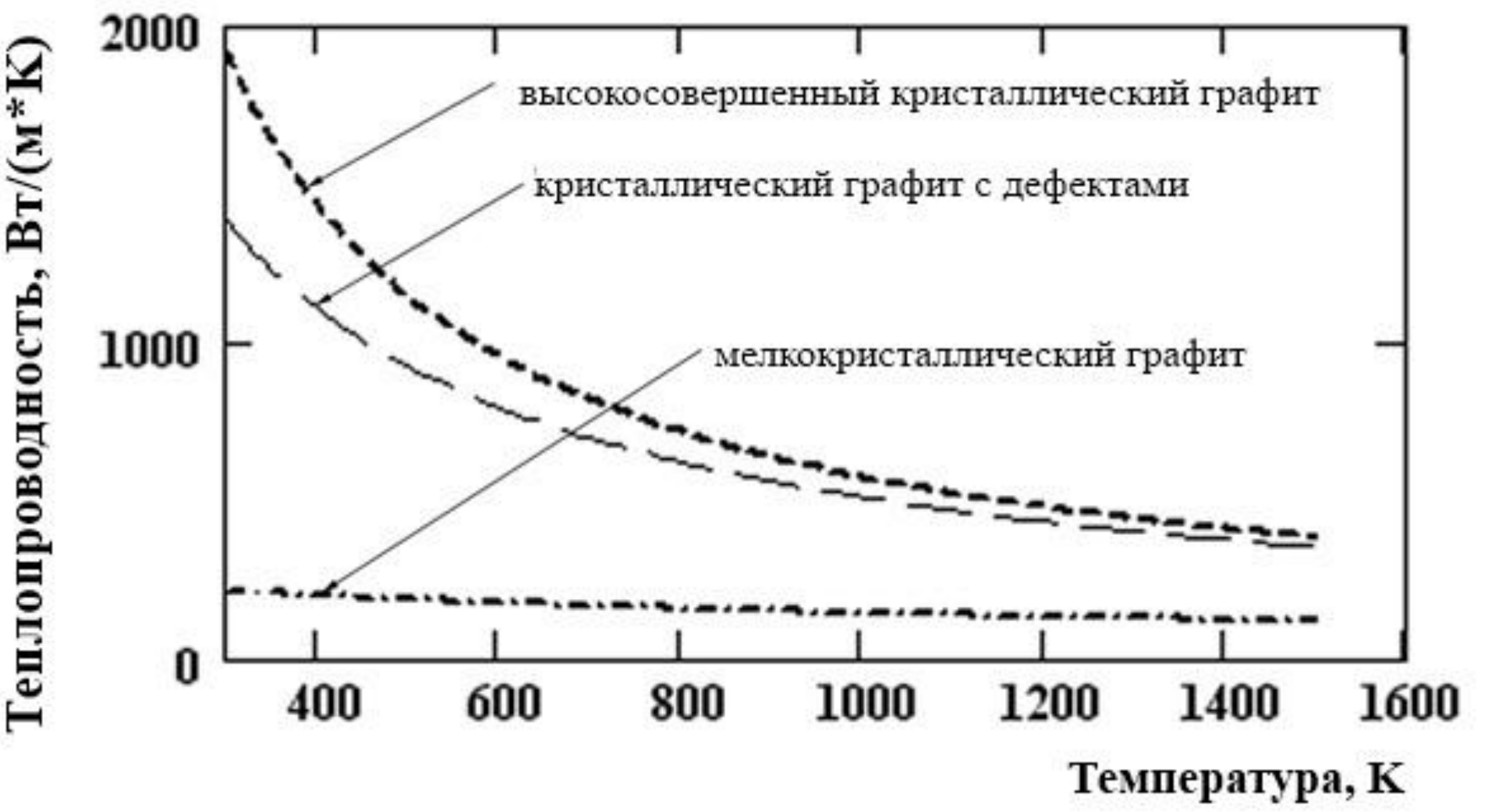

**Рис.5.10.** Теплопроводность, рассчитанная для кристаллического графита с учётом рассеяния на дефектах и границах зерна [152].

Это предполагает, что так же точно, как и в случае поликристаллического высокоупорядоченного графита фононный механизм теплопереноса является доминирующим во всём диапазоне температур.

Теплопроводность, связанная с фононами, может быть записана в виде [147, 154]:

$$\boldsymbol{\lambda}_{\mathrm{p}}=\Sigma_{\mathrm{j}} \int C_{\mathrm{j}}(\omega)\times \upsilon^{2}_{\mathrm{j}}(\omega) \times\tau_{\mathrm{j}}(\omega)\mathrm{d}\omega. \qquad (5.3),$$

где $j$ есть номер фононной ветви колебаний, куда входят две поперечные и одна продольная акустическая ветвь; $\upsilon_{\mathrm{j}}$ – есть групповая скорость фононов, которая часто может быть приблизительно равна скорости звука; $\tau_j$ есть время релаксации фононов; $\omega$ есть частота фононов и $C$ есть теплоёмкость. Длина свободного пробега фононов ($\Lambda$) связана со временем релаксации простым соотношением:

$$\boldsymbol{\Lambda=\tau\times\upsilon} \qquad (5.4).$$

В приближение времени релаксации, исходя из того, что различные механизмы рассеяния, которые ограничивают длину свободного пробега фононов $\boldsymbol{\Lambda}$ складываются:

$$\boldsymbol{\tau^{-1}=\Sigma} \qquad (5.5),$$

где $i$ нумерует механизм рассеяния.

Обычно акустические фононы, ответственные за теплоперенос в объёме, рассеиваются в результате фонон-фононного взаимодействия, на дефектах решётки, примесях, границах зёрен, а также на носителях заряда.

Клеменс [153] использовал дебаевскую скорость вместо групповой и параметр Грюнайзена $\gamma$ принял равным 2 (что соответствует экспериментально найденной для графита). Кроме того, используя для высоких температур величину теплоёмкости $C \sim \omega$ и полагая распределение фононов по частотам классическим, он оценил теплопроводность графитов, которая оказалась близка к экспериментальной для высокоориентированных графитов. Рис. 5.10 показывает теплопроводность, рассчитанную в приближении времени релаксации в трёхфононном процессе при учёте различных факторов рассеяния.

В ряде работ [158, 159] рассматривалось влияние изотопного рассеяния на теплопроводность графитов. Так, в работе [159] методами молекулярной динамики было показано, что в графеновом слое теплопроводность, связанная с фононами, может уменьшаться до 80% при случайном замещении атомов углерода $^{12}C$ на атомы изотопа $^{13}C$. Наибольшее уменьшение будет иметь место, если общая концентрация замещающих атомов изотопа $^{13}C$ составит примерно половину от общего количества атомов углерода в графеновом слое.

**Тепловое расширение** является одной из важнейших характеристик твердых тел. Наличие достоверных данных по температурной зависимости плотности особенно важно для конструкционных графитов, которые могут применяться при высоких температурах. Следует также иметь в виду, что данные по плотности используются при определении ряда других теплофизических характеристик и, возможно, она может служить тем интегральным параметром, который будет характеризовать структуру графитов при обобщении результатов измерений.

В работе [160] было проведено экспериментальное исследование теплового расширения композита на основе изотопа углерода $^{13}C$. На рис. 5.11 приведены данные по тепловому расширению графитов различных марок. Видно, что плотные поликристаллические графиты *POCO* и *МПГ-6* имеют близкие средние интегральные коэффициенты линейного расширения и практически совпадающие температурные коэффициенты. Отличие интегрального коэффициента линейного расширения (ИКЛР) от справочных данных для POCO AXM-5Q [161] не превышает $1{,}9\times10^{-7}$ К$^{-1}$ и $4{,}6\times10^{-7}$ К$^{-1}$ соответственно. Температурная зависимость ИКЛР углеродного композита на основе изотопа $^{13}C$ имеет существенно нелинейный характер и на 45-65% меньшие абсолютные значения.

Низкое значение коэффициента теплового расширения $^{13}C$ не вызывает удивления, если учесть различия в микроструктуре графитов. По данным просвечивающей электронной микроскопии размер области когерентного рассеяния в композитах на основе $^{13}C$ составляет величину порядка 10 нм, а по электрофизическим измерениям - 15 нм [132]. Поэтому средний размер микрокристаллитов $^{13}C$ может быть оценен величиной 10-15 нм. Открытая пористость в поликристаллическом графите МПГ-6, измеренная методом ртутной порометрии, составляет около $9\times10^{-2}$ см$^3$/г, при среднем радиусе преобладающих пор порядка 1 мкм [162]. Открытая пористость более рыхлого композита на основе углерода $^{13}C$ почти в четыреста раз превосходит суммарный объём пор композита МПГ-6, при этом максимум распределения макропор композита на основе $^{13}C$ приходится на диаметр 0,3-0,5 мкм.

В композите на основе $^{13}C$ имеются также ещё два пика: микропоры с диаметром около 2 нм и мезопоры со средним диаметром около 10 нм. Величина коэффициента термического расширения углеродного композита определяется двумя конкурирующими факторами: тепловым расширением микрокристаллитов и наличием пор, микротрещин и других нарушений структуры, которые способны компенсировать это расширение [2]. Из-за наличия большой пористости коэффициенты линейного и объёмного расширения турбостратных углеродных материалов всегда будет ниже таковых для монокристаллов и совершенных поликристаллов [36]. Кроме того, содержание в углеродном материале неупорядоченной аморфной фазы, расположенной между кристаллитами, также приводит к уменьшению коэффициента термического расширения, так как расширение этой фазы меньше, чем кристаллитов.

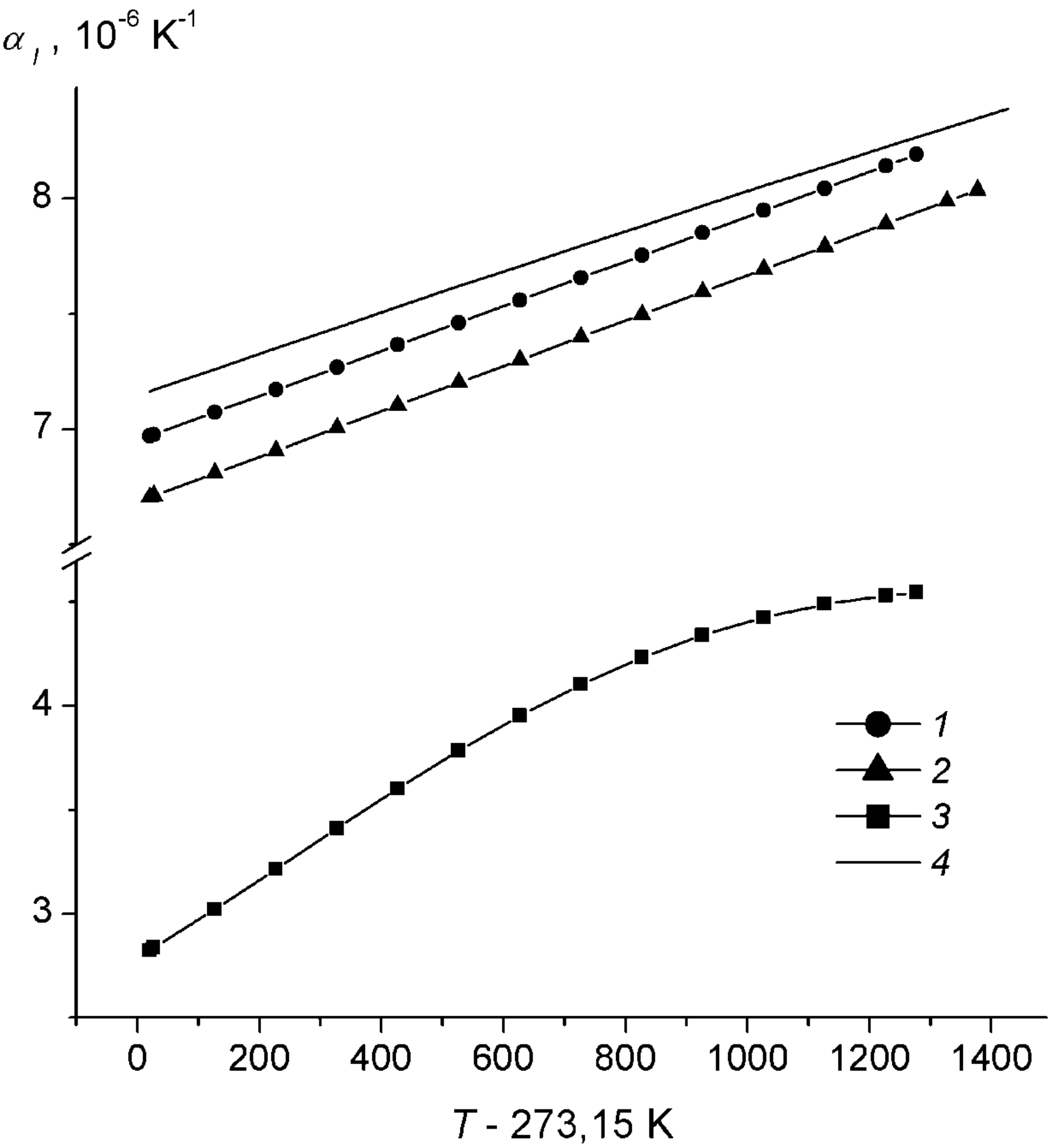

**Рис.5.11.** Средний интегральный коэффициент линейного расширения графитов. *1* - POCO AXF-5Q, *2* - МПГ-6, *3* - $^{13}$C, *4* - POCO AXM-5Q [160].

***

Таким образом, данные рентгеновской дифракции и просвечивающей электронной микроскопии показали, что образцы углеродных композитов на основе изотопа $^{13}$C имеют резко выраженную турбостратную структуру. В целом, внутренняя структура композитов, полученных на основе изотопа углерода $^{13}$C представляется достаточно сложной и состоящей из нескольких, заметно различающихся между собой морфологических форм углерода. Рентгеновские флуоресцентные CKα-спектры исходного порошка чистого изотопа $^{13}C$ и композитов на его основе заметно отличаются от спектра графита интенсивностью высокоэнергетического максимума. Квантово-химический расчет графена $C_{150}$ показал, что повышение плотности состояний обеспечивается электронами разорванных связей граничных углеродных атомов частиц размером ~20 Å.

Показано также, что температурная зависимость проводимости углеродного композита на основе $^{13}C$ обусловлена квантомеханическим эффектом двумерной слабой локализации носителей заряда. Данные по теплопроводности сопоставлены с полученными ранее рентгеноструктурными данными и результатами высокоразрешающей электронной микроскопии. Показано, что во всех случаях процессы теплопереноса в графитах обусловлены фононами. Эти данные могут быть использованы в тепловых расчётах конструкции нейтронного конвертора с использованием программы ANSYS.

# ГЛАВА 6

# ВЛИЯНИЕ ВЫСОКОДОЗНОГО НЕЙТРОННОГО ОБЛУЧЕНИЯ НА ИЗМЕНЕНИЕ ФИЗИЧЕСКИХ СВОЙСТВ РЕАКТОРНОГО ГРАФИТА

В настоящее время в России эксплуатируется 11 энергоблоков с реакторами большой мощности канального типа (РБМК) на трех АЭС – Ленинградской, Курской и Смоленской, вклад которых в общую выработку электроэнергии всеми АЭС России составляет около 50 %. Введенные в эксплуатацию в разное время (с 1973 по 1990 г.), они имеют 30-летний назначенный срок службы, и к настоящему моменту часть реакторов уже полностью выработала назначенный ресурс. Графитовая кладка (ГК), выполняющая роль замедлителя и отражателя нейтронов, является незаменяемым и ограниченно ремонтопригодным узлом реактора, и поэтому она определяет ресурс работы энергоблока в целом. Исследования отдельных кернов, выбуренных из графитовых блоков (ГБ) кладок реакторов после эксплуатации в течение 30 лет показали, что состояние кладки удовлетворительное и позволяет продолжать эксплуатацию реактора, в связи с чем возник вопрос об обосновании нового увеличенного ресурса кладки.

Тридцатилетний ресурс кладки был определен с использованием расчетных кодов на основе базы данных по радиационной стойкости реакторного графита ГР-280, содержащейся в «Нормах расчета на прочность типовых узлов и деталей из графита уран-графитовых канальных реакторов» (НГР) [163]. Следует отметить, что эта база относится к флюенсам нейтронов не более $2{,}2\cdot10^{26}$ м$^{-2}$ (здесь и далее приведен флюенс нейтронов с энергией больше 0,18 МэВ) и температурам в интервалах 350-450 и 500-600°С, что существенно ниже требуемых значений по флюенсу и температуре, а данные для предельных температур и флюенсов получены методом экстраполяции. Однако, ввиду того, что закономерности изменения свойств графита при таких параметрах облучения не изучены, а изменение свойств имеет немонотонный сложный характер, то детальная экстраполяция не является достаточно надежной в таком важном и сложном вопросе, как определение ресурса кладки, и заложенные экстраполированные кривые требуют своего экспериментального подтверждения.

Именно поэтому несколько лет назад концерн «Росэнергоатом» в рамках общей целевой программы Минатома РФ поставил задачу по обоснованию предельно достижимого срока службы графитовых кладок реакторов типа РБМК.

Основным этапом работ по этой проблеме является создание обновленной базы данных по радиационной стойкости графита ГР-280, адаптированной к параметрам облучения активной зоны реактора РБМК, чем и обусловлена высокая актуальность темы.

Для создания базы данных было необходимо проведение исследований образцов графита, облученных при условиях, перекрывающих интервалы флюенса и температуры, в пределах которых происходит обоснование ресурса, то есть до предельного флюенса около $3{\cdot}10^{26}$ $м^{-2}$ в интервале температур 450-650°С.

Кроме того, при создании базы данных необходимо использовать обоснованный набор методик. При расчете ресурса графитовых кладок используются два основных критерия: размерная стабильность и трещинностойкость ГБ, которые, в свою очередь, определяются напряженно-деформированным состоянием (НДС) графитовой кладки. Многочисленные работы показывают, что для расчета НДС блоков графита необходимо обладать зависимостями от флюенса и температуры таких свойств, как размеры, коэффициент теплопроводности, тепловой коэффициент линейного расширения (ТКЛР), модуль упругости, предел прочности, деформация радиационной ползучести. Такой же набор свойств использовался при создании базы данных, содержащейся в НГР [163].

## 6.1. Исследование радиационного формоизменения графита

Размерные изменения графитовых блоков при облучении являются одним из основных факторов, ограничивающих срок службы графитовой кладки. Это связано с тем, что по достижении определенного флюенса происходит исчерпание технологических зазоров между блоками и технологическими каналами, что приводит к росту напряжений в графитовых блоках, их растрескиванию и, в конечном счете, к потере работоспособности графитовой кладки. Кроме того, искривление графитовых колонн, вызванное размерными изменениями, может приводить к изгибу технологических каналов и заклиниванию тепловыделяющих сборок.

Зависимости относительного изменения длины и объема образцов от флюенса представлены на рис.6.1-6.2 [164]. На всех кривых условно можно выделить три стадии. На начальной стадии облучения происходит уменьшение размеров до определенного флюенса $Ф_0$, при котором размеры образцов достигают минимума. При увеличении флюенса выше $Ф_0$ размеры образцов увеличиваются с возрастающей скоростью и возвращаются к своему исходному значению при флюенсе $Ф_{кр}$, называемом критическим (вторая стадия). При дальнейшем увеличении флюенса выше $Ф_{кр}$ увеличение размеров образцов идет практически с постоянной скоростью $V_{кр}$ (третья стадия).

Таким образом, в первом приближении, зависимости изменения размеров образцов от флюенса нейтронов можно описать четырьмя величинами: максимальной усадкой $(\Delta l/l)_{max}$ и соответствующим ей флюенсом $Ф_0$; критическим флюенсом Фкр, при котором линейный размер либо объем образца возвращается к исходному значению; скоростью роста образцов при флюенсе, выше критического. На рис. 6.3 представлены зависимости изменения этих параметров от флюенса при заданной температуре облучения.

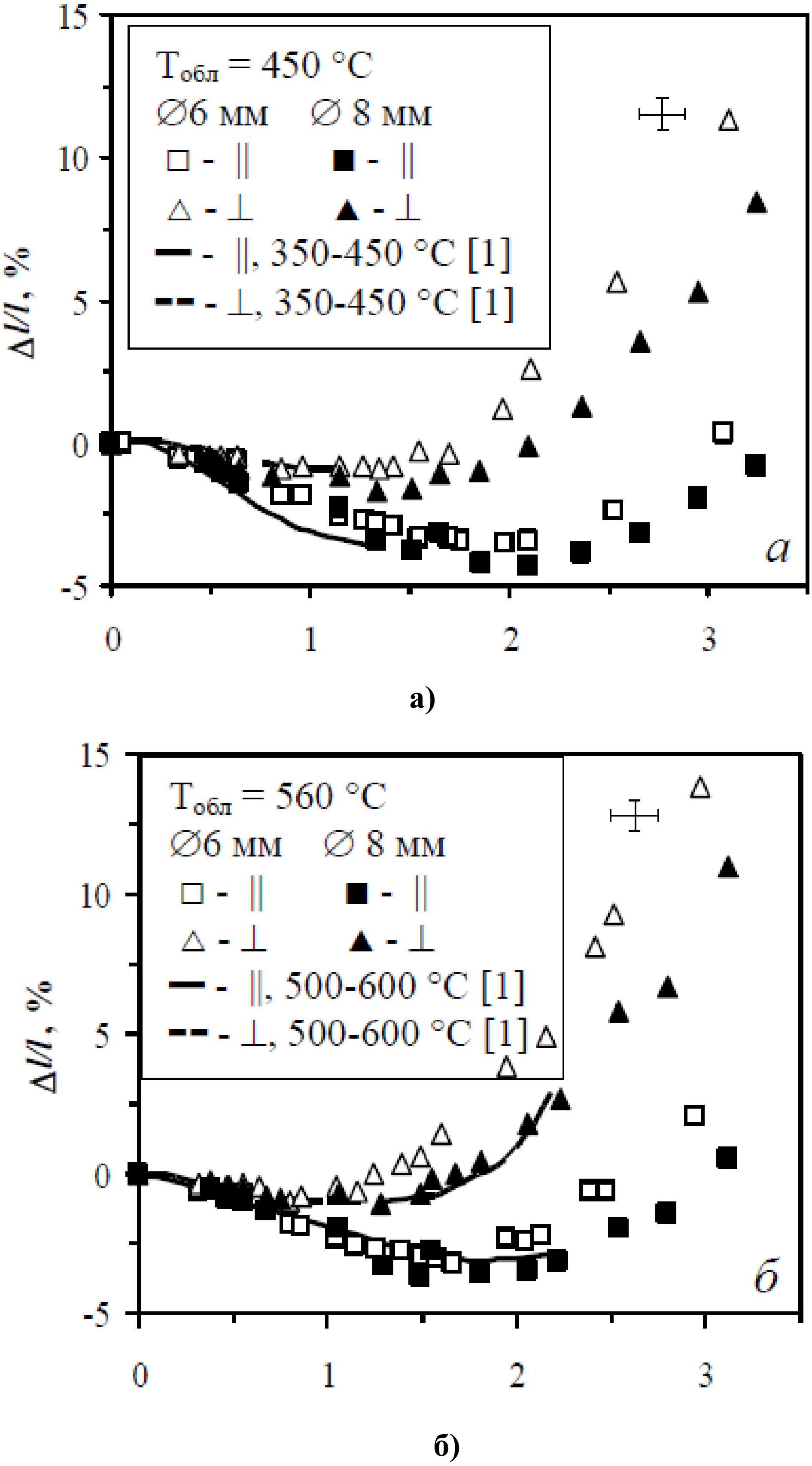

б)

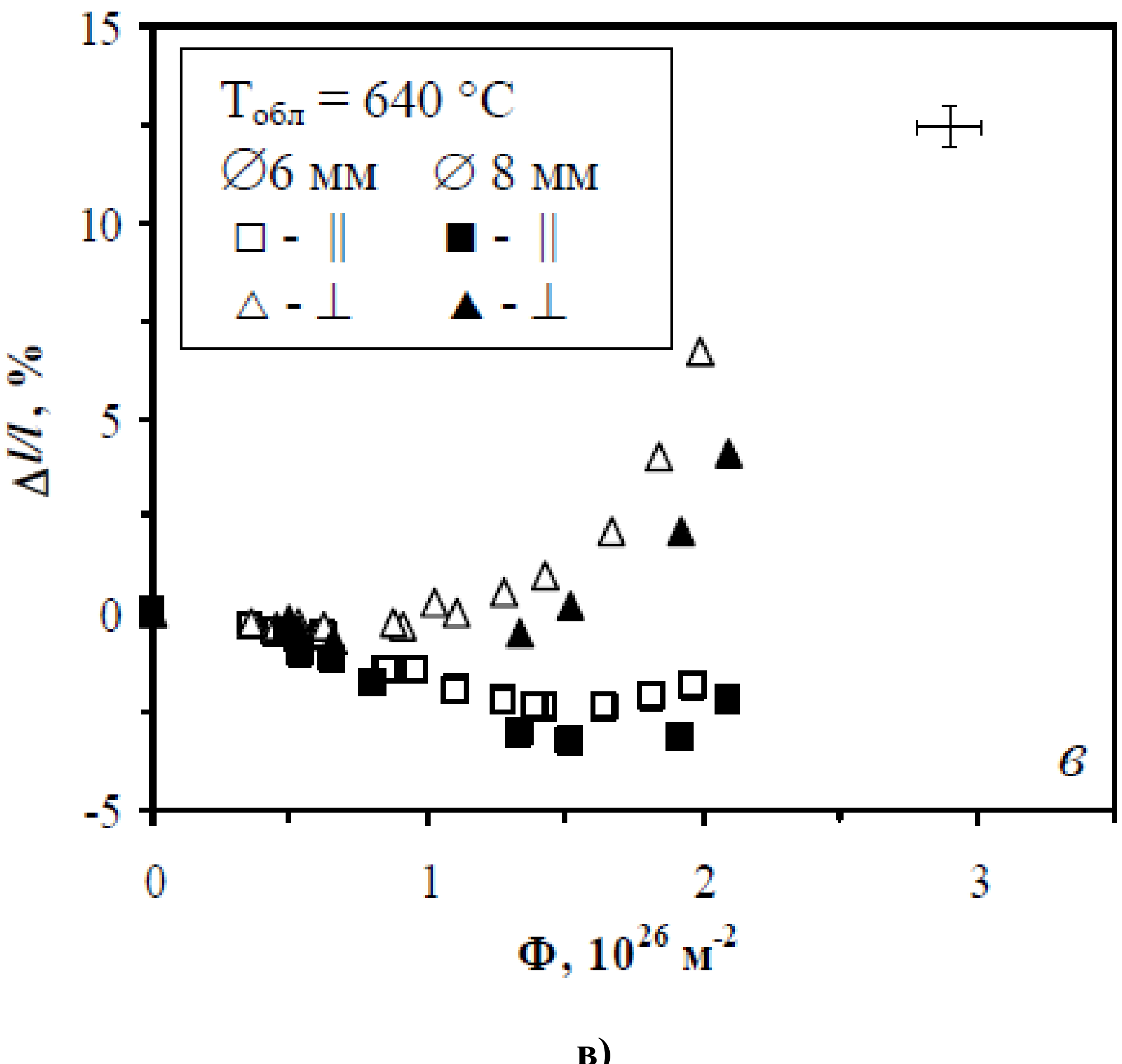

**в)**

**Рис.6.1.** Зависимости относительного изменения длины образцов диаметром 6 и 8 мм, облученных при температуре 450°С (а), 560°С (б) и 640°С (в), от флюенса нейтронов.

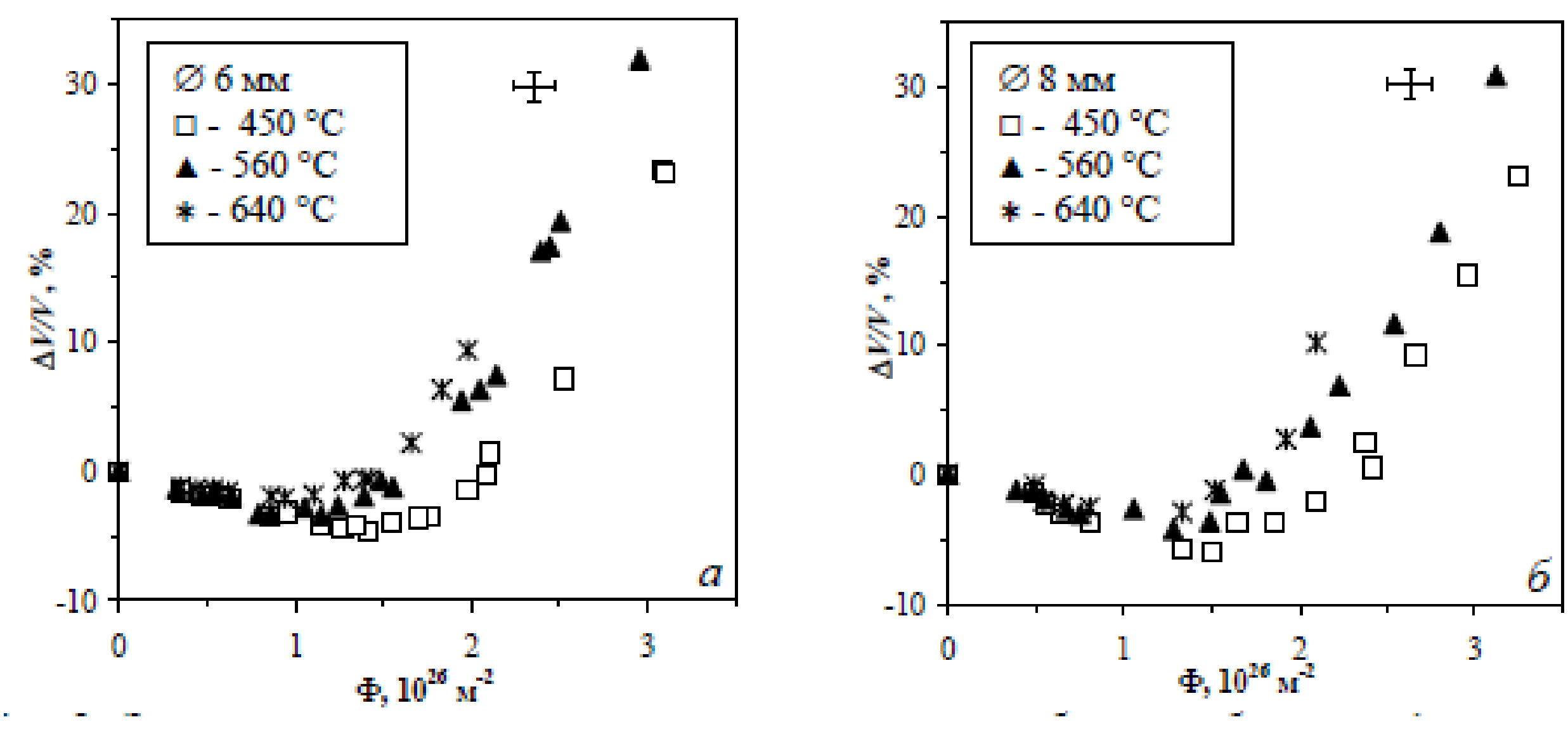

**Рис.6.2.** Зависимости относительного изменения объема образцов диаметром 6 мм (а) и 8 мм (б) от флюенса нейтронов.

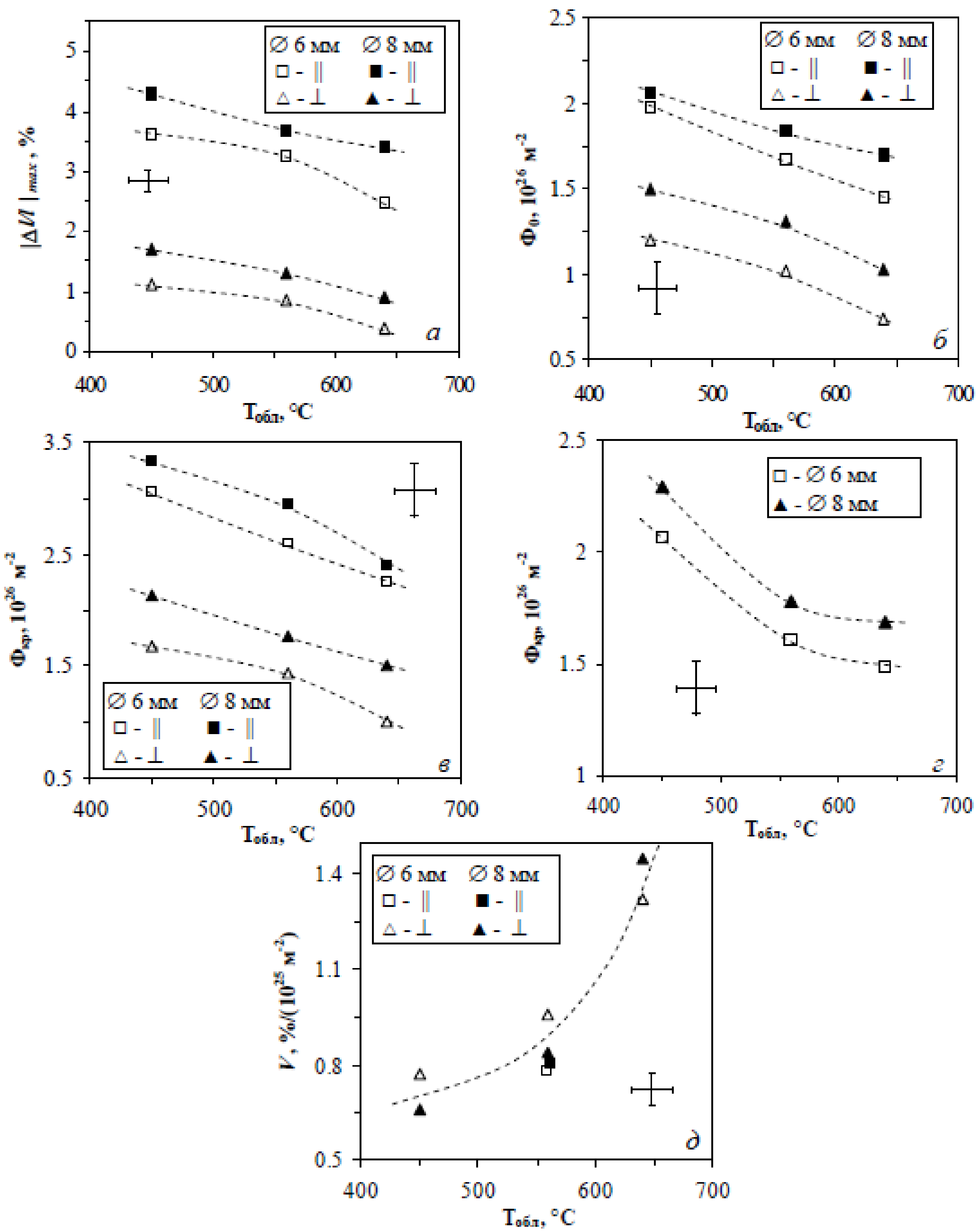

**Рис.6.3.** Зависимости максимальной усадки (а), соответствующего ей флюенса (б), критического флюенса, определенного по длине (в) и объему (г), и скорости роста на 3-ей стадии (д) от температуры облучения.

Параметры радиационного изменения размеров сильно зависят от направления вырезки образцов, так для образцов диаметром 8 мм значение $(\Delta l/l)_{max}$ в параллельном направлении примерно на 3 % больше, чем в перпендикулярном направлении, значения $Ф_0$ и $Ф_{кр}$ на 40-60 % больше в параллельном направлении, чем в перпендикулярном.

Различия значений параметров в параллельном и перпендикулярном направлениях объясняется преимущественной ориентацией кристаллитов базисными плоскостями вдоль оси формования блока и анизотропией формоизменения кристаллитов (при облучении размеры кристаллитов уменьшаются в направлении вдоль базисных плоскостей и увеличиваются в перпендикулярном направлении). Считается, что на первом этапе облучения формоизменение кристаллитов аккомодируется технологическими порами и трещинами, возникающими при остывании заготовок графита после их графитации, по исчерпании аккомодирующей способности пор усадка образцов сменяется их ростом [59, 165, 166].

Увеличение температуры облучения от 450°С до 640°С приводит к ускорению размерных изменений: значение $(\Delta l/l)max$ снижается примерно на 1 %, значения $Ф_0$ и $Ф_{кр}$ уменьшаются на 20-40 %, значение $V_{кр}$ увеличивается на 100 %.

Ускорение размерных изменений при увеличении температуры облучения объясняется увеличением скорости радиационного формоизменения кристаллитов, при этом ввиду того, что скорость размерных изменений зависит от размера кристаллита, причем тем сильнее, чем выше температура облучения, то возникают дополнительные внутренние напряжения, приводящие к образованию трещин при меньших значениях флюенса [167]. Кроме того, уменьшение флюенса и значения максимальной усадки связано со снижением эффекта упрочнения с ростом температуры облучения, в результате чего уменьшается величина напряжения, при котором начинается растрескивание [168].

Установлено существенное влияние исходных размеров образцов на их размерные изменения при облучении («масштабный фактор»): значения $(\Delta l/l)max$, $\Phi_0$ и $\Phi_{кр}$ значительно ниже для образцов диаметром 6 мм по сравнению с образцами диаметром 8 мм, так, например, значение $\Phi_{кр}$, определенное по объему, для образцов диаметром 6 мм на 10-15 % меньше, чем для образцов диаметром 8 мм.

Согласно одной из моделей [**61**] влияние исходных размеров на радиационное формоизменение объясняется меньшим сдерживанием радиационного формоизменения кристаллитов, располагающихся вблизи поверхности. Отмечено хорошее согласие зависимостей относительного изменения длины образцов диаметром 6 мм, облученных при температуре 450 и 560°С, с ограниченными по флюенсу литературными данными для интервалов температуры облучения 350-450 и 500-600°С [163] (рис.6.1).

Представленные в данной главе данные по радиационному формоизменению графита позволяют рассчитывать размерные изменения графитовых блоков при их эксплуатации до максимального флюенса $3{,}3\cdot10^{26}$ $м^{-2}$, соответствующего более чем 45 годам работы реактора. Кроме того, с использованием полученных в настоящей работе данных была разработана модель радиационного формоизменения графита [61], объясняющая влияние «масштабного фактора» на радиационное формоизменение и позволяющая корректно переносить результаты по размерным изменениям, полученным на образцах, на полномасштабные графитовые блоки.

## 6.2. Теплофизические свойства облучённого графита

Теплоемкость графита мало изменяется после облучения – наблюдается уменьшение теплоемкости не более чем на 10 %, что находится в пределах погрешности измерения, причем не отмечено влияния температуры облучения на теплоемкость [169].

На рис.3.4, а-в представлены зависимости относительного изменения коэффициента теплопроводности λ от флюенса нейтронов [170]. В исходном состоянии значения λ составляют 150 и 120 Вт/м·К в параллельном и перпендикулярном направлении соответственно. Различие значений λ связано с преимущественным расположением кристаллитов базисными плоскостями вдоль оси формования блока и анизотропией теплопроводности кристаллитов (в направлении вдоль базисных плоскостей значение λ на несколько порядков выше, чем в перпендикулярном направлении).

Наиболее резкое падение коэффициента теплопроводности λ происходит при относительно малом флюенсе. Так, при флюенсе около $0{,}1\cdot10^{22}$ см$^{-2}$ значение λ снижается на 60-70 % относительно исходного значения. Затем темп снижения коэффициента теплопроводности с увеличением флюенса замедляется, и в интервале $(0{,}7\text{-}1{,}5)\cdot10^{22}$ см$^{-2}$ изменение λ очень незначительно.

При флюенсе, близком к критическому, происходит ускорение темпов падения коэффициента теплопроводности. Резкое снижение значения λ на начальном этапе связано с уменьшением длины свободного пробега фононов за счет рассеяния фононов на возникающих радиационных дефектах [165, 171]. Вторичное снижение при флюенсе выше критического связано, по-видимому, с интенсивным трещинообразованием на стадии вторичного распухания. Относительное изменение λ не зависит от направления вырезки образцов (рис.6.4,а-в). В ряде работ, например [59], было показано, что так как значение λ кристаллитов в направлении вдоль базисных плоскостей на несколько порядков выше, чем в перпендикулярном направлении, то коэффициент теплопроводности поликристаллического графита в произвольном направлении будет пропорционален значению λ кристаллитов в направлении вдоль базисных плоскостей, и относительное изменение коэффициента теплопроводности не будет зависеть от текстуры. Таким образом, имея зависимость λ от флюенса для образцов с одной текстурой, можно построить аналогичные зависимости для образцов с другими текстурами, зная их исходный коэффициент теплопроводности.

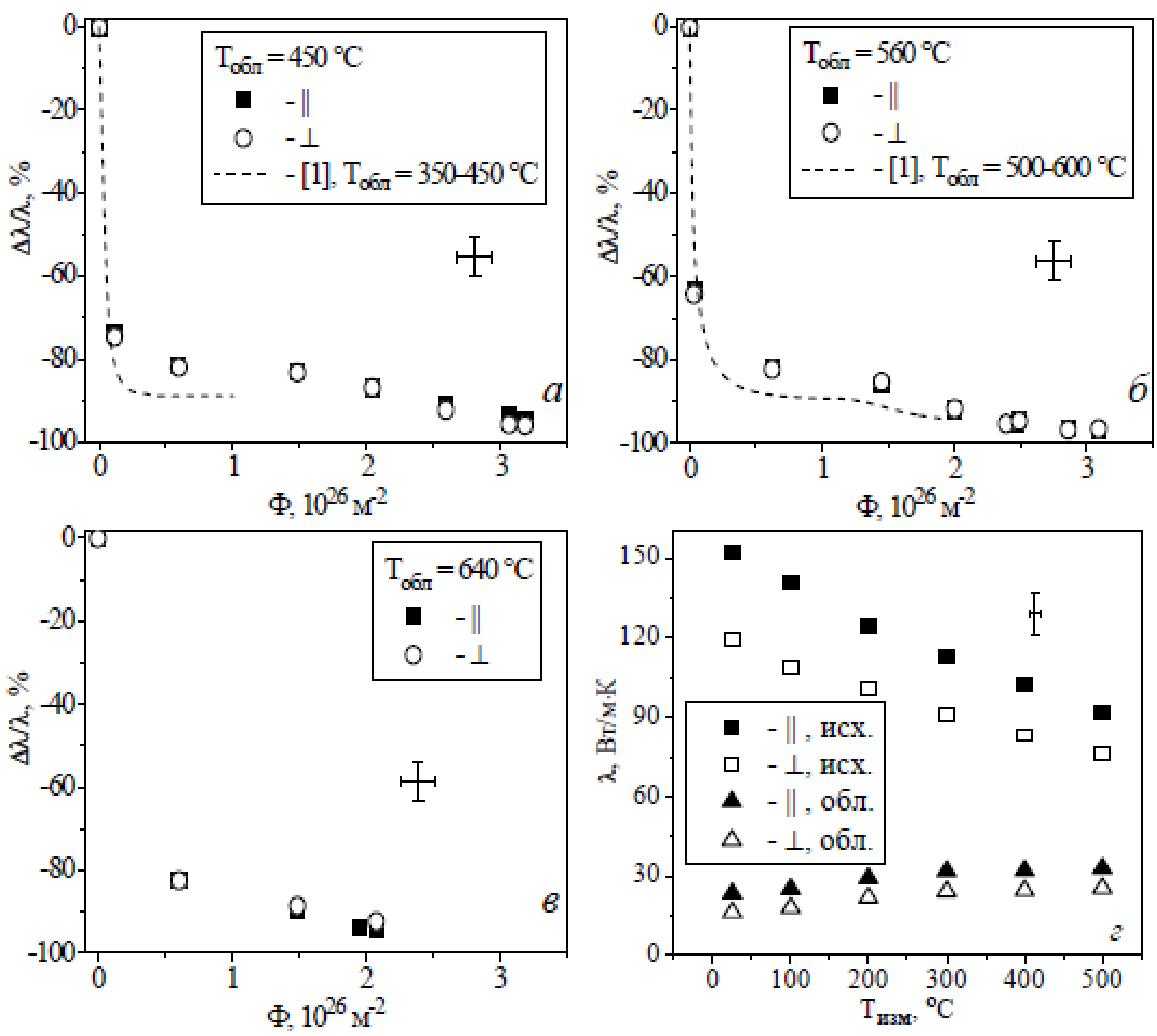

**Рис.6.4.** Зависимости относительного изменения коэффициента теплопроводности λ при 20°С образцов графита, облученных при 450 °С (а), 560 °С (б) и 640 °С (в), от флюенса и зависимость λ исходных и облученных при 560 °С до флюенса $1{,}5\cdot10^{26}$ м$^{-2}$ образцов от температуры измерения (г).

На рис.6.4, г показана зависимость λ от температуры для исходных образцов и образцов после облучения при температуре 560°С до флюенса нейтронов $1{,}5\cdot10^{22}$ см$^{-2}$. Видно, что при увеличении температуры с 20 до 500°С теплопроводность исходных образцов упала с 151 и 119 до 92 и 76 Вт/м·К в параллельном и перпендикулярном направлении соответственно.

Теплопроводность облученных образцов с повышением температуры, напротив, увеличилась с 23 и 16 до 33 и 26 Вт/м·К соответственно. Качественно данный эффект объясняется на основании теории Дебая. В исходном состоянии снижение коэффициента теплопроводности с повышением температуры связано с уменьшением длины свободного пробега фононов за счет интенсификации фонон-фононного рассеяния [172]. В облученном состоянии длина свободного пробега определяется рассеянием фононов на радиационных дефектах и практически не зависит от температуры, рост же значения λ с повышением температуры связан с увеличением теплоемкости. Наблюдается хорошее соответствие зависимостей коэффициента теплопроводности от флюенса, полученных в настоящей работе и приведенных в НГР [163] (рис.6.4, а, б).

Зависимости относительного изменения ТКЛР, усредненного в интервале температуры 25 – 400°С, от флюенса нейтронов представлены на рис.6.5. Исходные значения ТКЛР составляют $(2,5\text{-}3,5)\cdot10^{-6}$ $K^{-1}$ и $(3,5\text{-}4,5)\cdot10^{-6}$ $K^{-1}$ для образцов параллельной и перпендикулярной вырезки соответственно. Это различие связано с анизотропией ТКЛР кристаллитов (значение ТКЛР вдоль оси *c* кристаллитов ~ $24\cdot10^{-6}$ $K^{-1}$, вдоль оси ***a*** ~ $1\cdot10^{-6}$ $K^{-1}$) и преимущественной ориентацией кристаллитов базисными плоскостями вдоль оси формования блока.

Наблюдается сложный характер зависимости ТКЛР от флюенса, а именно: сначала происходит рост значения ТКЛР, не превышающий 20 %, до флюенса $(0,5\text{-}0,7)\cdot10^{22}$ $см^{-2}$, после чего значение ТКЛР уменьшается и далее при флюенсе выше $(1,5\text{-}2,0)\cdot10^{22}$ $см^{-2}$ снова начинает расти. Значение флюенса, при котором начинается вторичный рост ТКЛР, близко к значению критического флюенса.

Отмечено, что увеличение температуры облучения приводит к смещению положений максимального и минимального значений коэффициента α в сторону меньших флюенсов.

В работе [59] было показано, что значение ТКЛР кристаллитов практически не изменяется относительно их исходных значений при облучении при температуре выше 300°С, что связано с очень малыми изменениями параметров кристаллической решетки. Поэтому в нашем случае можно утверждать, что изменение ТКЛР связано с изменением микроструктуры поликристаллического графита (в основном пористой подсистемы), а не с изменением ТКЛР составляющих его кристаллитов.

Немонотонность изменения коэффициента $\alpha$ от флюенса нейтронов (наличие максимума и минимума) объясняется наличием нескольких конкурирующих процессов [173, 174].

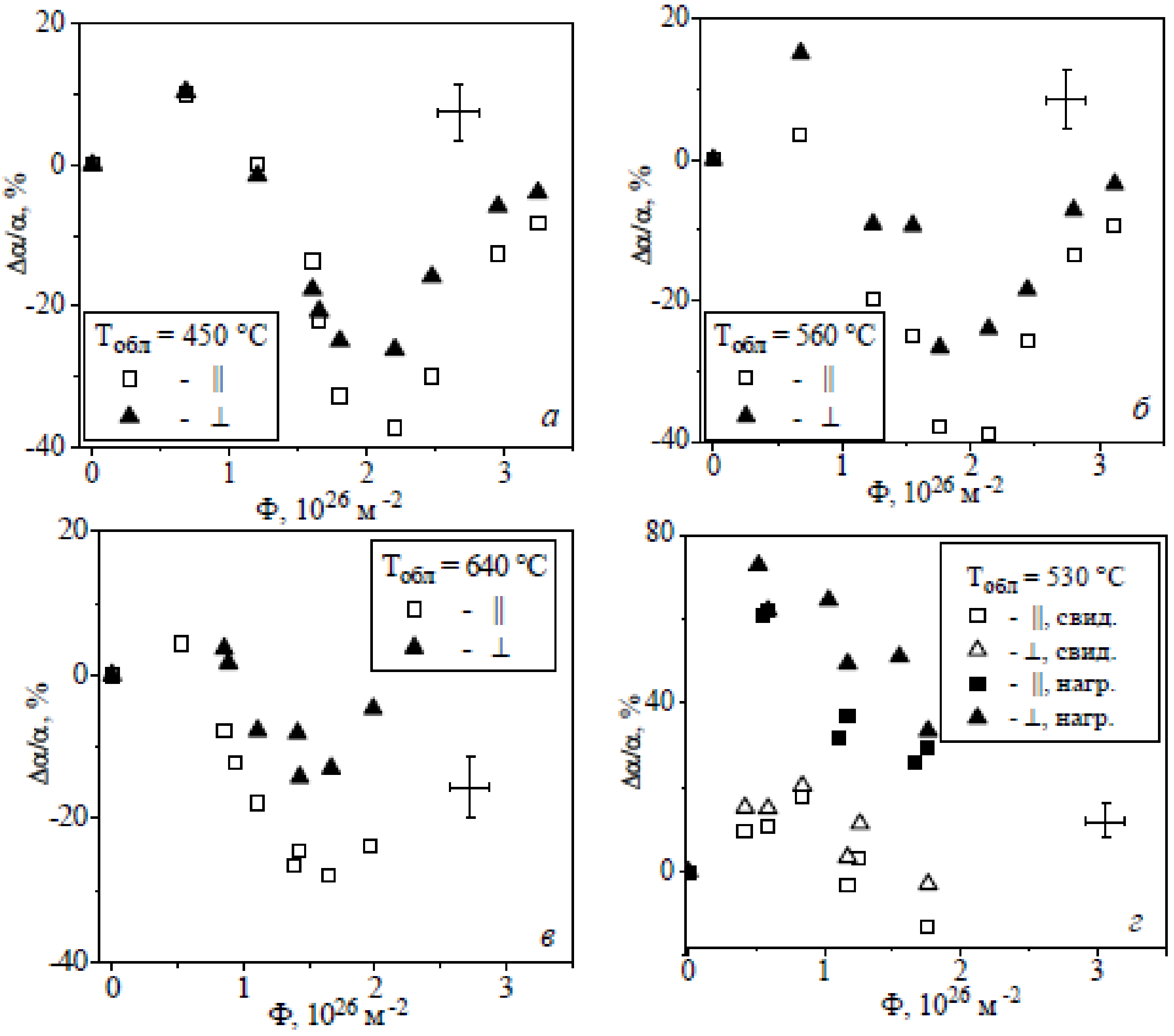

**Рис.6.5.** Зависимости относительного изменения ТКЛР образцов, облученных при температуре 450°С (а), 560°С (б), 640°С (в) и 530°С (г), от флюенса.

Радиационная деформация кристаллитов (рост вдоль кристаллографической оси ***c*** и сжатие в базисной плоскости) приводит к монотонному увеличению коэффициента $\boldsymbol{\alpha}$ поликристаллического графита ввиду анизотропии ТКЛР кристаллитов. Кроме того, деформация сопровождается заполнением межкристаллитных пор и снижением аккомодационной способности материала, что также приводит к увеличению ТКЛР. По исчерпании аккомодационного объема начинается механическое взаимодействие кристаллитов и их фрагментация. В результате развивается новая система трещин, и поэтому коэффициент $\alpha$ снижается. Вторичный рост значения ТКЛР при флюенсах выше критического может быть связан, во-первых, с ростом ТКЛР за счет радиационного формоизменения кристаллитов, и, во-вторых, с меньшим сдерживанием теплового расширения кристаллитов ввиду увеличения количества микротрещин на стадии вторичного распухания.

В данной работе выявлено сильное влияние нагрузки на изменение ТКЛР при облучении (рис.6.5,г). Образцы, облучавшиеся под сжимающим напряжением 15 МПа при температуре 530°С демонстрируют значительно более высокие значения ТКЛР по сравнению с ненагруженными образцами, особенно сильно этот эффект выражен для образцов перпендикулярной вырезки. Так, при флюенсе около $0{,}6\cdot10^{26}$ м$^{-2}$ ТКЛР нагруженных образцов перпендикулярной вырезки увеличивается почти на 70 % по сравнению с исходным значением, тогда как для ненагруженных образцов при том же флюенсе ТКЛР увеличился на 20 %. Данный эффект объясняется, по-видимому, ускоренным закрытием исходной пористости, происходящим за счет радиационной ползучести под воздействием сжимающей нагрузки.

Полученные результаты по радиационному изменению теплофизических свойств совместно с данными по изменению модуля упругости и предела прочности необходимы для оценки работоспособности графитовой кладки по критерию трещиностойкости, согласно которому величина напряжений в блоке не должна превышать предел прочности. Выявленное в настоящей работе сильное влияние напряжения на радиационное изменение ТКЛР позволяет выдвинуть требование о необходимости учета данного эффекта при расчетах НДС блоков, кроме того, этот эффект необходимо учитывать при определении параметров радиационной ползучести графита, что было сделано в настоящей работе.

## 6.3. Физико-механические свойства графита под облучением

Изменение динамического модуля упругости *E* при облучении при 560°С происходит в две стадии (рис. 6.6,а): на начальном этапе происходит увеличение значения Е до максимального прироста в 120 – 160 % при флюенсе $(1{,}5\text{-}2{,}0)\cdot10^{22}$см$^{-2}$, при дальнейшем увеличении флюенса наблюдается снижение значения модуля упругости. Зависимости модуля упругости при температуре облучения 450 и 640°С имеют аналогичный вид.

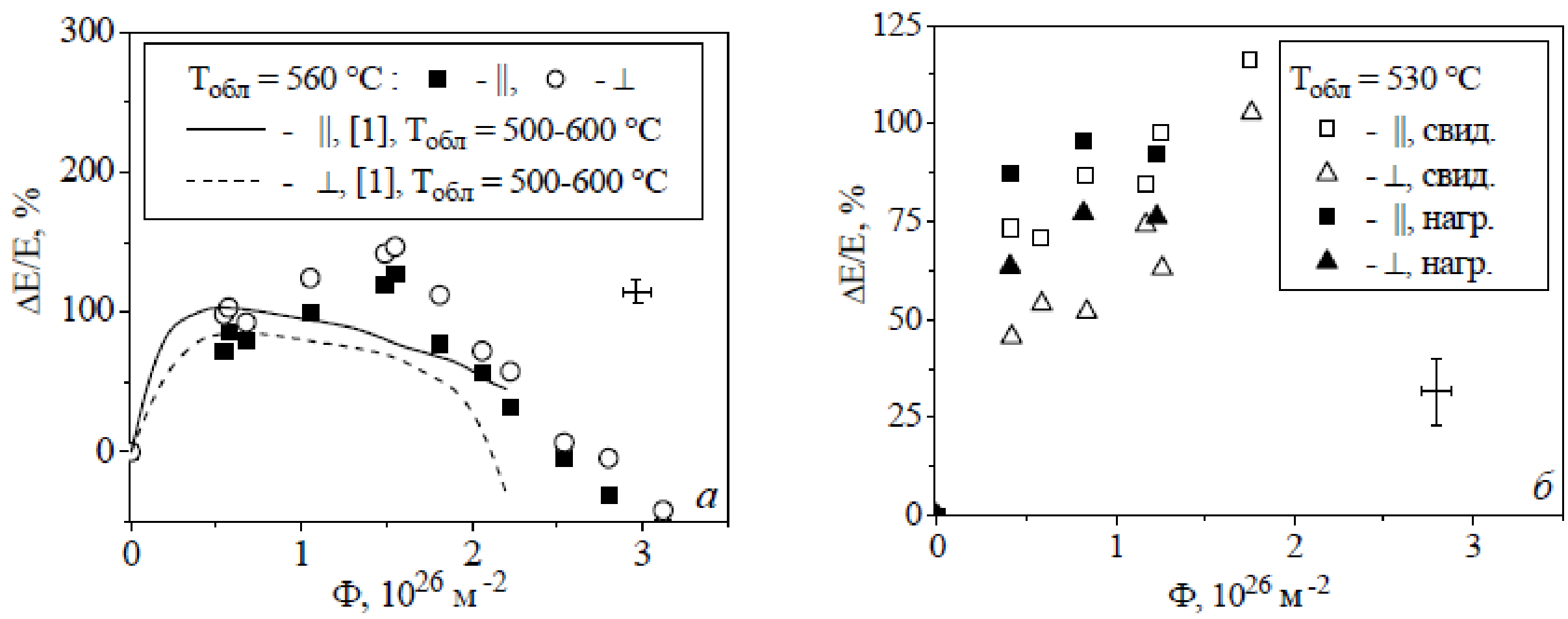

**Рис.6.6.** Зависимости относительного изменения динамического модуля упругости образцов диаметром 8 мм, облученных при температуре 560° С (а) и 530° С (б), от флюенса нейтронов.

Изменение модуля упругости поликристаллического графита при облучении обусловлено одновременным протеканием двух процессов: изменения упругих констант составляющих кристаллитов и изменения пористой подсистемы графита [175, 176]. На начальном участке модуль упругости поликристаллического графита растет за счет увеличения значения модуля сдвига кристаллитов $C_{44}$, связанного с закреплением подвижных дислокаций в базисных плоскостях на возникающих радиационных дефектах [59] и закрытием исходной пористости до флюенса, соответствующего максимальной объемной усадке, после чего происходит спад значения Е, вызванный растрескиванием графита.

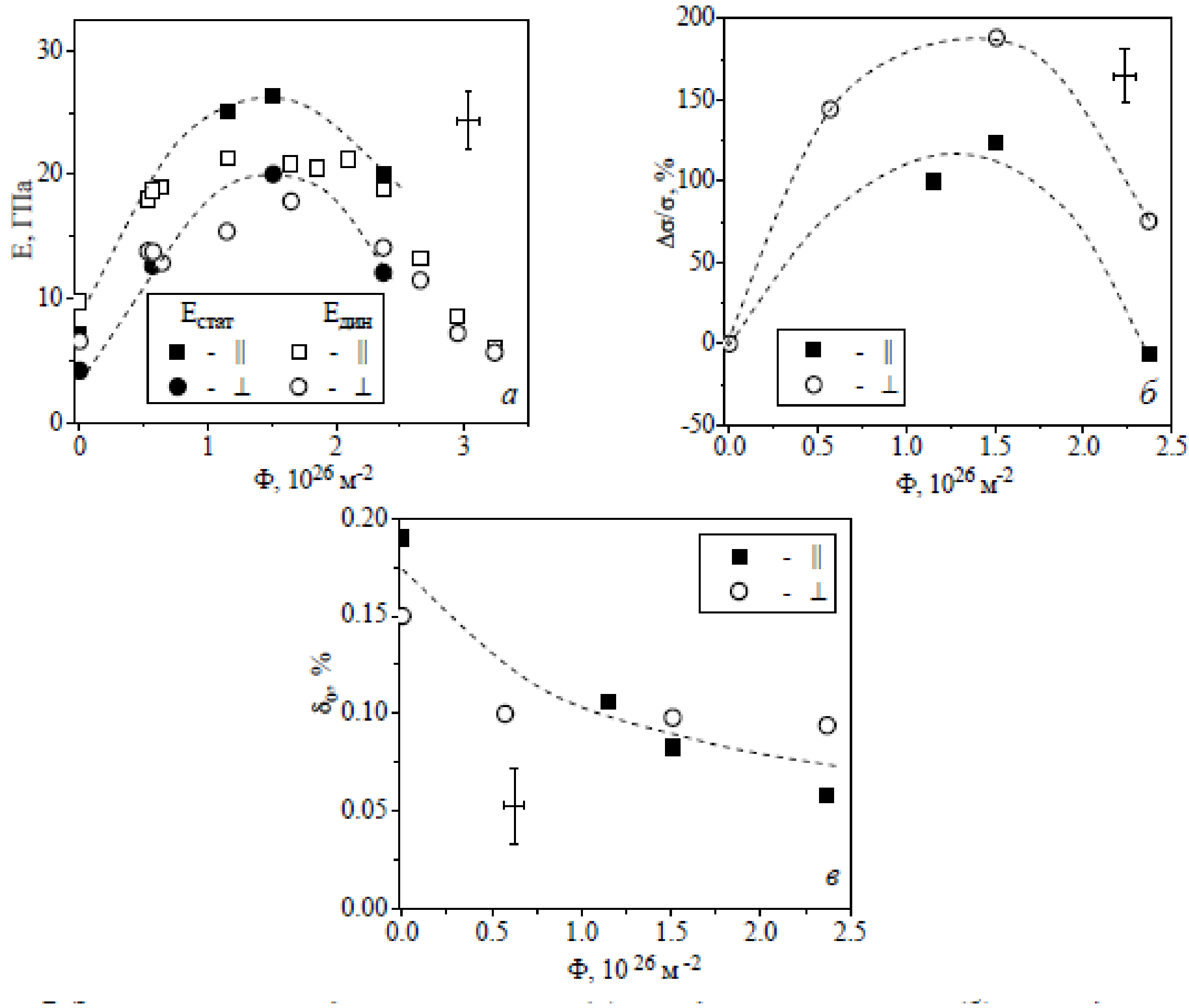

**Рис.6.7.** Зависимости модуля упругости (а), предела прочности (б) и предельной деформации (в) при испытаниях на растяжение образцов, облученных при температуре 450°С, от флюенса нейтронов.

Наблюдается небольшое расхождение зависимостей динамического модуля упругости от флюенса, полученных в настоящей работе и приведенных в НГР [163].

Ранее, в параграфе 6.2 было показано, что сжимающая нагрузка при облучении привела к значительному увеличению ТКЛР относительно ТКЛР облучавшихся без нагрузки образцов графита, при этом сжимающая нагрузка не вызвала заметного изменения динамического модуля упругости (рис.6.6,б).

Значения механических свойств при растяжении в исходном состоянии и соответствующие коэффициенты вариации (отношение среднеквадратичного отклонения к среднему значению) представлены в табл.1. Всего было испытано по 20 исходных образцов параллельной и перпендикулярной вырезки. Так как на диаграммах «напряжение – деформация» графита невозможно выделить начальный линейный участок, соответствующий области упругой деформации, условный модуль упругости определялся на начальном участке диаграммы в интервале удлинений 0 ÷ 0,04 %.

**Таблица 6.1**. Средние значения и коэффициенты вариации механических свойств при растяжении исходных образцов графита.

| Направление вырезки | $\sigma_p$, МПа | $v_\sigma$, % | $\delta_0$, % | $v_\delta$, % | E, ГПа | $v_E$, % |
|---|---|---|---|---|---|---|
| ǁ | 10,6 | 10 | 0,20 | 0,03 | 6,9 | 15 |
| ⊥ | 5,3 | 13 | 0,15 | 0,03 | 4,2 | 12 |

Средние значения предела прочности $\sigma_p$, общего удлинения $\delta_o$ и модуля *E* для образцов параллельной вырезки выше, чем для образцов перпендикулярной вырезки. Более высокое значение $\sigma_p$ в параллельном оси формования направлении связано с преимущественным расположением кристаллитов базисными плоскостями вдоль оси блока и тем, что разрушение графита при нагружении происходит за счет растрескивания кристаллитов вдоль базисных плоскостей. Более высокие значения *E* в параллельном направлении также объясняется текстурой графита и анизотропией упругих свойств кристаллитов (в направлении параллельном базисным плоскостям модуль упругости кристаллитов значительно выше).

На рис.6.7 приведены зависимости модуля упругости, предела прочности и предельной деформации от флюенса нейтронов для образцов, облученных при температуре 450°C. Зависимости механических свойств от флюенса для образцов, облученных при температуре 560°C, имеют аналогичный вид.

Для сравнения на рис.6.7,а приведены также зависимости динамического модуля упругости от флюенса. Можно отметить совпадение в пределах погрешности значений модуля упругости, определенного статическим и динамическим методами, что подтверждает возможность использования динамического метода для оценки статических модулей упругости.

Предел прочности растет приблизительно до флюенса $1{,}6\cdot10^{22}$ см$^{-2}$, после чего происходит его снижение, и при флюенсе около $2{,}4\cdot10^{22}$ см$^{-2}$ значение $\sigma_p$ становится равным исходному значению. Максимальный прирост предела прочности составляет 120-200 % в зависимости от направления вырезки. Облучение привело к уменьшению значения предельной деформации с 0,15-0,19 % до 0,06-0,10 % в зависимости от направления вырезки при флюенсе $2{,}4\cdot10^{22}$ см$^{-2}$.

Упрочнение графита на начальном участке может быть объяснено двумя механизмами: во-первых, упрочнением кристаллитов за счет закрепления дислокаций в базисных плоскостях и, во-вторых, уменьшением при облучении диаметра областей когерентного рассеяния $L_a$, так как в соответствии с теорией Гриффитса-Орована [166, 177] $\sigma_p \sim L_a^{-1/2}$. Снижение предела прочности при флюенсах $1{,}5\cdot10^{22}$ см$^{-2}$ связано, по-видимому, с интенсивным образованием трещин на стадии вторичного распухания.

Следует отметить, что флюенс, соответствующий максимальным значениям модуля упругости и предела прочности близок к значению флюенса максимальной объемной усадки, что является косвенным свидетельством того, что снижение модуля упругости и предела прочности на второй стадии происходит за счет образования трещин.

Полученные данные по радиационному изменению модуля упругости и предела прочности необходимы для расчета термических и радиационных напряжений в блоках, возникающих из-за градиента температур и неравномерности радиационного формоизменения по объему блока и, как отмечалось выше, для оценки работоспособности графитовой кладки по критерию трещиностойкости, согласно которому величина напряжений в блоке не должна превышать значения предела прочности.

## 6.4. Радиационная ползучесть графита

Радиационная ползучесть графита является одним из основных явлений, определяющих срок службы графитовой кладки, так как ползучесть приводит к дополнительным размерным изменениям блоков под воздействием внутренних и внешних напряжений. Кроме того, радиационная ползучесть способствует релаксации радиационных и термических напряжений, что благоприятно сказывается на стойкости графитовых блоков к возникновению трещин.

Радиационная ползучесть исследовалась на образцах графита Ø8×30 мм, величина сжимающего напряжения составляла около 15 МПа, облучение проводилось при температуре 530°С [178].

На рис.6.8,а представлены зависимости деформации ползучести $\varepsilon_c$ от флюенса, рассчитанной по стандартной формуле:

$$\varepsilon_c = (\Delta l / l)_{нагр} - (\Delta l / l)_{свид}, \qquad (6.1)$$

где $(\Delta l / l)_{нагр}$ и $(\Delta l / l)_{свид}$ - относительное изменение длины нагруженных образцов и образцов-свидетелей соответственно. Значение $\varepsilon_c$ увеличивается с ростом флюенса, причем деформация образцов перпендикулярной вырезки значительно выше, чем параллельной. Однако, приведенные на рис.6.8,а кривые демонстрирует значительное отклонение от ожидаемой на установившейся стадии линейной зависимости. С ростом флюенса происходит уменьшение скорости ползучести, особенно для образцов перпендикулярной вырезки, для которых она практически стремится к нулю при флюенсе выше $1{\cdot}10^{22}$ см$^{-2}$.

Для сравнения на рис.6.8,а приведены зависимости деформации ползучести, рассчитанные по эмпирическому уравнению, полученному авторами работы [179] после обработки данных по разным маркам графитов в широких интервалах флюенса и напряжений и температуры облучения 140-650°С:

$$\varepsilon_c = \left[1{,}63\cdot10^{-25}\frac{\sigma}{E_0}\Phi + \frac{\sigma}{E_0}\left(1-\exp\left(-2{,}84\cdot10^{-24}\Phi\right)\right)\right]\cdot100\%, \qquad (6.2)$$

где σ- напряжение, $E_0$ - модуль упругости необлученного графита, Φ - флюенс нейтронов, м$^{-2}$. Видно, что наблюдается большое расхождение между расчетными кривыми и экспериментальными данными.

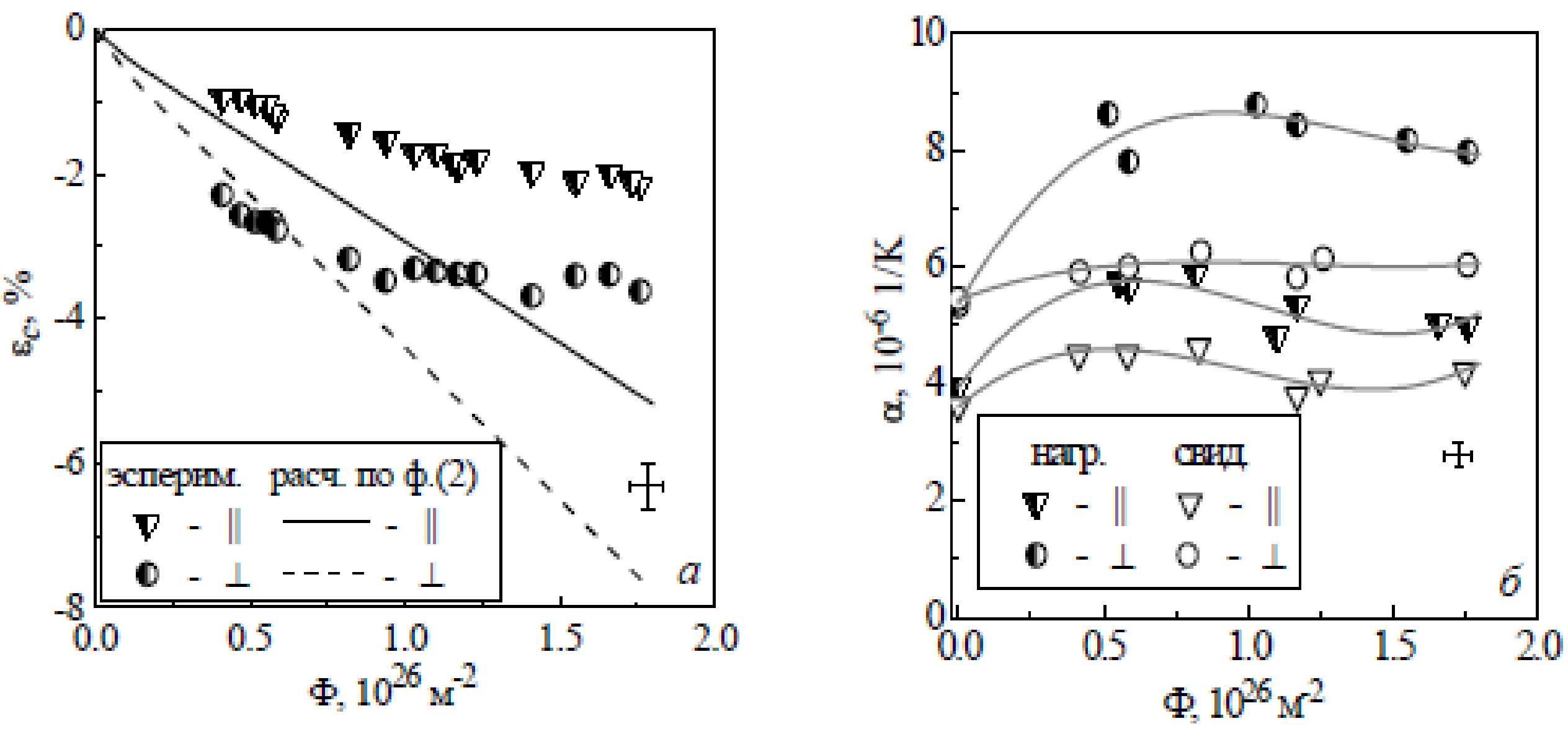

**Рис.6.8.** Зависимости экспериментальной и рассчитанной по формуле (6.2) деформации ползучести (а) и ТКЛР, измеренного при 530°С (б), от флюенса нейтронов.

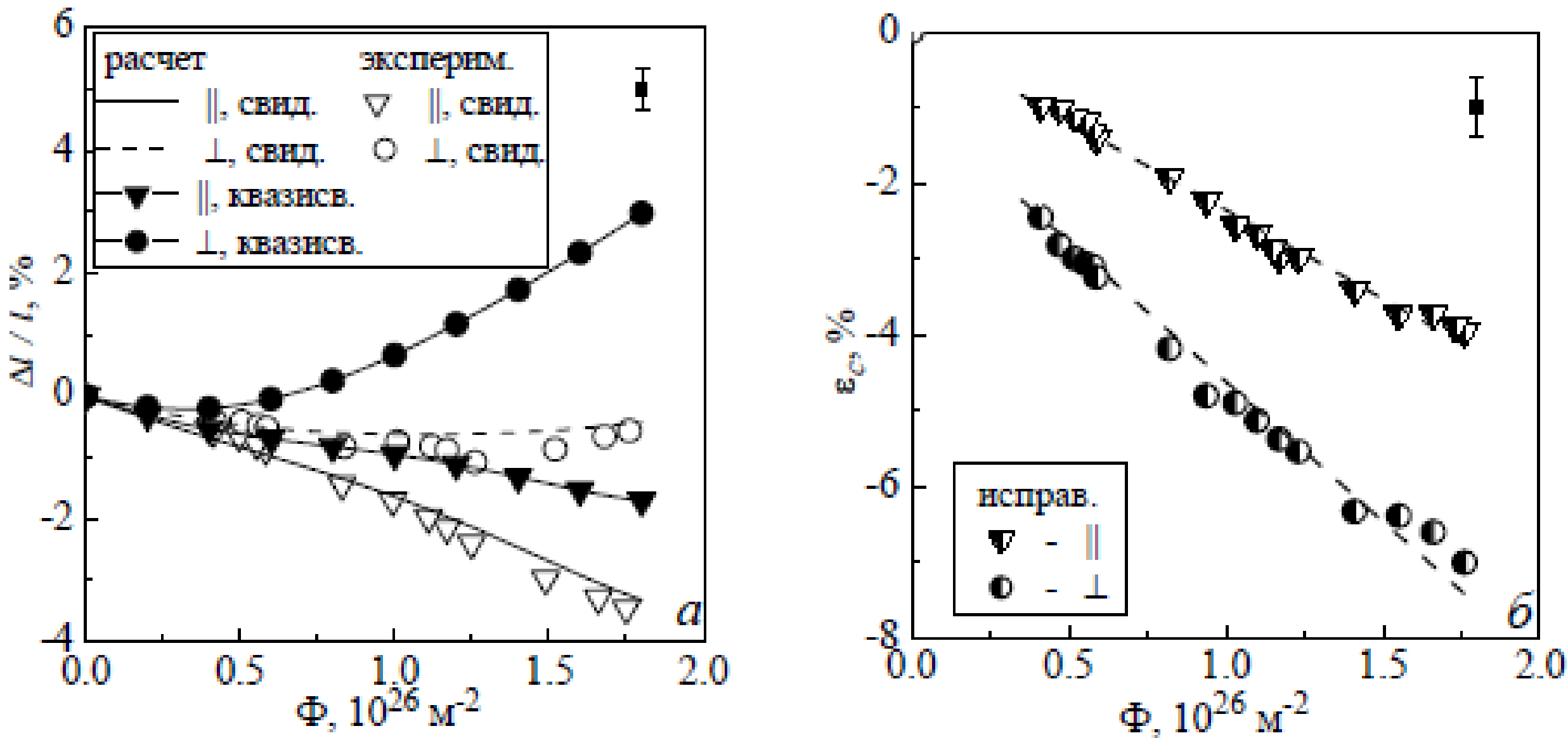

**Рис.6.9.** Зависимости относительного изменения длины образцов-свидетелей и квазисвободных образцов (а), рассчитанные по ф.(6.4) и (6.5), и зависимость исправленной деформации ползучести (б), рассчитанной по ф.(6.3), от флюенса нейтронов.

Отклонение от линейности зависимости деформации ползучести от флюенса наблюдалось авторами и других работ, например [180]. Ими было выдвинуто предположение о том, что при облучении под воздействием нагрузки существенно изменяется структура графита, прежде всего пористая подсистема, что приводит к дополнительным размерным изменениям. Поэтому деформация ползучести, определенная по формуле (6.1), будет включать в себя и деформацию, связанную с дополнительными структурными изменениями нагруженных образцов по сравнению с образцами-свидетелями. Для корректного определения деформации ползучести в формуле (6.1) усадка образцов-свидетелей $(\Delta l\ /\ l\ )_{свид}$ должна быть заменена на усадку неких идеализированных образцов, которые облучаются без нагрузки, но при этом их структура изменяется так же, как у нагруженных образцов (далее такие образцы будем называть квазисвободными и обозначать их размерные изменения $(\Delta l\ /\ l)_{квазисв}$):

$$\boldsymbol{\varepsilon}_{исп} = (\Delta l\ /l)_{нагр} - (\Delta l/\ l)_{квазисв}, \qquad (6.3)$$

Ввиду невозможности прямого экспериментального определения $\Delta l/l_{квазисв}$ она была рассчитана с использованием теории Симмонса [181, 182] по методике, предложенной в [59]. Теория Симмонса устанавливает связь между размерными изменениями и ТКЛР поли- и монокристаллического графита. Основу теории составляют уравнения:

$$\begin{cases} \alpha_x = A_x(\Phi)\alpha_c + [1 - A_x(\Phi)]\alpha_a, \\ \dfrac{d}{d\Phi}\left(\dfrac{\Delta l_x}{l_x}\right) = A_x(\Phi)\dfrac{d}{d\Phi}\left(\dfrac{\Delta X_c}{X_c}\right) + [1 - A_x(\Phi)]\dfrac{d}{d\Phi}\left(\dfrac{\Delta X_a}{X_a}\right); \end{cases} \qquad (6.4)$$

где $A_x(\Phi)$ – коэффициент, учитывающий текстуру и пористость и зависящий от флюенса; $\alpha_c$, $\alpha_a$ – ТКЛР монокристаллического графита в направлении осей $c$ и $a$; $\alpha x$ и $\Delta lx/lx$ - ТКЛР и размерные изменения поликристаллического графита в некотором направлении $x$; $\Delta Xc/Xc$ , $\Delta Xa/Xa$ - относительные размерные изменения кристаллитов в направлении осей $c$ и $a$.

Выражая $Ax(\Phi)$ из первого уравнения и подставляя во второе, получаем уравнения для определения скорости размерных изменений образцов свидетелей и квазисвободных образцов:

$$\frac{d}{d\Phi}\left(\frac{\Delta l_x}{l_x}\right)_{\text{свид}} = \frac{\alpha_x - \alpha_a}{\alpha_c - \alpha_a}\left[\frac{d}{d\Phi}\left(\frac{\Delta X_c}{X_c}\right) - \frac{d}{d\Phi}\left(\frac{\Delta X_a}{X_a}\right)\right] + \frac{d}{d\Phi}\left(\frac{\Delta X_a}{X_a}\right);$$

$$\frac{d}{d\Phi}\left(\frac{\Delta l_x}{l_x}\right)_{\text{квазисв}} = \frac{\alpha_x^{\text{нагр}} - \alpha_a}{\alpha_c - \alpha_a}\left[\frac{d}{d\Phi}\left(\frac{\Delta X_c}{X_c}\right) - \frac{d}{d\Phi}\left(\frac{\Delta X_a}{X_a}\right)\right] + \frac{d}{d\Phi}\left(\frac{\Delta X_a}{X_a}\right), \qquad (6.5)$$

где $\alpha_x^{\text{нагр}}$ - ТКЛР нагруженных образцов. С использованием описанной методики были рассчитаны зависимости усадок $(\Delta l/l)_{\text{квазисв}}$ и $(\Delta l/l)_{\text{свид}}$ от флюенса.

Зависимости ТКЛР при 530°С от флюенса представлены на рис.6.8,б, из которого видно, что для обоих направлений вырезки ТКЛР нагруженных образцов существенно больше таковых для образцов-свидетелей. Для расчетов полученные зависимости были аппроксимированы полиномами третьей степени, графики которых также приведены на рис.6.8,б.

В качестве размерных изменений кристаллитов $\Delta Xc/Xc$ , $\Delta Xa/Xa$ были использованы данные по пирографиту из работ [183, 184], интерполированные на температуру облучения 530°С. Значения ТКЛР кристаллитов были приняты постоянными $\alpha_c = 24\cdot10^{-6}$ К$^{-1}$, $\alpha_a$ = -1·10-6 К$^{-1}$, так как они не зависят от флюенса при температуре облучения выше 300°С [165].

После численного интегрирования уравнений (6.5) были получены зависимости $(\Delta l/l)_{\text{квазисв}}$ и $(\Delta l/l)_{\text{свид}}$ от флюенса нейтронов. Расчетные кривые для образцов-свидетелей хорошо согласуются с экспериментальными точками, что подтверждает корректность выбора исходных данных для расчета (рис.6.9,а). Также отмечено, что зависимости размерных изменений квазисвободных образцов лежат значительно выше соответствующих зависимостей образцов-свидетелей.

Зависимость исправленной деформации ползучести, определенной по формуле (6.3), значительно приблизилась к линейной зависимости (рис.6.9,б). По результатам линейной аппроксимации методом наименьших квадратов были определены скорости деформации ползучести для образцов параллельной и перпендикулярной вырезки, составившие $(1{,}6 \pm 0{,}2) \cdot 10^{-29}$ и $(2{,}5 \pm 0{,}3) \cdot 10^{-29}$ МПа$^{-1} \cdot$м$^{2}$ соответственно.

Следует отметить, что использование полученных в работе уточненных значений скорости радиационной ползучести в предварительных расчетах НДС графитовых блоков явилось одним из основных факторов, благодаря которым была показана возможность продления ресурса графитовых кладок

***

В ходе работы, изложенной в данной главе, ввиду большого объема получаемых экспериментальных данных возникла необходимость создания базы данных для их хранения и упрощения процесса их обработки и анализа. Она была построена по технологии клиент-сервер на основе свободно распространяемого SQL-сервера Firebird. Архитектура базы данных ориентирована на структуру эксперимента и позволяет отразить изменение свойств образцов, получаемых в результате измерений на различных этапах эксперимента. Все полученные в ходе экспериментов результаты были внесены в базу данных.

Для обработки первичных данных, помещаемых в базу, была разработана программа-клиент, которая позволяет выбирать и сортировать результаты экспериментов по различным критериям, проводить их простейшую статистическую обработку и выводить экспериментальные данные в виде зависимостей свойств от флюенса нейтронов и температуры. Созданная база войдет в состав новой версии «Норм расчета на прочность типовых узлов и деталей из графита уран-графитовых канальных реакторов», на основании которых будет уточнен ресурс графитовых кладок реакторов.

ПРИЛОЖЕНИЕ I

## КРЕПЛЕНИЕ ГРАФИТОВЫХ ОБРАЗЦОВ

Для крепления графитовых образцов были разработаны специальные зажимы из тантала, которые не только выдерживают высокие температуры без окисления, но и позволяют создать надежные токовводы, позволяющие исключить продольные нагрузки на образец (рис.I.1). На основе этих же креплений позднее были разработаны держатели, позволившие подавать регулируемую нагрузку на образец. Высокотемпературные испытания с повышенной нагрузкой показали заметное снижение времени жизни образца во время теста, в зависимости от величины приложенной нагрузки. Первые же пробы показали заметное снижение времени жизни образцов под нагрузкой во время высокотемпературных тестов, однако результаты этих тестов были сугубо предварительные, и анализ их не проводился.

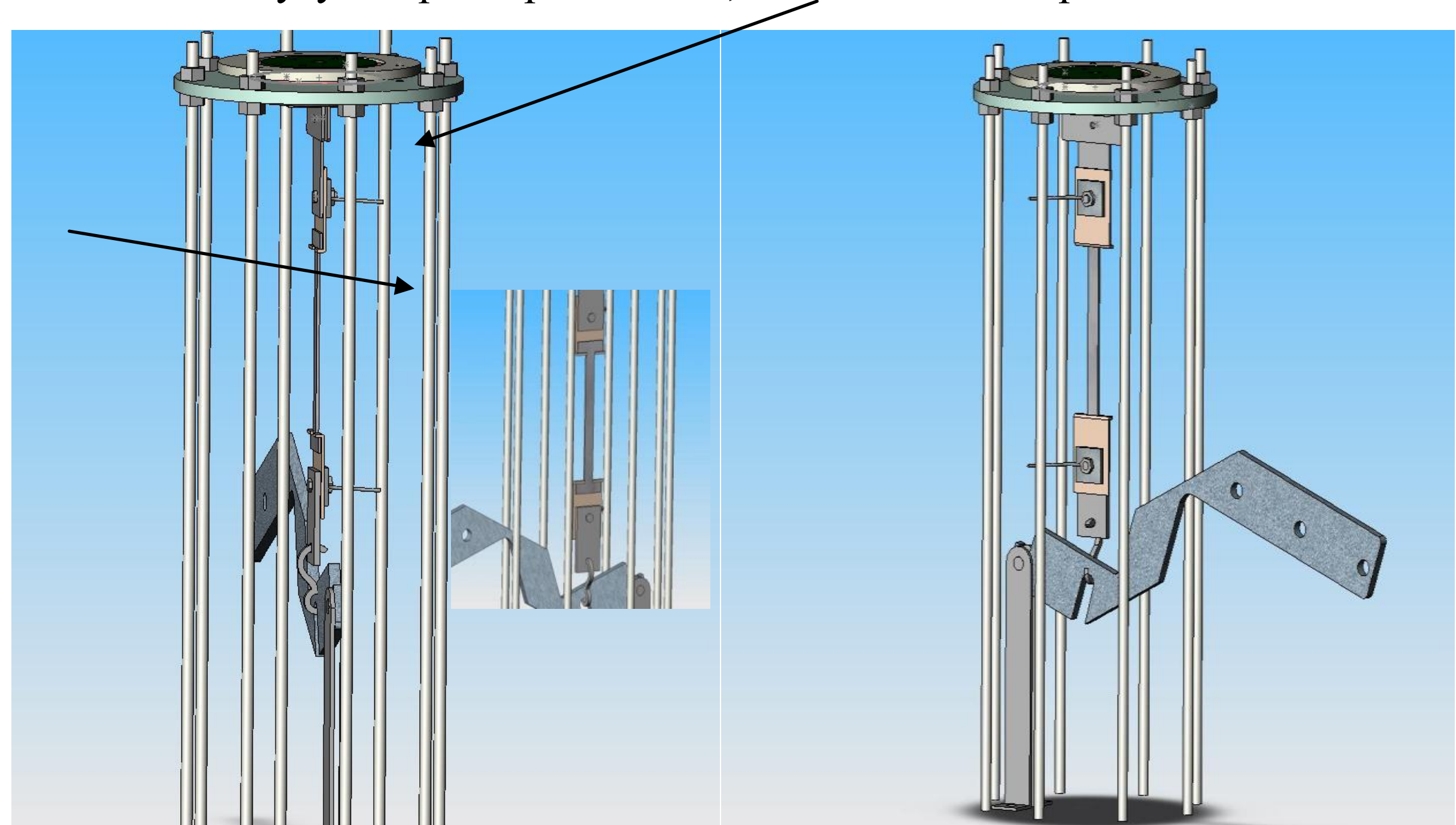

**Рис. I.1.** Держатель графитовых образцов (оригинальная разработка). Стрелками на верхнем чертеже указаны точки крепления, которые предоставляют свободу вращения на концах графитового образца, близкую к сферическим подшипникам (чертёж выполнен в 3D-графической программе). На врезке показано увеличенное изображение крепления графитового образца.

ПРИЛОЖЕНИЕ II

# КВАНТОВЫЕ ПОПРАВКИ К ПРОВОДИМОСТИ

Слабая локализация (СЛ) - совокупность явлений, обусловленных квантовой интерференцией электронов проводимости в проводниках с металлическим типом проводимости, т. е. обладающих остаточной проводимостью [82, 185]. Эффекты СЛ универсальны и проявляются в любых неупорядоченных системах - сильнолегированных полупроводниках, металлических стёклах, системах с двумерным электронным газом, тонких металлических плёнках и т. д. При температурах столь низких, что сопротивление проводника определяется рассеянием электронов на случайном потенциале, создаваемом, например, хаотически расположенными примесями, квантовая интерференция приводит к поправкам к классической электропроводности. Последнюю рассчитывают на основе кинетического уравнения Больцмана, при выводе которого предполагается, что между соударениями электрон движется по классической траектории и рассеяние на различных центрах происходит независимо. К слабой локализации приводит изменение скорости диффузии электронов за счёт интерференции электронных волн, многократно рассеиваемых дефектами кристаллической решётки.

Происхождение термина «Слабая локализация» объясняется тем, что интерференционные явления можно интерпретировать как предвестник андерсоновского перехода металл – диэлектрик, при котором благодаря достаточно сильному беспорядку происходит полная локализация электронных волн. Вдали от перехода квантовые поправки малы по параметру $\lambda/\boldsymbol{l}$, где $\lambda$ - длина волны электрона, $\boldsymbol{l}$ - длина его свободного пробега. Однако во многих случаях именно они определяют нетривиальные зависимости проводимости $\sigma$ от магнитного поля $\boldsymbol{H}$, температуры $\boldsymbol{T}$, частоты $\boldsymbol{\omega}$ переменных полей и размерности $\boldsymbol{d}$ образца.

Полное вычисление поправок производится с помощью методов квантовой теории поля. Однако их происхождение и основные свойства можно понять на основе следующих рассуждений.

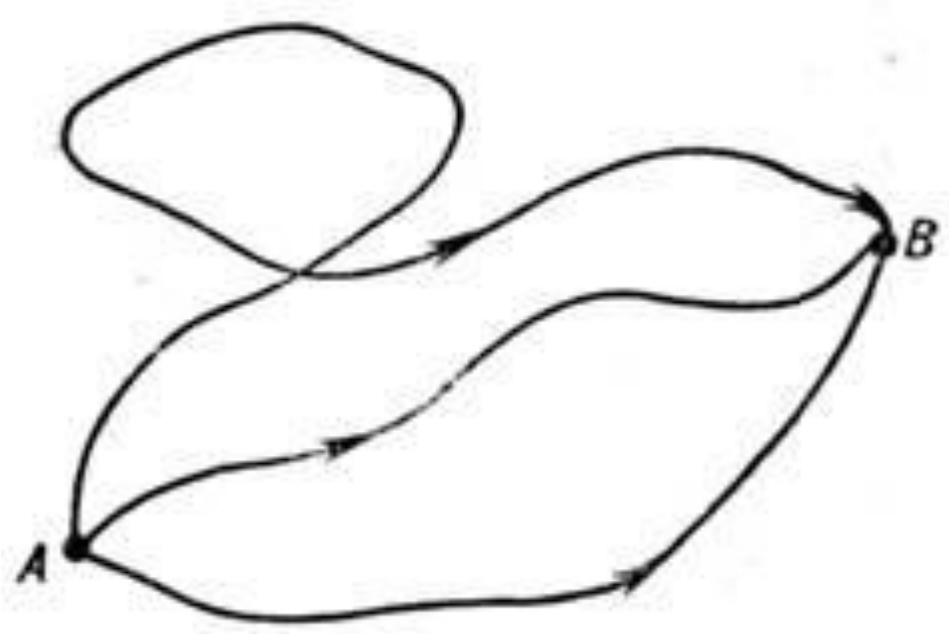

**Рис. II.1.** *Квантовая интерференция и эффект слабой локализации*

Рассмотрим проводник, в котором $l \gg \lambda$, и предположим, что за время $t$ электрон, испытывая рассеяние на примесях, переходит из точки A в точку B. При этом он может пройти по разным путям (рис.1). Согласно общим принципам квантовой механики, вероятность такого процесса W определяется выражением:

$$W = \left| \sum_i A_i \right|^2 = \sum_i |A_i|^2 + \sum_{i \neq j} A_i A_j. \qquad (1)$$

Здесь $A_i$ - амплитуда вероятности движения электрона вдоль j-го пути. Первое слагаемое в уравнении (1) описывает сумму вероятностей прохождения каждого пути, а второе - интерференцию разных амплитуд. Интерференция большинства амплитуд не даёт вклад в W, т. к. их фазы пропорциональны длине траектории и при суммировании взаимно погашаются. Исключение составляют траектории с самопересечением. Каждой такой траектории можно сопоставить две амплитуды $A_1$ и $A_2$, отвечающие различным направлениям обхода замкнутой петли. Эти две амплитуды когерентны друг другу, и поэтому их интерференцией нельзя пренебречь:

$$A_1A_2^* + A_2A_1^* = 2|A_i|^2.$$

Пренебрежение интерференцией отвечает классическому описанию (кинетическое уравнение Больцмана), а её учёт приводит к возникновению квантовых поправок.

Относительная величина вклада поправок в проводимость Δσ (она всегда отрицательна) пропорциональна вероятности самопересечения лучевой трубки с сечением $\lambda^{d-1}$ при диффузии за время $\tau_\varphi$ полного разрушения когерентности (сбоя фазы) из-за неупругих процессов или из-за рассеяния с переворотом спина. Оценка Δσ, полученная из приведённых рассуждений, по порядку величины совпадает с результатами точного расчёта и определяется выражением:

$$\Delta\sigma = -\frac{e^2}{\hbar}\begin{cases} L_\varphi, & d=1, \\ \ln L_\varphi/l, & d=2, \\ \mathrm{const} - L_\varphi^{-1}, & d=3. \end{cases} \qquad (2)$$

Здесь:

$$L_\varphi = \sqrt{D\tau_\varphi},$$

где D –коэффициент классической диффузии. Из уравнения (2) видно, что Δσ, хотя и мала по параметру **$\lambda/l$**, но определяет сингулярные зависимости проводимости от температуры $\tau_\varphi \sim T^{-1}$ или частоты поля.

Если доминирующим процессом сбоя фазы является неупругое рассеяние, то $\tau_\varphi$ растёт с понижением температуры и всё большее число петлеобразных участков траекторий с размерами $L \ll L\varphi$ даёт вклад в Δσ. При этом абсолютная величина Δσ увеличивается, а сама проводимость уменьшается согласно уравнению (2). Этим, в частности, объясняется появление минимума на температурной зависимости сопротивления металлических плёнок и вырожденных полупроводников. Рост сопротивления при понижении температуры есть результат совместного проявления поправок разной природы, возникающих как за счёт эффектов как слабой локализации, так и межэлектронного взаимодействия. Во внешнем магнитном поле амплитуды $A_1$ и $A_2$ приобретают дополнительный фазовый множитель:

$$\exp(\pm\pi\, \Phi/\Phi_0)$$

где Ф - поток магнитного поля через замкнутую петлю, а $\Phi_0 \equiv \pi c\hbar/l$ - квант магнитного потока, ± соответствует различным направлениям обхода петли. В результате у интерферирующих амплитуд возникает разность фаз $\Delta\varphi_H = 2\pi\Phi/\Phi_0$. Появление разности фаз $\Delta\varphi_H$ приводит к разрушению когерентности и уменьшению $\Delta\sigma$. Экспериментально это явление наблюдается в виде отрицательного магнетосопротивления, в относительно слабом магнитном поле.

# Литература

## Введение

## Глава 1

## **Глава 2**

**Глава 3**

**Глава 4**

## Глава 5

## Глава 6

## Приложение 2

## Некоторые условные сокращения и пояснения

**АО-** атомная орбиталь
**АЭС** – атомная электростанция
**АЭ** – акустическая эмиссия
**ГБ** – графитовые блоки
**ГК** – графитовая кладка
**КРФС** – кокс резольный фенол-формальдегидной смолы
**МНК** – метод наименьших квадратов
МО – молекулярная орбиталь
**НМ** – нанокристаллические материалы
**НГР** – нормы расчёта на прочность типовых узлов и деталей из графита
**НДС** – напряжённо-деформированное состояние графитовой кладки
**ОКР** - область когерентного рассеяния
**РБМК** – реактор большой мощности канальный
**ТКЛР** – тепловой коэффициент линейного расширения
**СНА** – число смещений на атом
**ВРЭМ (HRTEM)** – высокоразрешающая электронная микроскопия (High Resolution Electron Microscopy)

**Сведения об авторах**

Жмуриков Евгений Изотович, к ф-м н.

В 1977 г. с отличием закончил Томский институт автоматизированных систем управления и радиоэлектроники. Научные работы связаны с переносом заряда в конденсированных средах, а также в области взаимодействия пучков заряженных частиц с твёрдым телом. Автор и соавтор более 40 научных публикаций.

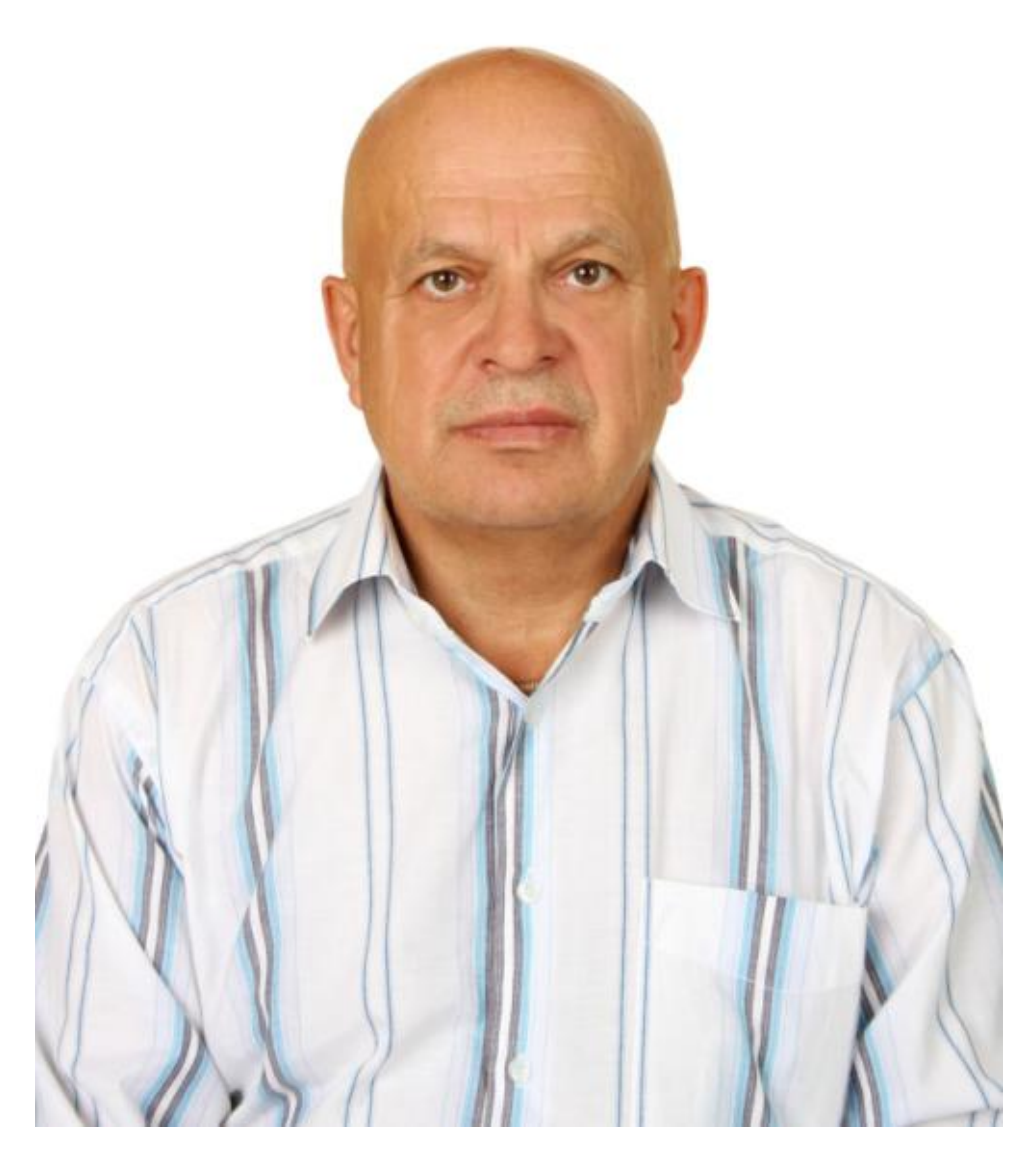

Бубненков Игорь Анатольевич, д.т.н.

Начальник отдела, д.т.н. В 1977 г. закончил с отличием Московский институт стали и сплавов. Научные интересы находятся в области разработки мелкозернистых и среднезернистых искусственных графитов; разработки специальных углеродных материалов для синтеза алмазов; углеродкарбидных материалов; в области исследования взаимодействие металлов с углеродом. Автор и соавтор 60 публикаций, 3 патента, 12 авторских свидетельств на изобретения, 1 монография.

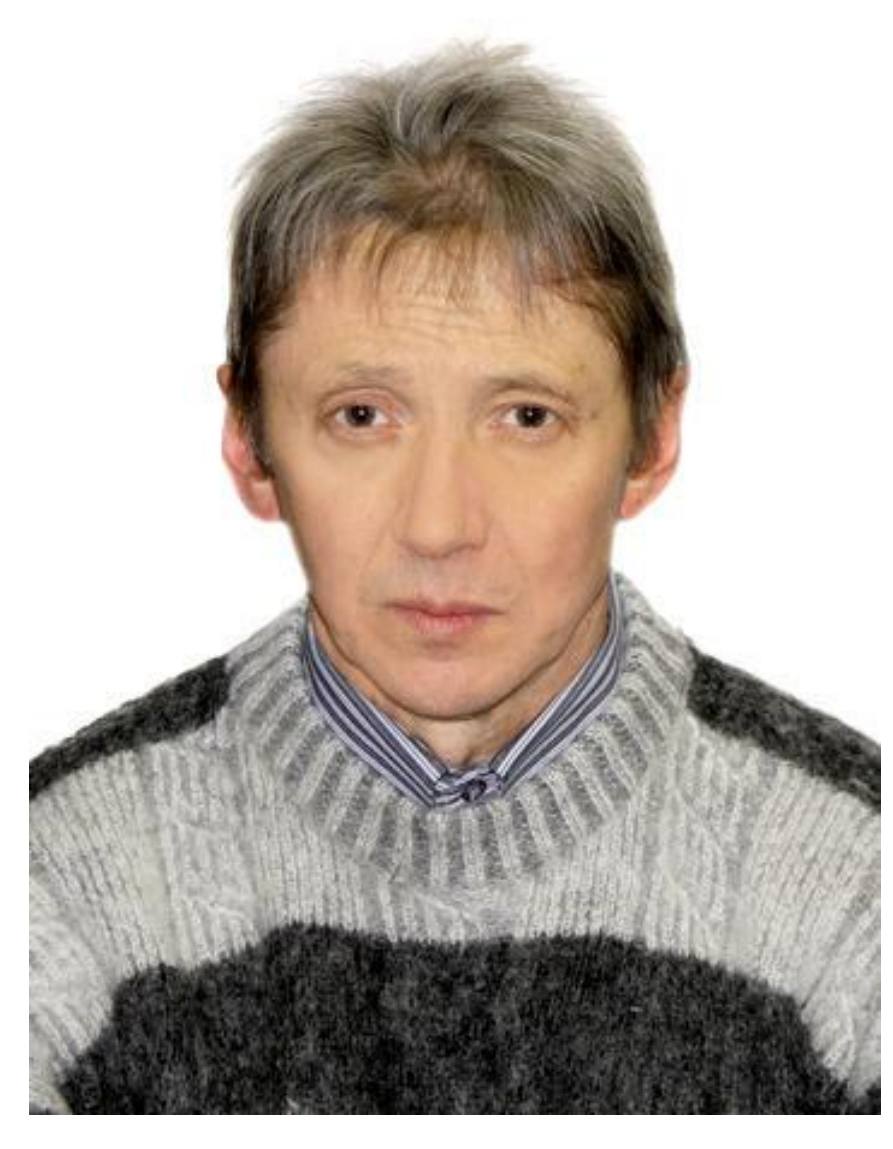

Дремов Владимир Владимирович, к.ф.-м.н.

Заместитель начальника отдела, к.ф.-м.н. В 1990 г. закончил физический факультет Челябинского государственного университета. Научные интересы: моделирование процессов высокоскоростной деформации (в том числе ударноволновых) и радиационных повреждений конструкционных материалов методами молекулярной динамики, а также в области уравнений состояния и расчета фазовых диаграмм веществ при высоких давлениях и температурах. Автор и соавтор 55 статей в реферируемых журналах.

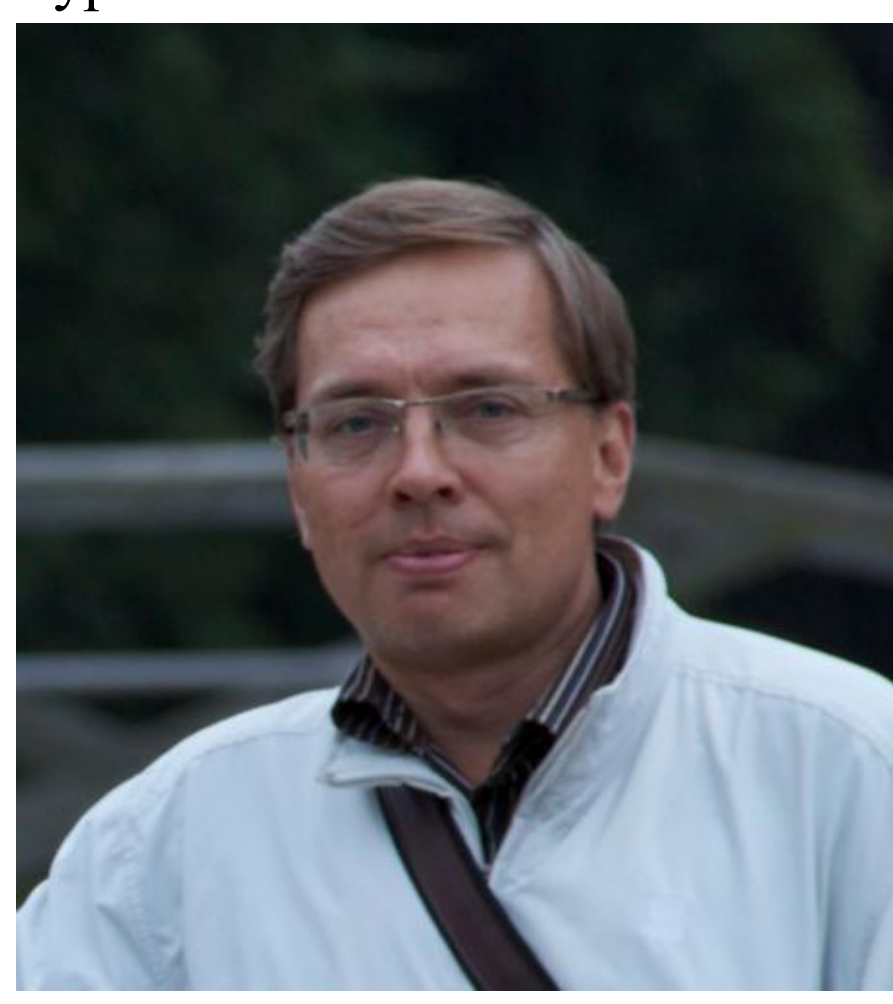

Самарин Сергей Иванович, к ф.-м.н..

Ведущий научный сотрудник, в 1990 г. закончил физический факультет Челябинского государственного университета. Научные интересы: моделирование переноса ионизирующего излучения, методы Монте-Карло. Автор и соавтор 20 статей в реферируемых журналах.

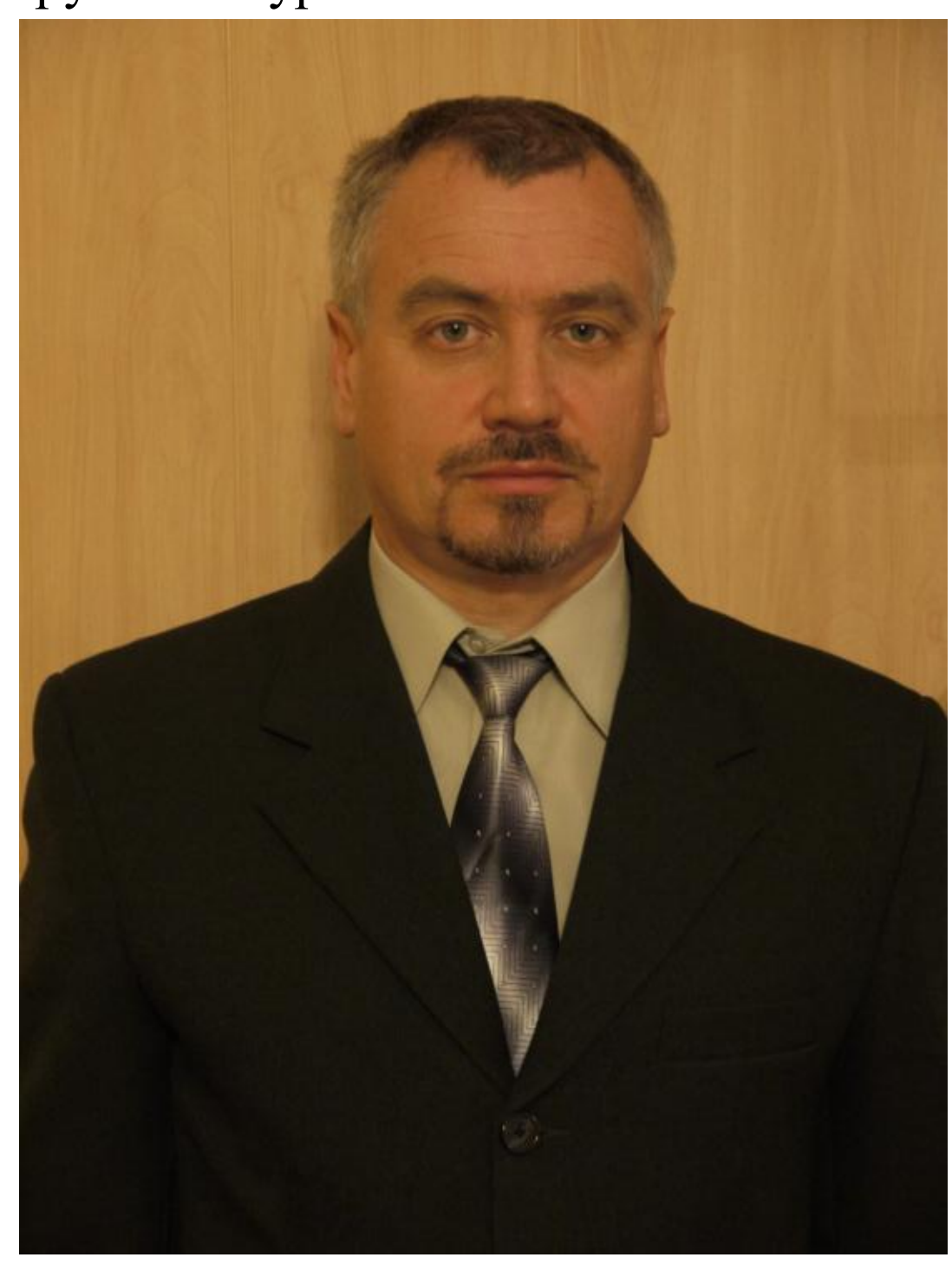

Покровский Александр Сергеевич, к.ф.-м.н.

Начальник лаборатории НИИ атомных реакторов. Закончил МИФИ в 1966 г. Старший научный сотрудник, специалист в области реакторного материаловедения. Занимался разработкой и испытаниями материалов для активных зон энергетических водоводяных реакторов ВВЭР и РБМК, а так же высокотемпературных газовых реакторов 5 поколения. Основные научные интересы лежат в области изучения влияния нейтронного облучения на изменение структуры и физико - механических свойств реакторных графитов и жаропрочных конструкционных сплавав и оценки их работоспособности в условиях эксплуатации.
Автор и соавтор 80 публикаций, 12 изобретений

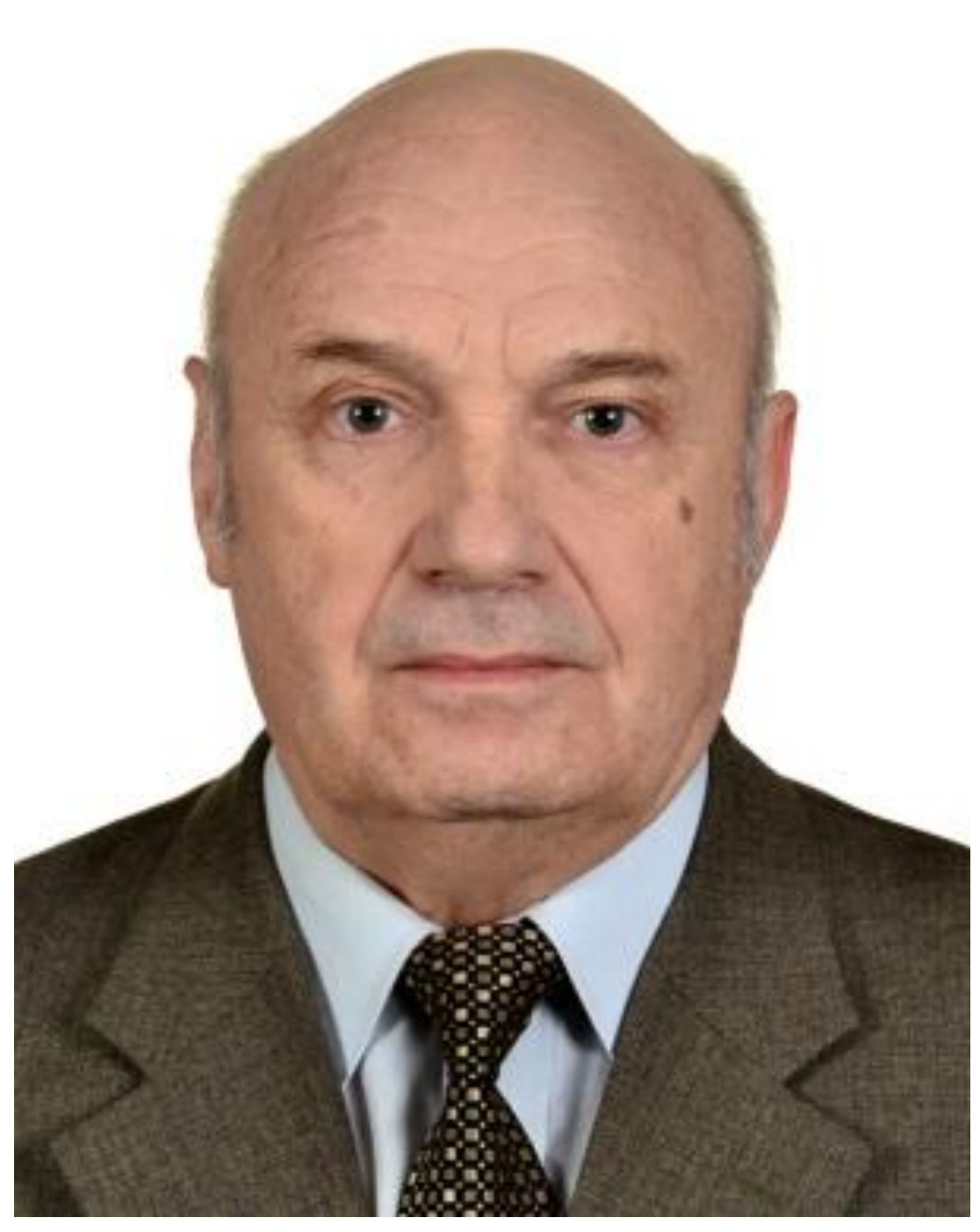

Харьков Дмитрий Викторович, к.т.н..

Старший научный сотрудник ОАО «ГНЦ НИИАР», к.т.н. Закончил физико-технический факультет Ульяновского государственного университета в 2002 г., специалист в области реакторного материаловедения. Автор и соавтор 15 публикаций.

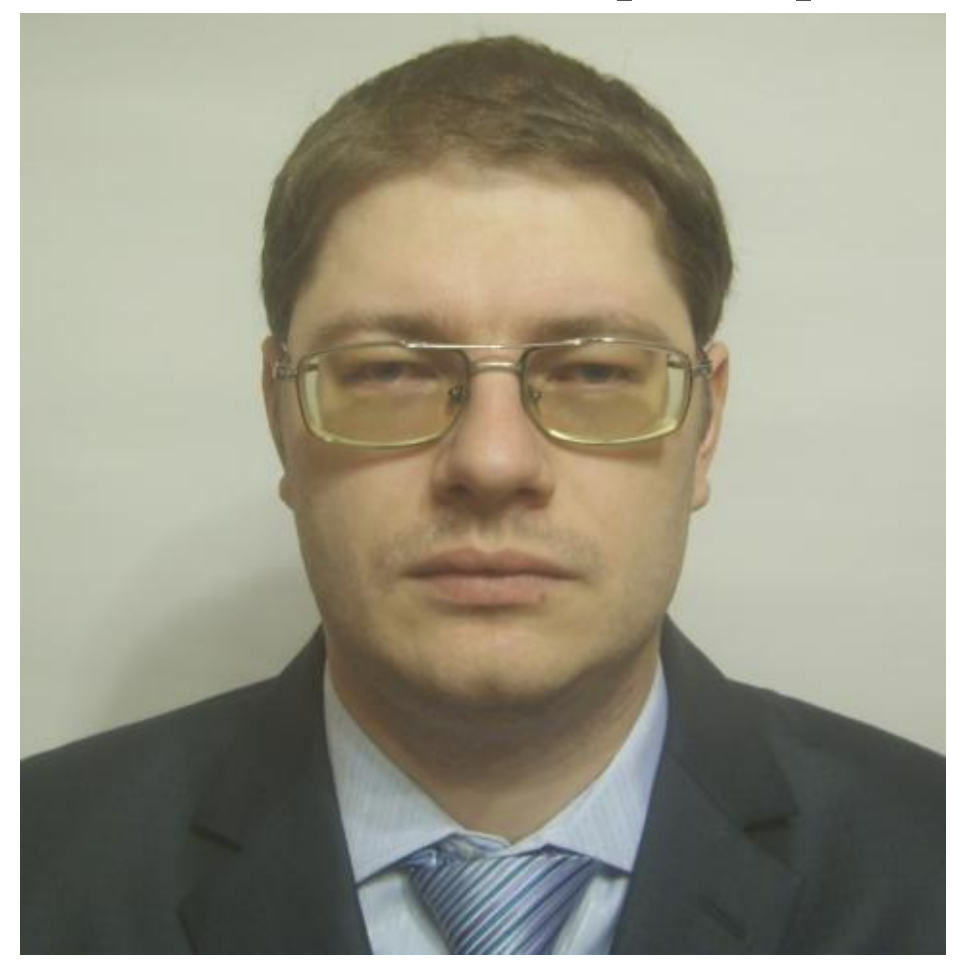